\documentclass[a4paper,12pt,twoside, openright]{report}
\usepackage{fullpage}
\usepackage{natbib}
\usepackage{fancyhdr}
\usepackage{subfigure}
\usepackage{graphicx}
\usepackage{longtable}
\usepackage{rotating}
\usepackage{array}
\usepackage{amssymb}
\usepackage{amsmath}
\usepackage{multirow}
\usepackage{lscape}
\usepackage{mathrsfs}
\usepackage[usenames]{color} 
\usepackage[top=2.5cm,       
            bottom=2.5cm,     
            right=2.4cm,     
            left=3.6cm,      
            headheight=1cm,  
            headsep=.5cm]{geometry}  

\usepackage[T1]{fontenc}

\definecolor{citecolor}{rgb}{0.,  0.,  .5}
\newcommand\cyt[1]{{\color{citecolor}{\textsc{#1}}}}

\newcommand\citepm[1]{{\color{citecolor}{\textsc{\citep{#1}}}}}
\newcommand\citem[1]{{\color{citecolor}{\textsc{\cite{#1}}}}}

\renewcommand{\=}{\!=\!}

\fancypagestyle{plain}{
\fancyhead{}

}

\newcommand{\grr}{g_{rr}}
\newcommand{\ghh}{g_{\theta\theta}}

\newcommand{\sss}{\scriptscriptstyle}
\newcommand{\planckh}{h}
\newcommand{\boltzk}{k_{\rm\sss B}}

\newcommand{\speedoflight}{c}
\newcommand{\fcolor}{f_{\rm c}}
\newcommand{\flimbdk}{\Upsilon}
\newcommand{\kappaES}{\kappa_{\rm es}}   
\newcommand{\mue}{\mu_{\rm e}}
\newcommand{\Mdot}{\dot{M}}
\newcommand{\dotm}{\dot{m}}
\newcommand{\Medd}{\dot{M}_{\rm Edd}}
\newcommand{\Ledd}{{L}_{\rm Edd}}
\newcommand{\Msun}{M_\odot}
\newcommand{\msun}{M_\odot}
\newcommand{\EE}[1]{\!\times\!\!{10}^{#1}\,}

\newcommand\AcA{Acta Astronomica}
\newcommand\actaa{Acta Astronomica}

\newcommand\apj{Astrophysical Journal}
\newcommand\apjl{Astrophysical Journal Letters}

\newcommand\apjs{Astrophysical Journal Suppl. Ser.}

\newcommand\aap{Astronomy \& Astrophysics}

\newcommand\aj{Astronomical Journal}
\newcommand\nat{Nature}
\newcommand\mnras{Monthly Notices of the Royal Astronomical Society}
\newcommand\arxiv{ArXiv Astrophysics e-prints}
\newcommand\pasp{PASP}
\newcommand\pasj{Publications of the Astronomical Society of Japan}

\def\tylda{~}
\newcommand\litr[7]{\item[#1] \textsc{\cyt{#2}} (#3)%
\def\REFARG{#4}\ifx\REFARG\tylda\else, {\it #4}\fi
\def\REFARG{#5}\ifx\REFARG\tylda\else, {\bf #5}\fi
\def\REFARG{#6}\ifx\REFARG\tylda\else, #6\fi
\def\REFARG{#7}\ifx\REFARG\tylda\else; {\it \textsf{ #7}}\fi
\addtocounter{nor}{+1}
}
\newcounter{nor}

\newcommand{\beq}{\begin{equation}
  \renewcommand{\int}{\intop\limits}
  \renewcommand{\oint}{\ointop\limits}}
\newcommand{\eeq}{\end{equation}}

\newcommand{\pder}[2]{\frac{\partial#1}{\partial#2}}
\newcommand{\der}[2]{\frac{{\rm d}#1}{{\rm d}#2}}
\newcommand{\derln}[2]{\ensuremath{\frac{{\rm d\,ln}\, #1}{{\rm d\,ln}\, #2}}}
\newcommand{\pderln}[2]{\ensuremath{\frac{\partial\,\rm ln\,#1}{\partial\,\rm ln\,#2}}}

\newcommand{\be}{\begin{equation}}
\newcommand{\ee}{\end{equation}}
\newcommand{\bea}{\begin{eqnarray}}
\newcommand{\eea}{\end{eqnarray}}

\newcommand{\mdot}{\dot M}

\definecolor{RedVioletDD}{rgb}{0.465,0.05,0.5}

\begin{document}

\pagestyle{empty}
\begin{center}
\vspace*{1cm}
{\Huge \textsc{ Slim accretion disks\\\vspace{.4cm} around black holes}}\\
\vspace{2cm}
{\Large \it{Aleksander S\k{a}dowski}}\\
\vspace{2.5cm}
{\large \textsc{Ph.D. Thesis under supervision of\\ \vspace{.2cm}prof. Marek Abramowicz}}\\
\vspace{6cm}
\includegraphics[width=3cm]{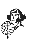}\\
\vspace{1cm}
{\large\textsc{Nicolaus Copernicus Astronomical Center\\Polish Academy
    of Sciences\\\vspace{.15cm}Warsaw 2011}}
\end{center}

\chapter*{Acknowledgements}

{\normalsize \noindent{\it I would like to thank my thesis supervisor, prof. Marek Abramowicz, for his support
 and patience. He was my invaluable guide on the paths of the disk accretion physics.  \\

\vspace{.3cm}
\noindent I am also thankful to my wife, Ewa, who gave me permanent support and valuable comments on this thesis.\\

\vspace{.3cm}
\noindent I am grateful to my colleagues, in the first place Wlodek Klu{\'z}niak, as well as Ramesh Narayan, Shin Mineshige, Agata R{\'o}{\.z}a{\'n}ska, Michal Bursa, Odele Straub, Akshay Kulkarni, Marek Sikora, Jean-Pierre Lasota, Mariusz Gromadzki, Krzysztof Nalewajko, Maciek Wielgus, Radek Wojtak, Krzysztof Hryniewicz and Mateusz Janiak for all the discussion we had which influenced this work.\\

\vspace{.3cm}
\noindent This work has been supported by the Polish Ministry of National Education Grant No. N203 3040351.   }}

\tableofcontents 

\chapter*{Summary}

In this thesis, I study hydrodynamical models of slim accretion disks --- advective, optically thick disks which generalize the standard models of radiatively efficient thin disks to all accretion rates. I start with a general introduction to the theory of accretion onto compact objects. It is followed by a derivation of the commonly-used standard models of thin disks. In the subsequent section I introduce the equations describing slim disks, explain the numerical methods I used to solve them and discuss  properties of such solutions. I also give a general derivation of non-stationary equations and present the time evolution of thermally unstable accretion disks. I introduce a state-of-the-art approach coupling the radial and vertical structures of an advective accretion disk and discuss the improvements it brings to vertically-averaged solutions. I also present a numerical model of self-illuminated slim accretion disks. Finally, I present and discuss applications of slim accretion disks: estimating of spin of the central black hole in LMC X-3 through X-ray continuum fitting basing on high-luminosity data, spinning-up of black holes by super-critical accretion flows and normalizing of magnetohydrodynamical global simulations. 

\vspace{2cm}

\noindent {\it This thesis presents and extends results included in the following papers:}\vspace{.3cm}\\
\citem{sadowski.slim} -- {\it Sections~\ref{s.stateq} --- \ref{s.stationarysolutions}},\\
\citem{leavingtheisco} --- {\it Section~\ref{s.inneredges}},\\
\citem{xueetal-11} -- {\it Chapter~\ref{chapter-nonstationary}},\\
\citem{sadowski-vertical} -- {\it Chapter~\ref{chapter-vertical}},\\
\citem{spinningup} --- {\it Section~\ref{s.spinningup}}.
\vspace{.3cm}\\
\noindent {\it Unpublished material is included in Section~\ref{s.slimbb} and in Chapter~\ref{chapter.selfirradiated}.}

\chapter{Introduction\label{C1}}
\pagestyle{fancy}
Microquasars and Active Galactic Nuclei, despite decades of research, are still not sufficiently understood. The basic theory of accretion disks gives us insight into physics involved in the process of accretion but is unable to explain the observed characteristic and time evolution of real accretion disks. Our understanding of the source of viscosity, spectral states, quasi-periodical oscillations, jets etc. is still far from satisfactory. New ideas and models are necessary to better understand one of the most powerful objects in the Universe.

\section{Origins of black hole physics\label{s.bhs}}

As early as in the end of 18th century John Michell\footnote{``If the semi-diameter of a sphere of the same density as the Sun were to exceed that of the
Sun in the proportion of 500 to 1, a body falling from an infinite height towards it would have
acquired at its surface greater velocity than that of light, and consequently supposing light to be
attracted by the same force in proportion to its vis inertiae (inertial mass), with other bodies, all
light emitted from such a body would be made to return towards it by its own proper gravity.''
\citepm{michell-1784}} and Laplace noted that according to the Newton's theory of gravity
and the corpuscular theory of light an object so massive and compact that even single photon
is not able to escape to infinity may exist. However, this reasoning found only few supporters.

In the early 1910's the general theory of relativity was established when Albert Einstein
published a series of papers on gravitational field \citepm{einstein-15}. These works were followed in a short time by Karl Schwarzschild's solution of Einstein's equations in the vicinity of a spherically symmetric mass \citepm{schwarzschild-16}. Both scientists appreciated this discovery but at that time they were not aware of the fact that it is the complete solution for spacetime surrounding a non-rotating, uncharged black hole (BH).

In 1930's Chandrasekhar discovered an upper mass limit for spherical stars built from degenerate matter \citepm{chandrasekhar-31}. A few years later \citem{baadezwicky-34} proposed that supernovae represent the transition of normal stars to neutron stars - compact objects supported by the pressure of degenerate neutrons. This discovery was followed by Eddington's remark that existence of the Chandrasekhar limit inevitably leads to a collapse of a sufficiently massive object to a BH \citepm{eddington-35}. However, he considered such objects unlikely and tried to modify the degenerate equation of state to obtain stable, non-collapsed, solutions for any mass of a star. Such an approach was common among astronomers at that time.

\citem{oppenheimersnyder-39} considered the collapse of a uniform, pressureless sphere of gas in the framework of the general theory of relativity. They analytically proved that an object resulting from the collapse is, at its particular stage, cut off from information exchange with the surroundings. For the first time an astrophysical process leading to the creation of a BH was proposed.

Due to the lack of further research BHs remained outside the interest of most physicists until 1960's when \citem{kerr-63} solved Einstein's equations in the vicinity of a rotating, uncharged compact object. This solution was generalized by \citem{newman-65} to charged objects. Today we know that the Kerr-Newman geometry perfectly describes the gravitational and electromagnetic fields of a stationary, solitary BHs. However, the importance of these discoveries were appreciated only a dozen or so years later.

Since that time a large number of papers devoted to different aspects of the BH theory have been published. The research was motivated by new discoveries: quasars \citepm{schmidt-63}, pulsars \citepm{hewish-68} and X-ray binaries \citepm{oda-71}. The observations of the Cygnus X-1 microquasar became first evidence for the existence of BHs in the Universe.

\section{Black holes in the Universe}

Astrophysical BHs originate mainly through the process of a gravitational collapse. It takes place when the gravity force acting on a massive object is not balanced by other forces e.g., gas or radiation pressure. Gravitational collapse has been common in the Universe. Such a process led to the creation of separated clouds of matter after the Big Bang, which later transformed into clusters of galaxies and ultimately into galaxies, stellar clusters, clusters and planets. 

Gravitational collapse is especially important in the stellar evolution. Protostellar clouds of matter collapse leading to the increase of the pressure and temperature, and, as a result, to the ignition of the nucleosynthesis in their centers. Stars collapse also in the end of their evolution once their internal sources of energy have run out. If the core contraction is stopped by the pressure of degenerate electrons, a white dwarf originates. This scenario is possible only for low mass stars. If the Zero Age Main Sequence (ZAMS) mass of a star exceeds $\sim 8\msun$ then the core exceeds the maximal mass of a white dwarf ($1.4\msun$, the Chandrasekhar mass) and continues its collapse. Only the pressure of degenerate neutrons can stop this process leading to a neutron star remnant. However, the cores of most massive stars ($>20\msun$) exceed the Tolman-Oppenheimer-Volkoff mass \citepm{oppenheimer-39} and even degenerated neutrons are not able to prevent further collapse. Such an objects continues its contraction and, once it has fallen inside its own event horizon, becomes a BH.

The scenario described in the previous paragraph leads to the creation of stellar-mass BHs. The resulting mass of the remnant depends mostly on the initial (ZAMS) mass of a star and its metallicity. According to the current evolutionary models (e.g., \citem{hurley-02}), masses of BHs resulting from evolution of single stars in the Galaxy range between $4$ and $12\msun$ depending on the initial stellar mass. In low-metallicity environments, e.g., globular clusters, their masses may be much higher, reaching $25\msun$, due to decreased efficiency of stellar winds.

During the supernova explosion, BHs resulting from the stellar evolution may remain in binary systems if the remnant does not obtain a kick sufficiently strong to disrupt a binary system or if the BH originates as a result of a direct collapse of the star. If the binary is close enough, the BH companion may fill its Roche lobe and start transferring mass onto the compact object. It may also happen that the companion's envelope embraces the BH. During such an event, called the common envelope, the system circularizes or, if there is not enough orbital energy to eject the envelope, merges giving rise to a more massive single BH or a Thorne-Zytkow object \citepm{thorne-zytkow-77}.

As it will be pointed out in detail in this thesis, matter falling onto a compact object loses its angular momentum in an accretion disk and emits radiation. In case of stellar mass BHs this emission falls into X-rays and is amenable to observations by X-ray satellites (e.g., XMM-Newton or RXTE). Detailed observations of light curves, together with spectroscopic research, provided estimates of BH masses in a number of systems (Table~\ref{t.gbhc}). In case of these objects, the lower mass estimates are higher than $2.5\msun$, which makes them likely BH candidates. So far, these objects have been the only galactic BHs known.

BHs of masses exceeding $10^5\msun$ are called supermassive and form another class of astrophysical BHs. They are observed indirectly in nuclei of many galaxies, in particular in the center of the Milky Way. It is assumed that the evolution of most, if not all, galaxies leads to the creation of a very massive BH in their nuclei.

The existence of supermassive BHs in active galactic nuclei (AGN) is supported by observations of visual/UV spectra of accretion disks which may be explained only under the assumption that they surround supermassive compact objects. In case of other than AGN types of galaxies we base on the Doppler measurements of water masers movement in the vicinity of nuclei. Their velocities reflect the existence of  massive, dark objects in nuclei of galaxies. According to our knowledge, only BHs are compatible with provided characteristics. Finally, the observed orbital motion of stars in the vicinity of the Milky Way center (Sgr A*) suggests the supermassive BH of mass $4.1\pm 0.6\cdot 10^6\msun$ \citepm{doelemanetaal-08}.

BHs of masses of the order of hundreds or thousands of solar masses form one more, so far only hypothetical, class of astrophysical compact objects - intermediate mass BHs. Observed Ultra-luminous X-ray Sources (ULXs) in nearby galaxies seem to suggest that BHs of such masses exist. However, the observed high luminosities may be explained by different processes e.g., by super-critical accretion onto stellar mass BHs. Doppler measurements of the velocities of objects gravitationally bound to intermediate mass BHs would provide evidence for their existence. However, so far no such observation has been obtained. It is also not clear how intermediate mass BHs would originate. They are too massive to result from the evolution of a single star of Population I or II. The regions where they are found differ significantly from galactic nuclei so the processes giving rise to supermassive BHs cannot be involved. Intermediate mass BHs could be a product of hierarchical mergers of massive, unevolved stars into very massive objects collapsing into a BH.

There is one more class of BHs that still remains hypothetical --- mini BHs, with masses smaller than any BH resulting from a stellar collapse. It is possible that such quantum primordial BHs were created in the high-density environment of the early Universe (or big bang), or possibly through subsequent phase transitions. They are expected to emit weak Hawking radiation, but direct detection of them will be impossible in the nearest future.

\begin{landscape}
\begin{longtable}{llccccc}
\caption{Galactic black hole candidates}\\
\hline\hline
Name& Other names  & Spectral type   & $P_{\rm orb}$ &$M_{\rm BH} /\msun$  &$M_{\rm sec}$ & Refs.\\
                &  & (O) &           &                 &      &  \\  
\hline   
\endfirsthead 
\multicolumn{7}{c}%
{{\tablename\ \thetable{} -- continued from previous page}} \vspace{.4cm} \\
\hline\hline
Name& Other names  & Spectral type    & $P_{\rm orb}$ &$M_{\rm BH} /\msun$  &$M_{\rm sec}$ & Refs.\\
                &  &   (O)&           &                 &      &  \\     
\hline
\endhead
\hline
\endfoot
\hline\hline\\
\multicolumn{7}{l}{Only objects with existing estimates of the BH mass are listed.}\\
\multicolumn{7}{l}{(O) - optical counterpart, $P_{\rm orb}$ - orbital period, $M_{\rm BH}$ - black hole mass,}\\
\multicolumn{7}{l}{$M_{\rm sec}$ - secondary mass or mass ratio $(M_{\rm sec}/M_{\rm BH})$ (in square brackets).}\\
\multicolumn{7}{l}{References: (1) \citem{ziolkowski-04}, (2) \citem{cowleyatal-95}, (3) \citem{oro+09}, (4) \citem{soriaetal-01}, }\\
\multicolumn{7}{l}{(5) \citem{hillwigetal-04}, (6) \citem{hynesetal-03}, (7) \citem{gelinoharrison-03}, (8) \citem{shahbazetal-94},}\\
\multicolumn{7}{l}{(9) \citem{shahbazetal-04}, (10) \citem{jonkeretal-05}, (11) \citem{maciasetal-11}, (12) \citem{2008AIPC.1010...82G},}\\
\multicolumn{7}{l}{(13) \citem{1997ApJ...489..272T}, (14) \citem{2004ApJ...613L.133C}, (15) \citem{1998ApJ...499..375O},
(16) \citem{2002ApJ...568..845O},}\\
\multicolumn{7}{l}{ (17) \citem{2004ApJ...616..376O}, (18) \citem{2011MNRAS.411..137H},
(19) \citem{1997ApJ...477..876O}, (20) \citem{2011arXiv1102.2102K},}\\
\multicolumn{7}{l}{(21) \citem{1996ApJ...459..226R}, (22) \citem{2007AAp...473..561S}, (23) \citem{2000ApJ...544..977H}, (24) \citem{2008AIPC.1053..183S},}\\
\multicolumn{7}{l}{(25) \citem{2003IAUS..212..365O}, (26) \citem{2001Natur.414..522G}, (27) \citem{2004AJ....127..481I}, (28) \citem{1994MNRAS.271L..10S}.}
\endlastfoot
                    &                      &           &             &               &              &            \\

Cyg X-1    \hspace{3cm}         & HDE 226868 (O),\hspace{1cm}        & O9.7 Iab  & $5^d 6$     & $13.5\div 29$     & $40\pm 10\msun$   &  (1)      \\
                    & V1537 Cyg (O)                     &           &             &               &              &            \\
LMC X-1             &                      & O7-9 III  &  $3^d 91$    & $10.91\pm1.41$ &$31.79\pm3.48$ &  (2,3)\\
LMC X-3             &                      & B3 V      & $1^d 70$    & $6\div 9$     &$[0.5\div0.91]$  &  (4)       \\
SS 433              & V1343 Aql (O)        & A3-7 I    & $13^d 1$    & $2.9 \pm 0.7$ &$10.9\pm 3.1\msun$& (5)           \\
GX 339-4            & V821 Ara (O)         & F8-G2 III & $1^d 756$   & $>6$          &$[<0.08]$     &  (6)          \\
GRO J0422+32        & V518 Per (O),          & M2 V      & $5^h 09$    & $3.97\pm 0.95$&$[0.116^{+0.079}_{-0.071}]$      &  (7)          \\
                    & XRN Per 1992                     &           &             &               &              &            \\
A 0620-00           & V616 Mon (O),             & K4 V      & $7^h75$     & $11\pm 2$     &$[0.074\pm0.006]$      & (8)           \\
                    & Mon X-1,                      &           &             &               &              &            \\
                    & XN Mon 1975                     &           &             &               &              &            \\
2S 0921-630         & V395 Cor (O)         & K0 III    & $9^d$       & $1.90^{+0.25}_{-0.25}\div 2.9^{+0.4}_{-0.4}$ &$[1.32\pm0.37]$       &  (9,10)          \\
GRS 1009-45         & MM Vel (O),               & K8 V      & $6^h 96$    & $\sim 8.5$ & $\sim0.5$       &          (11)  \\
                    & XN Vel 1993                     &           &             &               &              &            \\
XTE J1118+480       & KV Uma (O)           & K7-M0 V   & $4^h 1$     & $8.53\pm0.60$ &$0.09\div0.5\msun$      &  (12)          \\
GS 1124-683         & GU Mus (O),            & K0-5 V    & $10^h 4$    & $7.0\pm 0.6$  &$[0.128\pm0.04]$     &  (13)          \\                    
                    &XN Mus 1991                       &           &             &               &              &            \\
GS 1354-645         & BW Cir (O)           & G0-5 III  & $2^d 54$    & $>7.83\pm 0.50$&$1.07\div2.11\msun$     &  (14)          \\
4U 1543-47         & IL Lup (O)           & A2 V      & $1^d 12$    & $8.4\div 10.4$&$[0.25\div0.31]$     &   (15)         \\
XTE J1550-564       & V381 Nor (O)         & G8 IV-K4 III&$1^d 55$   & $9.7\div 11.6$&$[0.09\div0.15]$ &   (16)        \\
XTE J1650-500       &                      & K4V       & $0^d 3205$  & $2.73\div 7.3$&             &   (17)         \\
XTE J1652-453       &                      &           &             & $<30$         &                  & (18)           \\
GRO J1655-40        & V1033 Sco (O),            & F3-6 IV   & $2^d 62$    & $6.3\pm 0.3$  &$[0.40\pm0.03]$  &   (19)         \\
                    & XN Sco 1994                     &           &             &               &              &            \\
MAXI J1659-152      &                      &           & $2^h41$        & $20\pm 3$      &                 &   (20)         \\
H 1705-250          & V2107 Oph (O)        & K5 V      & $12^h 5$    & $5.7\div 7.9$ &$[0.014^{+0.019}_{-0.012}]$      &  (21)          \\
XTE J1817-330       &                      &           &             & $<6.0^{+4.0}_{-2.5}$        &              &  (22)          \\
XTE J1819-254       & V4641 Sgr (O)        & B9 III    & $2^d 817$   & $6.8\div 7.4$ &$[0.42\div0.45]$  &  (23)          \\
XTE J1856+053      &                      &           &              &$<4.2 $ &                  &   (24)         \\
XTE J1859+226       & V404 Vul (O)         & $\sim$G5  & $9^h 16$    & $7.6\div12$    &        &   (25)         \\
GRS 1915+105        & V1487 Aql (O),       & K-M III   & $33^d 5$    & $10\div 18$     &$>1.2\msun$          & (26)          \\
                    &  XN Aql 1992                      &           &             &               &              &            \\
GRS 2000+251         & QZ Vul (O),         & K5 V      & $8^h 3$     & $5.5\div 8.8$ &$0.16\div0.47$       &   (27)         \\
                    & XN Vul 1988                         &           &             &               &              &            \\
GRS 2023+338         & V404 Cyg (O),         & K0 IV     & $6^d 46$    & $10\div 13.4$ &$[0.060\pm0.005]$      &  (28)          \\
                    & XN Cyg 1989                       &           &             &               &              &            
\label{t.gbhc}
\end{longtable}
\end{landscape}

\section{Accretion onto compact objects}

In astrophysics the term \textit{accretion} is used to describe the process of acquiring matter, typically gaseous, by a massive body. As a result of the infall, the gravitational energy is extracted. Its amount is roughly proportional to the ratio of the mass and radius of the attracting object: $M/R$. It is clear that the more massive and compact the object is, the larger amount of energy may be released. Accretion onto compact objects (BHs, neutron stars or white dwarfs) is therefore an extremely important process in the astrophysical context, as it transforms the gravitational energy into radiation in many of the most luminous classes of objects in the Universe.

The spherically symmetric (zero angular momentum) accretion on a gravitating body is the simplest case one can imagine. It was first studied  by \citem{bondi-52} and is called \textit{the Bondi accretion}. It was shown that the solution describing such a process is unique and that the flow becomes transonic crossing the critical (sonic) radius. 
It is currently accepted that Bondi-type of accretion hardly occurs in the Universe because all potential sources of accreting matter (e.g., gas clouds or stars) have non-zero angular momentum.

As an example let us consider a molecular cloud collapsing under the gravity of the central compact object. As it becomes denser, the random gas motions that were originally present in the cloud, average out in favor of the direction of the nebula's net angular momentum. Conservation of angular momentum  causes the angular velocities to increase as the nebula becomes smaller. This rotation causes the cloud to flatten out and take the form of a disk.

Formation of a disk is required to support the accretion rate. A free particle with a given angular momentum will rotate around the central object in an eccentric orbit. 
As long as angular momentum is conserved, the test particle will not fall onto the gravitating body nor release its potential energy. In such a case there is no accretion and no energy is released. To support the accretion and form a luminous accretion disk the angular momentum must be transported outward enabling particles to flow inward. Such a process occurs due to the action of viscous torques within the disk. It may be shown that laminar viscosity cannot provide sufficiently high magnitude of torque which would lead to the observed disk luminosities. Turbulent motions are required to explain the angular momentum transport in accretion disks. The exact mechanism leading to turbulence is discussed in Section~\ref{s.viscosity}.

Accretion disks appear in many astrophysical objects. Protostellar clouds of molecular gas create protoplanetary disks surrounding newly formed stars. Stars in close compact binary systems transfer mass through accretion disks when they fill their Roche lobe or due to strong stellar winds. Supermassive BHs in AGNs are believed to accrete cold matter from their vicinity. According to the collapsar model, short $\gamma$-ray bursts accrete the ejecta at super-critical rates shortly after the outbursts. Other examples of accretion disks in the Universe may also be given. Accretion appears to be involved in most of the high-luminosity phenomena in the Universe.

\section{Spectral states}
\label{s.highsoft}
\begin{figure}
  \centering
\subfigure{\includegraphics[width=.7\columnwidth]{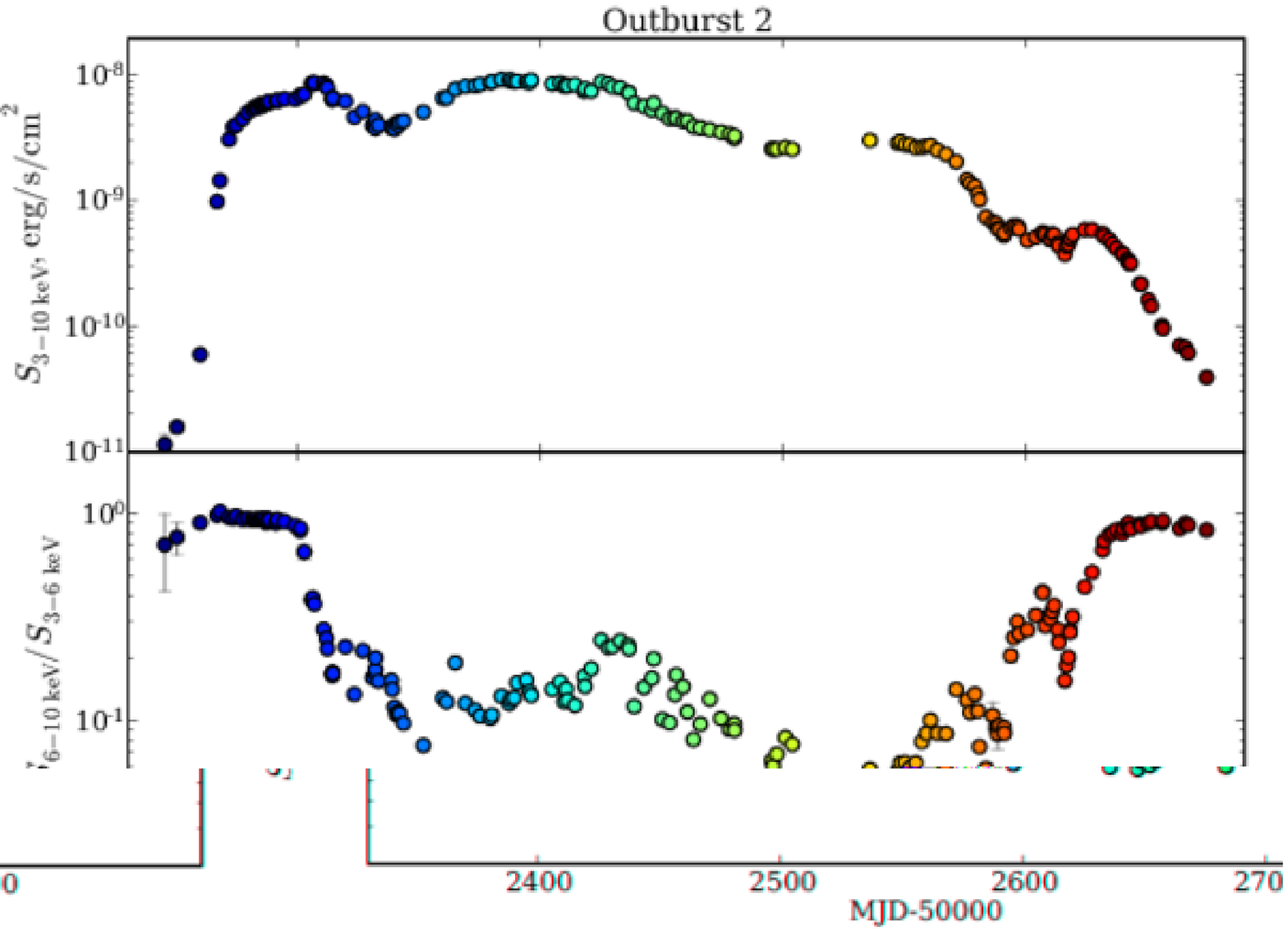}}\\
\subfigure{\includegraphics[width=.7\columnwidth]{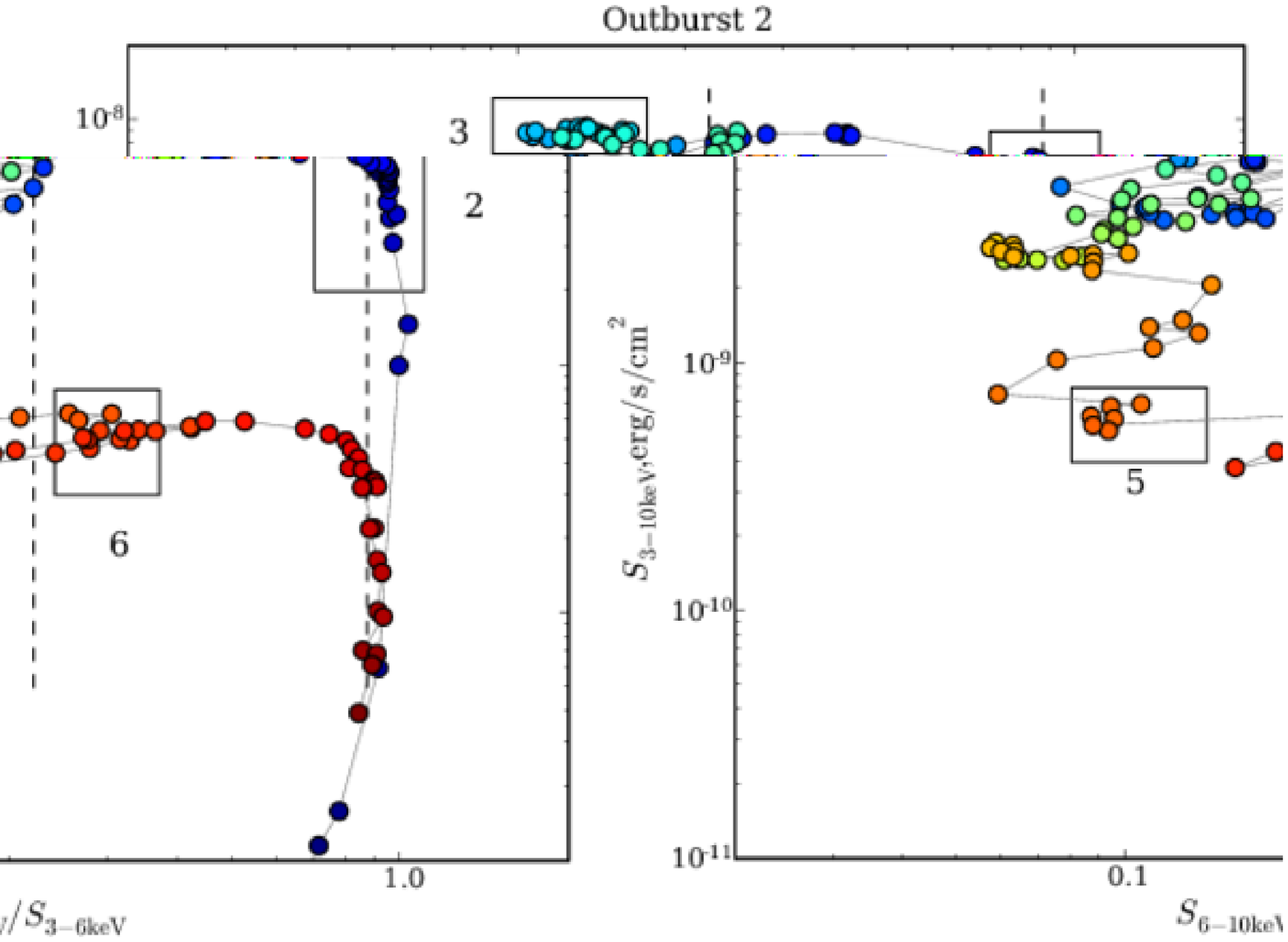}}
\caption{ Soft ($3-10 {\rm keV}$) and hard ($6-10 {\rm keV}$) flux components for GX 339-4 during one cycle of observations (top panel). Luminosity plotted against hardness ratio for the same set of observations (bottom panel). Colors denote particular observations. Figures after \citem{dunnetal-08}.
 }
  \label{f.dunnN.eps}
\end{figure}

The timescale for propagation of matter in accretion disks (the drift timescale) is much longer for disks around supermassive BHs than for disks in microquasars. For the latter, the changing mass supply from the companion may cause variability on the timescale of days. All of the observed microquasars exhibit this kind of variability proving that the process of accretion is not, in general, stationary. In the top panel of Fig.~\ref{f.dunnN.eps} we present, after \citem{dunnetal-08}, light curves in the hard and soft bands of microquasar GX 339-4 during the period March 2002 --- July 2003. Both fluxes vary over a few orders of magnitude. They are, however, anticorrelated which means that the shape of the spectrum has to change significantly. The bottom panel presents luminosity vs hardness ratio. The microquasar exhibits the behavior typical for all objects of this kind:  a given cycle starts with a hard spectrum of low luminosity (bottom right corner) which is followed by rapid increase of luminosity and softening of the spectrum leading to quasi-stationary state characterized by soft spectrum and high luminosity (top left). Close to the end of the cycle, the microquasar decreases its luminosity preserving, however, soft spectrum (left branch). Finally, the spectrum hardens again, but this time the change of hardness takes place at lower luminosity than during the beginning of the cycle.

\begin{figure}
  \begin{center}
\subfigure{\includegraphics[width=.35\columnwidth]{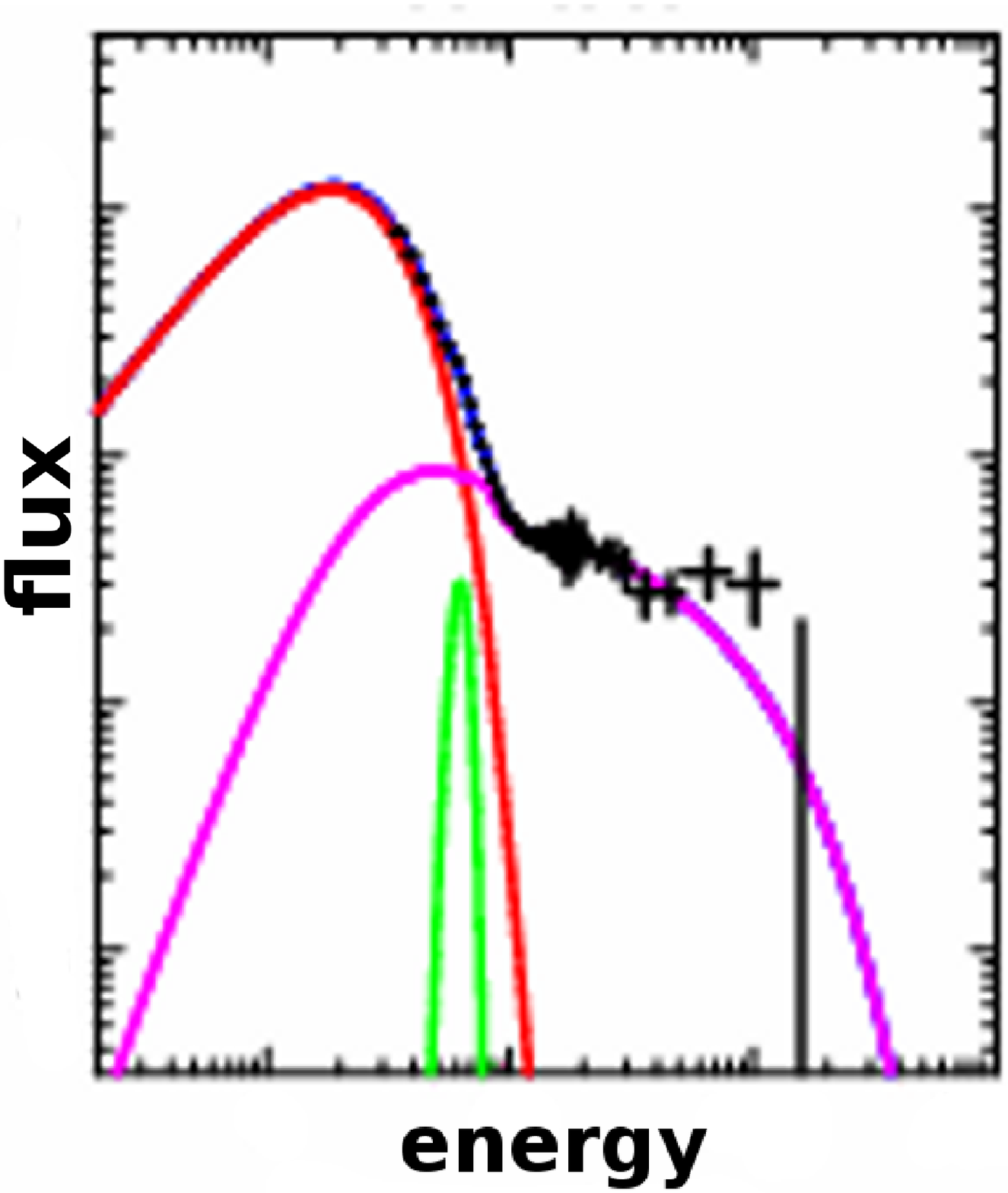}}\hspace{1cm}
\subfigure{\includegraphics[width=.35\columnwidth]{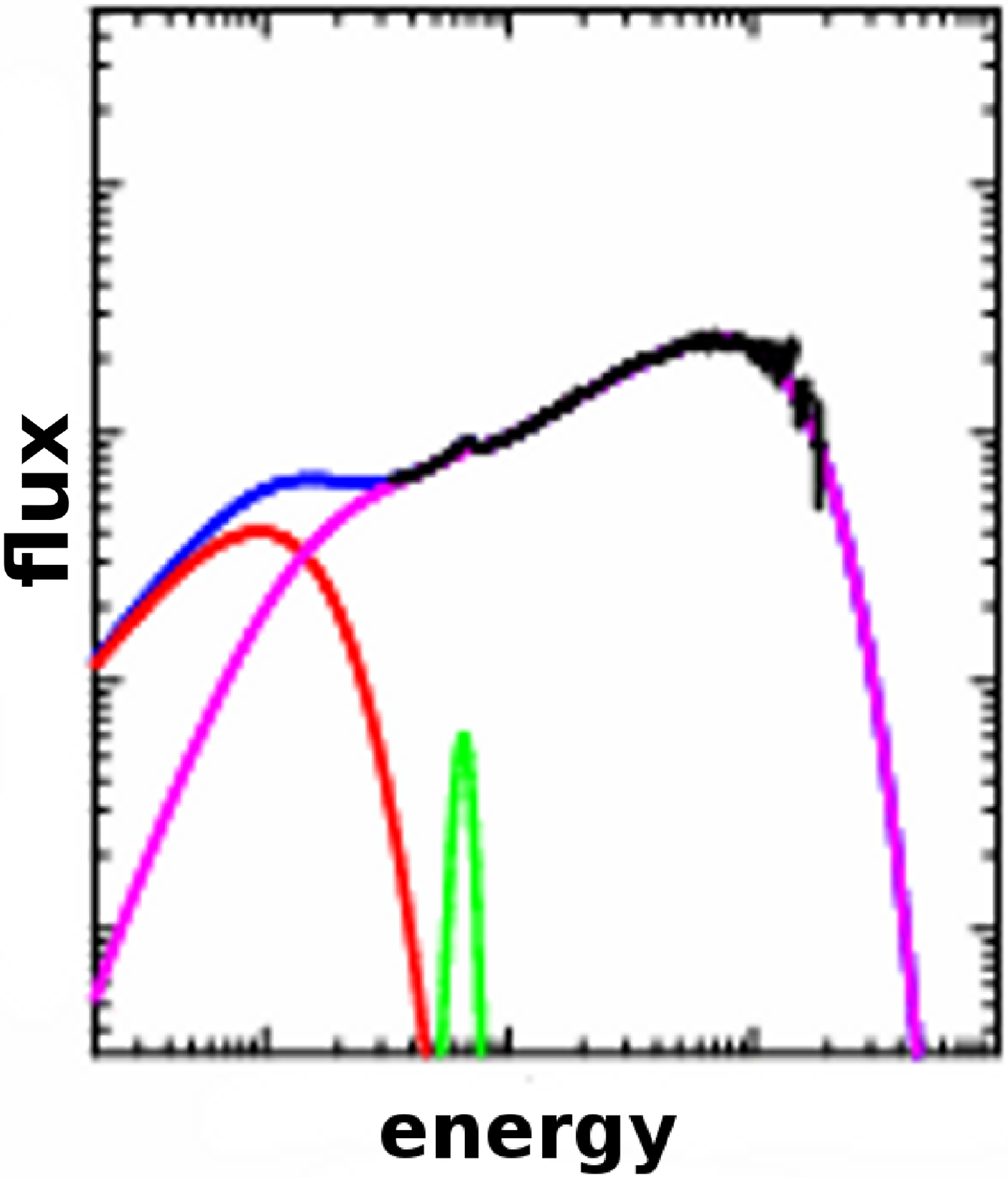}}
\end{center}
\caption{ Schematic plots presenting general features of microquasar spectra during high/soft (left) and
 low/hard (right) spectral states. The red curve corresponds to the disk component, the purple to the comptonized component, while the green curve fits the iron line at 6.4 keV. }
  \label{f.hardsoft}
\end{figure}

Two dominant spectral states can be distinguished: high/soft (soft spectrum and high luminosity) and low/hard (hard spectrum, low luminosity). Fig.~\ref{f.hardsoft} presents schematic pictures of the spectral energy distributions corresponding to these states on left and right panels, respectively. During the high/soft state the spectrum is dominated by the soft component which may be well modeled by 
superposition of black body radiation with different effective temperatures and magnitudes (multicolor black body) which corresponds to the emission predicted for an accretion disk (see e.g., Eq.~\ref{sh_flux1}).
The high energy tail comes most probably from photons up-scattered by 
hot electrons above the disk. The low/hard state is, on the contrary, dominated by the hard component, while the disk radiation contribute only to small fraction of the total emitted energy. 

The disk structure behind the high/soft state is relatively easy to resolve --- the spectrum can be explained by an optically thick accretion disk extending down to the BH horizon, which emits the viscosity-generated heat as multicolor black body radiation. The low/hard state, on the contrary, requires much more complicated disk structure. To obtain the observed relation between the hard and soft components one has to assume that the standard, optically thick accretion disk terminates far away from the BH, at radii of the order of hundreds of the gravitational radius. Inside this transition radius the flow has to be radiatively inefficient and hot so that it can up-scatter photons originating in the outer disk regions. 

A self-consistent model explaining the characteristic features of the low/hard spectral state has not been developed yet. The transition from the optically thick to thin disk may be a result of disk evaporation due to conductive heating from the corona (e.g., \citem{meyermodel}). The high/soft state, on the contrary, is well understood and consistent models of the underlying optically thick disks have been in use for many years. In this dissertation we address this state of accretion disks only.

\chapter{Accretion disk theory}
\pagestyle{fancy}

\section{Conservation laws}
\label{s.conslaws}
The flow of matter accreting onto a compact object through an accretion disk may be described by the equations of hydro- (or, more general, magnetohydro-) dynamics as the typical mean free path of collisions of microscopic particles in such medium is much shorter than the characteristic macroscopic length of the disk (e.g., its thickness). 

The first fundamental equation describing fluid dynamics is,
\be
\nabla_i T^i_k=0,
\ee
where the hydrodynamical stress-energy tensor has the general form\footnote{
In the case of magnetohydrodynamics, when the viscosity results directly from magnetorotational instability, as discussed in Section~\ref{s.mri}, the stress-energy tensor is,
\begin{equation}
 \label{e.cons2f}
  T^i_k=u^i u_k(p+\epsilon+b^2)+(p+b^2/2)\delta^i_k - b^ib_k + u^iq_k + u_kq^i.
\end{equation}
The contravariant fluid-frame magnetic field is given by $b^i$, and is related to the lab-frame three-field $B^i$ by,\be b^i=B^kh^i_k/u^t\ee
where $h^i_k=u^iu_k+\delta^i_k$ is a projection tensor.

Throughout this dissertation we use the $(-+++)$ signature.},
\begin{equation}
 \label{e.cons2}
  T^i_k=u^i u_k(p+\epsilon)+p\delta^i_k - t^i_k + u^iq_k + u_kq^i,
\end{equation}
where $u^i$ is the matter four-velocity, $p$ is the total pressure, $\epsilon$ is gas internal energy, $\delta^i_k$ is the Kronecker delta, $q^i$ is the radiative energy flux and the viscous stress tensor $t^i_k$ is given by,
\be
t^i_k=\nu\rho \sigma^i_k
\ee
($\nu$ is the kinematic viscosity coefficient and $\sigma^i_k$ is the shear tensor of the velocity field).

The second fundamental equation is that of the conservation of the baryonic number of particles,
\be
\nabla_i(\rho u^i)=0,
\ee
where $\rho$ is the gas density.

These general equations of hydrodynamics are used to construct all relativistic, hydrodynamical models of accretion disks. In Appendix~\ref{ap.nonstat} we give a detailed derivation of the non-stationary equations describing advective, optically thick, vertically-integrated, $\alpha$-disks. In this section we introduce and discuss, basing mostly on \citem{KatoBook}, the equations governing the stationary, axisymmetric accretion flow in the non-relativistic limit. 

Under this approximation the fluid dynamics is described by the Navier-Stokes equations:
\begin{eqnarray}
\label{e.navier1}
\pder\rho t+\nabla\cdot(\rho \vec u)&=&0,\\\label{e.navier2}
\rho\left(\pder{\vec u}t+(\vec u\cdot\nabla)\vec u\right)&=&-\rho \nabla \Phi-\nabla p + \nabla {\bf t},
\end{eqnarray}
where $\vec u$ is the fluid velocity, $\Phi$ is the gravitational potential and ${\bf t}$ is the viscous stress tensor.

The surface layers of accretion disks usually do not extend far from the equatorial plane. It is therefore possible to use the cylindrical system of coordinates ($r,\phi,z$). The equation of continuity takes the form,
\be
    \pder \rho t+\frac{1}{r}\frac{\partial}{\partial r}\left(\rho r u_r\right) + \frac{1}{r}\frac{\partial}{\partial \phi}\left(\rho u_\phi\right) + \frac{\partial}{\partial z}\left( \rho u_z\right) = 0. 
\ee
We neglect both time and azimuthal derivatives and, following the common approach, we integrate this equation vertically between the disk vertical boundaries at $z=\pm h$,
\be
\int_{-h}^{h}\frac1r\pder{}r\left(r\rho u_r\right){\rm d}z+\int_{-h}^{h}\pder{}z\left(\rho u_z\right){\rm d}z=0,
\ee
which is equivalent to,
\be
\frac1r\pder{}rr\int_{-h}^{h}\rho u_r{\rm d}z+\left.\rho u_z\right|_{-h}^{h}=0.
\ee
As the density $\rho$ is very low at the disk boundary, the second is negligible. Therefore, we may write,
\be
\label{e.cons.rad1}
\pder{}{r}\left(r\Sigma V\right) =0,
\ee
where
\be\Sigma=\int_{-h}^{h}\rho{\rm d}z,
\ee
\be
V=\frac{1}\Sigma\int_{-h}^{h}\rho u_r{\rm d}z,
\ee
are the surface density and density-weighted radial velocity, respectively. Finally, integrating Eq.~\ref{e.cons.rad1} we may introduce the integration constant $\Mdot$,
\be
\label{e.cons.rad2bis}
\Mdot=-2\pi r\Sigma V,
\ee
which is the accretion rate of gas. This equation describes the conservation of mass.

Taking the $\phi$ component of Eq.~\ref{e.navier2} we get, under the same assumptions: 
\begin{equation}
\rho u_r\frac{\partial}{\partial r}\Omega r^2=\frac{1}{r}\frac{\partial}{\partial r}(r^2 t_{r\phi}), 
\label{e.ang1}
\end{equation}
where $\Omega$ is angular velocity of the fluid and $t_{r\phi}$ is the only non-negligible component of the viscous
stress tensor. Introducing the angular momentum per unit mass ${\cal L}=\Omega r^2$ and integrating over $z$ and $r$ we obtain 
the angular momentum conservation law:
\begin{equation}
\frac{\dot{M}}{2\pi}({\cal L}-{\cal L}_{\rm in})=-2hr^2 t_{r\phi}.
\label{e.ang2}
\end{equation}
We have assumed that there is no torque at the inner boundary (i.e., BH horizon) so that the integration constant ${\cal L}_{\rm in}$ can be 
interpreted as the angular momentum of the matter at the inner edge. Eq. \ref{e.ang2} reflects the conservation of angular momentum which is transported outward by viscous stresses.

%

Energy is another conserved quantity. Three mechanisms of heating/cooling are involved for advective accretion disks: viscosity, advection and radiation. The energy balance may be formulated in the following way:
\be
Q^{\rm adv}=Q^{\rm vis} - Q^{\rm rad}
\ee
which reflects the fact that the heating or cooling rate by advection ($Q^{\rm adv}$) must be equal to the difference between the amount of heat generated by viscosity ($Q^{\rm vis}$) and the radiative cooling rate ($Q^{\rm rad}$).

\section{Source of viscosity}
\label{s.viscosity}

The angular momentum in accretion disks is transported by viscosity. In this section we will discuss this process in general, describe the $\alpha p$ formalism and discuss the magnetorotational instability which is considered the most likely source of turbulence in disks.

\begin{figure}
  \centering\resizebox{.8\textwidth}{!}{\includegraphics[angle=0]{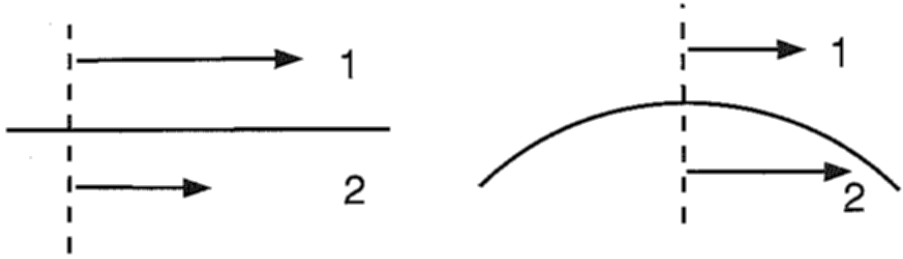}}
\caption{Parallel shear flow (left) and differentially rotating flow (right panel). Figure after \citem{KatoBook}.
 }
  \label{f.shear}
\end{figure}

Let us consider parallel shear flow (left panel of Fig.~\ref{f.shear}). As the upper layer moves with greater velocity the momentum transfer occurs from the top to bottom. As a result the upper layer will be decelerated while the bottom accelerated. Viscosity tends to reduce shear and produce uniform layers.

The viscous force exerted in the horizontal ($x$) direction on the unit surface of the interface plane is,
\be
t_{xy}=\eta\pder{v_x}{y},
\ee
where $y$ is the normal to the interface, $\eta=\rho\nu$ is the coefficient of dynamical viscosity and $\partial{v_x}/\partial{y}$ is the velocity gradient.

In the case of accretion disks we do not deal with parallel layers. Instead, we have differentially rotating annuli (right panel of Fig.~\ref{f.shear}). The rotational velocity increases in Keplerian flows inward. Therefore, the viscosity will transfer angular momentum from the inner annulus to the outer. 
In such a way, matter loses angular momentum and moves inward. If the disk was cut off at a certain outer radius and no external force acted on the disk, the outermost annuli would receive all the transported angular momentum and move outward to infinity. 

The viscous force per unit area exerted in the azimuthal direction of the interface plane between differentially rotating annuli is,
\be
\label{visc.stress2}
t_{r\phi}=\eta\left(\pder{v_\phi}{r}-\frac{v_\phi}{r}\right)=\eta r\pder{\Omega}{r}.
\ee

\subsection{$\alpha p$ prescription}
\label{standard-alpha-prescription}

In 1970s the understanding of viscous 
processes which take place in accretion disks was very poor. Nevertheless,
\citem{shakura-73} proposed an extremely
elegant and plausible formalism which has been used effectively for many years, making progress in accretion disk
theory possible. They assumed correctly that the source of viscosity in accretion disk is connected in some way to turbulence
in gas-dynamical flow. The nature of this turbulent motion was not known at that time. They assumed a simple expression for
kinematic viscosity coefficient:
\begin{equation}
 \nu\approx v_{\rm turb}l_{\rm turb},
\label{e.sh_nu}
\end{equation}
where $v_{\rm turb}$ and $l_{\rm turb}$ stand for typical turbulent motion velocity and length scale, respectively. Assuming that the velocity of turbulent elements cannot exceed the speed of sound $c_s$ and their typical size cannot
be larger that the disk thickness $h$ one gets
\be
\nu<c_s h.
\ee
Using the vertical equilibrium condition (Eq.~\ref{sh_vert2}), this expression may be put in the following form
\be
\nu<\frac{c_s^2}{\Omega}\approx \frac{p}{\rho \Omega}.
\label{visc.nuestim}
\ee
The stress is given by Eq.~\ref{visc.stress2}:
\be
t_{r\phi}=\eta r \pder\Omega r\approx- \rho \nu\Omega.
\ee
Taking Eq.~\ref{visc.nuestim} into account one gets
\be
|t_{r\phi}| < \rho\frac{p}{\rho \Omega}\Omega=p.
\ee
Therefore, a non-dimensional viscosity parameter $\alpha$ satisfying the condition $\alpha\leq 1$, may be introduced:
\begin{equation}
 t_{r\phi}=-\alpha p.
\label{e.sh_alphaP2}
\end{equation}
This expression for viscosity, proposed for the first time by \citem{shakura-73}, is called the \textit{$\alpha p$} or the \textit{$\alpha$ prescription} and has been widely used for many years in accretion 
disk theory. The most modern magnetohydrodynamical (MHD) simulations (e.g., \citem{hirose-09-b}) support application of the $\alpha$ formalism to \textit{mean}
accretion flow dynamics.

\subsection{Magnetorotational instability}
\label{s.mri}
For almost twenty years the source of the turbulent motion responsible for viscosity in accretion disks remained unknown. Only in 1991 \citem{balbushawley-91} identified the process which might lead to turbulent motion in accretion disks. They noted that in astrophysical disks a magnetorotational instability (MRI) may set in, amplifying the initial weak magnetic field and leading to the required angular momentum transport. Since that time a number of numerical methods have been developed to verify this possibility. They gave positive results and therefore magnetorotational instability is currently thought to be the most likely explanation of turbulent motion in accretion disks. 
However, it was suggested by \citem{umurhametal-07} on the
basis of perturbative weakly nonlinear analysis that the transport due to
the nonlinearly developed MRI may be very small in experimental setups
with the Prandtl number $P_m\ll1$. Several authors studied an
alternative picture of turbulence arousal in non-magnetized astrophysical
disks, related to their non-axisymmetric hydrodynamical disturbances,
which can, in principle, precipitate secondary instabilities \citepm{rebuscoetal-09}.

MRI was first noticed by \citem{velikhov-59} who studied stability of an ideally conducting liquid flowing between cylinders rotating in a magnetic field. His results were generalized by \citem{chandrasekhar-60}. Only in 1990's astrophysicists realized its paramount importance for triggering accretion in astrophysical objects.

The general idea of MRI may be illustrated in the following way. Let us consider a disk of fluid rotating with Keplerian angular velocities in a weak axially symmetric magnetic field. The magnetic tension acting on two fluid elements may be imagined as a spring connecting them. As the inner element orbits with higher angular velocity than the outer one, the spring stretches. In such a way the magnetic tension slows down the inner element and forces the other one to speed up. As a result, the inner loses its angular momentum while the angular momentum of the outer element increases moving the elements apart. In this way the angular momentum is transported outward making accretion possible. The crucial condition for arising of MRI is that the angular velocity of a magnetized fluid decreases with radius:
\be \der\Omega r<0,\ee
which is satisfied in the astrophysical context, e.g., for the Keplerian profile of rotation.

Most of research devoted to MRI is based on numerical simulations solving the equations of magnetohydrodynamics. They may be solved either globally in three-dimensions (e.g., \citem{pennaetal-10}) or locally in the shearing-box approximation (e.g., \citem{hirose-09-a}). So far, the radiative transfer has been neglected as it immensely increases the required computational power. Instead, different, purely numerical, methods of cooling are assumed. Only recently, first MHD simulations accounting for radiation, however axially symmetric, were performed \citepm{ohsuga-09}.

\begin{figure}
  \centering\resizebox{.7\textwidth}{!}{\includegraphics[angle=0]{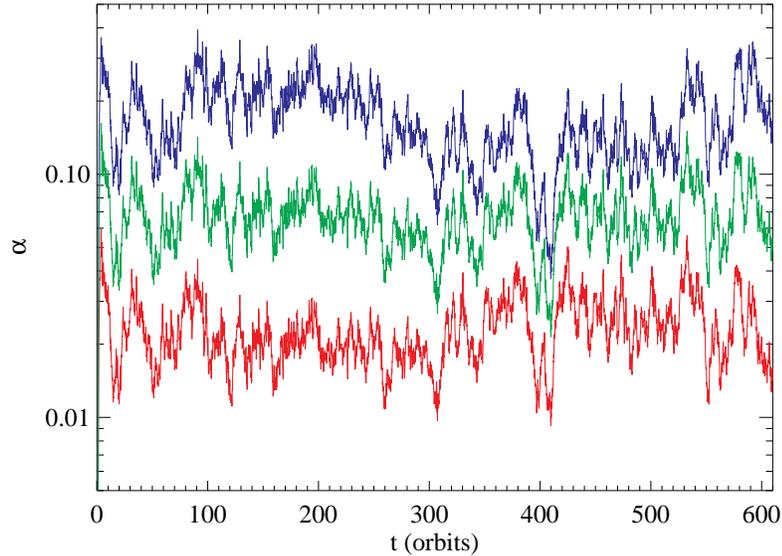}}
\caption{ Time dependence of the ratio $\alpha$ of the box-averaged $r\phi$ stress to
various box-averaged pressures in one of the simulations from \citem{hirose-09-a}. From top to bottom, the
stress is scaled with gas pressure, the geometric mean of gas and radiation
pressures, and with total (gas plus radiation) pressure, respectively. Figure after \citem{hirose-09-a}. }
  \label{f.hirose.f10.eps}
\end{figure}

The crucial question from the point of view of hydrodynamical $\alpha p$ models of accretion disks is whether the assumption made by Shakura \& Sunyaev remains valid facing the MHD physics. 
Fig.~\ref{f.hirose.f10.eps} shows time-dependence of the value of $\alpha$ (defined as the ratio of the box-integrated stress to the different box-integrated pressure quantities) for various stress prescriptions for simulation 1112a from \citem{hirose-09-a}. The short-time variability reflects the fluctuations in the $r\phi$ stress component. The total (gas plus radiation) case has the least variation of the $\alpha$ profile (by a factor of 4), while the largest (by a factor of 10) is for the gas pressure only case. Therefore, the stress component never follows the pressure precisely, but one may say that the $\alpha p$ prescription (with $p$ being the total pressure) quite well describes the mean dynamics of the flow.

\section{Efficiency of accretion}
\label{s.efficiency}
Let us consider the energetic balance of the accreted fluid between two radii: $r_{\rm in}$ and $r_{\rm out}$. 
The amount of generated energy (luminosity) between these two radii can be estimated from
\be
L_{r_{\rm in}\div r_{\rm out}}=  {\dot M}e_{r_{\rm out}} + ({\cal T}\Omega)_{r_{\rm out}}-{\dot M}e_{r_{\rm in}} - ({\cal T}\Omega)_{r_{\rm in}}
, \nonumber \\
\label{luminosity-estimate}
\ee
where ${\cal T}$ denotes the torque at a given radius, ${\cal T}\Omega$ is the outward energy transport due to viscosity and $e$ is the specific energy. If the disk extends to large radii, we may put $\Omega_{r_{\rm out}} \approx 0$ and $e_{r_{\rm out}} \approx 1$. Hence,
\begin{equation}
L_{r_{\rm in}} = {\dot M}\left(1 - e_{r_{\rm in}} - \chi ( {\cal L}\Omega )_{r_{\rm in}}
\right) \equiv \tilde\eta {\dot M},
\label{efficiency-estimate}
\end{equation}
where $\chi={\cal T}/\Mdot{\cal L}$ is the ratio of the viscous torque to the advective
flux of angular momentum (see Figs.~\ref{fig:inner-torque} and
\ref{fig:rstr}).

If the torque at the inner edge of the disk is negligible ($\chi \ll 1$),
the efficiency of accretion $\tilde \eta$ depends
mainly on the specific energy at the inner edge, $e_{r_{\rm in}}$, given by\footnote{Throughout this dissertation we use the geometrical ($G=c=1$) units. In these units mass is measured in meters, $M=r_G=GM/c^2$ and $\msun\approx 1477 \,{\rm m}$.}
\be
e_{r_{\rm in}}=\sqrt{1-\frac{2M}{3r_{\rm in}}}.
\label{e.etaorg2}
\ee
Assuming $\chi=0$ we get,
\be
\label{e.etaorg}
\tilde\eta=1-e_{r_{\rm in}}.
\ee
The farther away the inner edge from the innermost stable circular orbit\footnote{The innermost stable circular orbit (ISCO) is often called the marginally stable orbit.} (ISCO, Appendix~\ref{ap.isco}) is (and the closer to the
BH), the smaller the efficiency.

\begin{figure}
  \centering\resizebox{.7\textwidth}{!}{\includegraphics[angle=0]{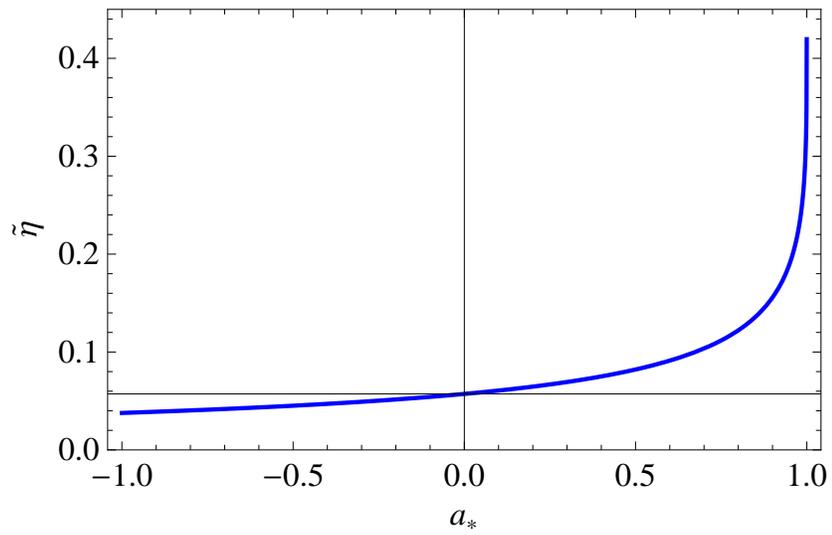}}
\caption{ Efficiency of accretion ($\tilde\eta=L/\dot Mc^2$) through a standard thin disk terminating at ISCO for a given value of BH spin.
 }
  \label{f.eff.eps}
\end{figure}

For accretion disks terminating at ISCO, we have,
\be
\tilde\eta=\left\{\begin{array}{lll}
1-\sqrt{\frac89}=0.057,&\quad&a_*=0,\\
1-\sqrt{\frac13}=0.420,&\quad&a_*=1.
\end{array}\right.
\label{e.efficiency}
\ee
Fig.~\ref{f.eff.eps} presents the dependence of the efficiency of accretion on the BH spin parameter $a_*$. For non-rotating BHs it equals $\eta=0.057$. The efficiency rapidly increases for BH spin values close to unity --- even small difference in angular momentum (e.g., $a_*=0.999$ vs $0.9999$) implies significant difference in the efficiency of extracting gravitational energy from the flow. 

\chapter{Classical models of accretion disks}

\begin{figure}[h]
\centering
\includegraphics[width=.75\textwidth]{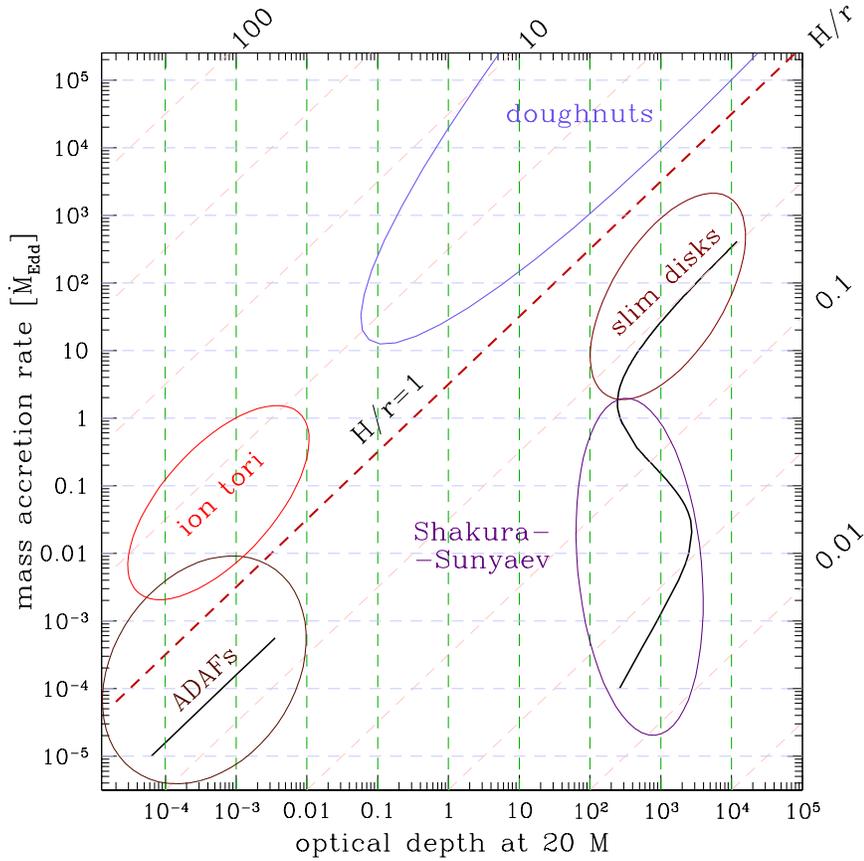}
\caption{This figure illustrates a few of the most well-known analytic and
semi-analytic solutions of the stationary black hole accretion
disks. Their location in the parameter space approximately
corresponds to viscosity $\alpha = 0.1$ and radius $r = 20~{\rm M}$.}
\label{fig:branches}
\end{figure}

Figure~\ref{fig:branches} shows the locations of analytic and semi-analytic
models of hydrodynamical accretion disks\footnote{For
detailed reviews of these solutions see, {\tt \tiny
http://www.scholarpedia.org/article/Accretion\_discs} or
\citem{KatoBook}.}, in the parameter space
described by the vertical optical depth $\tau$, dimensionless
vertical thickness and dimensionless accretion rate
${\dot m}= {\dot M}/{\dot M}_{\rm Edd} 
$ (defined in Eq.~\ref{e.mdotcritical}). In the following sections we
describe in detail Polish doughnuts and radiatively efficient disks \citepm{shakura-73,nt}. Slim disks are introduced and discussed in Chapters~\ref{chapter-stationary} --- \ref{chapter-applications}.

%
%

\section{Polish doughnuts}

One of the simplest and most useful models of geometrically thick accretion disks may be obtained adopting the following assumptions \citepm{abramowicz-78}: the matter is distributed uniformly and axisymmetrically, it orbits the BH along circular trajectories with given angular momentum ${\cal L}=u_\phi(r,\theta)$ and the stress energy tensor $T^i_k$ is given by:

\begin{equation}
 \label{e.paczki1}
  T^i_k=u^i u_k(p+\epsilon)+\delta^i_k p
\end{equation}
where $u^\mu$ is the matter four-velocity, $p$ the total pressure, $\epsilon$ the total energy density and $\delta^\mu_\nu$ is the Kronecker delta.

\begin{figure*}
  \centering

    \subfigure{\includegraphics[width=3.85cm]{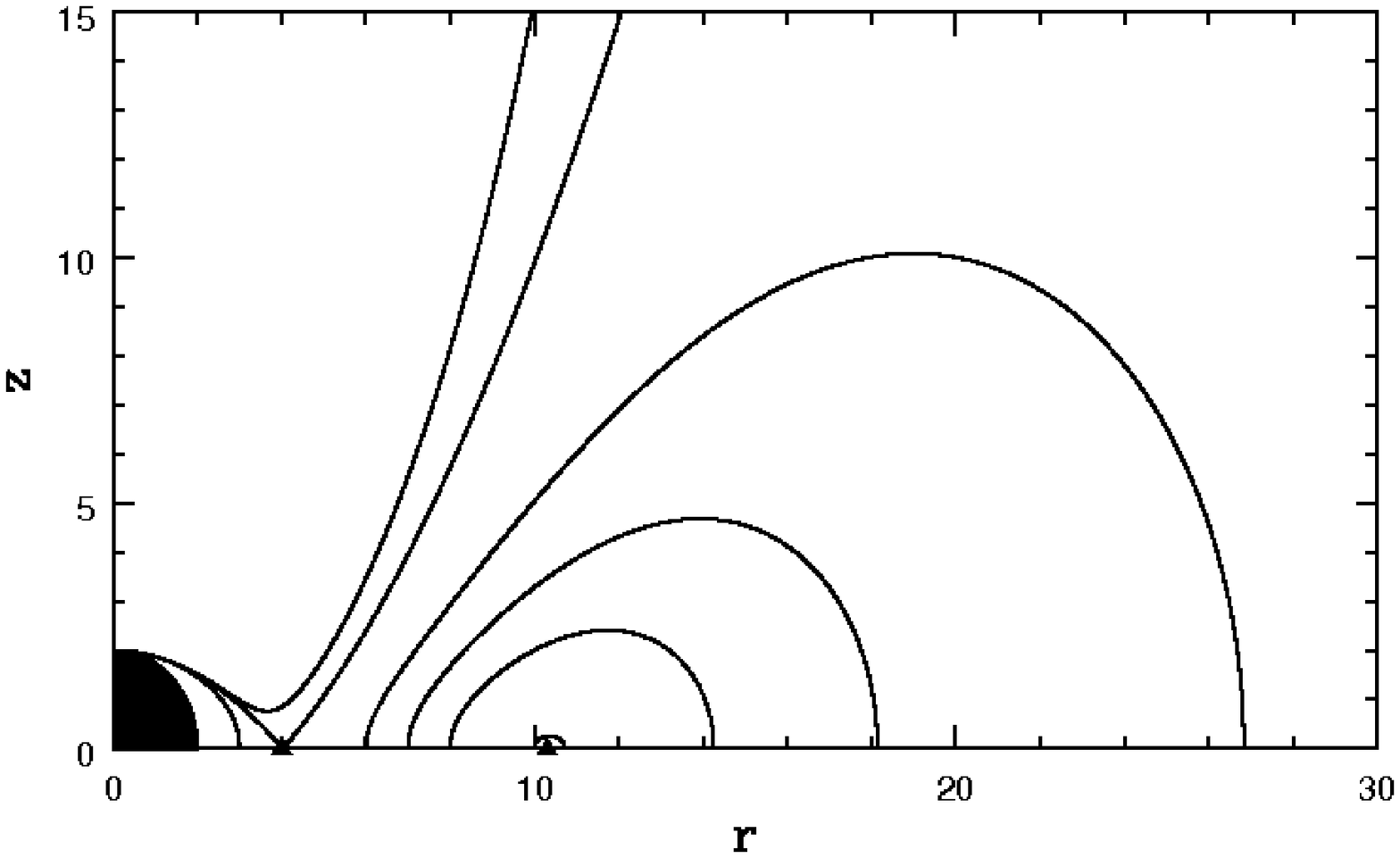}}\hspace{-.32cm}
   \subfigure{\includegraphics[width=3.85cm]{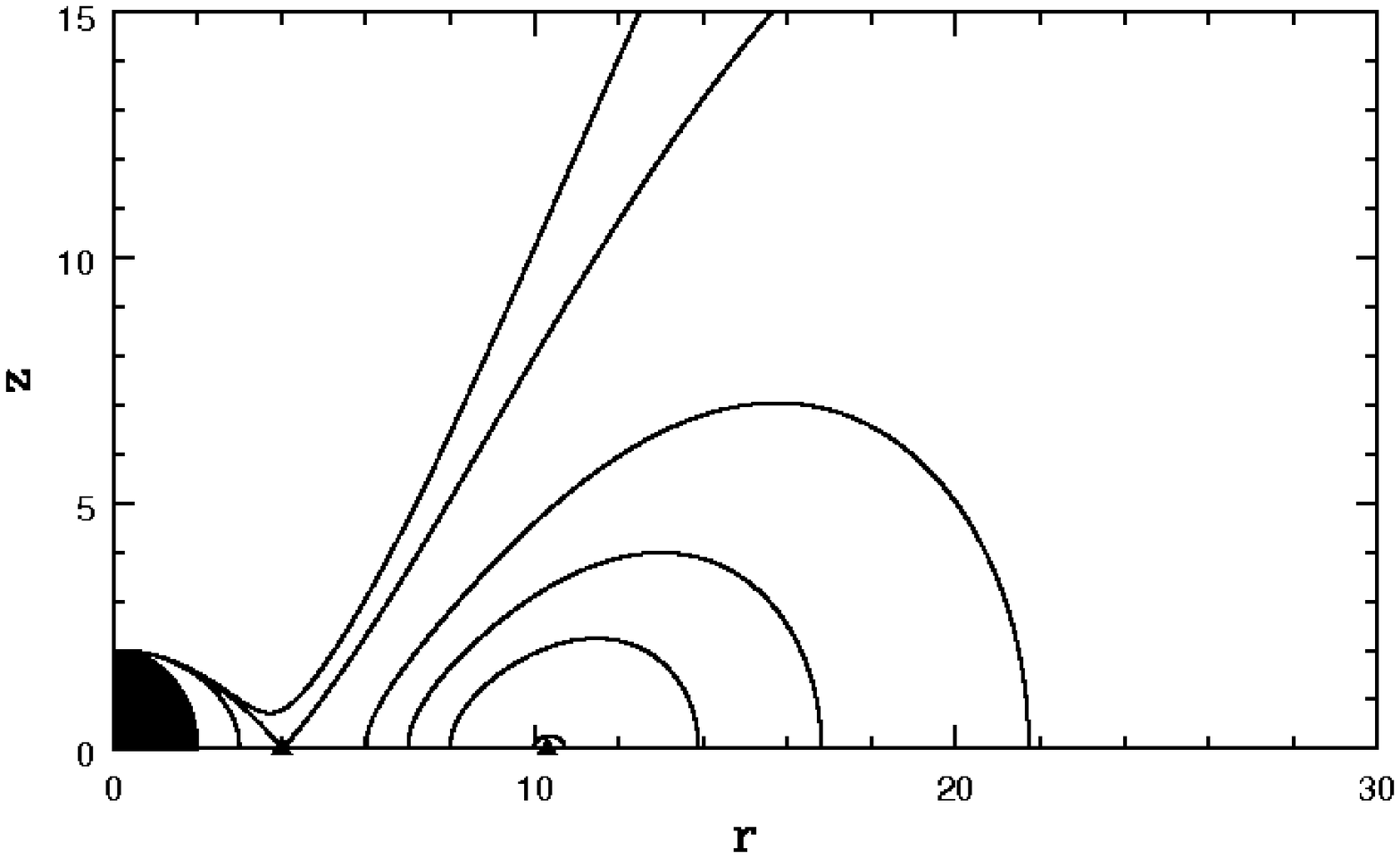}}\hspace{-.32cm}
    \subfigure{\includegraphics[width=3.85cm]{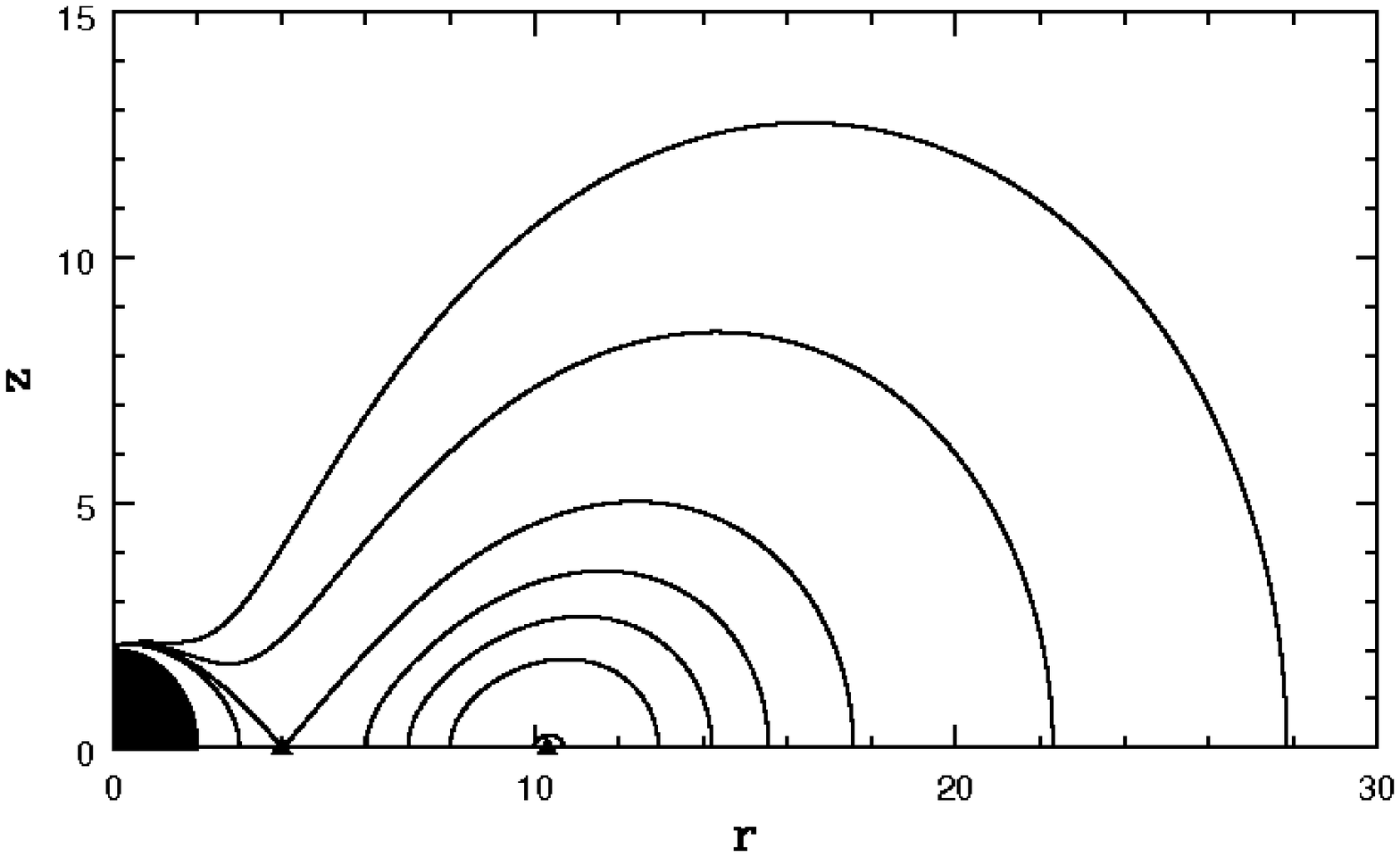}}\hspace{-.32cm}
    \subfigure{\includegraphics[width=3.85cm]{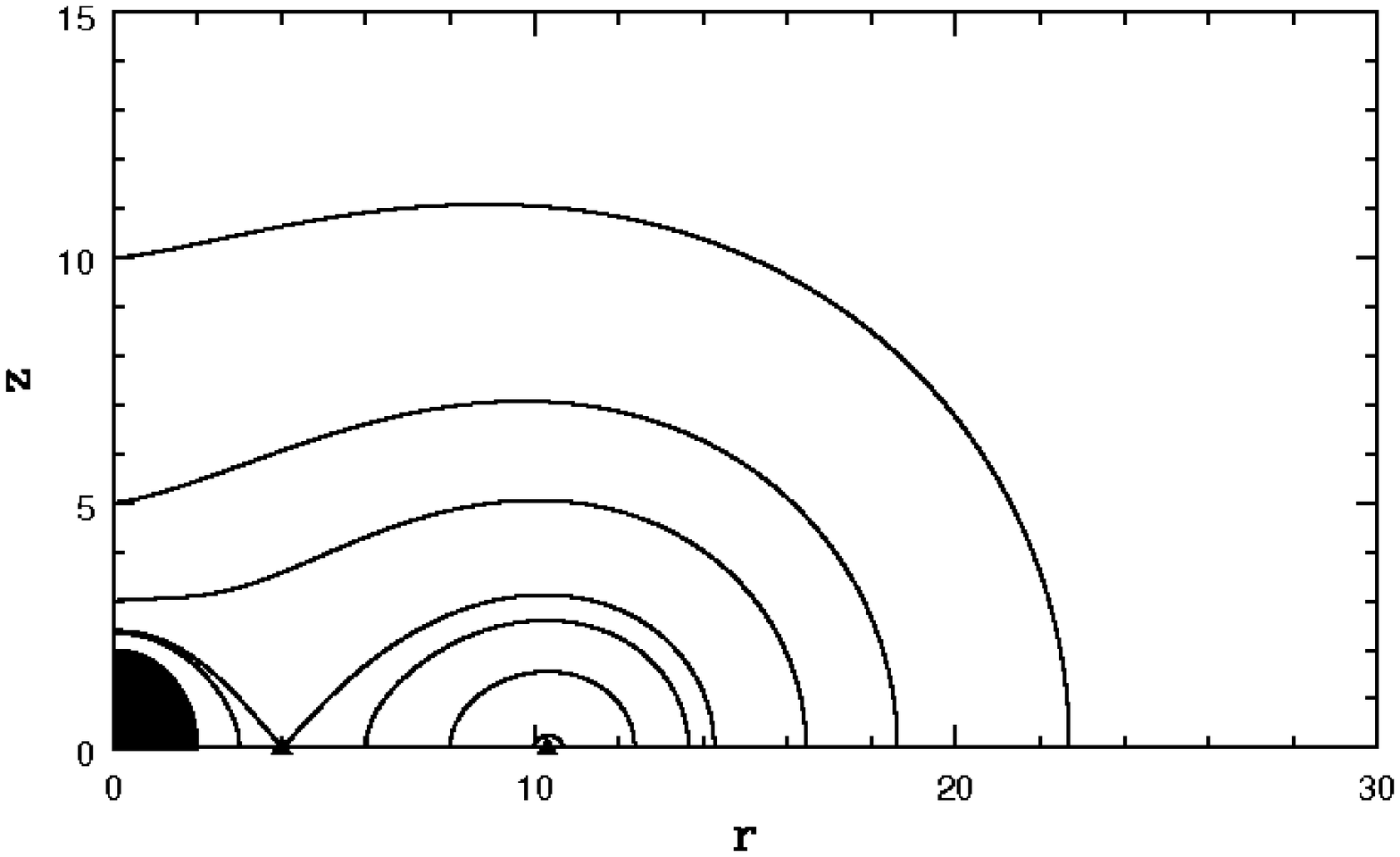}}\hspace{-.32cm}
    \subfigure{\includegraphics[width=3.85cm]{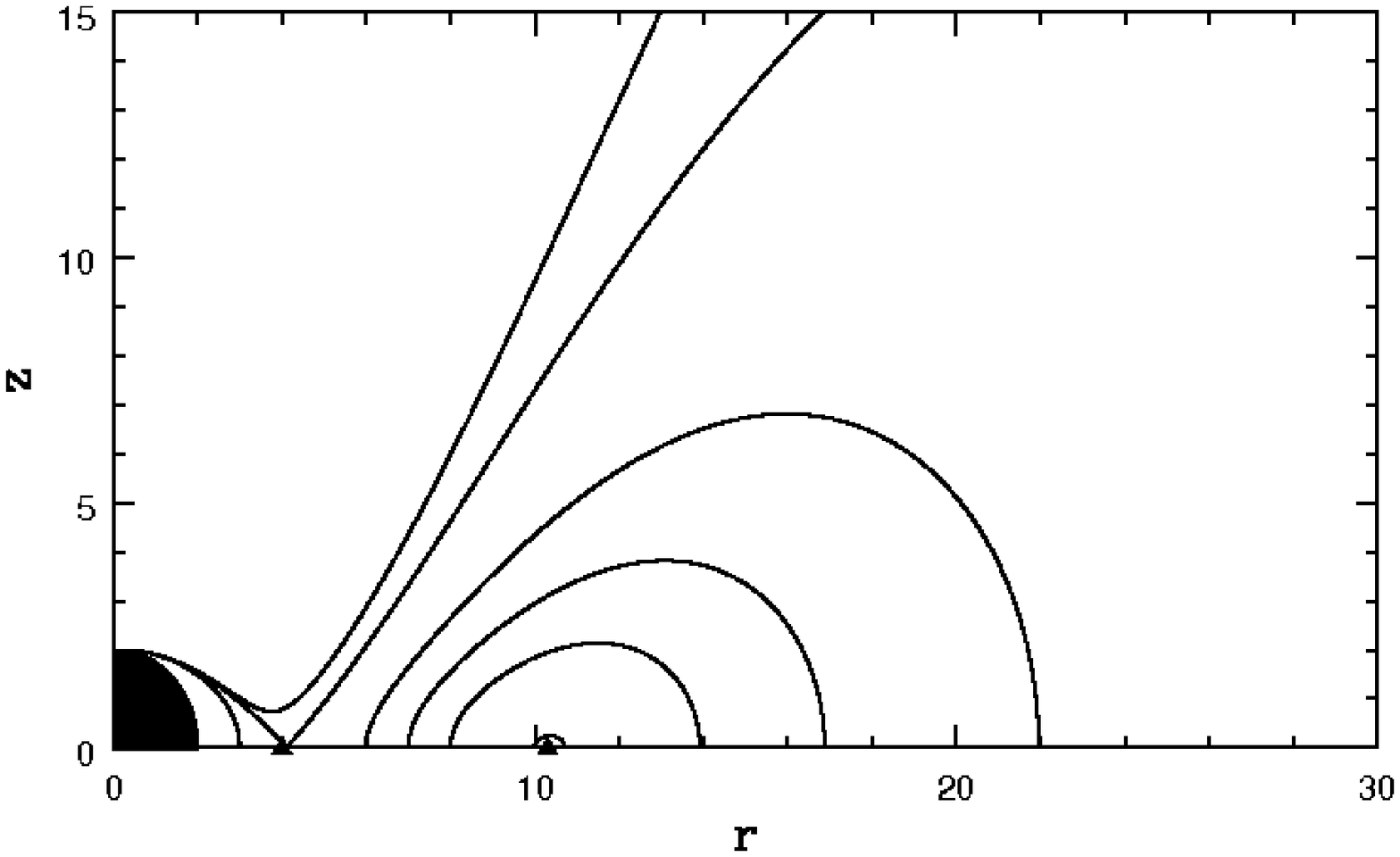}}\hspace{-.32cm}
    \subfigure{\includegraphics[width=3.85cm]{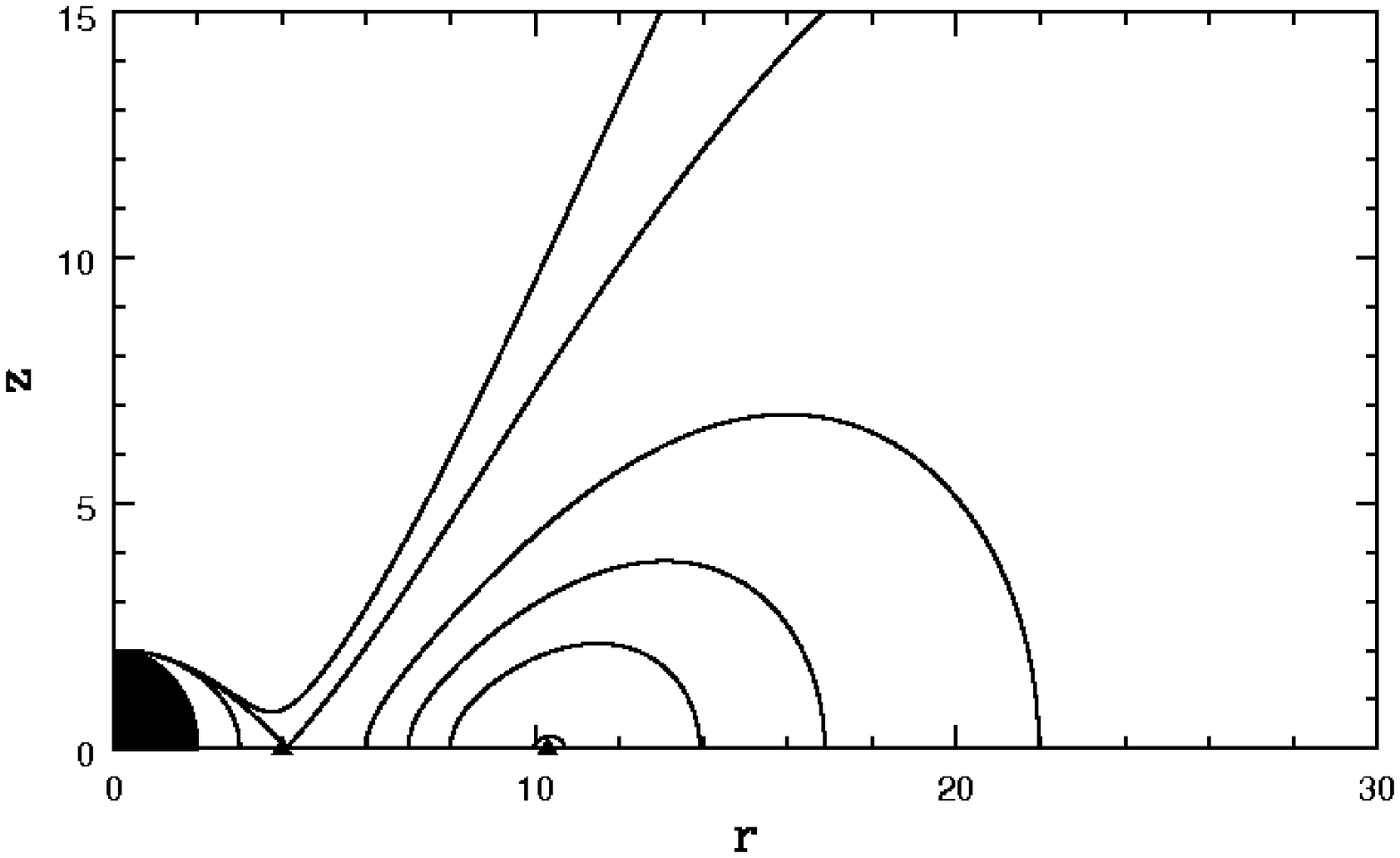}}\hspace{-.32cm}
    \subfigure{\includegraphics[width=3.85cm]{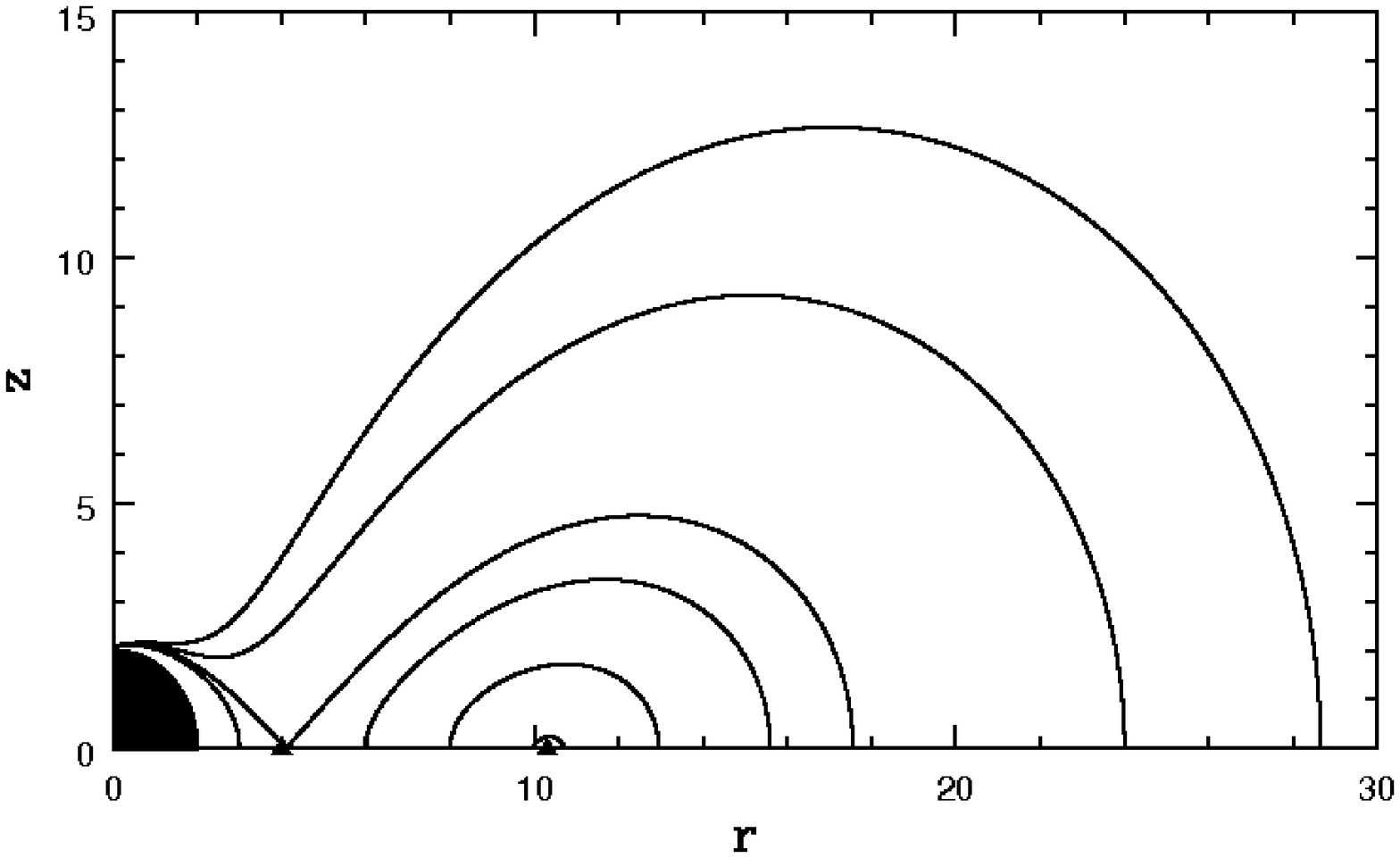}}\hspace{-.32cm}
    \subfigure{\includegraphics[width=3.85cm]{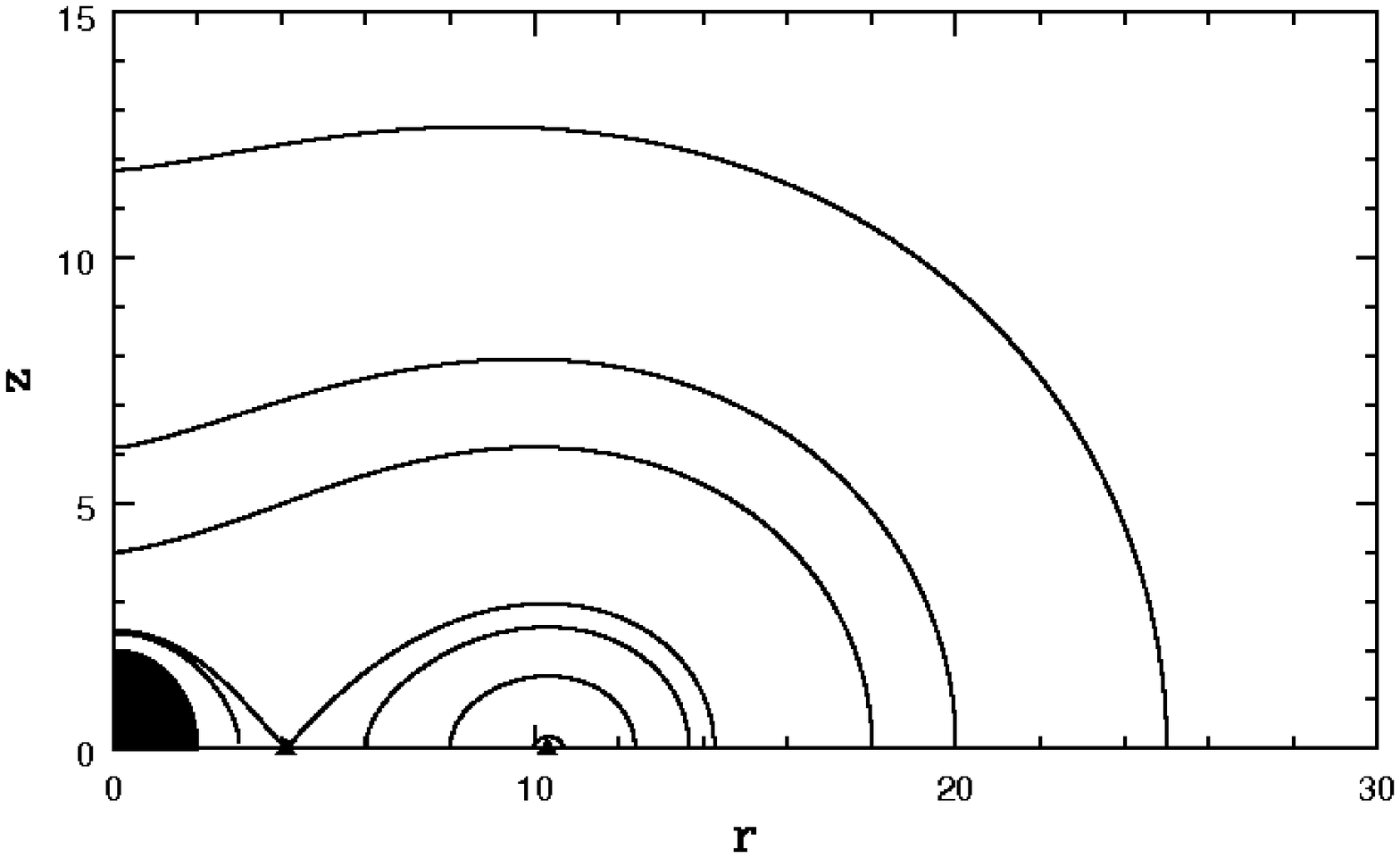}}\hspace{-.32cm}
    \subfigure{\includegraphics[width=3.85cm]{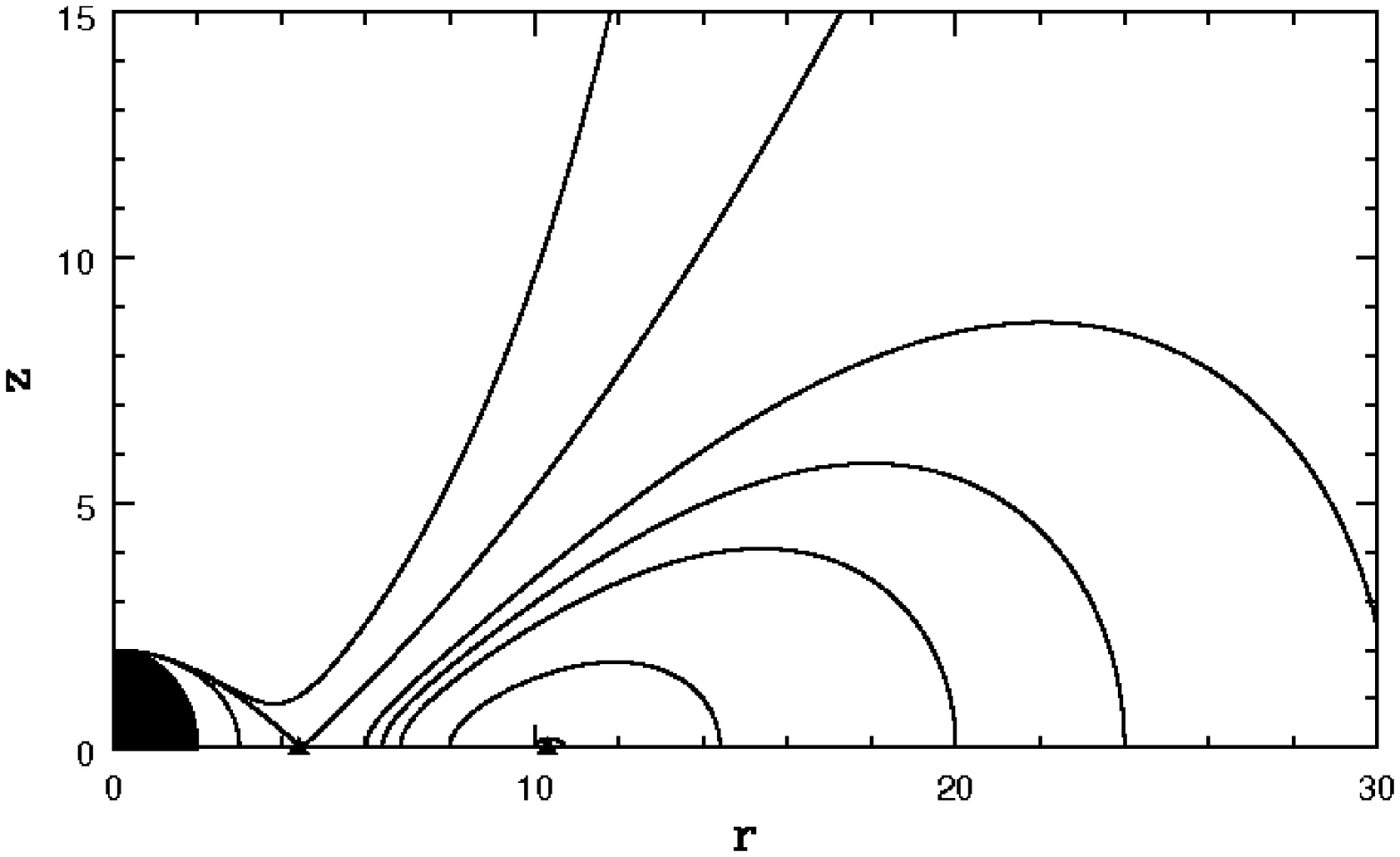}}\hspace{-.32cm}
    \subfigure{\includegraphics[width=3.85cm]{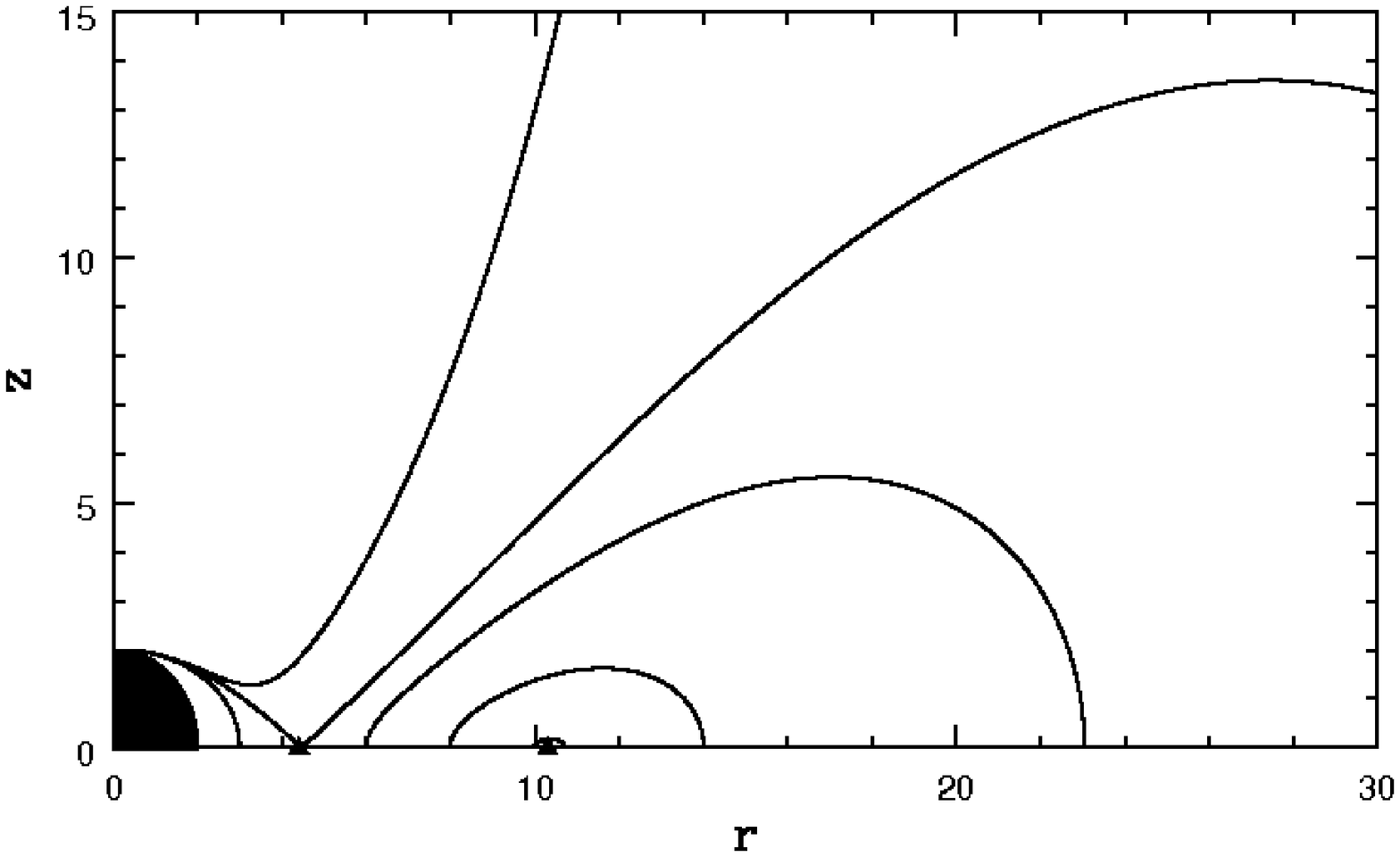}}\hspace{-.32cm}
    \subfigure{\includegraphics[width=3.85cm]{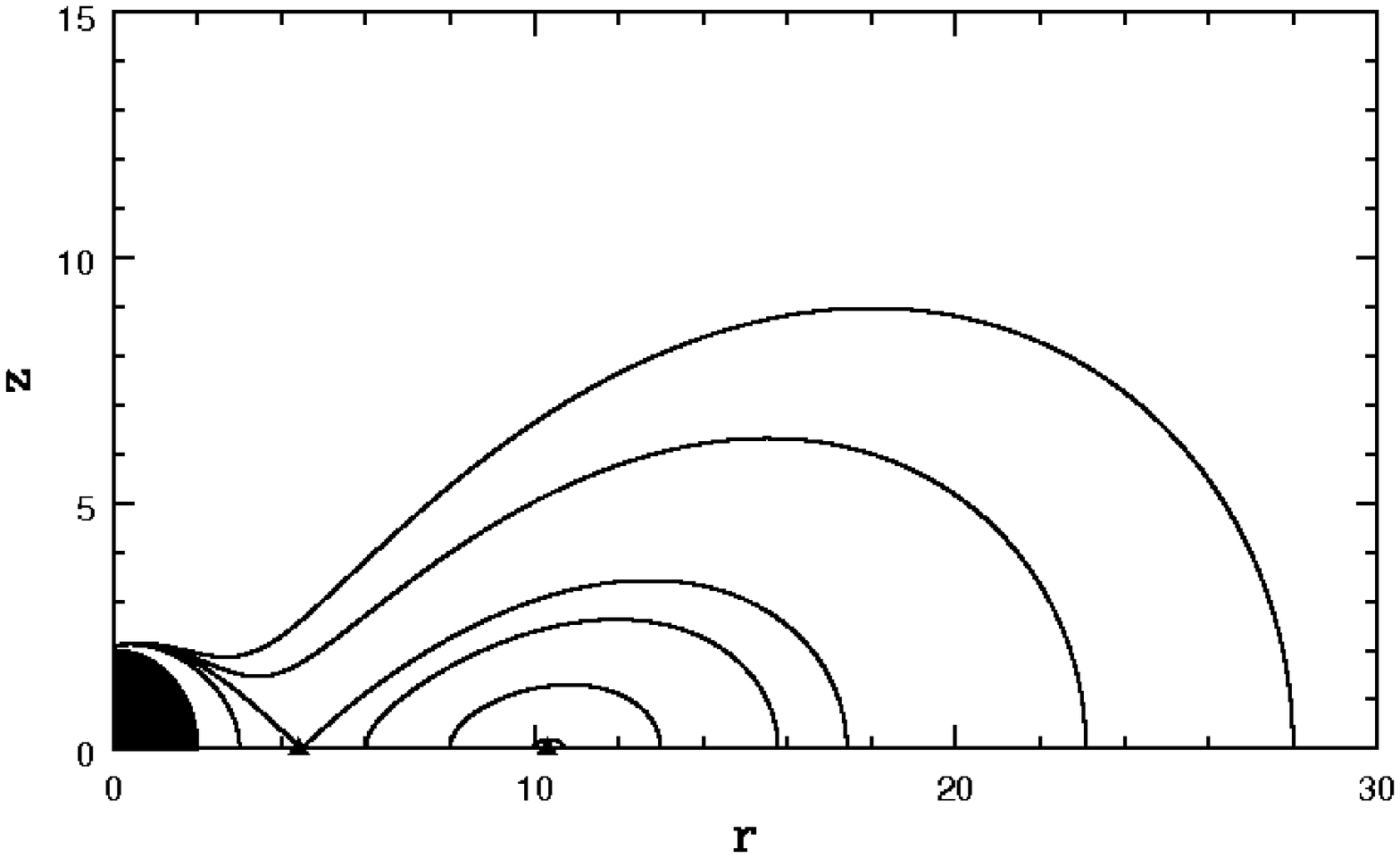}}\hspace{-.32cm}
   \subfigure{\includegraphics[width=3.85cm]{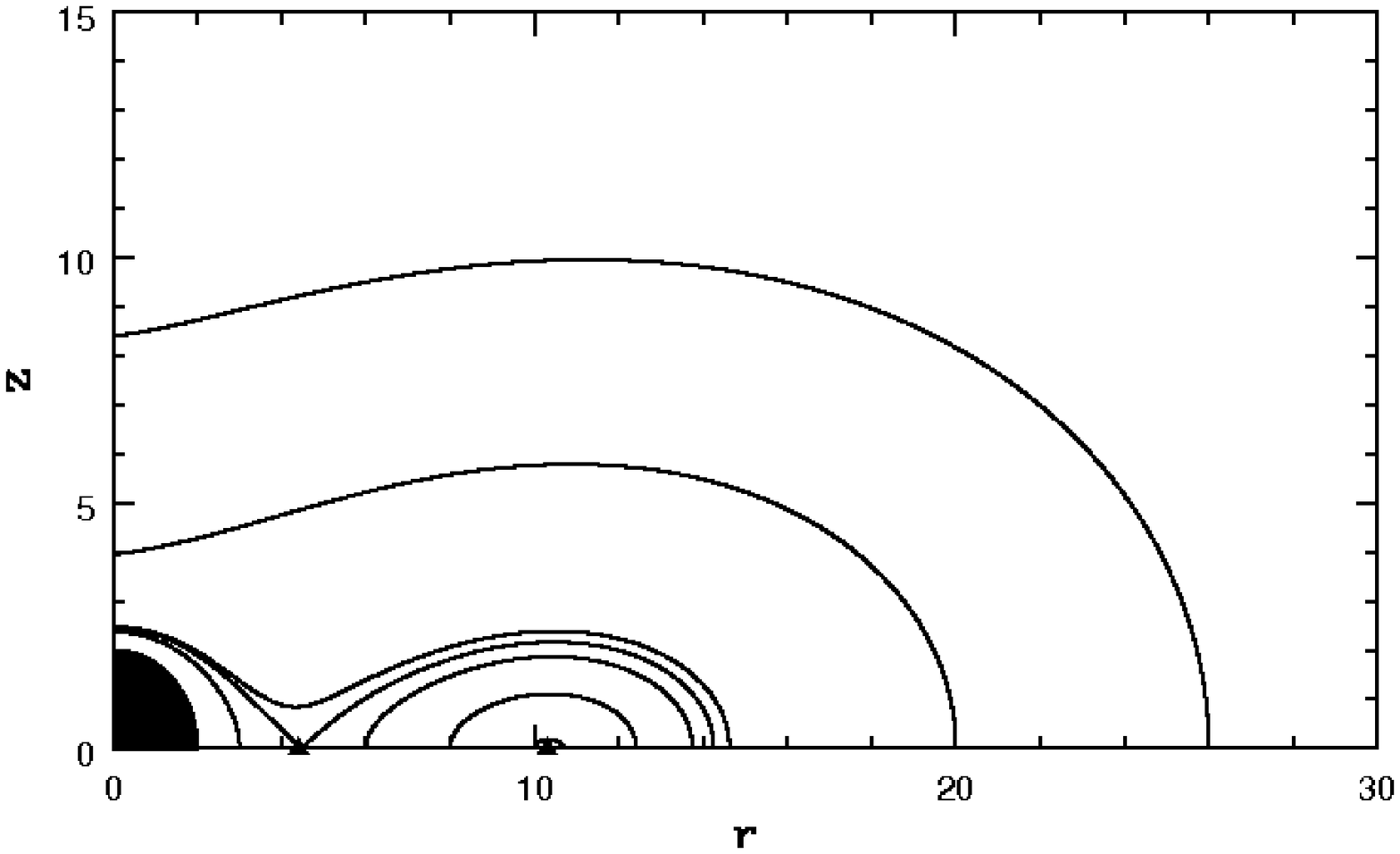}}\hspace{-.32cm}
    \subfigure{\includegraphics[width=3.85cm]{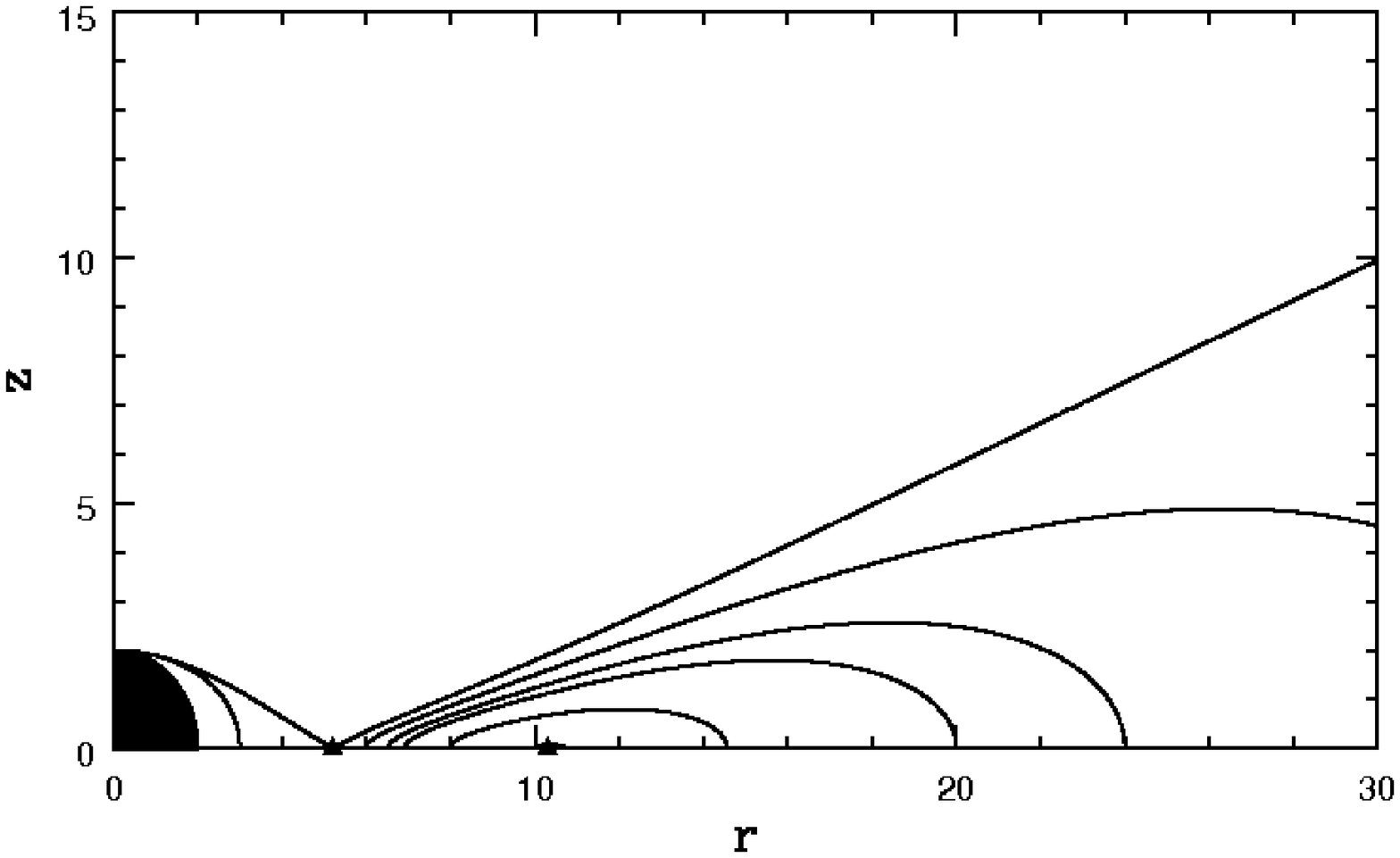}}\hspace{-.32cm}
    \subfigure{\includegraphics[width=3.85cm]{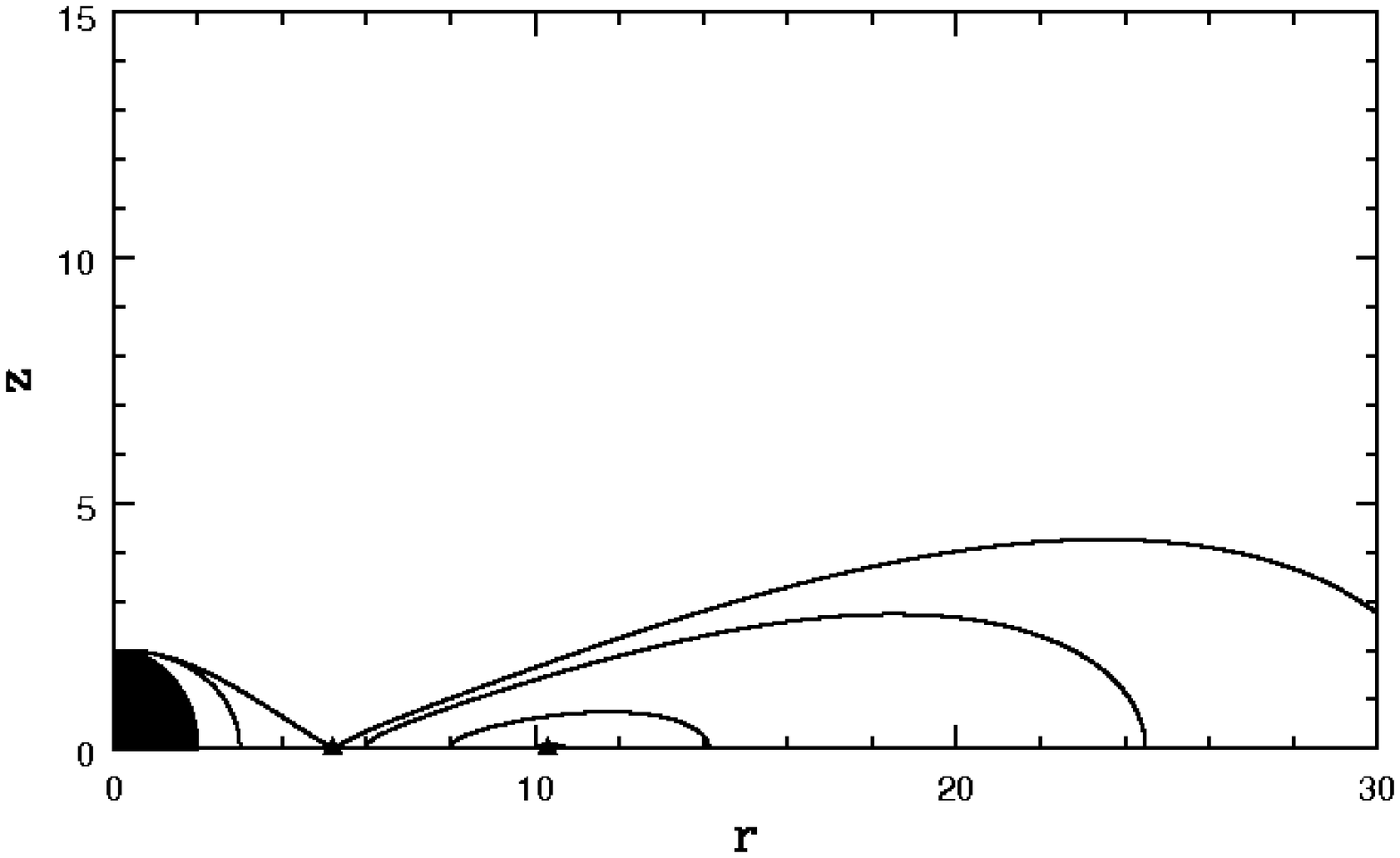}}\hspace{-.32cm}
    \subfigure{\includegraphics[width=3.85cm]{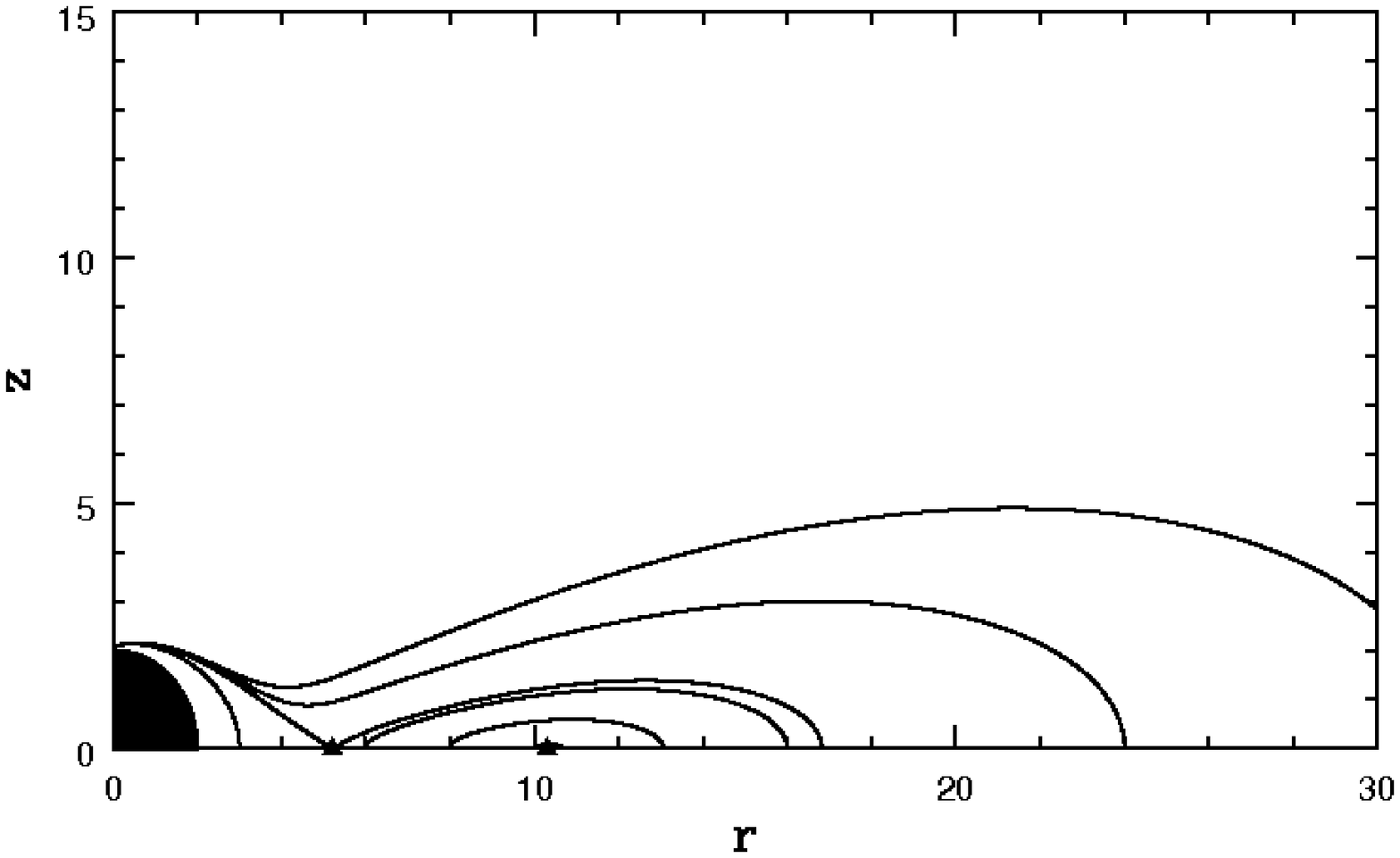}}\hspace{-.32cm}
    \subfigure{\includegraphics[width=3.85cm]{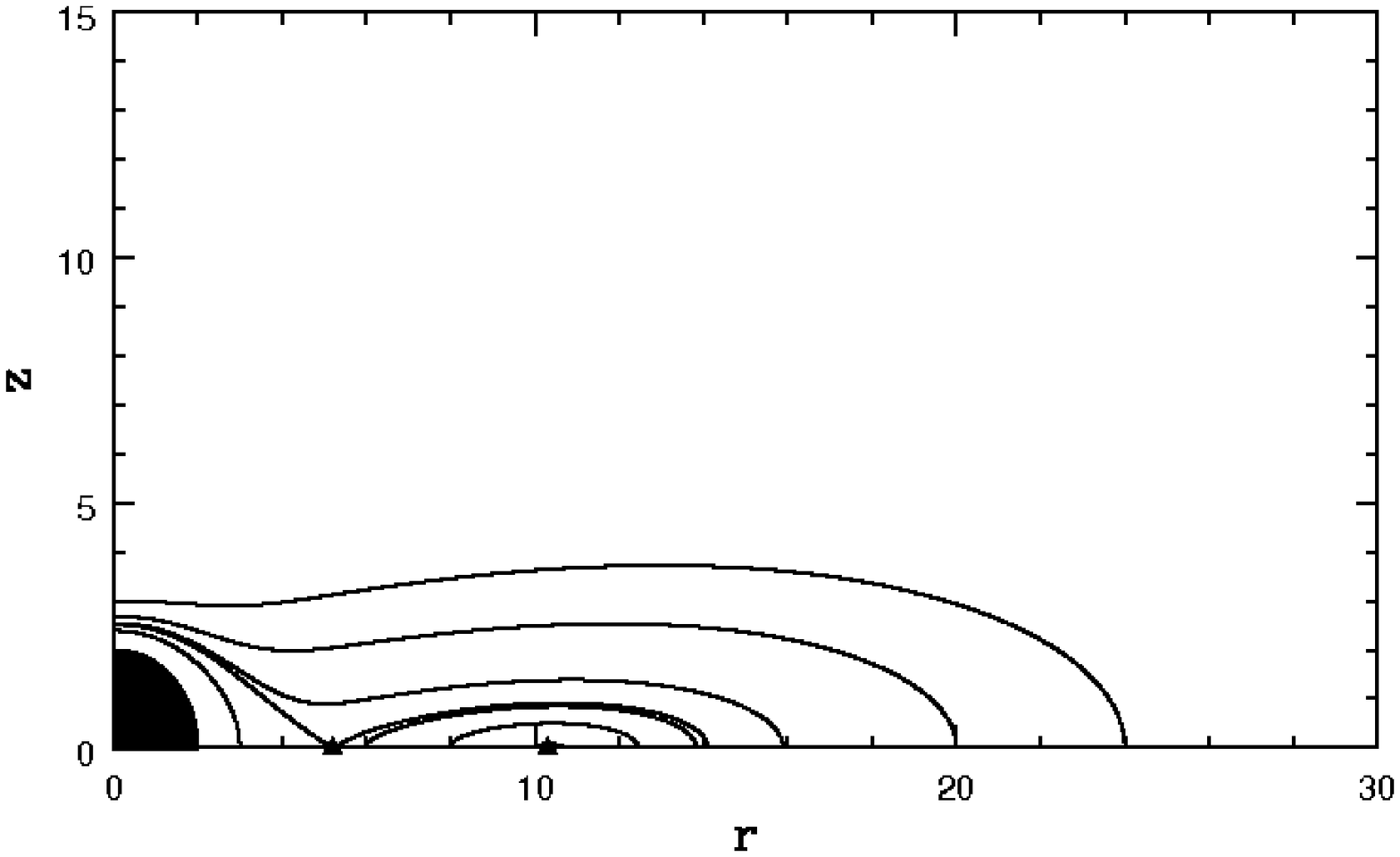}}\hspace{-.32cm}
    \subfigure{\hspace{-.3cm}\includegraphics[width=3.85cm]{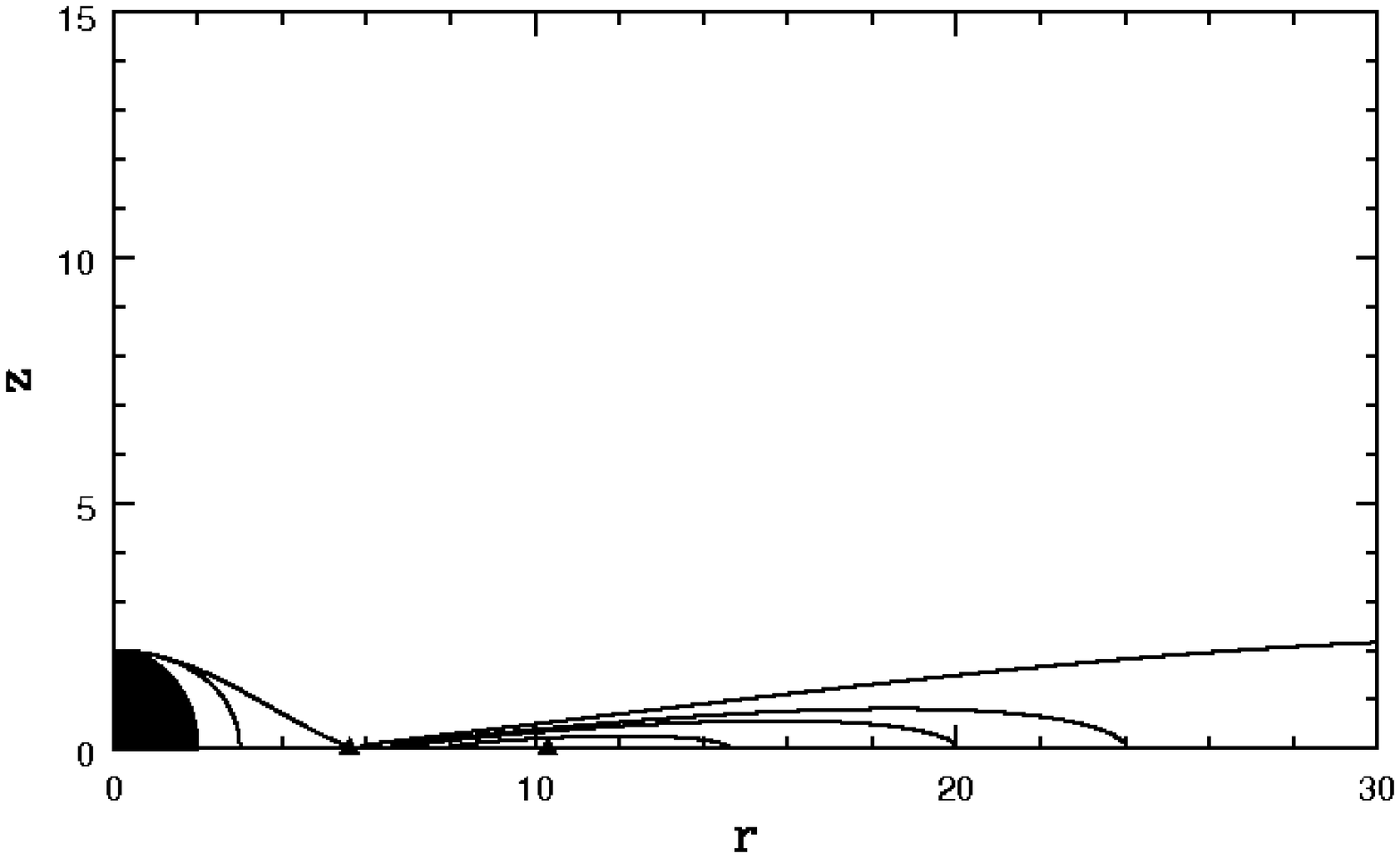}}\hspace{-.32cm}
    \subfigure{\includegraphics[width=3.85cm]{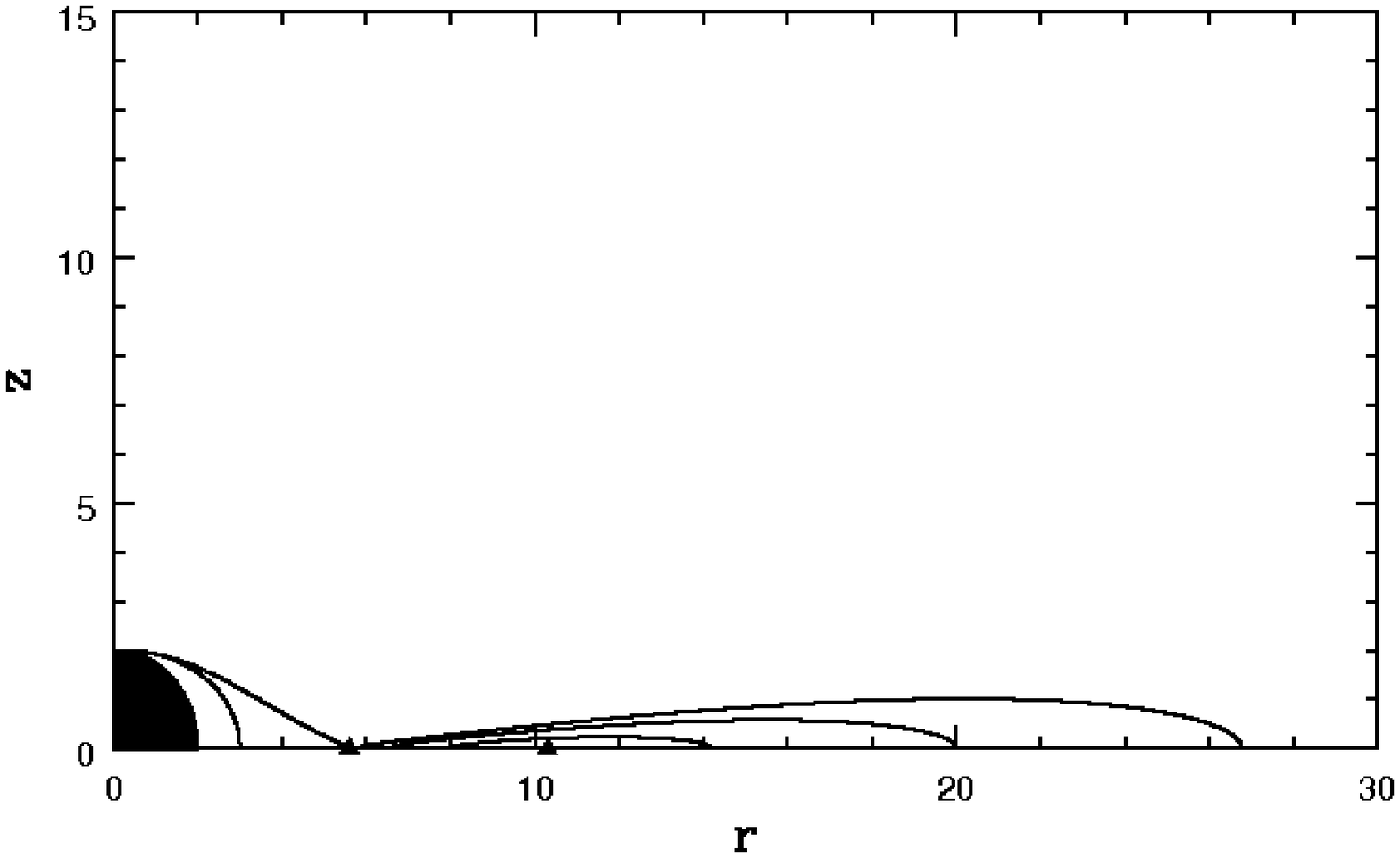}}\hspace{-.32cm}
    \subfigure{\includegraphics[width=3.85cm]{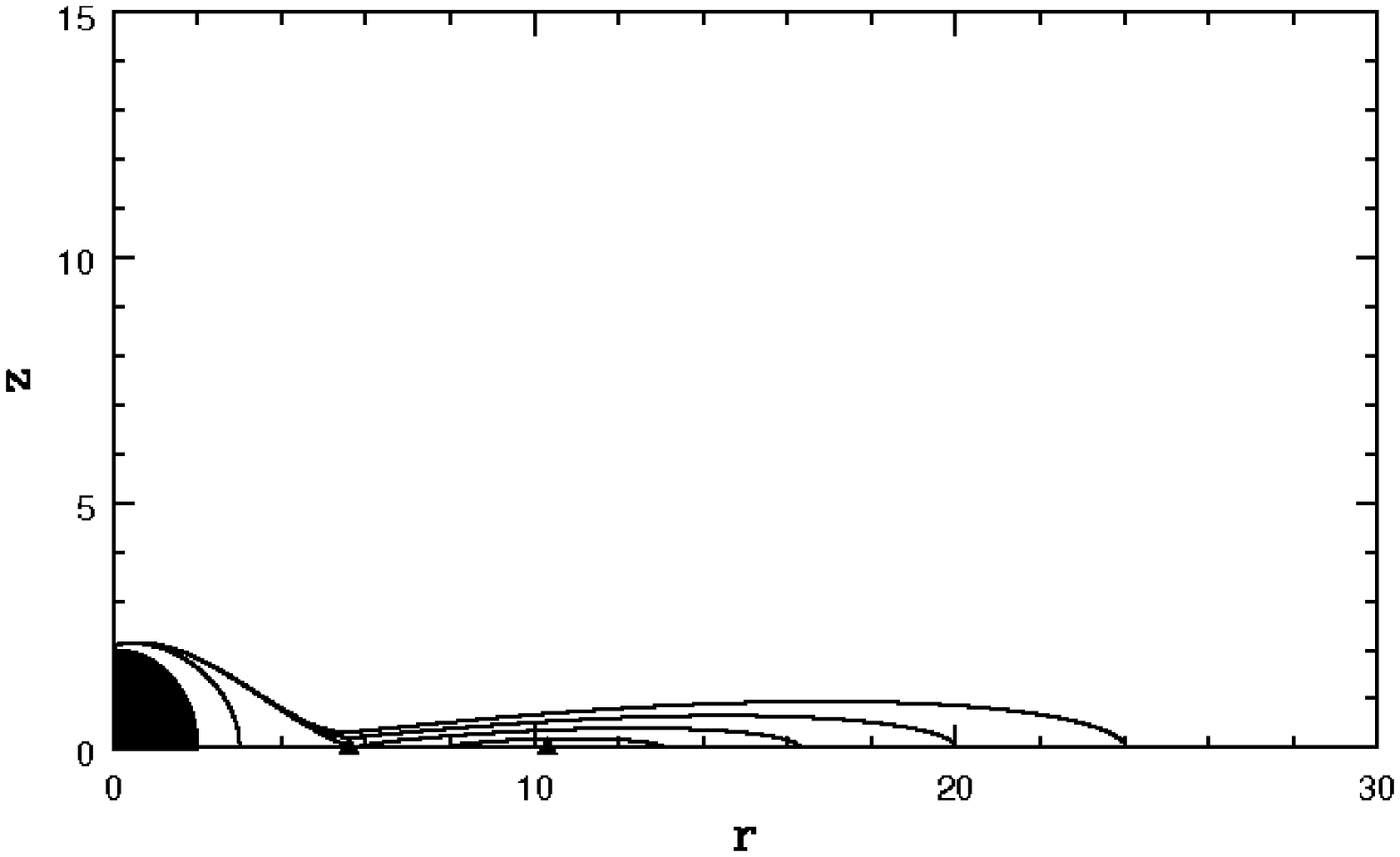}}\hspace{-.32cm}
    \subfigure{\includegraphics[width=3.85cm]{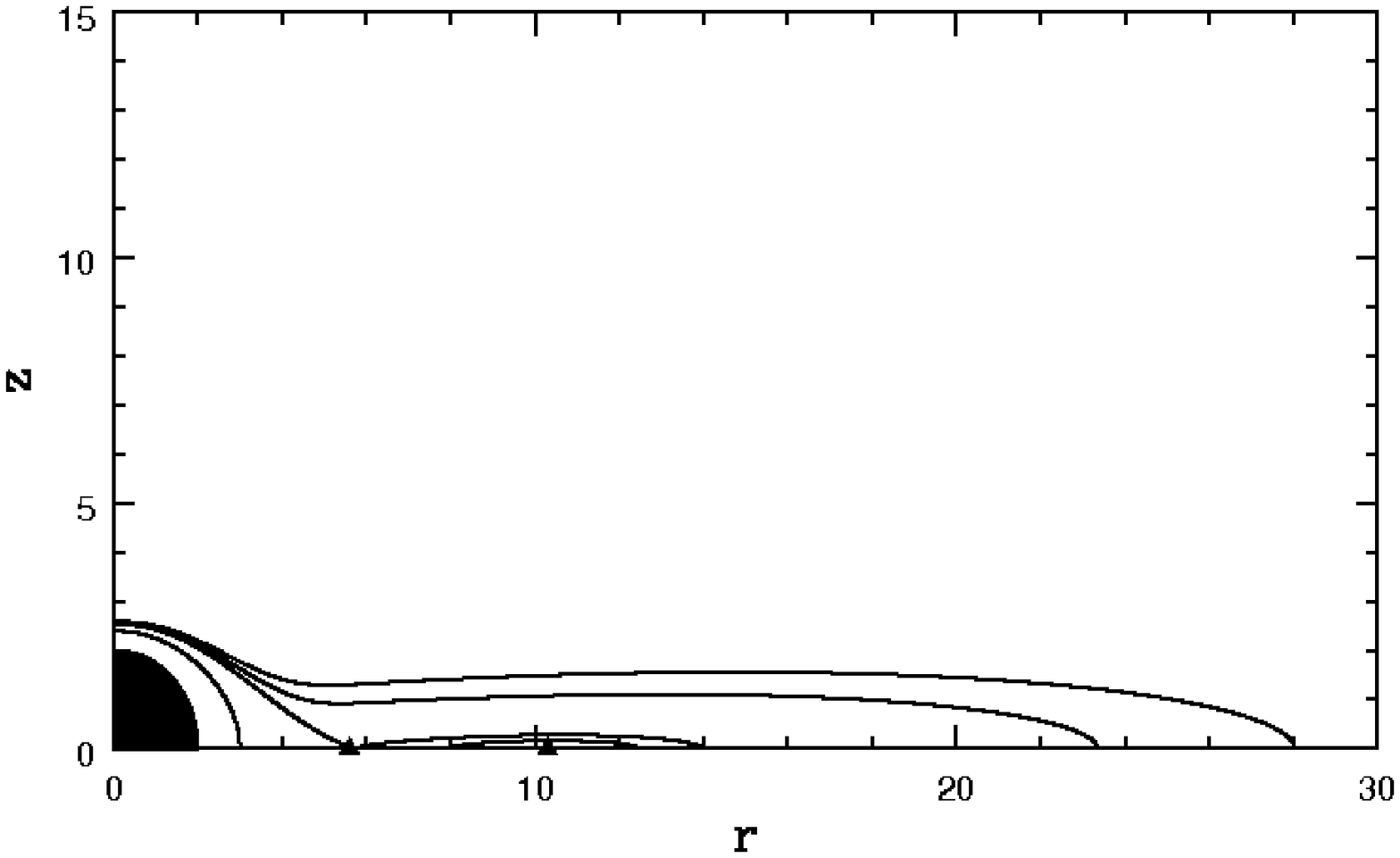}}\hspace{-.32cm}
%
   %
      \caption
      {Equipressure surfaces in (r,z) coordinates for a non-rotating BH and $\eta = 1.085$.
      Five rows correspond to $\beta = (0.0), (0.1), (0.5), (0.9), (0.99)$
      from the top to the bottom.
      Four columns correspond to $\gamma = (0.0), (0.1), (0.5), (0.9)$
      from the left to the right. The black area in the bottom left corner of each plot denotes the BH.
       Figures after \citem{qian-09}.
      }
         \label{sequence-beta-gamma}
   \end{figure*}

The four velocity of a particle on a circular orbit has the following form,
\be u^i=A\left(\eta^i+\Omega\xi^i\right),\ee
where $\eta^i$ and $\xi^i$ are the Killing vectors associated with the time and axis symmetry, respectively (Appendix~\ref{ap.killing}) and $\Omega=u^\phi/u^t$ is the angular velocity. The normalization constant $A$ comes from the $u^iu_i=-1$ condition,
\be -A^{-2}=g_{tt}+2\Omega g_{t\phi}+\Omega^2 g_{\phi\phi}.\ee
From the general equation $\nabla_i T^i_k=0$ it follows,
\begin{eqnarray}
0=\nabla_i T^i_k=\nabla_i\left(u^i u_k(p+\epsilon)+\delta^i_k p\right)=u_k\nabla_i\left((p+\epsilon)u^i\right)+u^i(p+\epsilon)\nabla_iu_k+\nabla_kp.
\end{eqnarray}
The first term vanishes as it reflects the conservation of number of particles. Thus, we have,
\begin{eqnarray}
-\frac{\nabla_kp}{p+\epsilon}=u^i\nabla_iu_k\equiv a_k.
\end{eqnarray}
The acceleration term may be expressed as follows, 
\bea
\label{e.paczki.ak}
a_k&=&Au^i\left(\nabla_i\eta_k+\Omega\nabla_i\xi_k+\xi_k\nabla_i\Omega\right)=Au^i\left(-\nabla_k\eta_i-\Omega\nabla_k\xi_i+\xi_k\nabla_i\Omega\right)=\\\nonumber
&=&-Au^i\nabla_k(\eta_i+\Omega\xi_i)+Au^i\left(\xi_i\nabla_k\Omega+\xi_k\nabla_i\Omega\right)=\\\nonumber
&=&-\frac12A^2\nabla_k\left(g_{tt}+2\Omega g_{t\phi}+\Omega^2 g_{\phi\phi}\right)+Au^i\left(\xi_i\nabla_k\Omega+\xi_k\nabla_i\Omega\right)=\\\nonumber
&=&-\frac12A^2\left(\nabla_kg_{tt}+2\Omega \nabla_kg_{t\phi}+\Omega^2\nabla_kg_{\phi\phi}\right)-\frac12A^2\left(2g_{t\phi}\nabla_k\Omega+2g_{\phi\phi}\Omega\nabla_k\Omega\right)+\\\nonumber
&+&Au^i\left(\xi_i\nabla_k\Omega+\xi_k\nabla_i\Omega\right)=-\frac12A^2\left(\nabla_kg_{tt}+2\Omega \nabla_kg_{t\phi}+\Omega^2\nabla_kg_{\phi\phi}\right)+
A^2\xi_k u^i\nabla_i\Omega.
\eea
The last term vanishes as $\Omega=\Omega(r)$. In the above derivation we have used the Killing equation ($\nabla_k\eta_i+\nabla_i\eta_k=0$). Finally, we get,
\be
\label{e.paczki0}
\frac{\nabla_k p}{p+\epsilon}=\frac12A^2\left(\nabla_kg_{tt}+2\Omega \nabla_kg_{t\phi}+\Omega^2\nabla_kg_{\phi\phi}\right).
\ee
By rewriting the $r$ and $\theta$ components of this equation and dividing them side by side we may get the following expression for 
the shape of the equipressure surfaces ($\nabla_k p = 0$),
\be
\label{e.paczki3}
\der \theta r = - \frac{\partial_r g_{tt}+2\Omega\,\partial_r g_{t\phi}+\Omega^2\,\partial_r g_{\phi\phi}}{\partial_\theta g_{tt}+2\Omega \,\partial_\theta g_{t\phi}+\Omega^2\,\partial_\theta g_{\phi\phi}}.
\ee
Knowing the angular velocity (or the specific angular momentum $\ell=\ell(r,\theta)=-u_{\phi}/u_{t}$) one can integrate the function given above and obtain the exact profiles of equipressure surfaces. \citem{qian-09} have recently made a comprehensive study of their shapes. They parametrized the angular momentum distribution by introducing the following function,
\be
\ell(r,\theta)=\left\{\begin{array}{lll}
\eta\ell_{\rm K}(r_{\rm ISCO})\left(\frac{\ell_{\rm K}(r)}{\eta \ell_{\rm K}(r_{\rm ISCO})}\right)^\beta\sin^{2\gamma}\!\theta & {\rm for} & r\ge r_{\rm ISCO},\\
\ell(r=r_{\rm ISCO})\sin^{2\gamma}\!\theta & {\rm for} & r<r_{\rm ISCO}, \\
\end{array}\right.
\label{e.paczki.ellfun}
\ee
where $r_{\rm ISCO}$ is the radius of the marginally stable orbit.
$\beta\rightarrow0$ corresponds to the limit of constant (in radius) angular momentum, while $\beta=1$ and $\eta=1$ gives the Keplerian profile ($\ell_{\rm K}(r)$) at the equatorial plane. The vertical distribution of angular momentum depends on the value of $\gamma$. Fig.~\ref{sequence-beta-gamma} presents profiles of the equipressure surfaces in the Polish doughnut model for a non-rotating BH. Rows correspond to increasing value of $\beta$ from $0$ (top) to $0.99$ (bottom). Columns show $\gamma$ ($0.0\div0.9$) dependence. The upper left panel shows a ``standard`` Polish doughnut. The lower right panel shows an almost Keplerian disk at the
      equatorial plane, surrounded by a very low angular momentum
      envelope.

\begin{figure}
  \centering
  \vskip 0.2truecm
   \includegraphics[width=8.0cm]{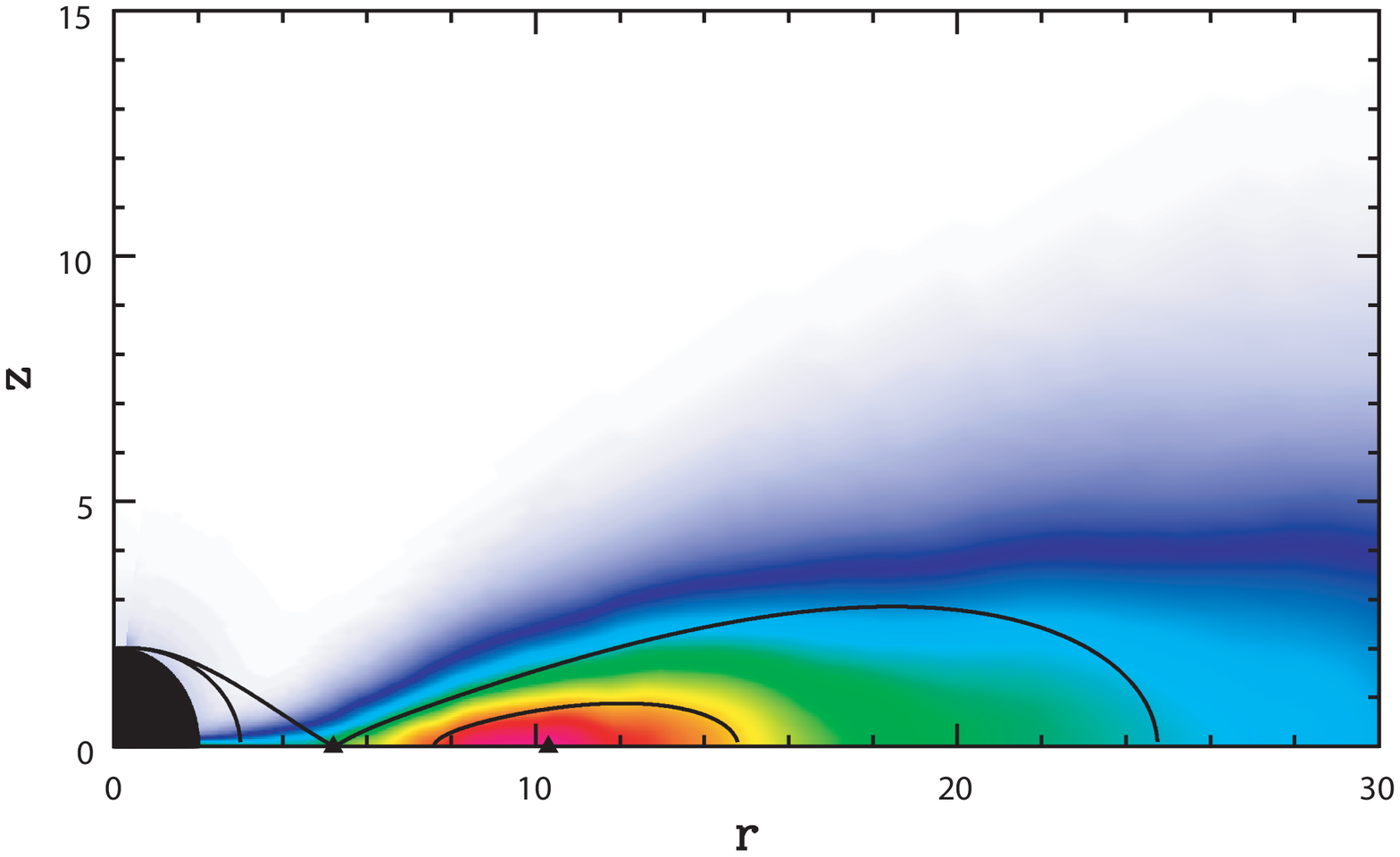}
   \vskip 0.8truecm
   \includegraphics[width=8.0cm]{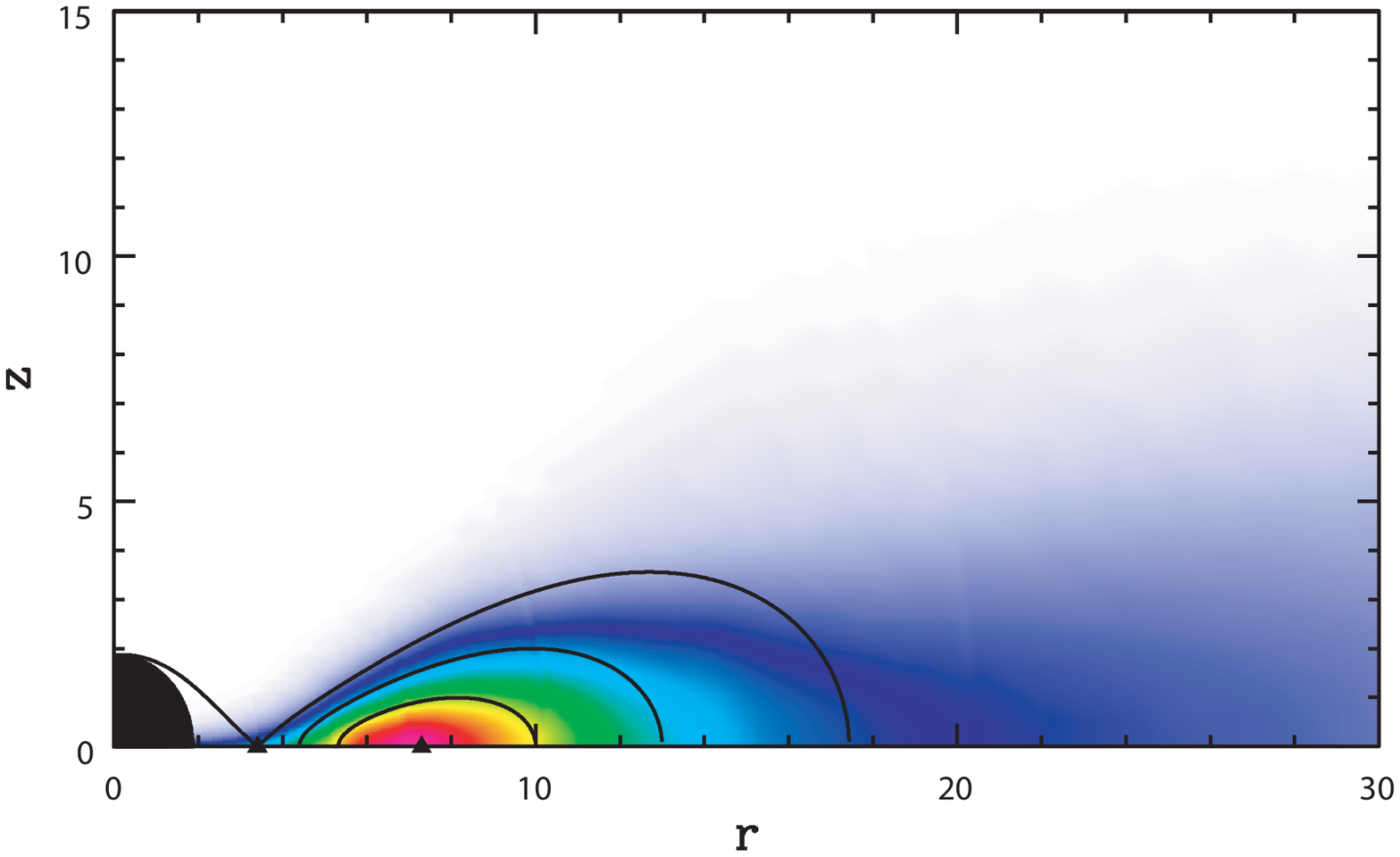}
      \caption
      {Comparison of pressure distributions between the analytic model ({\it dark lines})
      and numerical simulations ({\it colors}). The results of MHD
      simulations \citepm{fra-2007,fra-2008} have been
      time-averaged over one orbital period at $r=50 M$.
      {\it Upper panel:} Schwarzschild black hole ($a_*=0$); the analytic model
       parameters are $\eta=1.085$, $\beta=0.9$, and
       $\gamma=0.18$. {\it Lower panel:} Kerr black hole ($a_*=0.5$); the analytic model
       parameters are $\eta=1.079$, $\beta=0.7$, and
       $\gamma=0.2$. Figures after \citem{qian-09}.
      }
         \label{f.overlay}
   \end{figure}

Fig.~\ref{f.overlay} presents the comparison of equipressure surfaces profiles with results of a modern MHD simulation \citepm{fra-2007,fra-2008}. It is clear that appropriate choice of parameters describing the analytical model may reproduce most of the relevant features of the numerical results, including the locations of
the cusp and pressure maximum, as well as the vertical thickness
of the disk. The discrepancy inside the cusp (the numerical solution maintains constant vertical height while the analytical equipressure surfaces diverge) is due to significant radial velocities in this region which are not consistent with the adopted assumption of circular motion of fluid.

\section{Radiatively efficient, optically thick disks}

\subsection{Shakura \& Sunyaev (1973)}
\label{s.shakura}
When discussing the history of models describing disk accretion onto a compact object one has to mention the
outstanding, milestone work of \cyt{Shakura \& Sunyaev (1973)}. The importance of this work was mainly 
due to innovative approach to viscosity which was described in details in Section~\ref{standard-alpha-prescription}. The authors introduced several assumptions which are common for most of models describing disk accretion:
(1) the companion star in a close binary has negligible gravitational influence on the disk; (2) the self-gravitation
of the disk is not important; (3) disk is axisymmetric ($\partial_{\phi}=0$) (4) disk is 
thin ($h(r)\ll r$) (5) disk is in stationary state ($\partial_t=0$), (6) particles move around the compact 
object on Keplerian circular orbits and (7) disk is radiatively efficient (radiative cooling is the only cooling mechanism). 


The physics of disk accretion under the assumptions given above is described by four formulae: conservation of rest-mass and angular momentum, vertical equilibrium of forces and energy balance. First two have already been derived in Section~\ref{s.conslaws}:

(i) \textit{conservation of rest-mass} (Eq.~\ref{e.cons.rad2bis})
\begin{equation}
\dot{M}=-2\pi r\Sigma V,
\label{sh_cont} 
\end{equation}
where $\dot{M}$ stands for mass accretion rate (which is constant due to $\partial_t=0$ assumption), $\Sigma$ is surface
density defined as $\Sigma=\int_{-h}^h\rho\,dz$ and $V$ is the density-averaged radial velocity of matter. 

(ii) \textit{transport of angular momentum} (Eq.~\ref{e.ang2})
\begin{equation}
\frac{\dot{M}}{2\pi}({\cal L}-{\cal L}_{\rm in})=-2hr^2 t_{r\phi},
\label{sh_ang2}
\end{equation}
where, assuming that there is no torque at the inner edge of the disk, the integration constant ${\cal L}_{\rm in}$ can be 
interpreted as the angular momentum of the matter at the inner edge (${\cal L}_{\rm in}=\sqrt{GMr_{\rm in}}$ for the Newtonian potential), ${\cal L}=\Omega r^2$ is the angular momentum and $t_{r\phi}$ is the only non-negligible component of the viscous
stress tensor.

(iii) \textit{vertical 
momentum conservation} \\
According to the assumption that the disk is in the stationary state, it reduces to a simple hydrostatic equilibrium condition. Writing down the balance of the vertical component
of the gravitational force and the vertical pressure gradient we get:
\
\begin{equation}
\frac{1}{\rho}\frac{\partial p}{\partial z}=-\frac{GM}{r^2}\frac{z}{r}.
\label{sh_vert1}
\end{equation}
Following the standard approach, we replace the differentials with finite differences ($\Delta p\approx p(z=h)$ and 
$\Delta z\approx h$) obtaining the formula for disk semi-thickness $h$:

\begin{equation}
h\approx\sqrt{\frac{p}{\rho}\frac{r^3}{GM}}\approx\frac{c_s}{\Omega},
\label{sh_vert2}
\end{equation}
where $c_s$ is the sound speed in the midplane of the disk.

For a Newtonian accretion disk, the $t_{r\phi}$ component of the viscous stress tensor is given by (compare Eq.~\ref{visc.stress2}),
\begin{equation}
 t_{r\phi}=-\frac{3}{2}\rho\nu\Omega,
\label{sh_trphi1}
\end{equation}
where $\nu$ is the kinematic viscosity coefficient which has not been defined yet.
It can be shown (e.g., \citem{shapiroteukolsky}) that the heat 
is generated locally by viscosity at the rate:
\begin{equation}
 q^{\rm vis}\approx\frac{(t_{r\phi})^2}{\nu\rho}.
\label{sh_dotQ}
\end{equation}
Using Eq. \ref{sh_ang2} we can express $t_{r\phi}$ in the following way:
\begin{equation}
 t_{r\phi}=-\frac{\dot{M}}{4h\pi r^2}({\cal L}-{\cal L}_{\rm in}).
\label{sh_trphi2}
\end{equation}
Putting Eqs.~\ref{sh_trphi1} and \ref{sh_trphi2} into Eq.~\ref{sh_dotQ} we obtain the total heat generation rate
at radius $r$:
\begin{equation}
 Q^{\rm vis}=2h{q^{\rm vis}}=\frac{3\dot{M}}{4\pi r^2}\frac{GM}{r}\left(1-\sqrt{\frac{r_{\rm in}}{r}}\right).
\label{sh_heat}
\end{equation}
Assuming that the heat generated by viscosity is immediately radiated away we get the following expression for the outcoming
flux of energy from one face of the disk (top or bottom):
\begin{equation}
 F=\frac{1}{2}{Q^{\rm vis}}=\frac{3\dot{M}}{8\pi r^2}\frac{GM}{r}\left(1-\sqrt{\frac{r_{\rm in}}{r}}\right),
\label{sh_flux1}
\end{equation}
which is the \textit{energy conservation law} in the disk. It is of extreme importance to note that the expression for the
outcoming flux \textit{does not} depend on the viscosity prescription which we have not determined yet. This fact will
have major consequences in estimating BH spin basing on X-ray
spectra (see the appropriate section or \citem{shafee-06}). 

Let us now precise the treatment of radiation transport. The simplest thing one can do is to take the diffusive approximation (Section~\ref{s.radiative})
and replace differentials with finite differences (as we did when formulating the vertical equilibrium):
\begin{equation}
 F(r,z)=-\frac{16\sigma T^3}{3\kappa\rho}\frac{\partial T}{\partial z}\quad\Rightarrow\quad F=\frac{32\sigma T_C^4}{3\kappa\Sigma}
\label{sh_flux2}
\end{equation}
where $\kappa$ is the Rosseland-mean opacity coefficient and $T_C$ denotes the temperature at the equatorial plane.
In this thesis we often use the Kramer's formula:
\be
\label{e.kramers}
\kappa=\kappa_{\rm es}+\kappa_{\rm ff}=0.34+6.4\times 10^{22} \rho T^{-3.5}\,\,\,{\rm cm^2 / g}.
\ee


The total pressure of mixture of gas and radiation is given by,
\begin{equation}
\label{e.shak.pressure}
 p=NkT+\frac{1}{3}aT^4.
\end{equation}
Accretion disks are typically supported by the gas pressure with exception to the innermost regions where radiation pressure dominates. The formula given above assumes that the magnetic pressure is negligible.

To close this set of equations, the $(r,\phi)$ component of the viscous stress tensor has to be specified. \citem{shakura-73} introduced a non-dimensional viscosity parameter $\alpha$ satisfying the condition $\alpha\leq 1$ (Section~\ref{standard-alpha-prescription}), so that,
\begin{equation}
 t_{r\phi}=-\alpha p.
\label{sh_alphaP2}
\end{equation}
This expression for viscosity is called the \textit{$\alpha$ prescription} and has been widely used for many years in accretion 
disk theory.

The equations given above form a set which can be solved algebraically, e.g., for $\Sigma(r)$, $h(r)$ and $T(r)$. It turns out
that an accretion disk can be divided into three regions: (i) an outer region dominated by gas pressure and ''free-free``
absorptions, (ii) a middle region in which gas pressure and electron scatterings dominate and (iii) an inner region 
dominated by radiation pressure and electron scatterings (for low enough accretion rates, the middle and inner regions may not occur). Full formulae describing accretion disks in all three regimes may be found in the original work, as well as in, e.g., \citem{KatoBook}.

The seminal paper by \citem{shakura-73} laid the foundations of the theory of accretion disks. For the first time astrophysicists were able to
investigate their structure. The $\alpha p$ prescription, introduced by the authors, has
remained in wide use in astrophysics.

\begin{sidewaysfigure}
\centering
\includegraphics[height=6.0cm, angle=0]{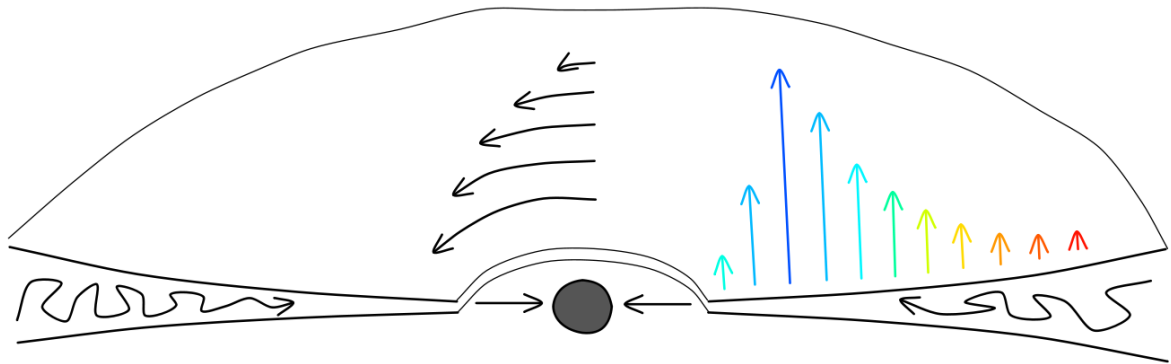}
\caption{A scheme of the standard thin disk model: Gas is accreted through a geometrically thin disk onto a BH. Viscosity results from turbulent motions. Orbits are nearly Keplerian. Disk terminates at the marginally stable orbit. Viscous heating is balanced by radiative cooling. Disk emits black body radiation with the effective temperature depending on radius.}
\label{f.thin.eps}
\end{sidewaysfigure}

In Fig.~\ref{f.thin.eps} we present a schematic picture of a radiatively efficient, geometrically thin accretion disk.

\subsection{Novikov \& Thorne (1973)}
\label{s.nt}
The obvious upgrade of the model given by \citem{shakura-73} was to write its equations in full general relativity. It was done a few months
later by \citem{nt}. They considered accretion in the equatorial plane of a BH in the framework of the Kerr spacetime metric. 
The authors introduced a convenient formalism of splitting formulae into Newtonian limits times relativistic 
corrections. However, to retain compatibility with other relativistic models we describe in this paper, we will present
\citem{nt} disk solutions using formalism proposed in
\citem{adafs}.

Throughout this paragraph we will use the Kerr metric in cylindrical coordinates
close to the equatorial plane (see Appendix \ref{ap.kerr}). We use geometrical units ($c=G=1$) and the following form of the stress-energy tensor of matter in the disk (compare Eq.~\ref{e.cons2}):
\begin{equation}
 T^{ik}=\rho u^iu^k+pg^{ik}-t^{ik}+u^kq^i+u^iq^k,
\label{stresstensor}
\end{equation}
where $\rho$ is the rest mass density, $t^{ik}$ is the viscous stress tensor and $q^i$ is the radiative energy flux. All 
physical values in this paragraph, if not stated otherwise, are averaged in the same sense as in Section~\ref{s.conslaws}. We
assume that the density of gas internal energy is much smaller than the density of gravitational binding energy.

The \textit{rest mass conservation law} requires that the covariant derivative of the product of rest mass density and four 
velocity disappears:
\begin{equation}
 \nabla_i(\rho u^i)=\frac{1}{\sqrt{-g}}(\sqrt{-g}\rho u^i)_{,i}=\frac{1}{r}(r\rho u^i)_{,r}=0.
\label{nt_cont1}
\end{equation}
Integration over $r$ introduces an integration constant which can be interpreted as the mass accretion rate:
\begin{equation}
 \dot M=-2\pi \Sigma V\Delta^{1/2},
\label{nt_cont2}
\end{equation}
where $\Sigma=\int_{-h}^{+h}\rho\,dz$ is disk surface density, $\Delta=r^2-2Mr+a^2$, and $V=\frac{r}{\Delta^{1/2}}u^r$ is now the fluid radial velocity
as measured in the local rest frame.

The general form of the \textit{angular momentum conservation law} is:
\begin{equation}
 \nabla_i(T^i_k\xi^k)=0,
\label{nt_ang1}
\end{equation}
where $\xi^k=\delta^k_{(\phi)}$ is the Kerr metric Killing vector along $\phi$ direction (Appendix~\ref{ap.killing}). After some algebra we derive:
\begin{equation}
 \frac{1}{r}\Sigma V\Delta^{1/2}\der{\cal L}{r}-\frac{1}{r}\der{}r\left(r\int_{-h}^{+h}t^r_\phi\,dz\right)+2F{\cal L}=0,
\label{nt_ang2}
\end{equation}
where ${\cal L}=u_\phi$ stands for the specific angular momentum per unit mass for circular orbit (please note that $\ell=-u_\phi/u_t$ is a constant   
of motion for perfect fluid) and $F$ is the outcoming flux of energy from one face of the disk. It can be 
shown \citepm{lasota94} that the $t^r_{\phi}$ component of the stress tensor is related to the comoving $t_{\bar r\bar\phi}$ by the 
following relation:
\begin{equation}
 t^r_\phi=\frac{A^{1/2}\Delta^{1/2}\gamma}{r^2}t_{\bar r\bar\phi}.
\label{nt_ang3}
\end{equation}
The Lorentz $\gamma$ factor is defined as: 
\be
\gamma^{-2}=1-(\Omega r)^2,
\ee
where
\be\Omega=\frac{u^\phi}{u^t}=\frac{M^{1/2}}{r^{3/2}+aM^{1/2}}\ee
is the angular velocity of the Keplerian orbits in the Kerr metric. Let us denote the integrated
shear stress $\int_{-h}^{+h}t_{\bar r\bar\phi}\,dz$ by $T_{r\phi}$. Now, the angular momentum equation takes the following form:
\begin{equation}
 \frac{1}{r}\Sigma V\Delta^{1/2}\der{\cal L}{r}-\frac{1}{r}\der{}{r}\left(\frac{1}{r}A^{1/2}\Delta^{1/2}\gamma{T_{r\phi}}\right)+2F{\cal L}=0.
\label{nt_ang4}
\end{equation}
Similarly like in the Newtonian case, one can prove the following
relation between shear and stress tensor components and the flux emerging from one side of the disk
\citepm{nt}:
\begin{equation}
 F\approx\frac{1}{2}\sigma_{\alpha\beta}^{\rm EG}\int_{-h}^{+h}t^{\alpha\beta}\,dz=\frac12\sigma_{\bar r\bar\phi}^{\rm EG}T_{r\phi},
\label{nt_ang5}
\end{equation}
where $\sigma_{\bar r\bar\phi}^{\rm EG}$  is the shear of the equatorial
geodesic orbits which is equal to \citepm{gammie}:
\begin{equation}
 \sigma_{\bar r\bar\phi}^{\rm EG}=\frac{1}{2}\frac{A\gamma^2}{r^3}\der\Omega r.
\label{nt_ang6}
\end{equation}
Putting the definition of flux from Eqs.~\ref{nt_ang5} and \ref{nt_ang6} into Eq.~\ref{nt_ang4} we finally obtain:
\begin{equation}
 \der{}r\left(\frac{A^{1/2}\Delta^{1/2}\gamma}{r}{T_{r\phi}}\right)=\frac{A\gamma^2}{r^2}\der\Omega r{\cal L}T_{r\phi}-\frac{\dot M}{2\pi}\der{\cal L}{r}.
\end{equation}
The only unknown function is ${T_{r\phi}}(r)$. Therefore, the above equation can be directly integrated. Originally, \citem{nt}
gave in their work a formula for $T_{r\phi}$ involving integrals. However, the following paper of \citem{pagethorne-74} gave explicit, algebraic
expression for this quantity. Once we know the integrated shear stress, we can use Eq. \ref{nt_ang5} to calculate outcoming flux. To resolve
radial dependencies of $V$, $\Sigma$, $T_C$ etc. we must introduce formulae for viscosity, vertical structure and opacity.

The authors followed \citem{shakura-73} and assumed
that the comoving $(r,\phi)$ component of the viscous stress tensor can be expressed as (Section~\ref{standard-alpha-prescription})
\begin{equation}
 t_{\bar r\bar\phi}=-\alpha p,
\label{nt_visc1}
\end{equation}
where $\alpha<1$ and $p$ is the total pressure. This assumption leads to,
\be
{T_{r\phi}}=-\alpha P,
\ee
where $P=\int_{-h}^{+h}p{\rm d}z$ is the vertically integrated pressure.

The \textit{vertical hydrostatic equilibrium} is constructed assuming that the pressure gradient is balanced by the 
vertical component of the gravitational force which corresponds to the vertical epicyclic frequency $\Omega_\perp$,
\begin{equation}
 \frac1\rho\frac{\partial p}{\partial z}=- \Omega_\perp^2 z=-\frac{M}{r^3}\frac{\cal H}{\cal C} z,
\label{nt_vert1}
\end{equation}
where ${\cal H}$ and ${\cal C}$ are relativistic correction factor defined in Eq.~\ref{e.relcorr}.
Once again, we replace differentials by finite differences to obtain:
\begin{equation}
 \frac{P}{\Sigma}=\frac{M}{r^3}\frac{\cal H}{\cal C}h^2.
\label{nt_vert2}
\end{equation}

To close this set of equations we need to describe the \textit{energy balance}. Similarly like in the previous section we assume
that radiative transport is the only cooling mechanism. Introducing finite differences into the diffusive approximation we get:
\begin{equation}
F=\frac{32\sigma T_C^4}{3\kappa\Sigma},
\label{nt_rad1}
\end{equation}
where $\tau$ is the total optical depth.

Taking all the equations given above together we get a set which can be solved in algebraic way, like in the Newtonian case. Similarly,
we can distinguish three regions in the accretion disk depending on the dominating pressure and opacity sources. Explicit expressions
for outgoing flux, surface density, disk semi-thickness, central temperature etc. are given in \citem{nt} with additional formulae in \citem{pagethorne-74}.

\chapter{Relativistic, stationary slim accretion disks}
\label{chapter-stationary}

Standard models of thin disks \citepm{shakura-73,nt} 
assume that all the heat generated at a given
radius by viscosity is immediately radiated away,
meaning that the thermal equilibrium is determined
by the local balance of viscous heating and
radiative cooling, $Q^{\rm vis} = Q^{\rm rad}$.
No other cooling mechanisms are considered
in the standard model. This
assumption is fairly consistent and correct
only for sufficiency small accretion rates.
For high accretion rates
it fails as the simple
argument below shows.

The ``advective'' radial flux of heat, $Q^{\rm adv}$,
is a consequence of the radial motion of hot
matter. It can be estimated as \citepm{KatoBook},
\begin{equation}
Q^{\rm adv} \approx - \frac{1}{r} V P = \frac {\dot M}{2\pi r^2}\frac{P}\Sigma.
\end{equation}
Assuming $P=P_{\rm rad}=2H\frac{4\sigma} {3} T_C^4$ and taking the rate of radiative cooling, $Q^{\rm rad}=\frac{64\sigma T_C^4}{3 \kappa_{\rm es}\Sigma}$, we may estimate this expression in the following way,
\be
Q^{\rm adv} \approx \frac {\dot M}{2\pi r^2}\frac{P}\Sigma\times Q^{\rm rad}\left/\frac{64\sigma T_C^4}{3 \kappa_{\rm es}\Sigma}\right. = \frac{ \dot M \kappa_{\rm es} h}{16\pi r^2}Q^{\rm rad}.
\ee
It is then obvious, that for $\dot M \gtrsim \frac{16\pi r^2}{\kappa_{\rm es}h}$ ($h$ increases with $\dot M$), advective cooling predicted by the \citem{shakura-73}
model is larger than the radiative cooling $Q^{\rm rad}$,
and that the standard thin accretion
model cannot be adequate at such accretion rates. Detailed calculations show that advection starts to modify the disk structure at accretion rates corresponding to $L\approx0.6L_{\rm Edd}$ (Section~\ref{s.inneredges}).

\citem{slim} constructed optically
thick ``slim`` accretion disk model which accounts for the advective cooling and does not impose the Keplerian angular momentum. Therefore, it may be applied to disks at any accretion rate.

Advection
appears also at much smaller (sub-Eddington)
accretion rates, but for optically very thin
disks. These disks, today known as advection-dominated accretion flows (ADAFs), were propheted
in a paper by \citem{ichimaru-77} (see also \citem{reesetal-82}).
Only after their rediscovery in the mid 1990s by
\citem{narayanyi-94,narayanyi-95} and by \citem{abramowiczetal-95}, ADAFs started to be intensely
studied by many authors (see reviews in \citem{lasota-99a, lasota-99b} and \citem{narayanmcclintock-08}).

In this doctoral thesis, we study optically thick
slim disks only, and this Chapter is devoted to the
discussion of their relativistic, stationary model.

\section{Equations}
\label{s.stateq}

\subsection*{Assumptions}

We assume an axisymmetric, stationary fluid configuration in the
Kerr metric with fixed values of BH mass $M$ and BH spin $a$ parameters\footnote{
$a$ is the angular momentum per unit mass, $a=J/M$, while $a_*=a/M=J/M^2$ is called the dimensionless spin parameter.}.
The disk is symmetric  under reflection in the equatorial plane. Matter is supplied at a steady rate, $\dot M$, through a
boundary ``at infinity'' and angular momentum is removed through the
same boundary, whereas zero torque is assumed at the BH
horizon. We include no self-irradiation of the disk, and we neglect
the magnetic pressure, as well as the angular
momentum carried away by radiation.  Dissipation and angular momentum transport
are given by the $\alpha$ prescription
(Section~\ref{standard-alpha-prescription}), with a constant value of $\alpha$ in radius. We assume that
the dissipation rate is proportional to the total pressure, 
$p=p_{\rm gas}+p_{\rm rad}$.

We take $G=c=1$ and put $r$ for the
cylindrical radius.  We also define the following
expressions involving the BH spin:
\begin{eqnarray}
\label{e.relcorr}
\Delta&=&M^2(r_*^{2}-2r_*+a_*^2),\\\nonumber
A&=&M^4(r_*^{4}+r_*^2a_*^2+2r_*a_*^2),\\\nonumber
\cal C&=&1-3r_*^{-1}+2a_*r_*^{-3/2},\\\nonumber
\cal D&=&1-2r_*^{-1}+2a_*^2r_*^{-2},\\\nonumber
\cal H&=&1-4a_*r_*^{-3/2}+3a_*^{2}r_*^{-2}, \nonumber
\end{eqnarray}
with $a_*=a/M$ and $r_*=r/M$.
The radial gas velocity, $V$, as measured by the observer co-rotating with the
fluid at a fixed value of $r$, is given by the relation \citepm{adafs},
\be
\label{e.vasmeasured}
V/\sqrt{1-V^2}=u^r g_{rr}^{1/2}=\frac{r u^r}{\Delta^{1/2}}.
\ee

We adopt an equation of state corresponding to the choice
$p_{\rm gas}=k\rho T/(\mu m_{\rm p})$, and $p_{\rm rad}= aT^4/3$,
with $k$ the Boltzmann constant, $m_{\rm p}$ the proton mass,
and $a$ the radiation constant
(no confusion with the spin parameter may arise). The mean molecular weight is
taken to be $\mu=0.62$ corresponding to $X=0.7$ and $Y=0.3$.

Only for the purpose of integrating the vertical structure, we assume the polytropic equation of state with the polytropic
index $N$: $p=K\rho^{1+1/N}$ ($N=3$ and $K=\rm const$ correspond to a mixture of gas and radiation with constant pressure ratio). The vertical integration of the hydrostatic 
equilibrium formula (Eq.~\ref{nt_vert1}), gives
\be
\rho=\rho_0\left(1-\frac{z^2}{H^2}\right)^N.
\label{rho.polytropic}
\ee
From the previous equation and the polytropic equation of state, it follows
\be
T=T_0\left(1-\frac{z^2}{H^2}\right)\label{T.polytropic},
\ee
where $\rho_0$ and $T_0$ denote density and temperature at the equatorial plane, respectively.

\subsection*{Equations}
\label{s.sslim.eq}

In this section we present equations describing relativistic, stationary slim disks.
They were derived originally by \citem{lasota94} and improved 
by several other authors including \citem{adafs}, \citem{gammie} and \citem{KatoBook}. 
More general derivation for the non-stationary case is given in Appendix~\ref{ap.nonstat}.

(i) The mass conservation:
\begin{equation}
 \dot M=-2\pi \Sigma\Delta^{1/2}\frac{V}{\sqrt{1-V^2}},
\label{eq_poly_cont2}
\end{equation}
where $\Sigma=\int_{-h}^{+h}\rho\,dz$ is disk surface density.

(ii) The radial momentum conservation:
\begin{equation}
\frac{V}{1-V^2}\frac{dV}{dr}=\frac{\cal A}{r}-\frac{1}{\Sigma}\frac{dP}{dr},
\label{eq_poly_rad3}
\end{equation}
where
\begin{equation}
{\cal
A}=-\frac{MA}{R^3\Delta\Omega_k^+\Omega_k^-}\frac{
(\Omega-\Omega_k^+)(\Omega-\Omega_k^-)}{1-\tilde\Omega^2\tilde R^2}
\label{eq_rad4}
\end{equation}
and $\Omega=u^\phi /u^t$ is the angular velocity with respect to the stationary
observer, $\tilde\Omega=\Omega-\frac{2Mar}{A}$ is the angular velocity with respect to
the inertial observer, $\Omega_k^\pm=\pm M^{1/2}/(r^{3/2}\pm aM^{1/2})$ are the
angular frequencies of the co-rotating and counter-rotating Keplerian orbits,
$\tilde R=A/(r^2\Delta^{1/2})$ and $P=\int_{-h}^{+h}p\,dz$ is vertically integrated pressure.

(iii) The angular momentum conservation:
\begin{equation}
 \frac{\dot{M}}{2\pi}({\cal L}-{\cal
L}_{\rm in})=\frac{A^{1/2}\Delta^{1/2}\gamma}{r}\alpha P,
\label{eq_ang6}
\end{equation}
where ${\cal L}=u_\phi$, ${\cal L}_{\rm in}$ is the angular momentum at the disk
inner edge, and
\be
\gamma=\left(\frac1{1-V^2}+\frac{{\cal L}^2r^2}{A}\right)^{1/2}
\ee is the Lorentz factor.

(iv) The vertical equilibrium:

\noindent Integration of Eq.~\ref{nt_vert1}, taking into account Eq.~\ref{rho.polytropic}, leads to
\begin{equation}
\label{eq_vertbalance}
h^2\Omega_\perp^2=(2N+3)\frac P\Sigma,
\end{equation}
where \be
\label{eq.omegatilde}
\Omega_\perp=
\sqrt{\frac{M}{r^3}\frac{\cal H}{\cal C}}
\ee is the vertical epicyclic frequency and $h$ corresponds to the disk thickness. Herein, we always assume $N=3$.

(v) The energy conservation: 

\noindent The advective cooling is defined, in terms of the vertically integrated quantities, as \citepm{KatoBook}
\be
Q^{\rm \rm adv}=\frac1r\der{}{r}(r u^r(E+P))-u^r\der Pr-\int_{-h}^{h}u^z\pder pz\,dz,
\label{eq.advcooling}
\ee
where $E$ is the height-integrated energy density,
\be
E=\int_{-h}^{h}\left(\frac{p_{\rm gas}}{5/3-1}+3p_{\rm rad}\right)\,dz.
\ee

Using the mass conservation (Eq. \ref{eq_poly_cont2}), hydrostatic equilibrium (Eq.~\ref{nt_vert1}), and the following derivation,
\begin{eqnarray}\nonumber
&&-\int_{-h}^{h}u^z\pder pz\,dz=\Omega^2_\perp\int_{-h}^{h}u^z\rho z\,dz=\\\nonumber
&&=-\frac12\Omega^2_\perp\int_{-h}^{h}\pder{}z(u^z\rho) z^2\,dz=\\\nonumber
&&=\frac1{2}\Omega^2_\perp\int_{-h}^{h}\frac1r\pder{}r(r\rho u^r) z^2\,dz\\\nonumber
&&=-\frac{\dot M}{2\pi r}\Omega^2_\perp\der{}r\left(\frac1\Sigma\int_{0}^{h}\rho z^2\,dz\right)=\\\nonumber
&&=-\frac{\dot M}{2\pi r}\Omega^2_\perp\der{\eta_4}r,
\end{eqnarray}
$Q^{\rm adv}$ may be expressed as
\begin{eqnarray}\nonumber
\label{eq.qadv}
Q^{\rm \rm adv}&=&-\frac{\mdot}{2\pi r^2}\left(\eta_3\frac{P}\Sigma\derln{P}{r} - (1+\eta_3)\frac{P}\Sigma\derln\Sigma r+\right.\\
&+&\left. \eta_3\frac P\Sigma\derln{\eta_3}{r}+\Omega^2_\perp\eta_4\derln{\eta_4}{r}\right).
\end{eqnarray}
Formulae for coefficients $\eta_3$ and $\eta_4$ are given below.

The amount of heat advected $Q^{\rm adv}$ is equal to the difference between viscous heating and radiative cooling:
\be
\label{e.radial.Qadv}
Q^{\rm adv}=Q^{\rm vis}-Q^{\rm rad}=-\alpha P\frac{A\gamma^2}{r^3}\der\Omega r - \frac{64\sigma T_C^4}{3\Sigma\kappa},
\ee
where $T_C$ is the temperature at the equatorial plane.
The opacity coefficient $\kappa$ is calculated using the Kramer's approximation (Eq.~\ref{e.kramers}). 

(vi) The equation of state (in terms of vertically integrated quantities)
\be
P=\eta_2\frac{k}{\mu m_{\rm p}}\Sigma T_C+\frac23\eta_1aT_C^4.
\ee
The radial derivative of the height-integrated pressure takes the form
\begin{eqnarray}
\nonumber
\label{e.dPdr}
\der{\ln P}r&=&(4-3\beta)\der{\ln T_C}r+\beta\der{\ln\Sigma}r+\\
&+&(1-\beta)\der{\ln \eta_1}r+\beta\der{\ln \eta_2}r,
\end{eqnarray}
with $\beta=\eta_2 (k/\mu m_{\rm p})\Sigma T_c/P$.

The coefficients $\eta_1$ to $\eta_4$ are given by, and equal to (assuming Eqs.~\ref{rho.polytropic} and \ref{T.polytropic}),
\begin{eqnarray}\label{def.eta1}
\eta_1&\equiv&\frac1{T_0^4}\int_0^hT^4\,dz=I_4 h,\\
\eta_2&\equiv&\frac2{\Sigma T_0}\int_0^h\rho T\,dz=I_{N+1}/I_N,\\
\eta_3&\equiv&\frac1P\left(\frac1{\gamma-1}\frac{k}\mu\frac{I_{N+1}}{I_N}\Sigma T_C+2I_4aT^4_Ch\right),\\\label{def.eta4}
\eta_4&\equiv&\frac1{\Sigma}\int_0^h\rho z^2\,dz=J_N h^2,
\end{eqnarray}
where\footnote{$I_4=\frac{128}{315}$; $\quad I_4/I_3=\frac{8}{9}$; $\quad J_3=\frac1{18}$.}
\begin{equation}
I_N=\frac{\sqrt{\pi}}2\frac{\Gamma(1+N)}{\Gamma(3/2+N)}\overset{N\in\mathbb{N}}{=}\frac{(2^NN!)^2}{(2N+1)!},\\
\label{e.IN}
\end{equation}
\be
J_N=\frac14\frac{\Gamma(3/2+N)}{\Gamma(5/2+N)}=\frac1{6+4N}.
\ee

\section{Numerical methods}
\label{sect.numerical}

The equations given above form a two-dimensional set of ordinary differential
equations with a critical (sonic) point. One may choose e.g., $V$ and $T_C$ as the dependent variables.
By a series of algebraic manipulations of 
Eqs.~\ref{eq_poly_cont2}, \ref{eq_poly_rad3}, \ref{eq_ang6},  \ref{eq.qadv},
and \ref{e.dPdr},
 we obtain the following set of ordinary differential equations for $V(r)$ and $T_C(r)$\footnote{For clarity and compatibility with Chapter~\ref{chapter-vertical}, we leave the radial derivatives of $\eta_1$ to $\eta_4$ in Eqs.~\ref{eq_derT} and \ref{e.NN}. They are reduced to the derivatives of $V$ and $T_C$, but the ultimate formulae are long and unreadable.}:
\be
\frac{1}{1-V^2}\der{\ln V}{\ln r}=\frac{\cal N}{\cal D},
\label{eq_derV}
\ee
\begin{eqnarray}\nonumber
(4-3\beta)\der{\ln T_C}{\ln r}=\left(\frac{\cal N}{\cal D}+\frac{r(r-M)}\Delta\right)(\beta-\tilde\Gamma_1)&&\\\nonumber
-\frac{2\pi r^2}{\dot M\eta_3}\frac\Sigma P Q^{\rm adv}-(1-\beta)\derln{\eta_1}r-\beta\derln{\eta_2}r&&\\
-\derln{\eta_3}r-\frac{\Omega_\perp^2\Sigma}{P}\frac{\eta_4}{\eta_3}\derln{\eta_4}r,
\label{eq_derT}
\end{eqnarray}
with $\cal N$, $\cal D$ and $\tilde \Gamma_1$ given by
\begin{eqnarray}\nonumber\label{e.NN}
{\cal N}&=&{\cal A}+\frac{2\pi r^2}{\dot M\eta_3}Q^{\rm adv}+
\frac{P}\Sigma
\left(\frac{r(r-M)}\Delta\tilde\Gamma_1+\right.\\
&+&\left.\derln{\eta_3}r\right)+\Omega^2_\perp\frac{\eta_4}{\eta_3}\derln{\eta_4}r,\\
{\cal D}&=&V^2-\tilde\Gamma_1\frac P\Sigma,\\
\tilde\Gamma_1&=&1+\frac 1{\eta_3}.
\end{eqnarray}

To obtain the regular solution one has to solve these equations together with outer boundary conditions given at some large radius $r_{\rm out}$ and the following regularity conditions at the sonic radius $r_{\rm son}$:
\be
\label{num_regcond.0}
\left.{\cal N}\right|_{r=r_{\rm son}}=\left.{\cal D}\right|_{r=r_{\rm son}}=0
\ee
which provides additional condition for calculating ${\cal L}_{\rm in}$ which is the eigenvalue of the problem. The location of the sonic point is not known \textit{a priori}. A mathematical problem defined in such a way can be classified as a \textit{two point boundary value problem} which we solve applying the relaxation technique \citepm{numericalrecipes}. To start the relaxation process one has to provide a \textit{trial solution}. The convergence strictly depends on the quality of such initial guess. In this work we apply the following method of searching for the proper initial conditions for relaxation.

The angular momentum at the inner edge of the disk, ${\cal L}_{\rm in}$, is the eigenvalue of the problem and must be chosen properly to satisfy the regularity conditions given by Eq.~\ref{num_regcond.0}. The value of ${\cal L}_{\rm in}$ determines the shape of a solution. It turns out that the topology of slim disk solutions with ${\cal L}_{\rm in}$ higher and ${\cal L}_{\rm in}$ lower than the proper value ${\cal L}_{\rm in,0}$ is different. The former branch of solutions terminates as soon as the denominator $\cal D$ vanishes (the regularity condition is not satisfied). The latter does not cross the sonic point at all ($\cal D$ is always positive). The self-consistent solution is expected to be the common limit of these two branches. A trial solution can be achieved in a few iteration steps. We start integration at $r>1000M$ assuming the \citem{nt} boundary conditions and we use the implicit Runge-Kutta method of the 4th order.

\begin{figure}
\centering \includegraphics[width=.65\textwidth]{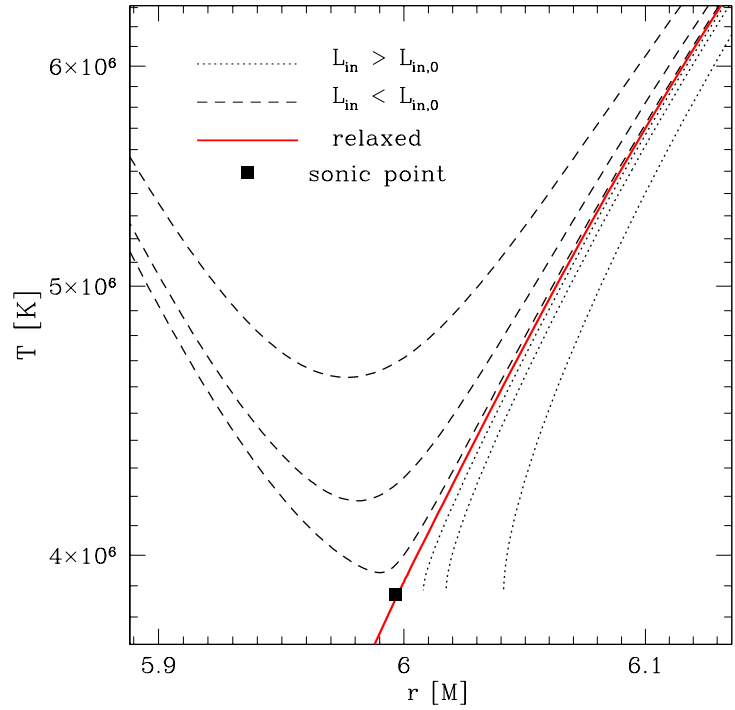}
\caption{
Temperature profiles in the vicinity of the sonic point for a few iterations leading to the trial model used as initial condition in the relaxation procedure (for details see Section~\ref{sect.numerical}). Solutions with too high value of ${\cal L}_{\rm in}$ terminate before reaching the proper sonic point while solutions with too low ${\cal L}_{\rm in}$ go through the sonic point radius but follow an improper branch. The relaxed solution and the location of the sonic point are also presented. Models calculated for a non-spinning BH with $\Mdot=0.1\Medd$.
}
\label{f.mtrial}
\end{figure}

Once the trial solution is found one can start relaxation process between $r_{\rm out}$ and the estimated position of the sonic radius $r_{\rm son}$. The standard approach has to be modified due to the fact that we expect singularity at the inner boundary. Thus, we treat the problem as a free boundary problem and introduce one more variable describing the position of the critical point \citepm{numericalrecipes}. In this work we usually use 100 mesh points spaced logarithmically in radius between the sonic point and the outer boundary.

A relaxed solution is obtained in a few iteration steps and is then used as the initial condition for relaxation when looking for the solution of a problem with one parameter (e.g., mass accretion rate $\Mdot$ or BH spin $a$) slightly modified. For the new system parameters we look for the outer boundary conditions at $r_{\rm out}$ by integrating the equations from $1.1r_{\rm out}$ (assuming the same outer boundary conditions as before) until we reach $r_{\rm out}$. In such a way a full spectrum of system parameters can be achieved effectively.

Once the solution outside the sonic point is found we numerically estimate the radial derivatives of $V$ and $T_C$ at the sonic point using values given at $r>r_{\rm son}$. Taking them into account we make a small step from the innermost mesh point (which corresponds to the sonic point) inward. Then we start integrating using standard Runge-Kutta method until we get close enough to the horizon.

\section{Solutions}
\label{s.stationarysolutions}
\subsection*{Definitions}

We introduce the following critical (Eddington) luminosity,
\be
    \Ledd = \frac{4 \pi \speedoflight G M}{\kappaES} = 1.25\EE{38}
    \frac{M}{\Msun} {\;\rm erg\;s^{-1}}\;,
    \label{eq:lum-edd}
\ee
and mass accretion rate\footnote{Many authors use a different definition, ${\dot M}_{\rm Edd} =
L_{\rm Edd}/c^2$.},
\be
    \Medd = 16\times\frac\Ledd{c^2}=16\times\frac{4 \pi M G}{\speedoflight\,\kappaES} = 2.23\EE{18}
    \frac{M}{\Msun} {\;\rm g\;s^{-1}}\;,
    \label{eq:mdot-edd}
\ee
where $\kappaES\!=\!0.20(1+X)\,{\rm cm^2/g}$ is the electron scattering opacity 
and here we take $X\!=\!1$ being the hydrogen mass fraction. We also define 
the efficiency of accretion as
\be
    \eta = 16\times\frac{L}{\Mdot\,c^2} = \frac{L/\Ledd}{\Mdot/\Medd}\;.
    \label{eq:efficiency}
\ee
The factor $16$ in Eqs.~\ref{eq:mdot-edd} and \ref{eq:efficiency} resembles the fact that the efficiency 
of accretion in the Pseudo-Newtonian potential for a thin, Shakura-Sunyaev disk, is
$1-u_{t,\rm in}=1/16$ (compare Section~\ref{s.efficiency}). In such a way we may consider $\Medd$ as an accretion rate for 
which, in case of a non-rotating BH only, the disk luminosity 
is close to $\Ledd$. We also introduce a dimensionless accretion rate
$\dot m$,
\be
\label{e.mdotcritical}
\dot m\equiv\frac {\Mdot}{\Mdot_{\rm Edd}}.
\ee

\subsection*{Disk appearance}
\label{s.diskappearance}

In this paragraph we discuss most important properties of slim disk solutions for the purpose of spectra modeling.

\begin{figure}
\centering
  \includegraphics[width=.7\textwidth,angle=0]{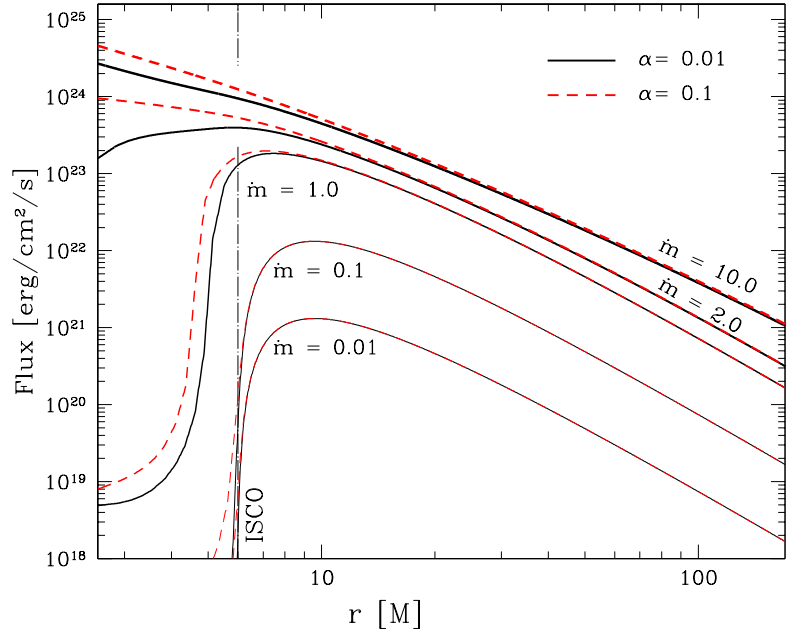}
\caption{Flux profiles for different mass accretion rates in case of a non-rotating BH
and two values of $\alpha$: $0.01$ (black solid), $0.1$ (red dashed lines).
For each value five lines for the following mass accretion rates: 
$0.01$, $0.1$, $1.0$, $2.0$ and $10.0 \dot M_{\rm Edd}$ 
are presented. BH mass is $10\Msun$.}
  \label{f.flux}
\end{figure}

In Fig.~\ref{f.flux} we present profiles of the emitted
flux for a Schwarzschild BH and a range of accretion rates. Profiles
for two values of the $\alpha$ parameter are shown: $0.01$ (solid
black) and $0.1$ (red dashed lines).
In the limit of low accretion rates the flux profiles coincide
with the Novikov \& Thorne solution - no flux emerges from inside
the marginally stable orbit. The higher the accretion rate, 
the higher departure from the standard profile as the advection
of heat becomes more important. For accretion rates exceeding $\Mdot_{\rm Edd}$
the disk emission is significantly shifted inward.
The emission 
profile in the inner disk regions for super-Eddington accretion rates
has a different slope ($F\propto r^{-0.5}$)
instead of $\propto r^{-0.75}$ for radiatively efficient
disks. 
Such behavior is valid for all
values of the viscosity parameter $\alpha$. However, the higher
the value
of $\alpha$, the earlier advection affects the disk emission.

\begin{figure}
\centering
  \includegraphics[width=.7\textwidth,angle=0]{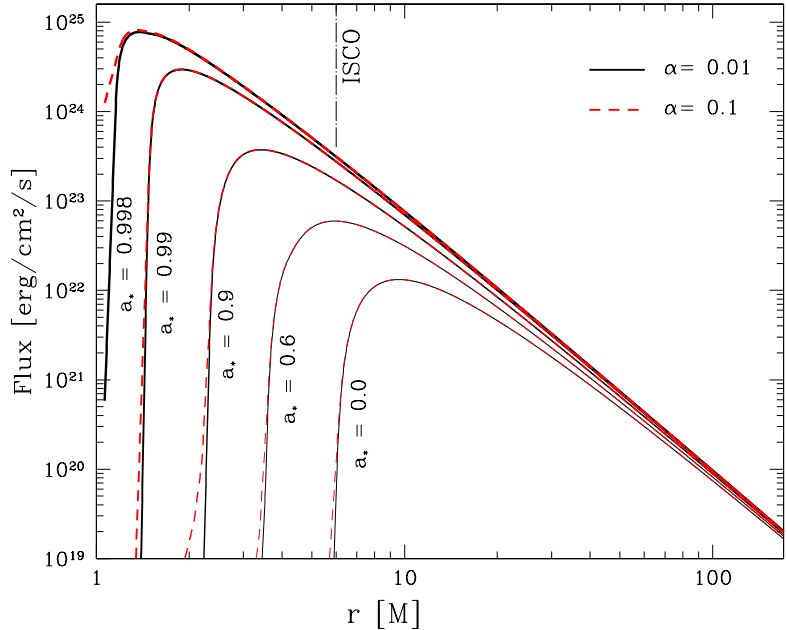}
\caption{Flux profiles for $\dot m=0.1$  and different values of BH spin.
Results for two values of $\alpha$: $0.01$ (black solid), $0.1$ (red dashed lines) are presented. BH mass is $10\Msun$.}
  \label{f.flux.aN.rad}
\end{figure}

Fig.~\ref{f.flux.aN.rad} presents profiles of emission from a disk with $\dot m=0.1$ for a few values of BH spin. They all coincide at large radii, where the impact of BH rotation is negligible. The larger BH spin, the closer to the horizon the marginally stable orbit is, and, as a result, the disk emission extends much closer to the horizon, well below $r=6\rm M$. It is clear, that the luminosity of a disk increases with BH spin. This increase of luminosity corresponds exactly to the increase of the efficiency of accretion (Eq.~\ref{e.etaorg}) for low accretion rates.

\begin{figure}
\centering
 \subfigure
{
\includegraphics[width=.7\textwidth]{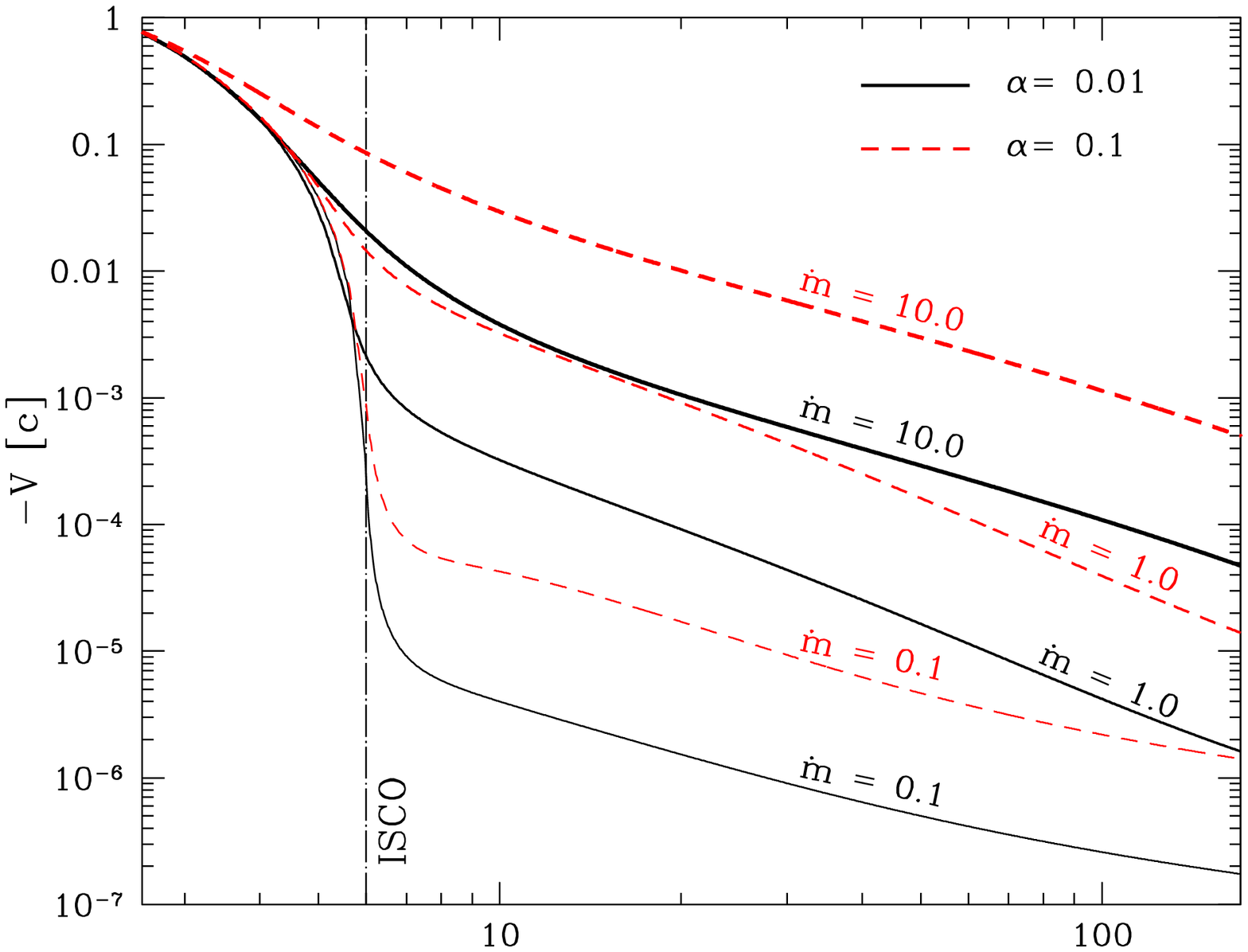}
}
\\
\vspace{-.07\textwidth}
 \subfigure
{
\includegraphics[width=.7\textwidth]{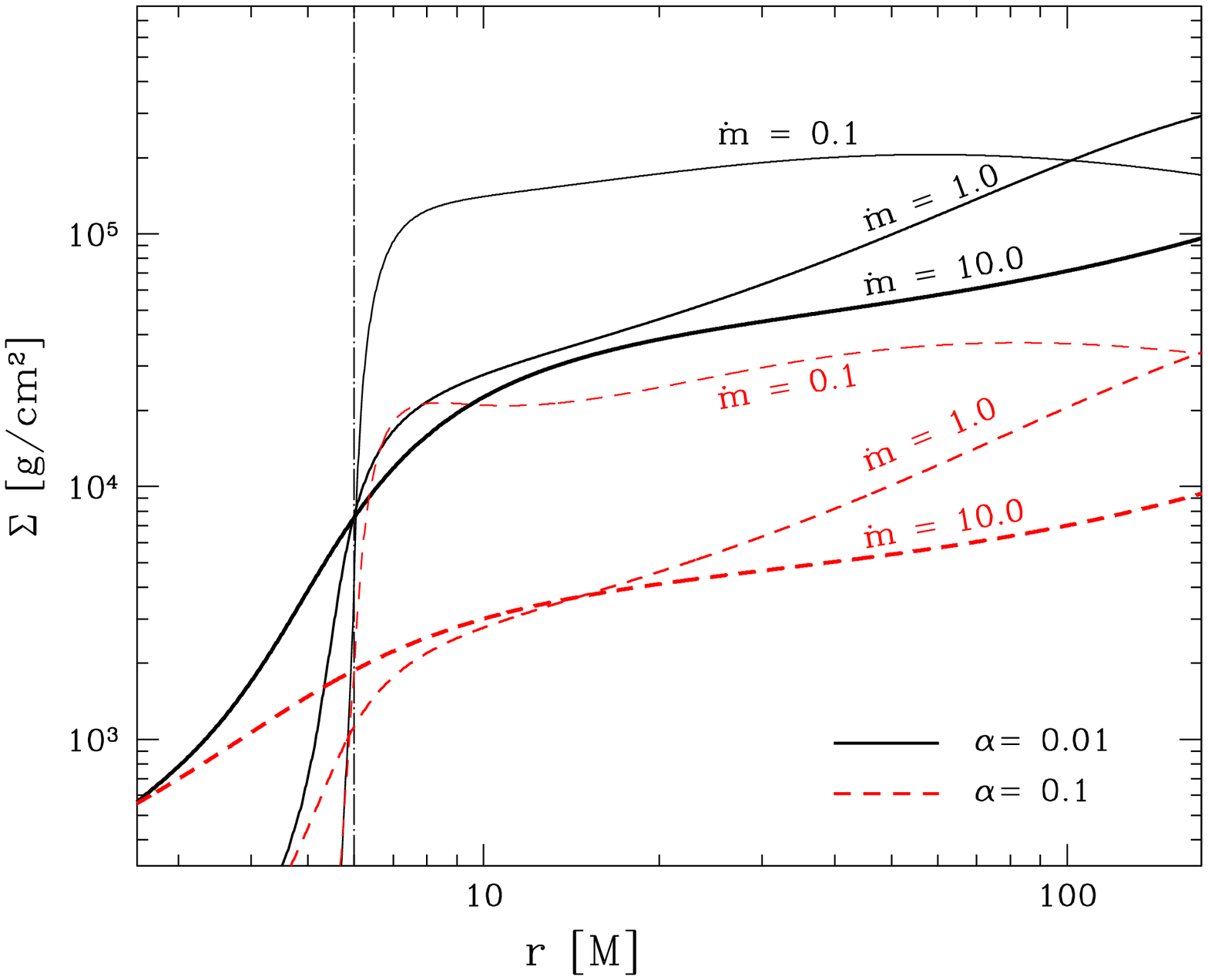}
}
\caption {Profiles of the radial velocity (top panel) and surface
density (bottom panel) for slim disk solutions with different accretion
rates in case of a non-rotating BH with $M_{\rm BH}=10\Msun$. Solid black
lines present solutions for $\alpha=0.01$ while red dashed lines for $\alpha=0.1$.
}
 \label{f.vrSigma}
\end{figure}

Profiles of the radial velocity ($V$) and surface density ($\Sigma$) are shown
in the top and bottom panels of Fig.~\ref{f.vrSigma}, respectively.
The black lines present solutions for $\alpha=0.01$, while the red
dashed lines for $\alpha=0.1$.
The radial velocity for large radii coincides with values given 
by \citem{nt} 
which are assumed as the outer boundary conditions, and 
approaches the speed of light when
 getting close to the horizon. At a given radius the 
radial velocity increases with mass 
accretion rate. The radial velocity is directly related to the surface
density through the mass conservation equation (Eq.~\ref{eq_poly_cont2}).
For large radii, where the disk is
gas pressure dominated, the surface density increases with accretion
rate (this region corresponds to the lower branch of solutions on the 
($T$,$\Sigma$) plane, (see e.g., \citem{slim}) while for moderate 
radii the relation is opposite (the middle branch). For the highest
accretion rates ($\dot m\gtrsim 10$) the solutions enter the upper, advection dominated branch. As the surface density is expected to be inversely proportional to $\alpha$, its values for $\alpha=0.1$ are roughly ten times smaller than for $\alpha=0.01$.

\begin{figure}
\centering
  \includegraphics[width=.7\textwidth,angle=0]{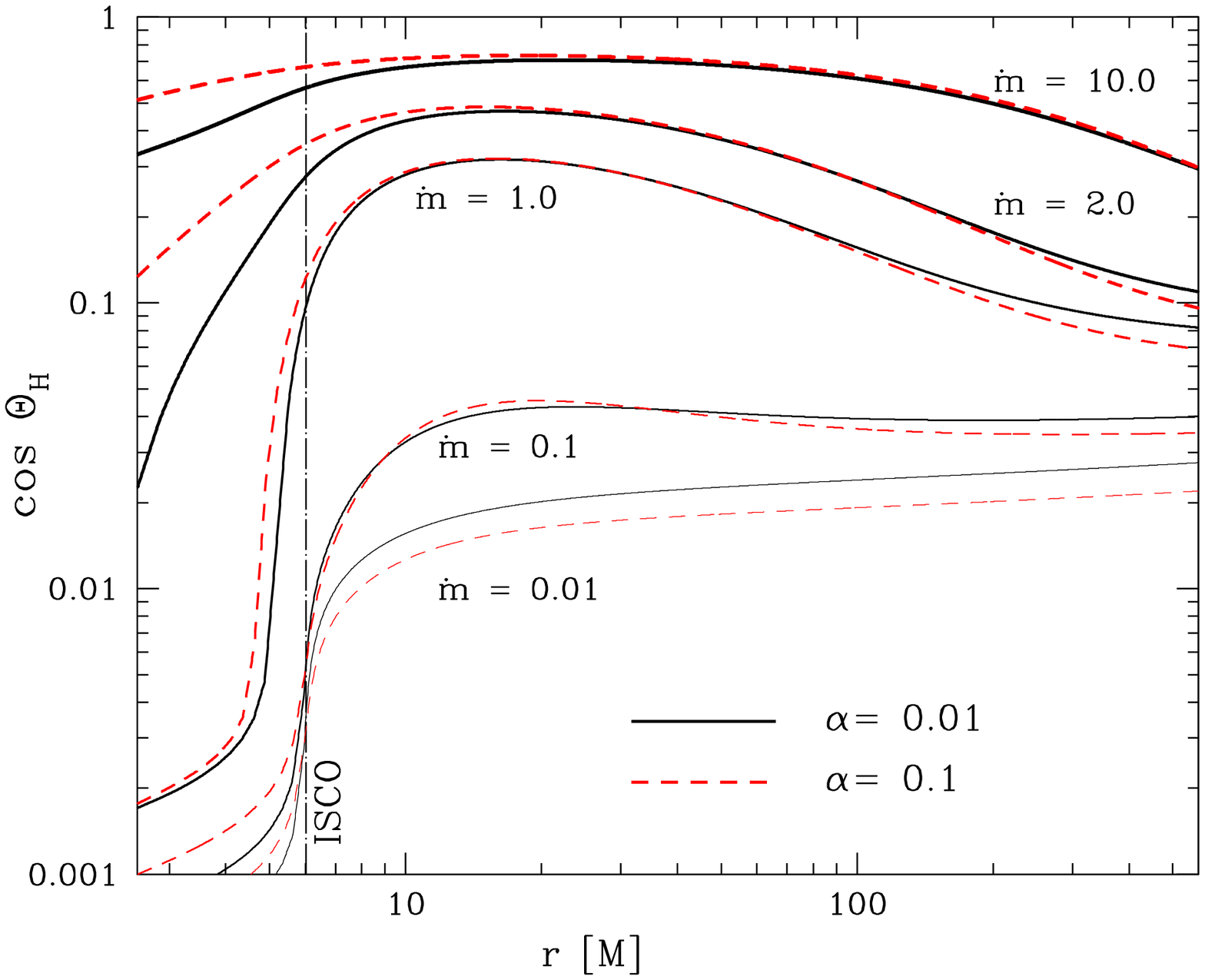}
\caption{Profiles of the disk photosphere in terms of $\cos\Theta_H=H_{\rm phot}/r_{\rm sph}$ 
for a number of accretion rates and two values of $\alpha=0.01$ and $0.1$. $M_{\rm BH}=10\Msun$ and 
$a_*=0$. }
  \label{f.hr}
\end{figure}

It has been proven \citepm{sadowski.photosphere} that the disk photosphere
location should be taken into account when performing ray-tracing of photons
emitted from an accretion disk. It becomes of major importance when the
accretion rates are high and disks are no longer geometrically thin. 
In Fig.~\ref{f.hr} we plot profiles of the photosphere (disk height, $\cos \Theta_H=H_{\rm phot}/r_{\rm spherical}$) for 
different accretion rates and a non-rotating BH. It is clearly visible
that for accretion rates $\dot m>0.1$ the inner 
regions are thicker due to the radiation pressure. For 
$\Mdot \ge 1.0\Medd$ the highest value of $\cos \Theta_H$ ratio 
exceeds $0.3$. In Fig.~\ref{f.sir.height} similar profiles are plotted using the same scale on both axes.

\begin{sidewaysfigure}
\centering

\includegraphics[width=.7\textwidth]{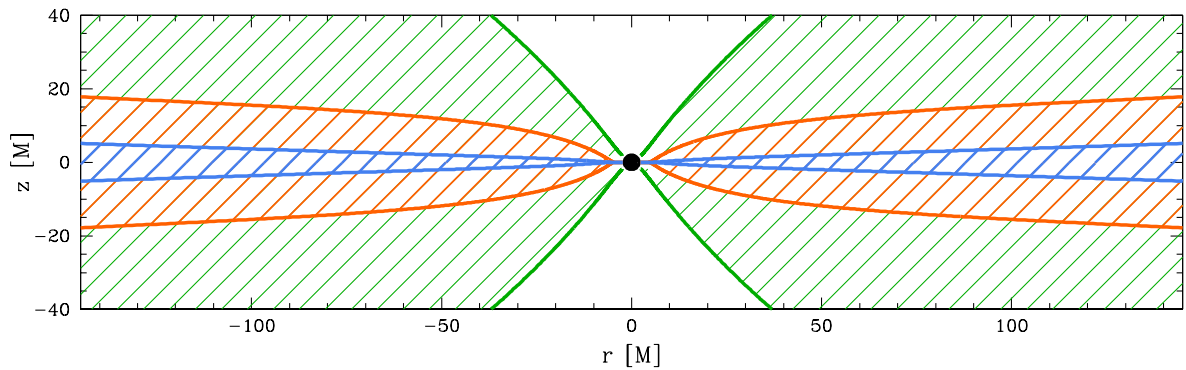}

\caption{
Disk height profiles of slim disks with three accretion rates: $0.1$ (blue), $1.0$ (orange) and $10.0\Medd$ (green line), and $\alpha=0.1$. The black circle denotes the central BH.
}
\label{f.sir.height.stand}
\end{sidewaysfigure}

\begin{figure*}
\centering
 \subfigure
{
\includegraphics[height=.55\textwidth]{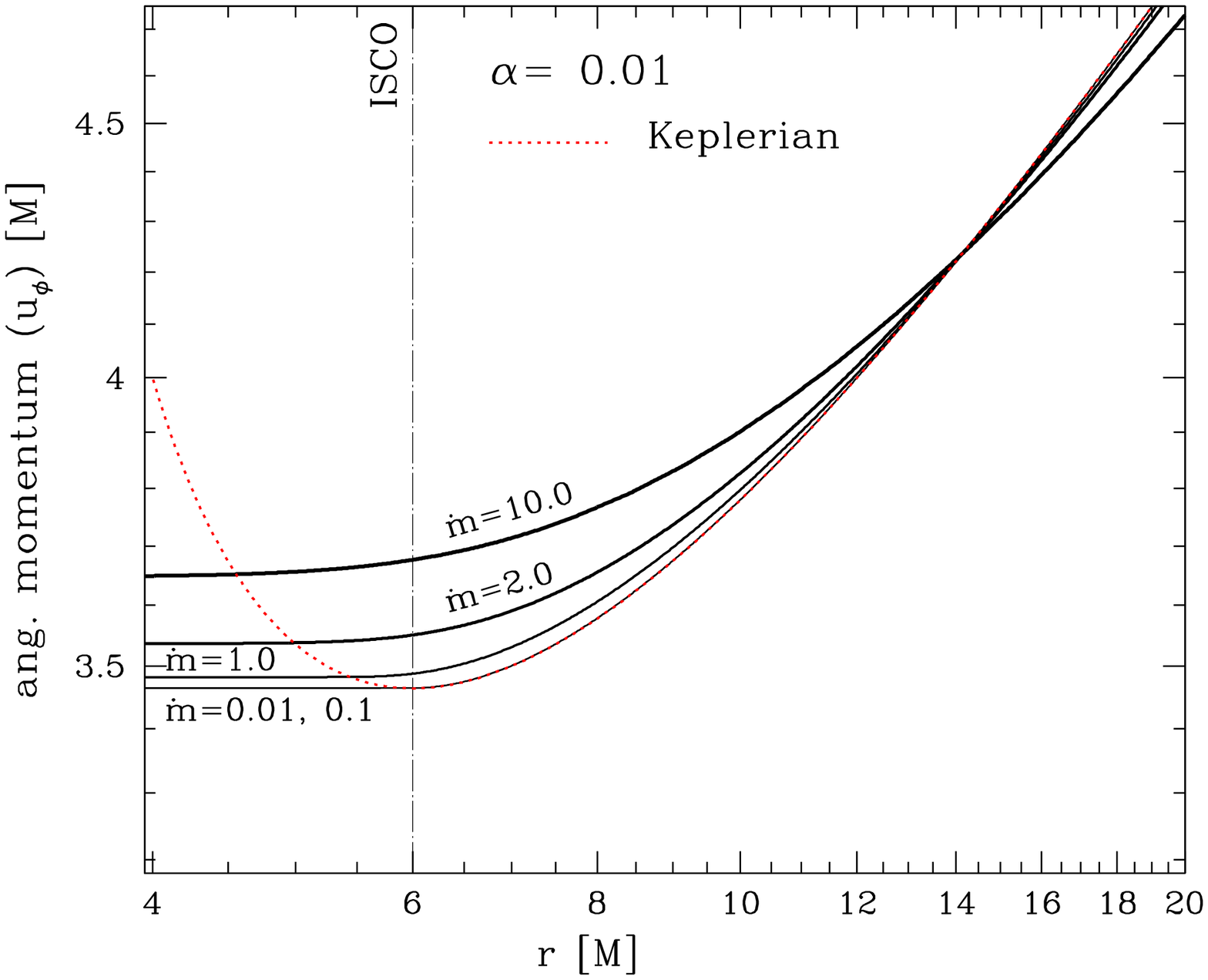}
}
\hspace{-.07\textwidth}
 \subfigure
{
\includegraphics[height=.55\textwidth]{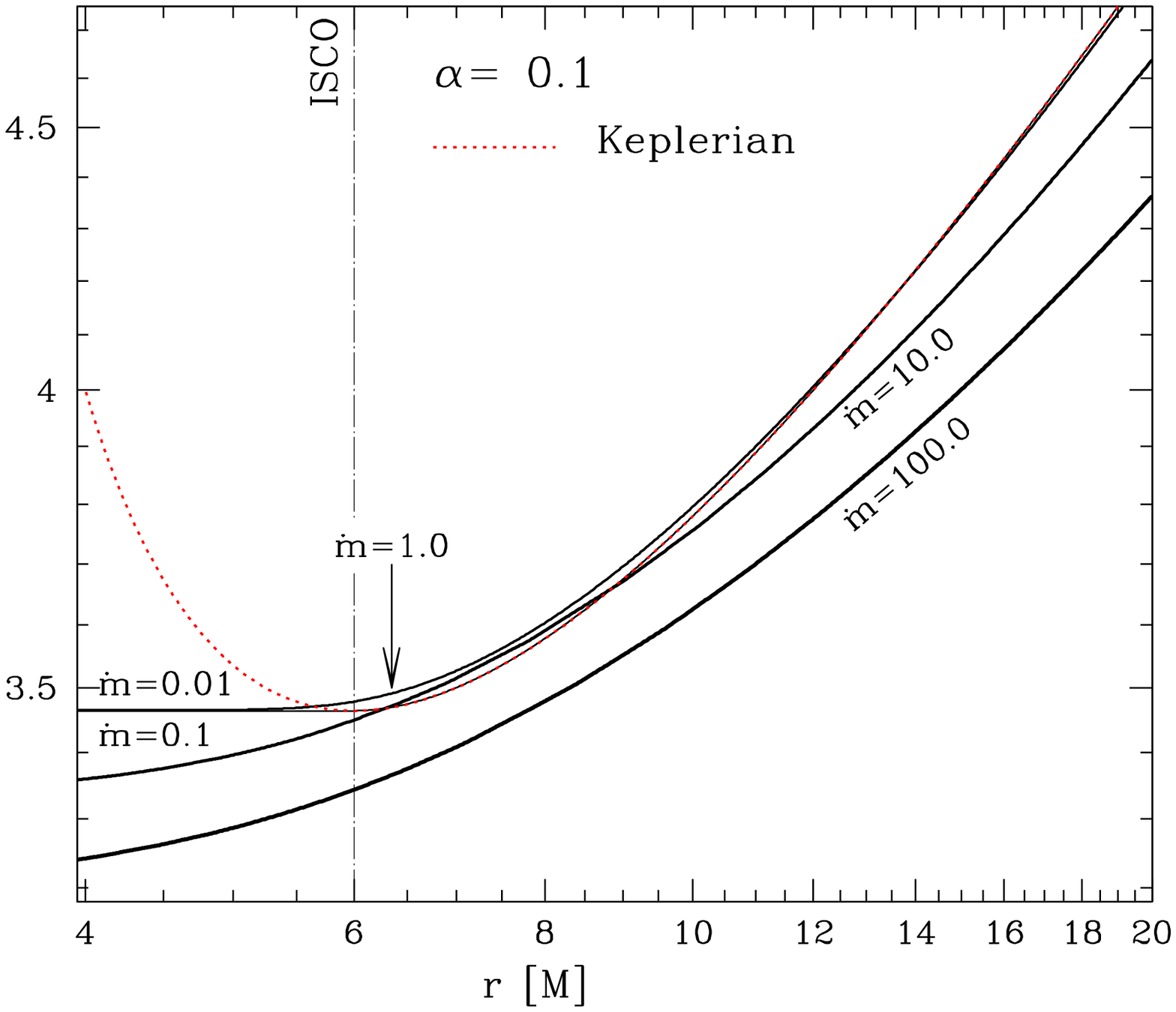}
}
\caption{ Profiles of the disk angular momentum ($u_\phi$) for $\alpha=0.01$ (top) 
and $\alpha=0.1$ (bottom panel) for different accretion rates. The spin
of the BH is $a_*=0$.}
\label{f.l.a0}
\end{figure*}
\begin{figure}[h]
\centering
\includegraphics[height=.7\textwidth,
angle=-90]{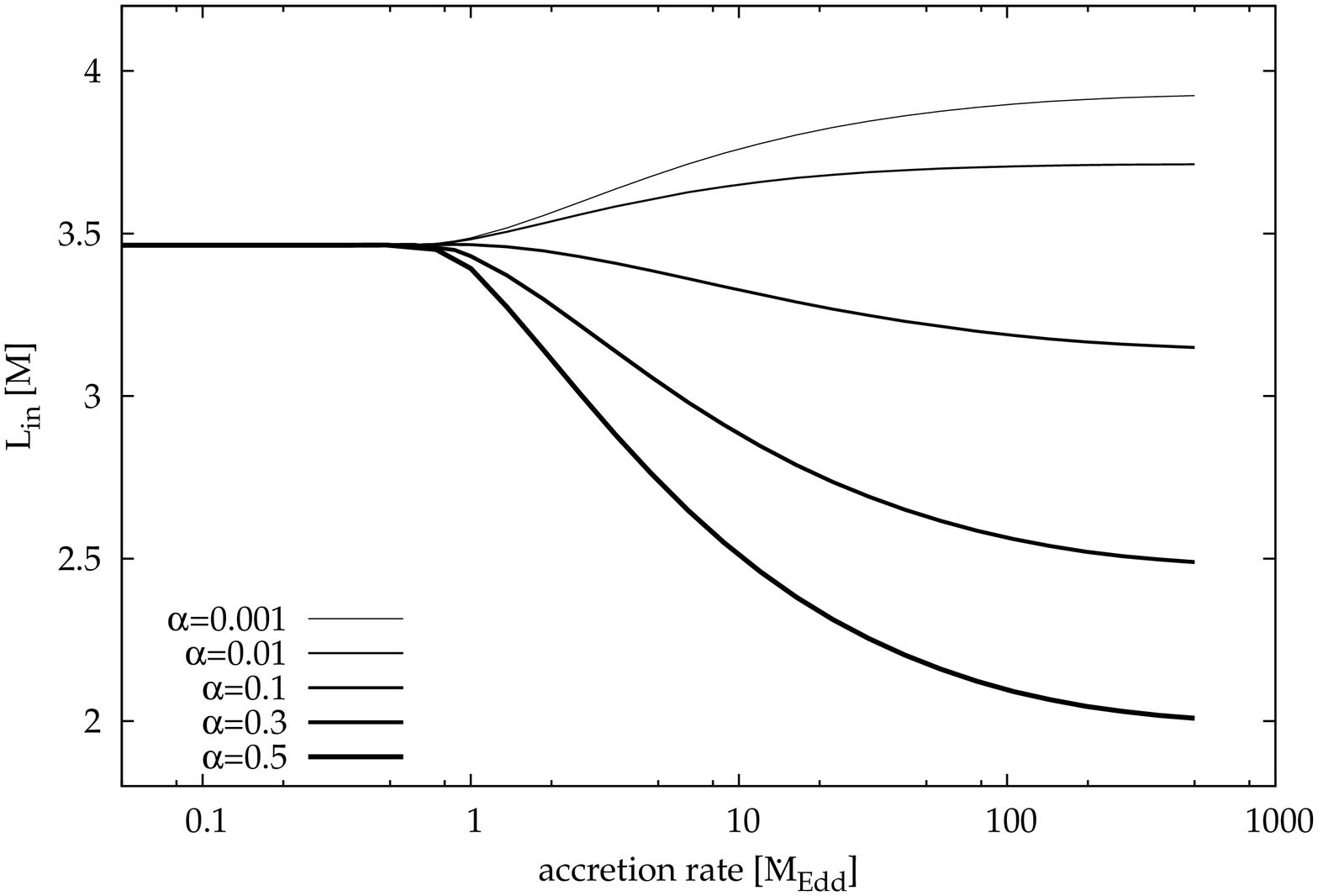}
\caption{Dependence of the angular momentum at the horizon on accretion
rate for solutions with different values of $\alpha$ for $a_*=0$.}
\label{fig:horizon-angular-momentum}
\end{figure}

In Fig.~\ref{f.l.a0} we present profiles of the 
angular momentum for two representative values of $\alpha$ 
($0.01$ and $0.1$ on left and right panel, respectively)
and a few accretion
rates. For low accretion rates the angular momentum follows the
Keplerian profile (down to the marginally stable orbit) for both cases. For higher accretion rates
it departs from the Keplerian flow. For the lower value of $\alpha$ the
flow is super-Keplerian for moderate radii.
The higher mass accretion rate the more significant deviation 
from the Keplerian profile. For the higher $\alpha$ ($0.1$) the
behavior is quantitatively different as the angular momentum 
at the horizon (${\cal L}_{\rm in}$) falls below the Keplerian value at
the marginally stable orbit. The higher the accretion rate the smaller
the value. The impact of $\alpha$ on the slim disk solution topology
has been extensively studied in \citem{leavingtheisco}. Figure~\ref{fig:horizon-angular-momentum} shows how 
${\cal L}_{\rm in}$ depends on both the accretion rate and the $\alpha$
viscosity parameter.


%
%

\begin{figure}
\centering
  \includegraphics[width=.7\textwidth,angle=0]{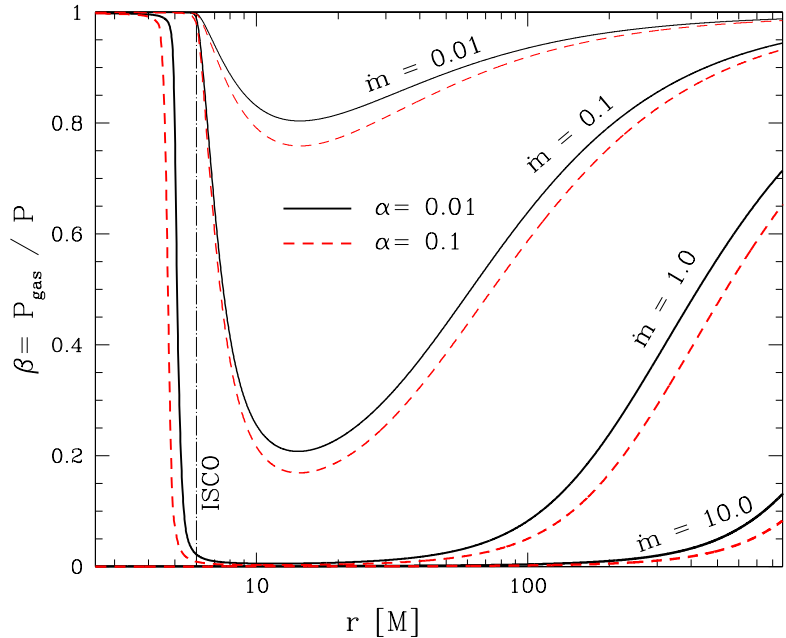}
\caption{Gas pressure to total pressure ratio for slim disk solutions with
  $\dot m=0.01$, $0.1$, $1.0$ and $10.0$. Profiles for $\alpha=0.01$ 
  and $0.1$ are presented with solid black and dashed red lines, respectively.
}
  \label{f.beta}
\end{figure}

Accretion disks are expected to be gas-pressure dominated in the limit of low accretion rates \citepm{shakura-73}. However, the temperature, and corresponding radiation pressure, increases with accretion rate. As a result, for accretion rates higher than a particular value (e.g., $0.1\Mdot_{\rm Edd}$ for $a_*=0$), there is a radiation-pressure dominated region, the extent of which increases with further grow of accretion rate. This behavior is presented in Fig.~\ref{f.beta} which shows the gas pressure to total pressure ratio ($\beta$) for different accretion rates and two representative values of $\alpha$. For $\dot m=0.1$, the region between $8\rm M$ and $50\rm M$ is dominated by radiation pressure. The pressure ratio drops below $0.4$ which is the critical value for triggering the thermal instability according to the standard theory (for detailed discussion see Section~\ref{s.thermalinstability}). For the Eddington accretion rate ($\dot m=1.0$) the unstable region extends between $5\rm M$ and $200\rm M$. For super-Eddington accretion rates the disk is radiation-pressure supported down to BH horizon.

\begin{figure}
\centering
  \includegraphics[width=.7\textwidth,angle=0]{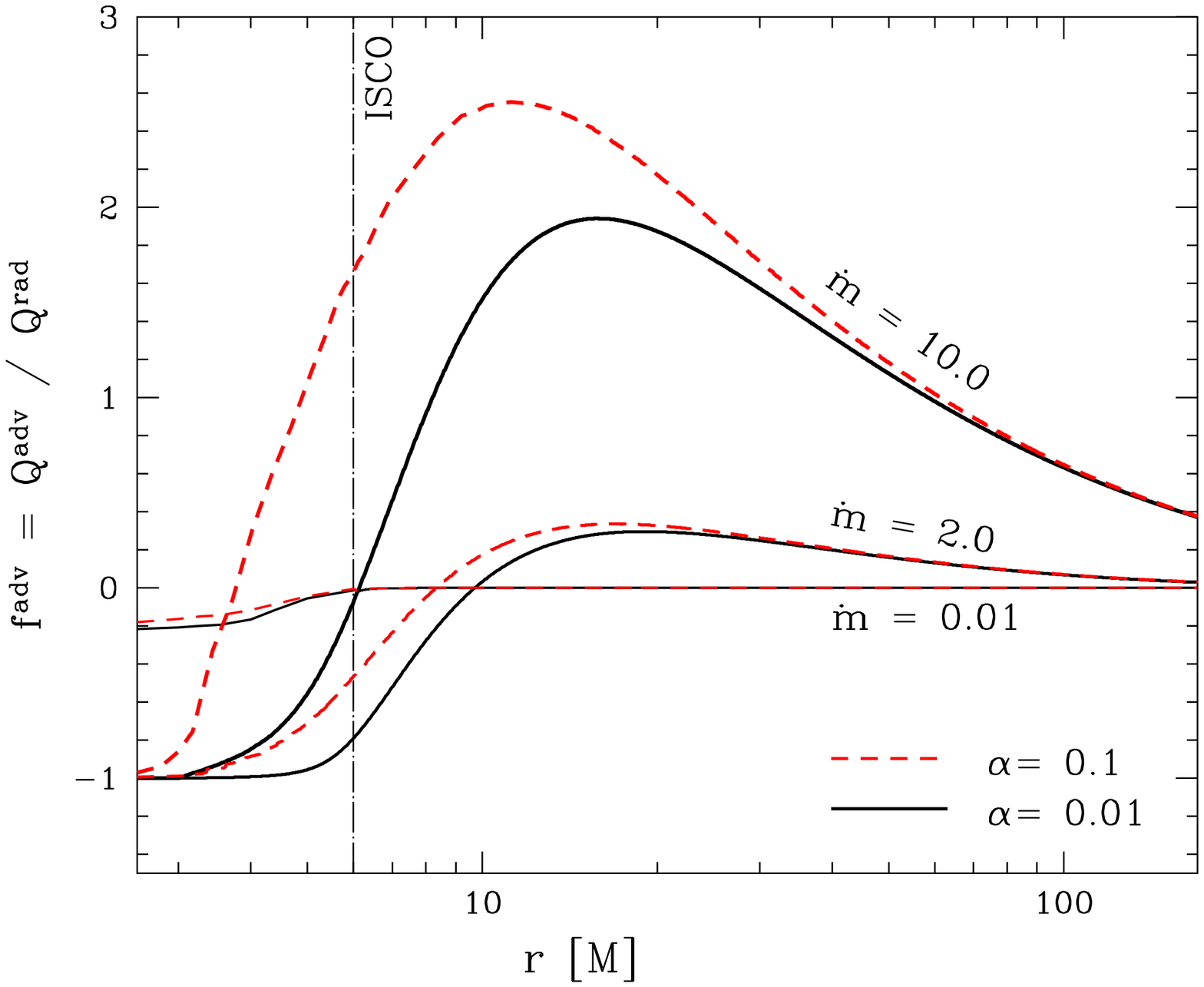}
\caption{The advection factor profiles for
  $\dot m=0.01$, $2.0$ and $10.0$. Profiles for $\alpha=0.01$ 
  and $0.1$ are presented with solid black and dashed red lines,
  respectively. The
  fraction $f^{\rm adv}/(1+f^{\rm adv}$) of heat generated by viscosity
  is accumulated in the flow. In regions with $f^{\rm adv}<0$ the advected
  heat is released.
}
  \label{f.fadv}
\end{figure}

Standard thin accretion disks are radiatively efficient --- viscous heating is balanced by radiative cooling. This is not the case for accretion rates close to, and higher than the Eddington one. Such disks are thick and are characterized by high radial velocity of gas. As a result, additional mechanism of cooling triggers in: advection. In Fig.~\ref{f.fadv} we plot the advection coefficient $f^{\rm adv}$ defined as the ratio of the advective to radiative cooling rates. Values close to zero denote radiatively efficient disk regions, higher than unity --- regions where advective cooling dominates over radiative cooling, negative --- regions where heat advected from outer disk annuli is released (advective heating). Disk solutions for the lowest accretion rates are in fact radiatively efficient ($f^{\rm adv}\approx 0$). For accretion rates close to $\Medd$ the region $r>10\rm M$ is significantly affected by advection, e.g., the advective cooling equals up to 30\% of the radiative cooling for $\dot m=2.0$. The region inside $r=10\rm M$ is characterized by negative advection coefficient. Radiative cooling balances both viscous and advective heating. The heat advected into this region is the only heating mechanism inside ISCO ($f^{\rm adv}\approx -1$), causing the emission from inside this radius visible in Fig.~\ref{f.flux}. For strongly super-Eddington accretion rates (e.g., $\dot m =10$) the advective cooling rate overwhelms the rate of radiative cooling outside ISCO, while advection heating still dominates close to BH horizon.

\begin{figure}
\centering
  \includegraphics[width=.7\textwidth,angle=0]{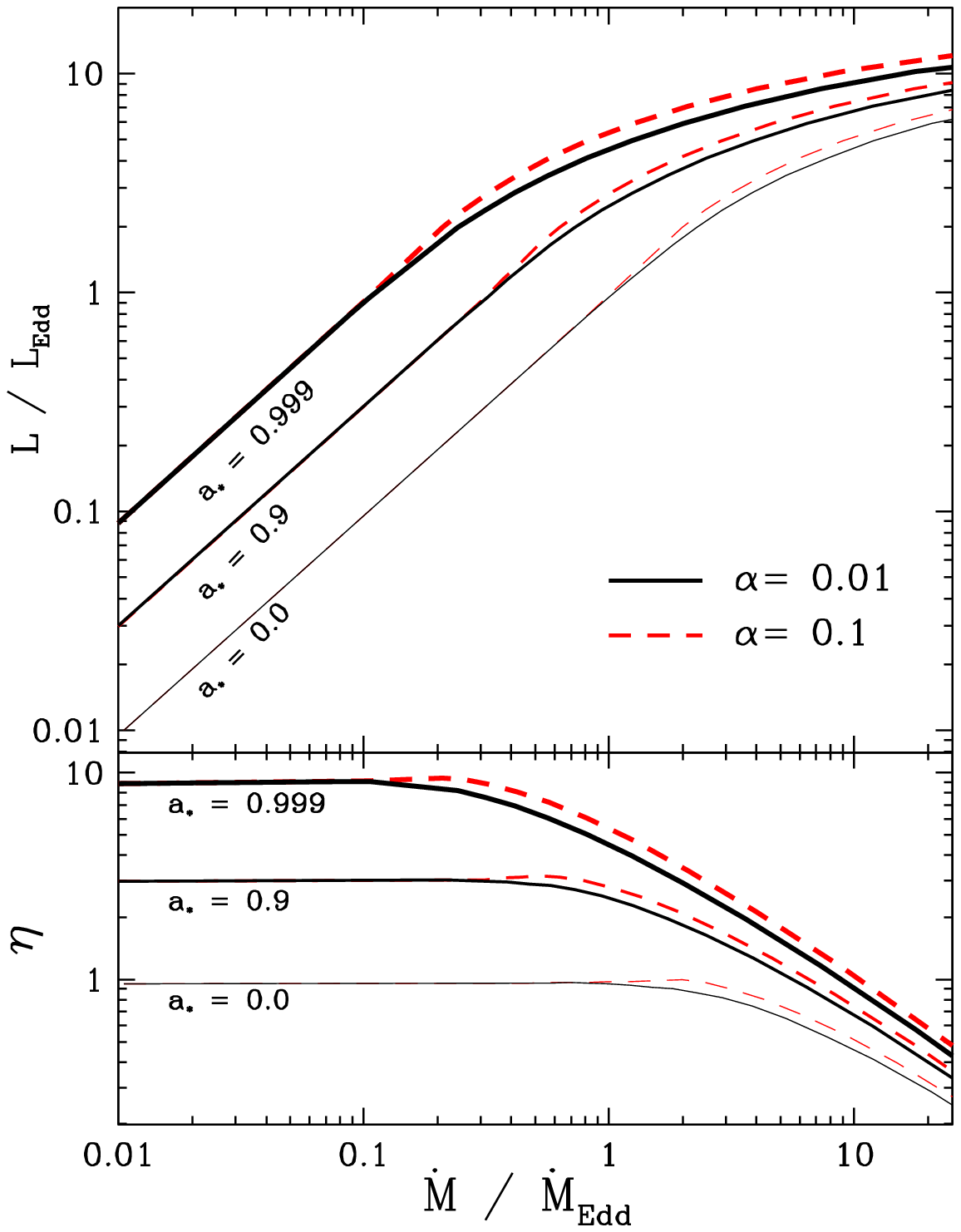}
\caption{Top panel: Luminosity vs accretion rate for three values of BH spin
  ($a_*=0.0$, $0.9$, $0.999$) and two values of $\alpha=0.01$ (black) and
  $0.1$ (red line). Bottom panel: efficiency of
  accretion $\eta=16\, L/\mdot c^2$. The luminosity $L$ was integrated locally, not taking into account the relativistic g-factor (see Section~\ref{s.gfactor}), and therefore only roughly corresponds to $\tilde\eta$ defined in Eq.~\ref{e.etaorg}.
}
  \label{f.eff}
\end{figure}

The advection of heat in slim disks significantly changes disk
properties for luminosities $L\gtrsim L_{\rm Edd}$. This fact is clearly
visible in Fig.~\ref{f.eff}. In the top panel we plot the relation between disk
luminosity and mass accretion rate for three values of BH spin. 
For sub-Eddington regime the luminosities are proportional to the
accretion rates as the efficiency of accretion ($\eta=16\, L/\mdot c^2$,
plotted in the bottom panel) remains constant. For higher accretion
rates, however, the efficiency drops down as some amount of heat
generated by viscosity is captured in the flow and advected onto the
BH. The higher the accretion rate, the lower the efficiency of accretion. For
$a_*=0.9$ it decreases by an order of magnitude for $\dot m\approx 10$.

\begin{sidewaysfigure}
\centering
\includegraphics[height=6.0cm, angle=0]{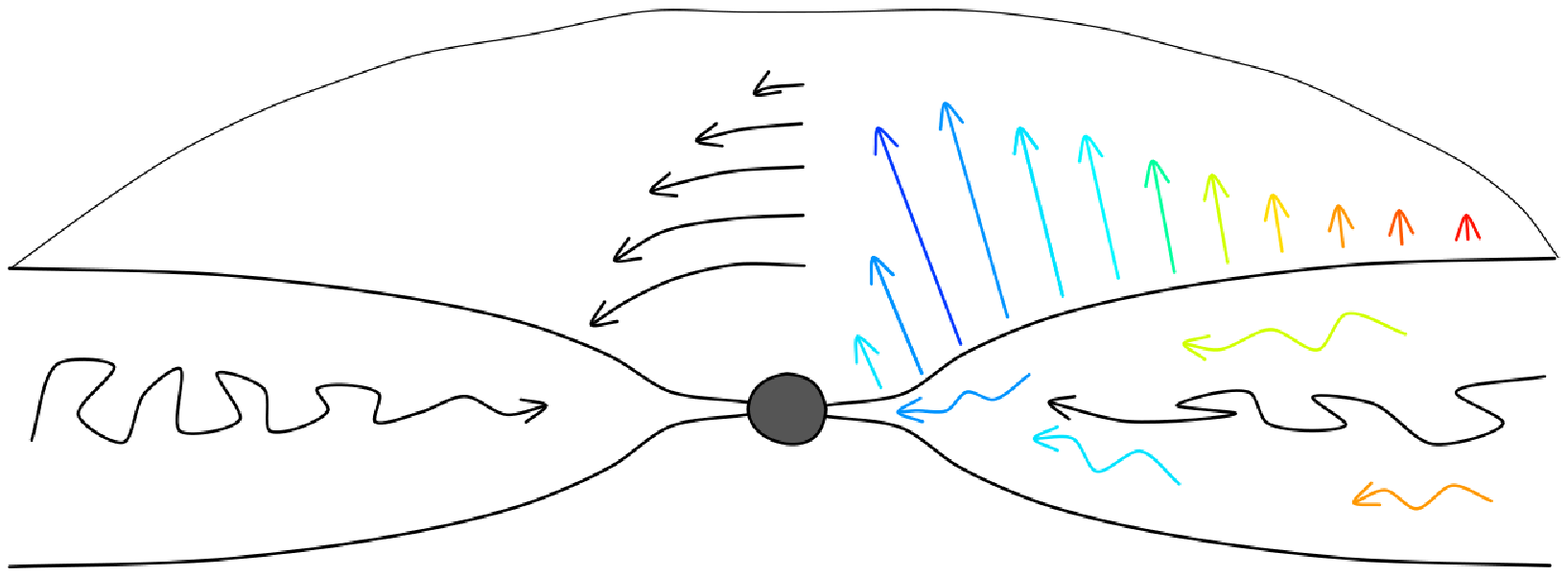}
\caption{A scheme of an accretion disk at super-Eddington accretion rate: Gas is accreted onto a BH through a disk that significantly extends above the equatorial plane (geometrically slim). Viscosity results from turbulent motion. The profile of rotation departs from the Keplerian one. Disk extends down to BH horizon. Viscous heating is balanced by radiative and advective cooling mechanisms. The disk emission does not terminate at ISCO --- significant emission comes from the vicinity of the BH, resulting from heat advected from outer regions.
}
\label{f.slim.eps}
\end{sidewaysfigure}

In Fig.~\ref{f.slim.eps} we present a schematic picture of a slim accretion disk with super-Eddington accretion rate.

\section{Inner edge(s)}
\label{s.inneredges}
Accretion flows onto BHs must change character before
matter crosses the event horizon, for two reasons. First, matter must cross the
black-hole surface at the speed of light as measured by a local
inertial observer \citepm{gj-06}, so that if the flow
is subsonic far away from the black-hole (in practice it is
always the case) it will have to cross the sound barrier (well)
before reaching the horizon. This is the property of all realistic
flows independent of their angular momentum. This sonic surface can be considered as the inner edge of the accretion
flow.

The second reason is related to angular momentum. Far from the
hole many (most probably most) rotating accretion flows adapt the
Keplerian angular momentum profile. Because of the existence of the
innermost stable circular orbit (ISCO), these flows must stop to be
Keplerian there. At high accretion rates when pressure gradients
become important, the flow may extend below the ISCO but the
presence of the innermost bound circular orbit (IBCO) defines
another limit to a circular flow (the absolute limit being given
by the circular photon orbit, the CPO). These critical circular
orbits provide another possible definition of the inner edge of the
flow, in this case of an accretion disk.

The question is: what is the relation between the accretion flow
edges? In the case of geometrically thin disks the sonic and
Keplerian edges coincide and one can define the ISCO as the inner
edge of these disks. \citem{paczynski-2000} showed rigorously that
independent of e.g., viscosity mechanism and the presence of magnetic fields, the ISCO is the universal inner disk's edge for not too-high
viscosities. The case of thin disks is therefore
settled\footnote{\citem{pennaetal-10} studied the
effects of magnetic fields on thin accretion disk (the disk
thickness $h/r \le 0.07$. They found that to within a few percent the
magnetized disks are consistent with the \citem{nt}
model, in which the inner edge coincides with the ISCO.}.

However, this is not the case for non-thin accretion disks, i.e.,
of medium and high luminosities. The problem of defining
the inner edge of an accretion disk is not just a formal exercise.
\citem{afshordi-2003} explored several reasons which made
discussing the precise location of inner edge $r = r_{\rm in}$ of the
BH accretion disks an interesting and important issue. One
of them was,\\
\par \hangindent=0.5cm {\it Theory of accretion disks is several
decades old. With time ever more sophisticated and more diverse
models of accretion onto BHs have been introduced.
However, when it comes to modeling disk spectra, conventional
steady state, geometrically thin-disk models are still used,
adopting the classical ``no torque'' inner boundary condition at
the marginally stable orbit.}\\

\citem{kro-2002} proposed several ``empirical'' definitions of the
inner edge, each serving a different practical purpose (see
also the follow-up investigation by \citem{bec-2008}). We add to these a few more
definitions. The list of the inner edges considered in this section
consists of\footnote{Krolik \& Hawley defined the inner edges no. 4, 5 and 6
and in addition a seventh edge 7, {\it the turbulence edge}, where
flux-freezing becomes more important than turbulence in
determining the magnetic field structure. Magnetic fields are not
considered for slim accretion disks, and we therefore do not discuss
7.},

\begin{enumerate}
 \item  {\it The potential spout edge}~$r_{\rm in} = r_{\rm
pot}$, where the effective potential forms a self-crossing Roche
lobe, and accretion is governed by the Roche lobe overflow.

\item  {\it The sonic edge}~$r_{\rm in} = r_{\rm son}$,
where the transition from subsonic to transonic accretion occurs.
Hydrodynamical disturbances do not propagate upstream a supersonic
flow, and therefore the subsonic part of the flow is ``causally''
disconnected from the supersonic part.

\item  {\it The variability edge}~$r_{\rm in} = r_{\rm
var}$, the smallest radius where orbital motion of coherent spots
may produce quasi-periodic variability.

\item  {\it The stress edge}~$r_{\rm in} = r_{\rm str}$,
the outermost radius where the Reynolds stress is small, and
plunging matter has no dynamical contact with the outer accretion
flow.

\item  {\it The radiation edge}~$r_{\rm in} = r_{\rm rad}$,
the innermost place from which significant luminosity emerges.

\item  {\it The reflection edge}~$r_{\rm in} = r_{\rm
ref}$, the smallest radius capable of producing significant
fluorescent iron line.
\end{enumerate}

In the next six subsections, we discuss these six edges one by one basing on the solutions of the relativistic slim disk model presented in the previous section. This discussion is similar to the study of \citem{leavingtheisco} but is based on a slim disk model with more consistent (i.e., polytropic) treatment of the vertical structure. Qualitative behavior is exactly the same. However, the impact of advection sets on at a higher ($0.6$ versus $0.3\Medd$) accretion rate. We also point out here that locations of some of these inner edges are sensitive to the assumed value of the $\alpha$ parameter. Solutions presented in the previous Secion are based on the $\alpha=\rm const$ assumption. This assumption, however, may not be satisfied inside the marginally stable orbit, as some of the most recent GRMHD simulations show (e.g., \citem{pennaetal-10}). Results presented in this Section should be therefore considered qualitative only.

\subsection{Potential spout edge}
\label{section-potential-spout}
%

\begin{sidewaysfigure}
\centering
\includegraphics[height=4.5cm, angle=0]{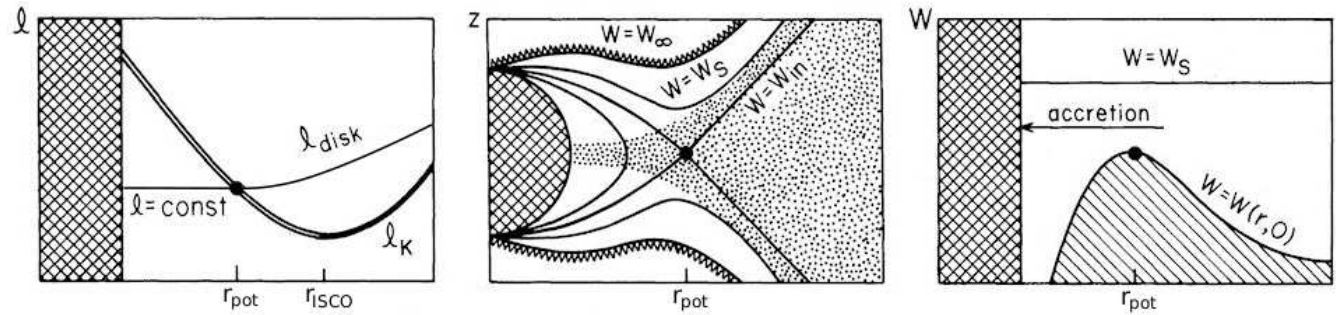}
\caption{An illustrative visualization of the Roche lobe overflow.
The leftmost panel schematically presents disk angular momentum
profile and its relation to the Keplerian distribution. The middle
panel shows the equipotential surfaces. The dotted region denotes
the volume filled with accreting fluid. The rightmost panel
presents the potential barrier at the equatorial plane ($z=0$) and
the potential of the fluid ($W_S$) overflowing the barrier. The
figure is taken from {\tt
http://www.scholarpedia.org/article/Accretion\_discs}.
  }
\label{figure-roche-lobe-ill}
\end{sidewaysfigure}

The idea of the ``relativistic Roche lobe overflow'' governing accretion
close to the BH was first explained by Paczy{\'n}ski
(see \citem{koz-1978}). It was later explored in detail by many
authors analytically (e.g., \citem{abr-1981, abr-1985}) and by
large-scale hydrodynamical simulations (e.g., \citem{igu-1997}).
It became a standard concept in BH accretion theory.
Figure \ref{figure-roche-lobe-ill} schematically illustrates
the Roche lobe overflow mechanism.
The leftmost panel presents a demonstrative profile of disk angular momentum,
which reaches the Keplerian value at the radius corresponding to
the self-crossing of the equipotential surfaces presented in the
middle panel. To flow through this ``cusp'', matter must have potential
energy higher than the value of the potential at this point, i.e., the
``potential barrier'' is crossed only when the matter overflows
its Roche lobe. 

%
%

The potential difference between the horizon and the spout is
infinite, and therefore no external force can prevent the matter located
there from plunging into the BH. At radii greater that
$r_{\rm pot}$, the potential barrier at $r = r_{\rm pot}$ retains
the matter inside this radius. We note, that because the dynamical equilibrium is
given (approximately) by $\nabla_i {\cal U}_{\rm eff} = \nabla_i
P/\rho$, with $\rho$ being the density, one may also say that it
is the pressure gradient (the pressure stress) that holds the
matter within $r_{\rm pot}$.

The specific angular momentum in the Novikov-Thorne model is {\it
assumed} to be Keplerian. Slim disk models do not a priori assume
an angular momentum distribution, but self-consistently calculate
it from the relevant equations of hydrodynamics (see Section~\ref{s.sslim.eq}). These calculations
indicate that the type of angular momentum distribution depends on
whether the accretion rate and viscosity constrain the flow to be either
disk-like or Bondi-like type.

In the {\it Bondi-type} accretion flows, the angular momentum is
everywhere sub-Keplerian, ${\cal L} < {\cal L}_K$. These flows are
typical of high viscosities and high accretion rates, as the
case of $\alpha = 0.1$ and $\dot m = 100$ shown in the bottom panel of
Fig.~\ref{f.l.a0}. This is the only Bondi-like
flow in this figure.

In the {\it disk-like} accretion flows, the angular momentum of
the matter in the disk is sub-Keplerian everywhere, except the
strong-gravity region $r_{\rm pot} < r < r_{\rm cen}$, {\it where
the flow is super-Keplerian}, ${\cal L} > {\cal L}_K$. The radius
$r_{\rm cen}
> r_{\rm ISCO}$ corresponds to the ring of the maximal pressure in
the accretion disk. This is also the minimum of the effective
potential. The radius $r_{\rm pot} < r_{\rm ISCO}$ delineates a saddle
point for both pressure and effective potential; this is also the
location of the ``potential spout inner edge'', $r_{\rm in} =
r_{\rm pot}$.

We note that in the classic solutions for spherically accretion flows
found by \citem{bondi-52} the viscosity is unimportant and the
sonic point is saddle, while in the ``Bondi-like'' flows discussed
here, angular momentum transport by viscosity is essentially
important and the sonic point is usually nodal. Therefore, one
should keep in mind that the difference between these types of
accretion flows is also determined by the relative importance of pressure
and viscosity. For this reason, a different terminology is often
used. Instead of ``disk-like'', one uses the term
``pressure-driven'', and instead  of ``Bondi-like'', one uses
``viscosity-driven'' (see e.g., \citem{mat-1984, KatoBook}).

From the above discussion, it is clear that the location of this
particular inner edge $r_{\rm pot}$ is formally given as the
smaller of the two roots, $r_{\pm} = (r_+, r_-)$, of the equation
\begin{equation}
\label{definition-potential-edge}
{\cal L}(r) - {\cal L}_K(r)= 0.
\end{equation}
The larger root corresponds to $r_{\rm cen}$. Obviously, Eq.~\ref{definition-potential-edge} has always a solution for the
disk-like flows, and never for the Bondi-like flows.

\begin{figure}[h]
\centering
\includegraphics[width=.7\textwidth,
angle=0]{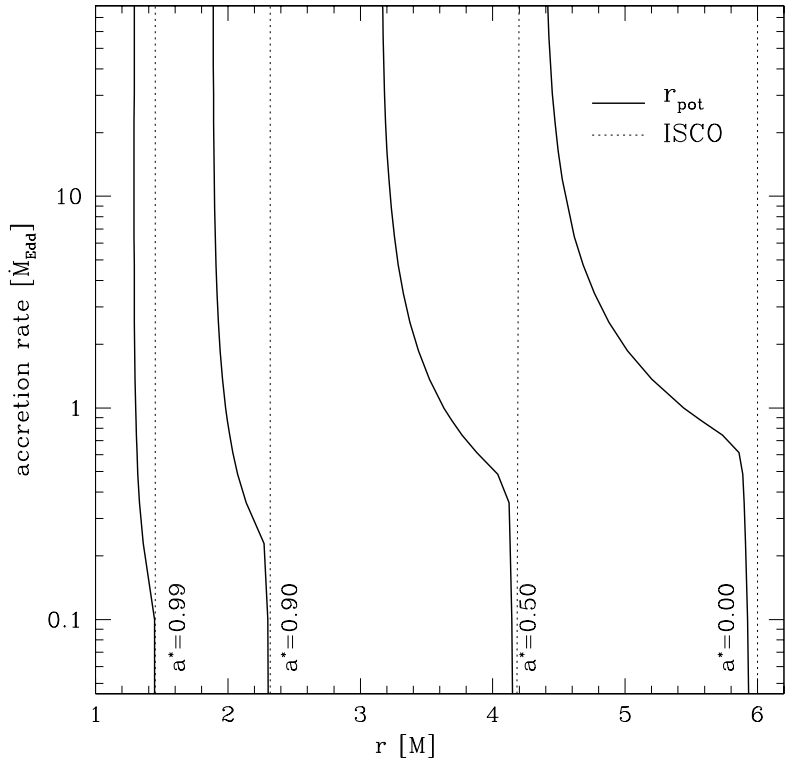}
\caption{Location of the potential spout inner edge $r_{\rm pot}$
for viscosity $\alpha=0.01$ and $a_*=0$. Solid lines show the exact location of
$r_{\rm pot}$ given by Eq.~(\ref{definition-potential-edge}).}
\label{figure-potential-edge}
\end{figure}

The location of the potential spout inner edge $r_{\rm pot}$ is
shown in Fig.~\ref{figure-potential-edge} for $\alpha=0.01$.
We note that for low accretion rates, (e.g., ${\dot m} \lesssim 0.6$ for $a_*=0$),
the location of the potential spout inner edge coincides with ISCO. At
${\dot m} \approx 0.6$, the location of the potential spout jumps
to a new position, which is close to the radius of the innermost
bound circular orbit, $r_{\rm IBCO}=4\rm M$. This behavior has long been recognized first by \citem{koz-1978} for Polish
doughnuts, and then by \citem{slim} for slim disks.
%
%
%
%
%

\subsection{Sonic edge}

\label{section-sonic}
%

As it has already been discussed, by a series of algebraic manipulations, one may reduce the slim disk
equations into a set
of two ordinary differential equations for two dependent
variables, e.g., the radial velocity $V$ and the
central temperature $T_C$,
\bea
\der{\ln V}{\ln r}&=&\frac{{\cal N}_1(r,V,T_C)}{{\cal D}(r,V,T_C)},\\\nonumber
\der{\ln T_C}{\ln r}&=&   \frac{{\cal N}_2(r,V,T_C)}{{\cal
D}(r,V,T_C)}.
\label{eq_reg}
\eea
%
%
\begin{figure}[h]
\centering
\includegraphics[height=.7\textwidth, angle=-90]{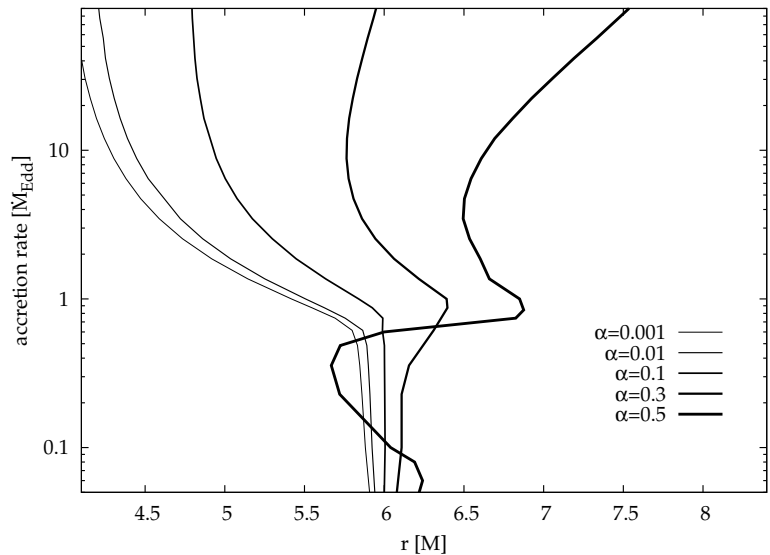}
\caption{Location of the sonic point as a function of the
accretion rate for different values of $\alpha$, for a
non-rotating black hole, $a_* = 0$. }
\label{fig:sonic-point-location}
\end{figure}

For a non-singular physical solution, the numerators ${\cal N}_1$
and ${\cal N}_2$ must vanish at the same radius as the denominator
${\cal D}$. The denominator vanishes at the sonic edge (or sonic
radius) where the Mach number is close to unity, i.e.
\begin{equation}
{\cal D}(r,V, T_C)|_{\,r = r_{\rm son}} = 0.
\label{definition-sonic}
\end{equation}
For low mass accretion rates, lower than about $0.6{\dot
M_{\rm Edd}}$ in the case of $a_*=0$, the sonic edge $r_{\rm son}$ is
close to ISCO, independently of the viscosity $\alpha$, as
Fig.~\ref{fig:sonic-point-location} shows. At about $0.6{\dot
M_{\rm Edd}}$, a qualitative change occurs, resembling a ``phase
transition'' from the Shakura-Sunyaev behavior to a very
different slim disk behavior.

For higher accretion rates, the location of the sonic point
significantly departs from ISCO. For low values of $\alpha$, the
sonic point moves closer to the horizon down to $\sim4M$ for
$\alpha=0.001$. For $\alpha>0.3$, the sonic point moves outward
with increasing accretion rate reaching values as high as $7.5M$ for
$\alpha=0.5$ and $100\dot M_{\rm Edd}$. This effect was first
noticed for low accretion rates by \citem{muchotrzeb-1986} and
later investigated for a wide range of accretion rates by
\citem{slim}, who explained it in terms of the disk-Bondi
dichotomy. The dependence of the sonic point location on the
accretion rate in the near-Eddington regime is more complicated
and is related to, for this range of accretion rates,
the transition from the radiatively efficient disk
to the slim disk occurring close to the sonic radius. 
It is also sensitive to the coefficients of the vertical integration introduced in Eqs.~\ref{def.eta1} --- \ref{def.eta4}, as comparison with similar plots in \citem{sadowski.slim} and \citem{leavingtheisco} shows.

The topology of the sonic point is important, because physically
acceptable solutions must be of the saddle or nodal type, the
spiral type being forbidden. The topology may be classified by the
eigenvalues $\lambda_1, \lambda_2, \lambda_3$ of the Jacobi
matrix
\be
{\cal J} = \left[
\begin{array}{ccc}
 \pder{\cal D}r      &\pder{\cal D}V      &\pder{\cal D}{T_C}\\\\
 \pder{{\cal N}_1}r  &\pder{{\cal N}_1}V  &\pder{{\cal N}_1}{T_C}\\\\
 \pder{{\cal N}_2}r  &\pder{{\cal N}_2}V  &\pder{{\cal N}_2}{T_C}
\end{array}
\right].
\ee
Because ${\rm det}({\cal J}) =0$, only two eigenvalues $\lambda_1,
\lambda_2$ are non-zero, and the quadratic characteristic equation
that determines them takes the form,
\be
2\,\lambda^2 - 2\,\lambda\,{\rm tr}({\cal J}) - \left({\rm
tr}({\cal J}^2) - {\rm tr}^2({\cal J})\right) = 0.
\ee
The nodal type is given by $\lambda_1\lambda_2 > 0$ and the saddle
type by $\lambda_1\lambda_2 < 0$. For the lowest values of $\alpha$, only
the saddle-type solutions exist. For moderate values of $\alpha$
($0.1\le\alpha\le0.4$), the topological type of the sonic point
changes at least once with increasing accretion rate. For the
highest $\alpha$, solutions have only nodal-type critical points.

\subsection{Variability edge}

\label{section-QPO-variability}
%

Axially symmetric and stationary states of slim accretion disks
are, obviously, theoretical idealizations.
Real disks are non-axial and non-steady. In particular, one
expects transient coherent features at accretion disk surfaces
--- clumps, flares, and vortices. The orbital motion of these features
could quasi-periodically modulate the observed flux of radiation,
mostly by means of the Doppler effect and the relativistic beaming.
We define $\Pi$ to be the ``averaged'' variability period, and $\Delta \Pi$
a change in the period during one period caused by the radial motion of a
spot.
%
%
\begin{sidewaysfigure}
\centering
\includegraphics[height=7.0cm, angle=0]{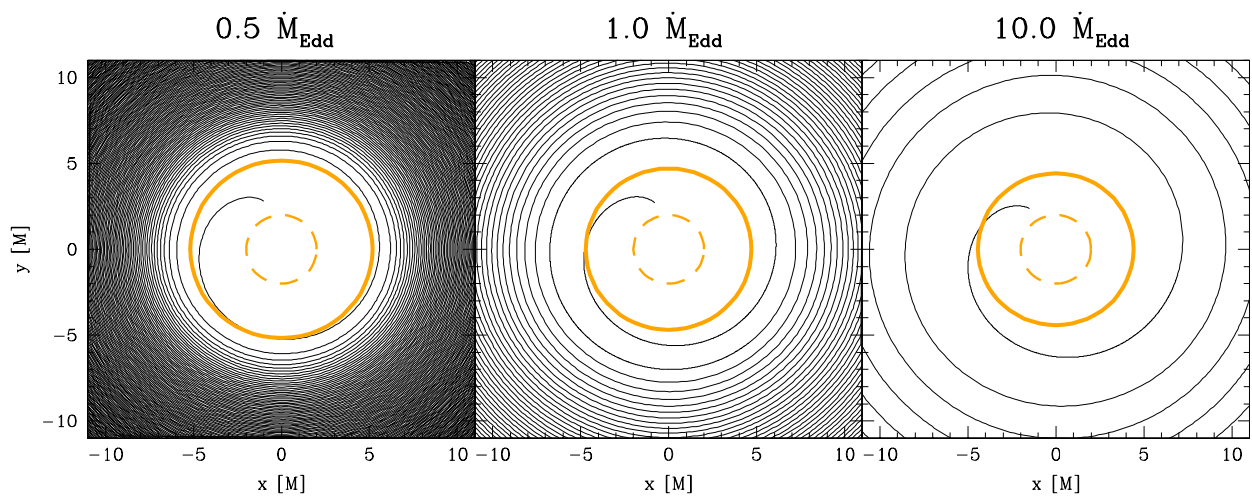}
\caption{The fluid
flow trajectories in slim accretion disks shown by thin solid
lines for different accretion rates. Locations of $r_{\rm pot}$
and the location of
black hole horizon are shown by thick orange solid and broken lines, respectively.
For low accretion
rates, the pattern of trajectories consists of very tight spirals
(almost circles) for $r > r_{\rm pot}\approx r_{\rm ISCO}$ and very wide spirals
(almost a radial fall) for  $r < r_{\rm pot}$. In this case,
there is a sharp transition from almost circular motion to almost
radial free-fall that clearly defines the variability edge as
$r_{\rm var} = r_{\rm ISCO}.$ For higher accretion rates, the
fluid trajectories are wide open spirals in the whole inner region
of the flow and the variability edge makes no sense.}
\label{figure-spirals}
\end{sidewaysfigure}
The variability quality factor $Q$ may be estimated by,
\begin{equation}
\frac1Q=\frac{\Delta
\Pi}{\Pi}=
\frac{\Delta\Omega}{\Omega}=\frac1{\Omega}\der\Omega{r}\Delta
r
= 2\pi\frac1{\Omega^2}\der\Omega{r}\frac{u^r}{u^t}
\label{quality-factor-definition},
\end{equation}
where $u^r/u^t = dr/dt$ and $u^r$ and $u^t$ are contravariant
components of the four velocity. The period is given in terms of the orbital
angular velocity by $\Pi = 2\pi/\Omega$. Using the relations (see
Eqs.~\ref{e.relcorr} for the explanation of the notation used)
\begin{eqnarray}
u^r &=& \frac V{\sqrt{1-V^2}}\frac{\sqrt{\Delta}}r,\nonumber \\
u^t &=& \frac{\gamma\sqrt A}{r\sqrt\Delta}=\frac{\sqrt
A}{r\sqrt\Delta}\frac1{\sqrt{(1-V^2)(1-({\tilde V}^\phi)^2)}}
\label{contravariant-velocities},
\end{eqnarray}
where $V$ is the radial velocity measured by an observer
co-rotating with the fluid, one obtains
\begin{equation}
Q = \frac{1}{2\pi}\left\vert\der{\log\Omega}{\log
r}\right\vert^{-1} \left\vert\frac{{\bar
V}^\phi}{V}\right\vert\,f^*(a_*, r),
\label{quality-final}
\end{equation}
where
\begin{eqnarray}
f^*(a_*, r) &\equiv& \frac{r^3}{\sqrt{\Delta A}}
= \left( 1 - X - X^2a_{*}^2\,(a_{*}^2 + 1) - X^5\,a_{*}^4
\right)^{-1/2}, \\
{\bar V}^{\phi} &=& \frac{V^{\phi}}{\sqrt{1-(V^\phi-\frac{2Mar}A\tilde
R)^2}},\nonumber \\
V^\phi&=&\tilde \Omega \tilde R,\nonumber 
\label{quality-two-functions}
\end{eqnarray}
and $X=2M/r$. From their definitions, it is clear that $\Delta A > 0$ outside the
BH horizon. We note that in the Newtonian limit it is $X \ll 1$
and one has $f^*(a_*, r)=1$. In this limit, $V$ and ${\bar V}^{\phi}$
are the radial and azimuthal components of velocity, and Eq.~\ref{quality-final} takes its obvious Newtonian form.
%
%
%
\begin{figure}[h]
\centering
\includegraphics[width=.7\textwidth,
angle=0]{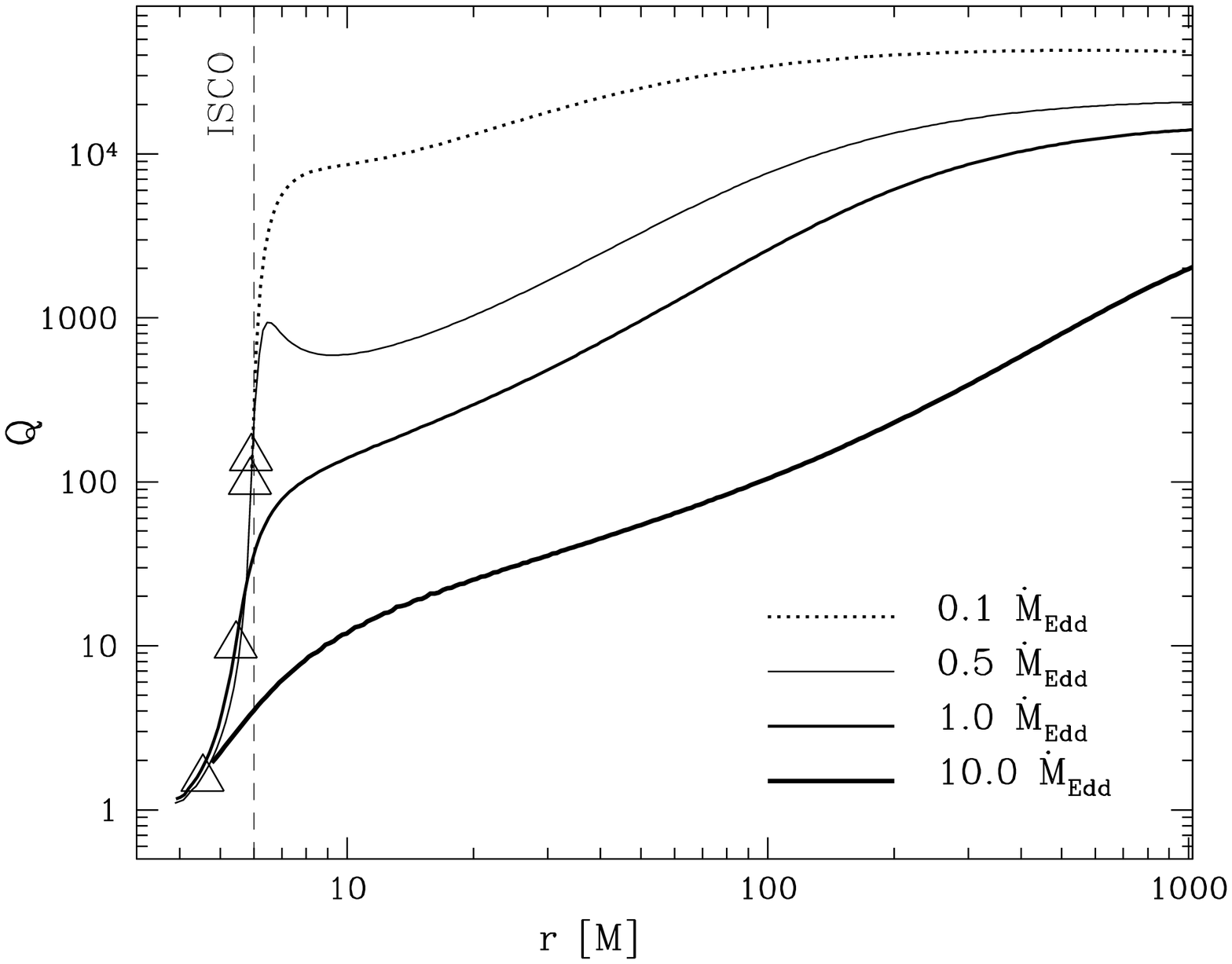}
\caption{The quality factor $Q$ profiles for different accretion
rates and $\alpha=0.01$. Triangles show $r_{\rm pot}$ for each rate.
The vertical dashed line denotes the location of ISCO.}
\label{figure-quality-factor-Q}
\end{figure}

The behavior of the quality factor $Q$ is shown in Fig.~\ref{figure-quality-factor-Q}. 
Profiles for four accretion rates
are drawn. As Fig.~\ref{figure-spirals} shows, the lower accretion
rate, the smaller radial velocity component, hence the
quality factor $Q$ in general increases with decreasing accretion
rate. For the lowest values of $\dot m$, a rapid drop is visible
at ISCO corresponding to the change in the nature of the flow
(gas enters the free-fall region below ISCO). For higher accretion
rates, this behavior is suppressed as the trajectories become
wide open spirals well outside ISCO.

We note that our definition given in Eq.~\ref{quality-factor-definition} of the
quality factor $Q$, essentially agrees with a practical definition
of the variability quality factor $Q_0$ defined by observers with
the help of the observationally constructed Fourier variability
power spectra, $I(\nu)$. Here $I(\nu)$ is the observed variability
power (i.e. the square of the observed amplitude) at a particular
observed variability frequency $\nu$. Any observed {\it quasi}
periodic variability of the frequency $\sim \nu_0$ can clearly be seen in the
power spectrum as a local peak in $I(\nu)$, centered on a particular
frequency $\nu_0$. The half-width $\Delta \nu$ of the peak defines
the variability quality factor by $Q_0 = \nu/\Delta \nu_0$.

Quasi-periodic variability of kHz frequencies, called kHz QPO,
is observed from several low-mass neutron star and BH
binaries. In a pioneering and important piece of research, \citem{barr-2005}
carefully measured the quality factor for a particular source in
this class (4U 1608-52) and found that $Q_0 \sim 200$, i.e. that
the kHz signals are very coherent. They argued that $Q_0 \sim 200$ cannot
be due to kinematic effects in the orbital motion of hot spots, clumps,
or other similar features located at the accretion disk surface,
because these features are too quickly sheared out by the
differential rotation of the disk (see also \citem{bath-1974,
pringle-1981}). They also argued that although coherent vortices
may survive much longer times at the disk surface
(e.g., \citem{abr-nature-1992}), if they participate in the inward
radial motion, the observed variability $Q_0$ cannot be high. Our
results shown in Fig.~\ref{figure-quality-factor-Q} illustrate
and strengthen this point. We also agree with the conclusion
reached by \citem{barr-2005} that the observational evidence
against orbiting clumps as a possible explanation of the
phenomenon of kHz QPO, seems to indicate that this phenomenon is
probably caused by the accretion disk global
oscillations\footnote{\citem{barr-2005} also found how $Q_0$ varies
in time for each of the two individual oscillations in the
``twin-peak QPO''. This places strong observational constraints on
possible oscillatory models of the twin peak kHZ QPO; see also
\citem{boutelier-2010}.}. For excellent reviews of the QPO
oscillatory models, we refer to \citem{wagoner-1999} and \citem{kato-2001}.

Although clumps, hot-spots, vortices or magnetic flares cannot
explain the coherent kHz QPOs with $Q_0 \sim 200$, they
certainly are important in explaining the continuous, featureless Fourier
variability power spectra (see e.g., \citem{abr-bao-1991, schnittman05, pech-2008} and references quoted
there). Our results shown in
Fig.~\ref{figure-quality-factor-Q} indicate that: (i) at low
accretion rates, a sharp high-frequency cut-off in $I(\nu)$ may be
expected at about the ISCO frequency; (ii) at high accretion rates,
there should be no cut-off in $I(\nu)$ at any frequency; and (iii) the
logarithmic slope $p = (d\ln I/d\ln \nu)$ should depend on ${\dot
m}$.


\subsection{Stress edge}
\label{section-stress}
%

The Shakura-Sunyaev model {\it assumes} that there is no torque at
the inner edge of the disk, which in this model coincides with
ISCO. Slim disk model {\it assumes} that there is no torque at the
horizon of the BH. It makes no assumption about the torque at
the disk inner edge, but calculations {\it prove} that the torque
is small there.

\begin{figure}
\centering
 \subfigure
{
\includegraphics[height=.7\textwidth, angle=270]{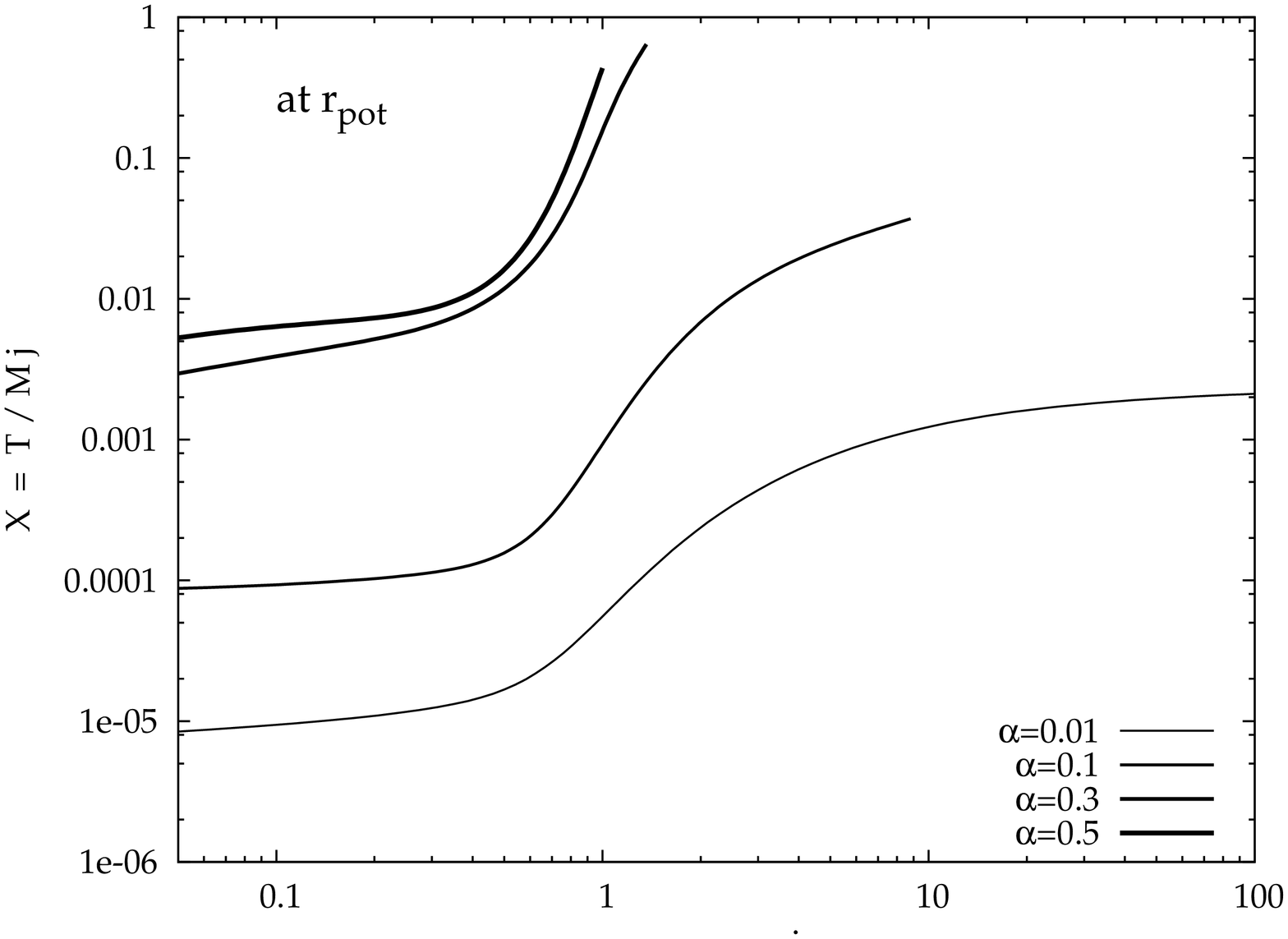}
}
\\
\vspace{-.02\textwidth}
 \subfigure
{
\includegraphics[height=.7\textwidth, angle=270]{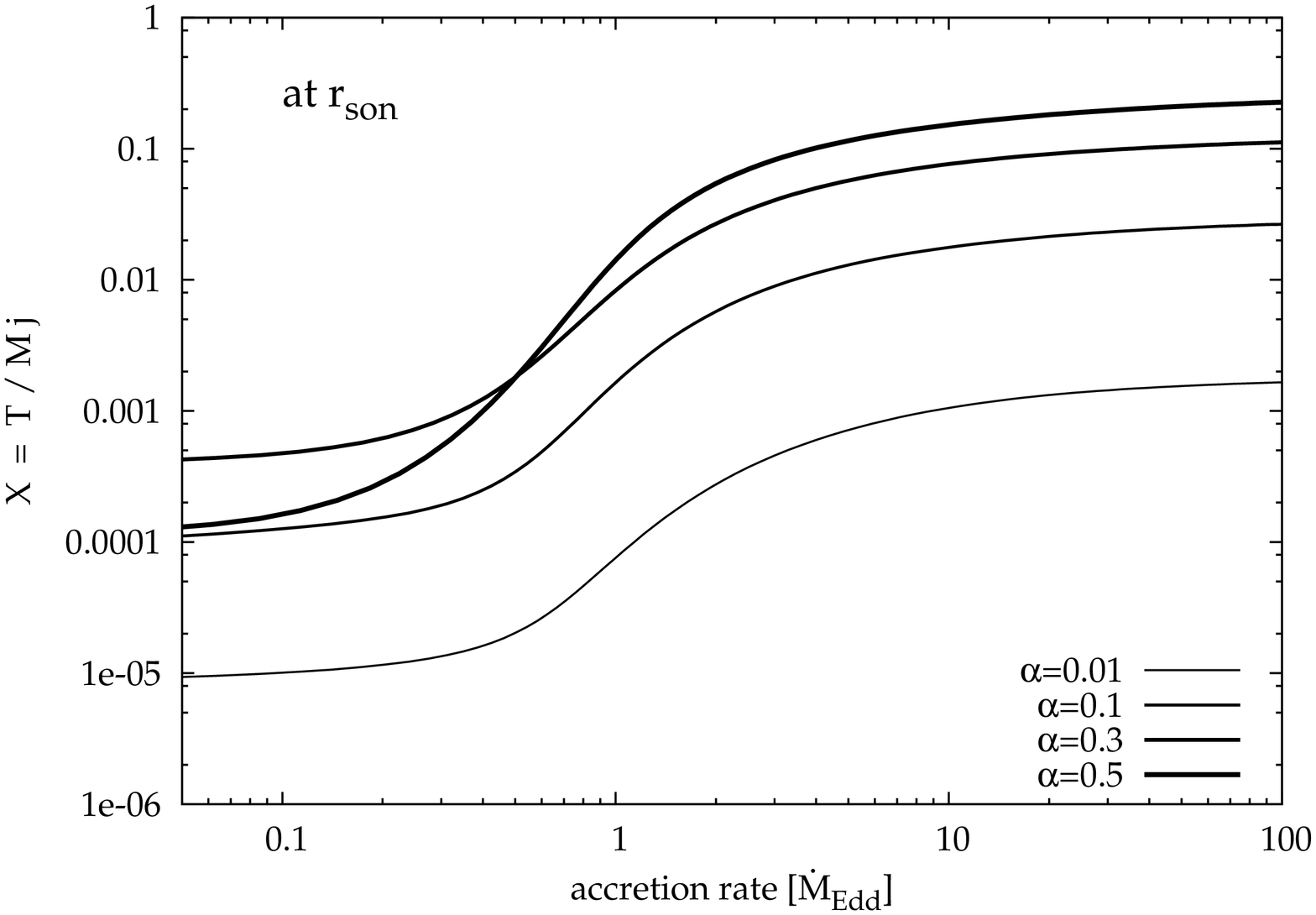}
}
\caption{Ratio of the angular momentum flux caused by torque to
the flux caused by advection calculated at $r_{\rm pot}$ (top) and
$r_{\rm son}$ (bottom panel) versus mass accretion
rate for a number of values of $\alpha$ and $a_*=0$. The $r_{\rm pot}$
profiles for high viscosities terminate when the disk enters the
Bondi-like regime.}
\label{fig:inner-torque}
\end{figure}

The zero-torque at the horizon is consistent with the small torque
at the inner edge of slim disks, as Fig.~\ref{fig:inner-torque}
shows.
%
%
The Figure presents the relative importance of the torque ${\cal
T}$ by comparing it with the ``advective'' flux of angular
momentum ${\dot M}j$ ($\dot J=\dot Mj+{\cal T}$). For
the viscosity parameter $\alpha$ smaller than about $0.01$, the
ratio $\chi = {\cal T}/{\dot M}j$ at both $r_{\rm pot}$ and $r_{\rm son}$
is smaller than $0.01$ even for highly super-Eddington accretion
rates, and for low accretion rates the ratio is vanishingly
small, $\chi \approx 10^{-5}$. For high viscosity, $\alpha = 0.5$,
the ratio is small for low accretion rates, $\chi <
10^{-2}$ and smaller than $0.1$ for
super-Eddington accretion rates (calculated at the sonic radius, as
the disk enters the Bondi-like regime at these high accretion
rates). For two highest viscosity models ($\alpha =0.3$ and $0.5$) the $\chi$ parameter
calculated at $r_{in}$ exceeds $0.1$ for accretion rates close to the disk-Bondi limiting value what
may suggest the no-stress at the disk inner edge assumption is violated for such high viscosities.

%
%
\begin{figure}[h]
\centering
\includegraphics[width=.7\textwidth, angle=0]{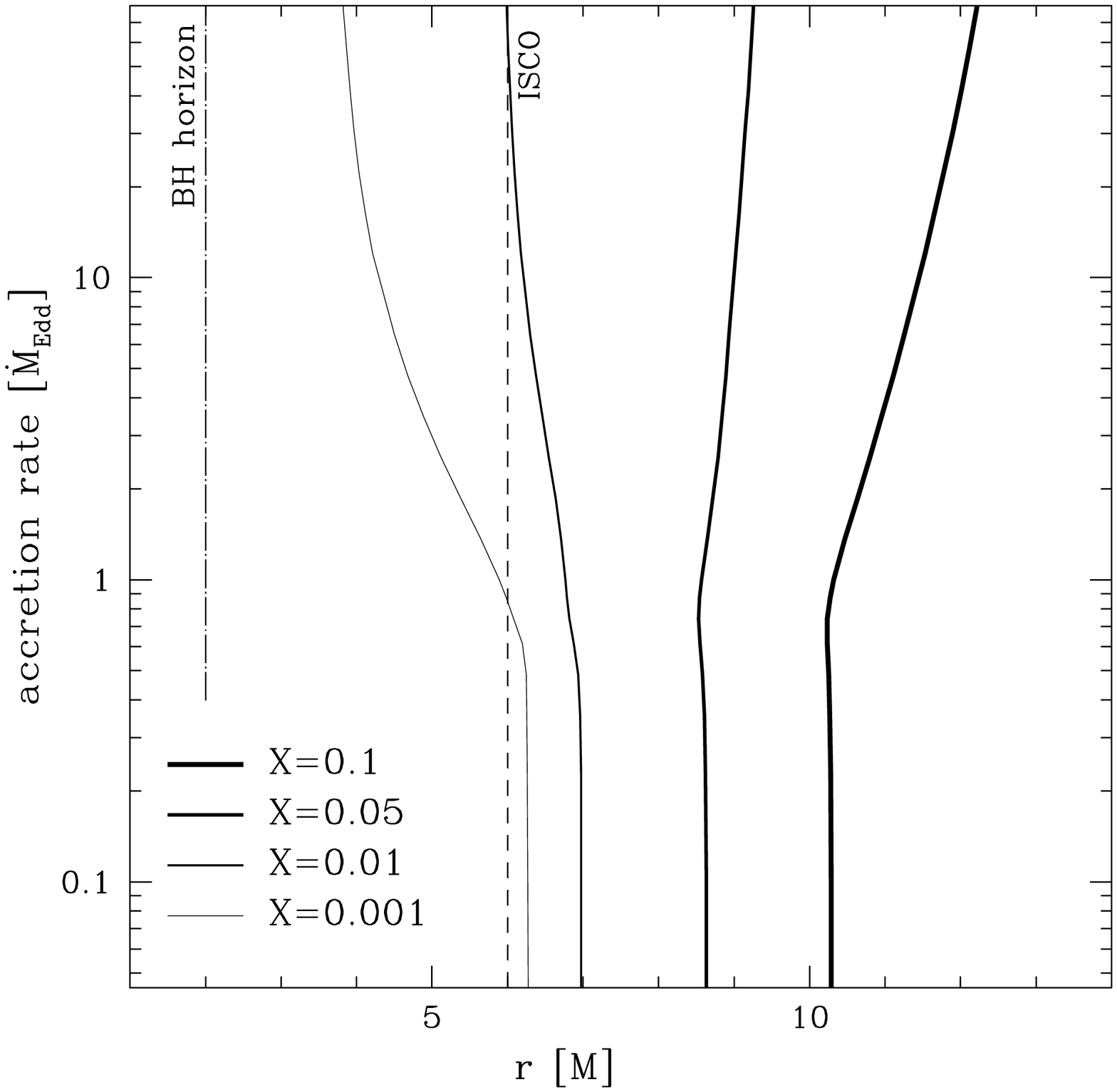}
\caption{Profiles of $r_{\rm str}$ defined as the radius with given
value of the torque parameter $\chi$ for $\alpha=0.01$. BH horizon and ISCO are also
shown with dot-dashed and dashed lines, respectively.}
\label{fig:rstr}
\end{figure}

To define the \textit{stress inner edge} $r_{\rm str}$, one has to
specify the characteristic value of the torque parameter ${\chi}$.
Profiles of $r_{\rm str}$ for a few values of ${\chi}$ and $\alpha=0.01$ are shown in
Fig.~\ref{fig:rstr}. The stress edge for ${\chi} \rightarrow 0$ is
located at ISCO for low accretion rates. When the accretion rate
exceeds $\sim 0.6\dot M_{\rm Edd}$, the edge departs from ISCO and moves
closer to BH approaching its horizon with increasing $\dot m$.
The behavior of $r_{\rm str}$ profiles for higher ($\gtrsim 0.1$) values
of ${\chi}$ is different - they move away from the BH as the
angular momentum profiles become flatter with increasing accretion
rates (compare Fig.~\ref{f.l.a0}).

By pushing the MHD numerical
simulations to their limits, \citem{shafee-06} and \citem{pennaetal-10}
 calculated a thin,
$H/r \lesssim 0.1$, disk-like accretion flow, and demonstrated that its 
inner edge torque is in fact small.
%
\subsection{Radiation edge}
\label{section-radiation}
\begin{figure}[h]
\centering
\includegraphics[width=.7\textwidth, angle=0]{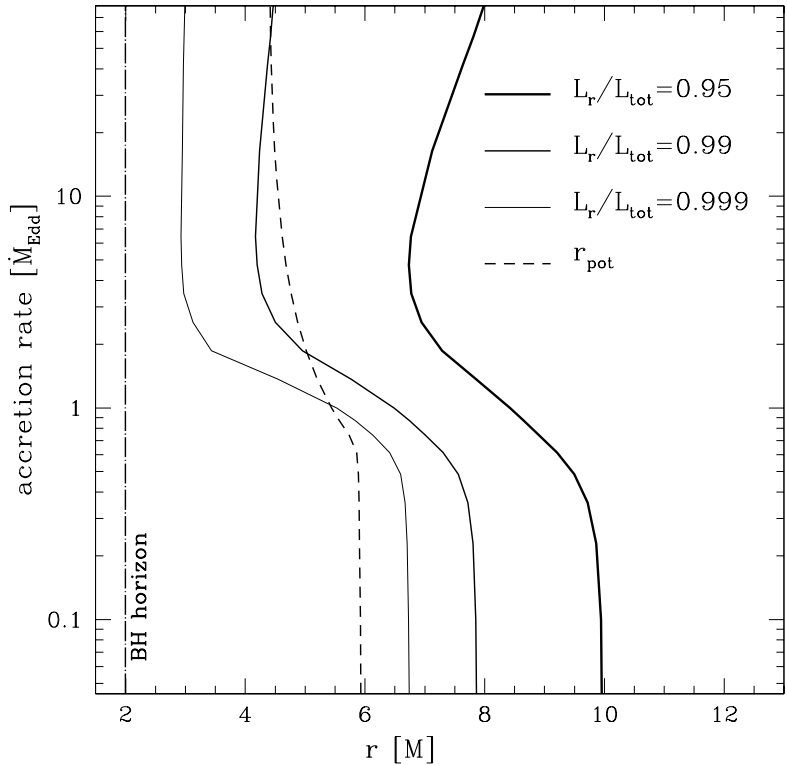}
\caption{``Luminosity edges'' defining the inner radii of the area emitting
a given amount of the total disk radiation. The lines are drawn
for 95\%, 99\%, and 99.9\% of the total emission. The dashed line
shows the location of the \textit{potential spout} inner edge $r_{\rm pot}$.
The gravitational suppression of the radiation has been taken into
account.}
\label{fig:radiation-radius}
\end{figure}

%
As discussed in the previous section, the torque at $r_{pot} <
r_{\rm ISCO}$ is small, but non-zero and therefore there is also
orbital energy dissipation at radii smaller than ISCO. Thus,
some radiation does originate in this region. Moreover, the heat advected
with the flow may be radiated away in this region. As a result, the radiation edge $r_{\rm rad}$
is not expected to coincide with ISCO. In Fig.~\ref{fig:radiation-radius}, we present profiles
of $r_{\rm rad}$ defined as the radii limiting area emitting a given
fraction of the disk total luminosity. For low accretion rates
($<0.6\dot M_{\rm Edd}$), the disk emission terminates close to ISCO as the
classical models of accretion disks predict. The locations of the
presented $r_{\rm rad}$ are determined by the regular Novikov \&
Thorne flux radial profile. For higher accretion rates, the disk
becomes advective and the maximum of the emission moves
significantly inward. As a consequence of the increasing rate of
advection (and resulting inward shift of $r_{\rm pot}$), the efficiency
of accretion drops down.

We emphasize that the location of the radiation edge is
{\it not} determined by the location of the stress edge (as some
authors seem to believe), but by the significant
advection flux bringing energy into the region well below ISCO.

%
%

\subsection{Reflection edge}
\label{section-reflection}
%

The iron K$_{\alpha}$ fluorescent line is an
observed characteristic feature of many sources with BH
accretion disks \citepm{Miller-2006,Remillard-2006}. The intensity and the shape
of this line depends strongly on the physical conditions close to
the inner edge.
This has been discussed by many authors, including
\citem{rey-2008} who gave three conditions for line formation: (i)
the flow has to be Thomson-thick in the vertical direction; (ii)
disk has to be irradiated by an external source of X-rays (hard X-ray
irradiation plays a
crucial role in the process of fluorescence and changes the
ionization degree of matter); and (iii) the ionization
state should be sufficiently low (iron cannot be fully ionized).

Nevertheless, since fluorescent iron-line emission depends on many aspects, such as the
energy distribution of illuminating photons, temperature, ionization state, and density 
of the emitting matter as well as iron abundance, there is no 
obvious condition for the reflection edge (defined as the minimal radius where the
reflection features originate). Additional computations of reflection models 
show that for some set of parameters
iron fluorescent line may arise even from Thomson-thin matter \citepm{dumont-2002}.
In this paper, we assume that the reflection edge is connected to the condition 
\begin{equation}
\label{effective-depth}
\tau_{\rm eff} = \sqrt{\tau_{\rm abs} (\tau_{\rm abs}+\tau_{\rm es})} > 1.
\end{equation}
However, one has to keep in mind that the effective optical depth at the
iron line band may be much larger than the above, frequency-averaged value.

\begin{figure}
\centering
 \subfigure
{
\includegraphics[height=.65\textwidth, angle=270]{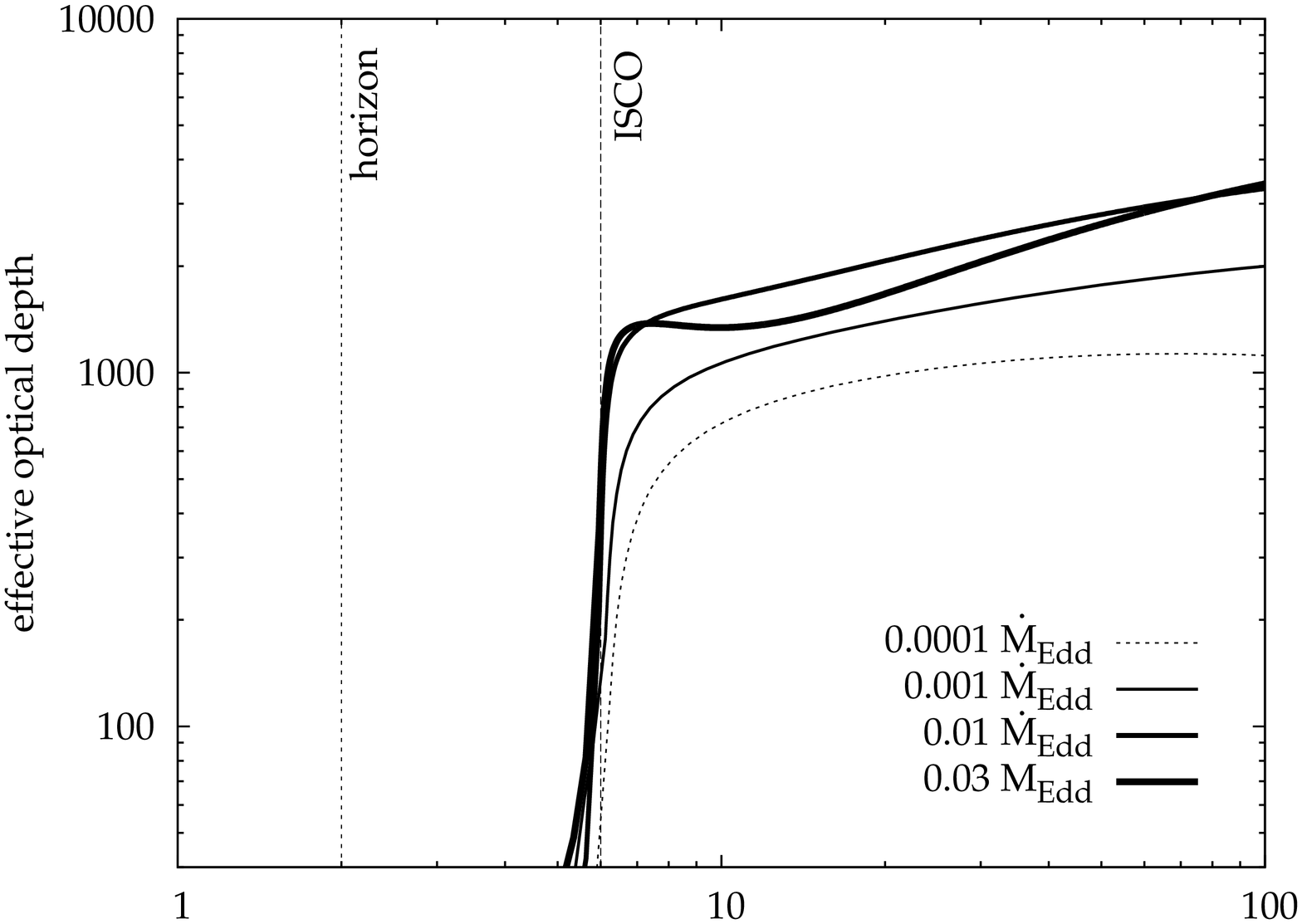}
}
\\
\vspace{-.05\textwidth}
 \subfigure
{
\includegraphics[height=.65\textwidth, angle=270]{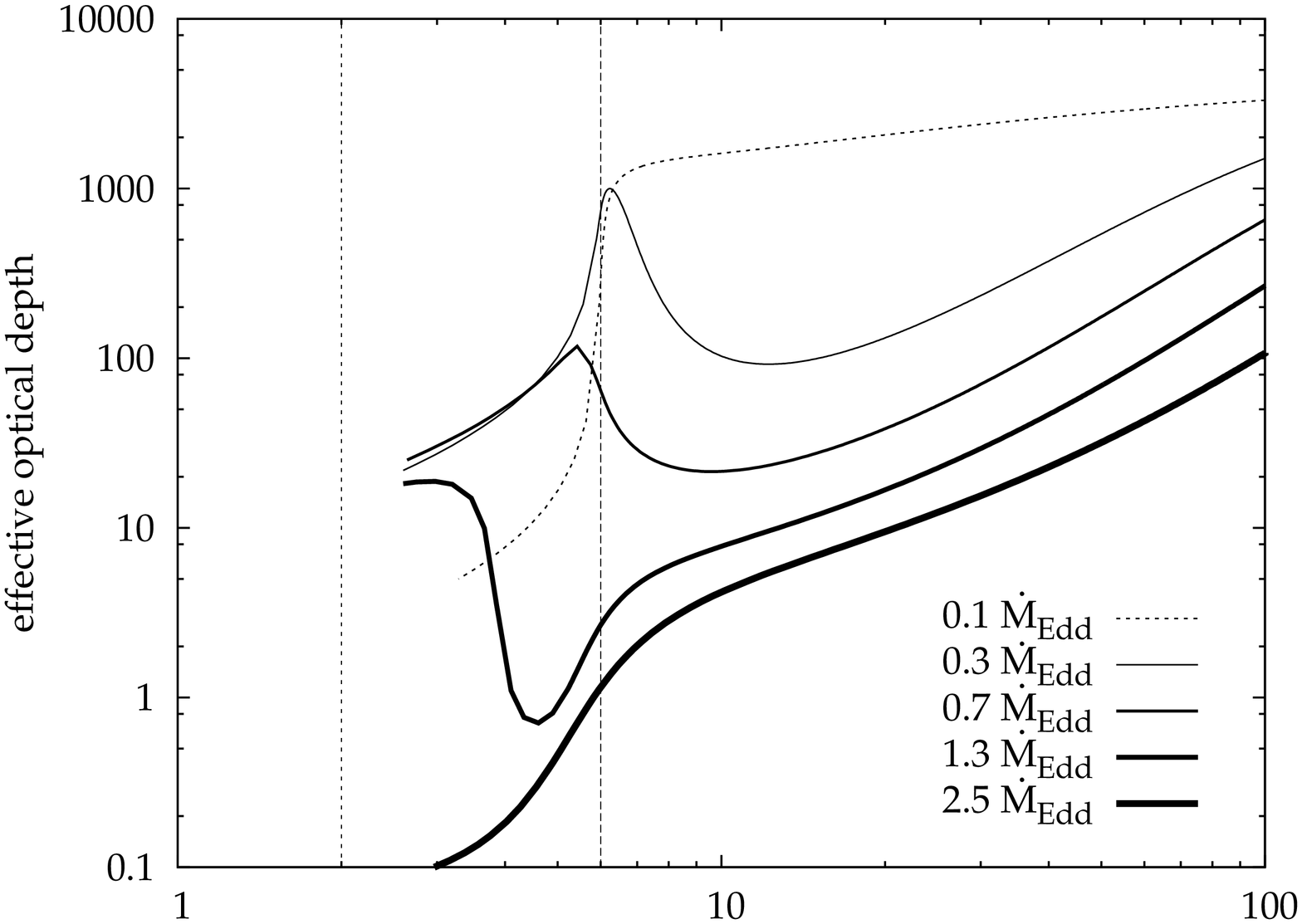}
}
\\
\vspace{-.05\textwidth}
 \subfigure
{
\includegraphics[height=.65\textwidth, angle=270]{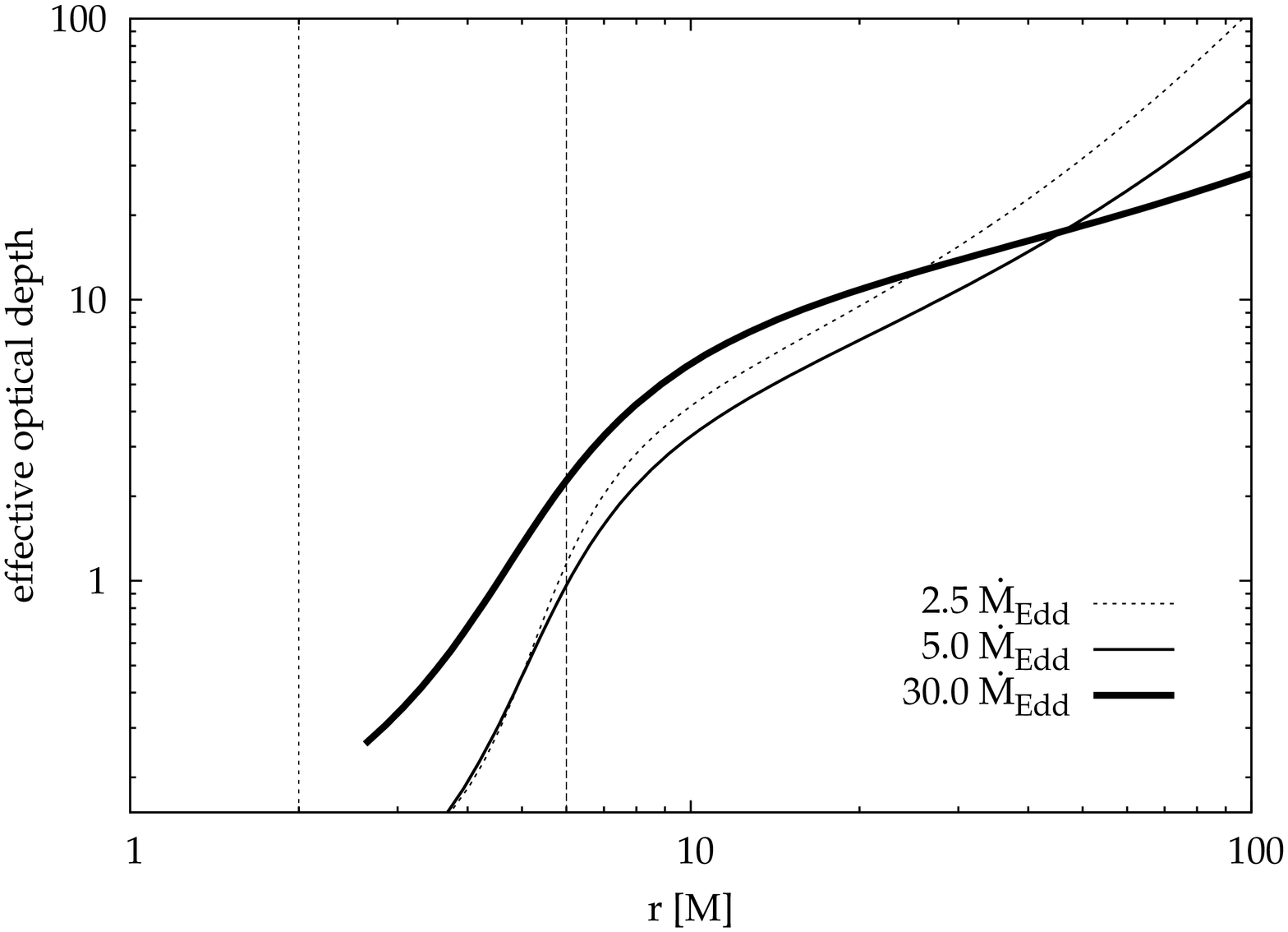}
} \caption {Profiles of the effective optical depth $\tau_{\rm eff}$
for $\alpha=0.01$ and $a_* = 0$ in three different regimes of
accretion rate. Vertical lines denote the locations of the BH horizon
(dotted) and ISCO (dashed). Three types of behavior of
$\tau_{\rm eff}$ can be seen: {\it sharp drop} at ISCO at the lowest
accretion rates, {\it maximum} near ISCO for moderate accretion
rates, and {\it monotonic} at all radii for the highest accretion rates.
}
\label{fig:tau_a0}
\end{figure}

In Fig.~\ref{fig:tau_a0}, we
show profiles of the effective optical depth $\tau_{\rm eff}$
in different regimes of accretion rates for $\alpha=0.1$ and $a_*
= 0$. Three characteristic types of their behavior are shown:
\textit{sharp drop}, \textit{maximum} and \textit{monotonic} at
the top, middle, and bottom panels, respectively. The behavior of
different values of $\alpha$ and $a_*$ is qualitatively similar
(but not quantitatively as in general $\tau_{\rm eff}$ increases
with decreasing $\alpha$). The top panel of the figure, corresponding to the lowest
accretion rates, shows a {\it sharp drop in} $\tau_{\rm eff}$ near
ISCO. The same behavior was noticed previously e.g., by
\citem{rey-2008}. The drop may clearly define the inner
reflection edge $r_{\rm ref}\approx r_{\rm ISCO}$, limiting the radii
where formation of the fluorescent iron line is prominent. The
middle panel, corresponding to moderate accretion rates, shows a
{\it maximum in} $\tau_{\rm eff}$ near ISCO. The non-monotonic
behavior is caused by the regions of moderate radii
outside ISCO being both radiation pressure and scattering dominated.
We note, that the top of the maximum of $\tau_{\rm eff}$ remains close to
ISCO for a range of accretion rates, but for accretion rates
higher than $0.6\dot M_{\rm Edd}$, it moves closer to the black
hole with increasing ${\dot m}$ as the disk emission profile
changes due to advection. The bottommost panel corresponds to
super-Eddington accretion rates. The profiles are {\it monotonic
in} $\tau_{\rm eff}$ and define no characteristic inner reflection
edge. Close to the BH, these flows are effectively optically thin
reaching $\tau_{\rm eff}=1 $ at about a few tens of gravitational
radii.

When the effective optical depth of the flow becomes less then unity,
opacities becomes invalid. In these
cases, full radiative transfer through accretion disk atmospheres
has to be considered
(e.g., \citem{davisomer05,rozanskamadej08}). Nevertheless,
our results allow us to estimate roughly how far from the black
hole the iron line formation is most prominent, assuming that the disk
is uniformly illuminated by an exterior X-ray source. For accretion rates
lower than $0.6\dot M_{\rm Edd}$, the reflection edge is located
very close to ISCO and we may identify the shape of the iron line
with the gravitational and dynamical effects of the ISCO. In the case of higher but
sub-Eddington accretion rates, the maximum of the effective
optical depth is located inside the ISCO, which may possibly allow us to study
extreme gravitational effects on the iron line profile. However, the
assumption that the line is formed at the ISCO is no longer satisfied.
The super-Eddington flows have smooth and monotonic profiles
of effective optical depth. Therefore, the reflection edge cannot
be uniquely defined and no relation between the shape of the
fluorescent lines and ISCO exists. Finally, we note that these lines may be
successfully modeled by clumpy absorbing material and have nothing to do
with relativistic effects (see e.g., \citem{Milleretal-2009} and references therein).
The role of the ISCO in determining the
shape of the Fe lines was also questioned in the past (based on different
reasoning) by \citem{ReynoldsBegelman-97}, whose arguments were then refuted by
\citem{Youngetal-98}.

%
%

\subsection{Summary}

We have addressed the inner edge issue by discussing the behavior of six
differently defined ``inner edges'' of slim accretion disks around
a Kerr BH. We have found that the slim disk inner edges
behave very differently than the corresponding \citem{shakura-73} and
\citem{nt} ones. The differences are qualitative and become 
important for the same range of luminosities independently of the BH spin.
Even at
moderate luminosities, $L/L_{\rm Edd} \gtrsim 0.6$,
there is no unique inner edge. Differently defined edges are located at
different places. For nearly Eddington luminosities, the
differences are huge and the notion of the inner edge loses all
practical significance. We summarize the properties and locations of the six inner edges
in Table \ref{table:1}. 
%
\begin{sidewaystable*}

\label{table:1}      
\centering                           
\begin{tabular}[width=1.0\textheight]{l m{4.cm} m{2.5cm} m{2.cm} m{3.cm} m{3.cm} m{3cm} m{.0001cm}}        
\hline\hline                 
  & \centering$r_{\rm pot}$ & \centering$r_{\rm son}$ & \centering$r_{\rm var}$ & \centering$r_{\rm str}$ & \centering$r_{\rm rad}$ & \centering $r_{\rm ref}$ &\\
\hline                        
&&&&&&\\
 $L/L_{\rm Edd}\lesssim 0.6$
&
 \multicolumn{6}{c}{
$ r_{\rm in} \approx
r_{\rm pot} \approx
r_{\rm son} \approx
r_{\rm var} \approx
r_{\rm str} \approx
r_{\rm rad} \approx
r_{\rm ref} \approx
r_{\rm ISCO}
$
}
\\&&&&&&\\
\hline                        
 $L/L_{\rm Edd}\gtrsim 0.6$ &
\begin{center}
for $\alpha\lesssim0.1$ moves inward with increasing $\dot m$ down to $\sim r_{\rm mb}$;

~

for $\alpha\gtrsim0.1$ and sufficiently high $\dot m$ disk enters the Bondi regime --- undefined
\end{center}&
\begin{center}
departs from ISCO;

~

for $\alpha\ll0.1$:

$r_{\rm son}\approx r_{\rm mb}$;

~

for $\alpha\gtrsim0.2$:

$r_{\rm son}>r_{\rm ISCO}$ \end{center}
&
\begin{center}
undefined\end{center}
&
\begin{center}
moves inward with increasing $L$ down to BH horizon.\end{center}
&
\begin{center}
moves inward with increasing $L$ down to BH horizon.\end{center}
&
\begin{center}
for $0.6\lesssim L/L_{\rm Edd}\lesssim1.0$: $r_{\rm ref}<r_{\rm ISCO}$

~

for $L/L_{\rm Edd}\gtrsim1.0$:

undefined\end{center}
&
\\
\hline                                    
\end{tabular}

\caption{Inner edges of a slim accretion disk: 
$ r_{\rm pot}$ - potential spout,
$r_{\rm son}$ - sonic radius,
$r_{\rm var}$ - variability edge,
$r_{\rm str}$ - stress edge,
$r_{\rm rad}$ - radiation edge,
$r_{\rm ref}$ - reflection edge.}             
\end{sidewaystable*}

\section{Spectra}

The rate of viscous heating at given radius depends (in the $\alpha$ approximation) on the disk pressure and the rate of shear. These parameters change with radius what implies that the heating (and the corresponding cooling rate) significantly depends on radius too (profiles of the emitted flux of energy are presented e.g., in Fig.~\ref{f.flux}). In the standard framework (assuming e.g., large optical thickness), the gas is assumed to radiate locally as a blackbody with the characteristic effective temperature corresponding to the amount of energy released in the form of the radiative flux. Since this quantity changes with radius, the outcoming spectrum of an accretion disk is a superposition of the black body spectra with a wide range of effective temperatures (multicolor blackbody).

Let us consider the spectrum of a radiatively efficient disk. The outcoming flux of energy is given by the corresponding formulae of \citem{shakura-73} and \citem{nt}. For the non-relativistic case it is given by (Eq.~\ref{sh_flux1}),
\be
F=\sigma T^4_{\rm eff}=\frac{3GM\dot M}{8\pi r^3}\left(1-\sqrt{\frac{r_{\rm in}}{r}}\right).
\label{e.spectra.F}
\ee
It is assumed that at given radius disk radiates the blackbody radiation corresponding to the effective temperature $T_{\rm eff}$:
\be
B_\nu(r)=\frac{2h}{c^2}\frac{\nu^3}{e^{h\nu/k_BT_{\rm eff}(r)}-1}.
\label{e.Bnu}
\ee
The observed spectrum may be calculated by integrating the specific intensities ($I_\nu=B_\nu$) over the whole surface of the disk\footnote{This formula neglects the gravitational g-factor. The precise way of calculating spectrum is given in Section~\ref{s.slimbb}}:
\be
S_\nu(r)=\int_{\rm disk} I_\nu d\Omega=\frac{\cos i}{D^2}\int_{r_{\rm in}}^{r_{\rm out}}B_\nu 2\pi r dr,
\ee
where $i$ is the inclination angle and $D$ is the distance to the observer.

\begin{figure}[h]
\centering
\includegraphics[width=.6\textwidth, angle=0]{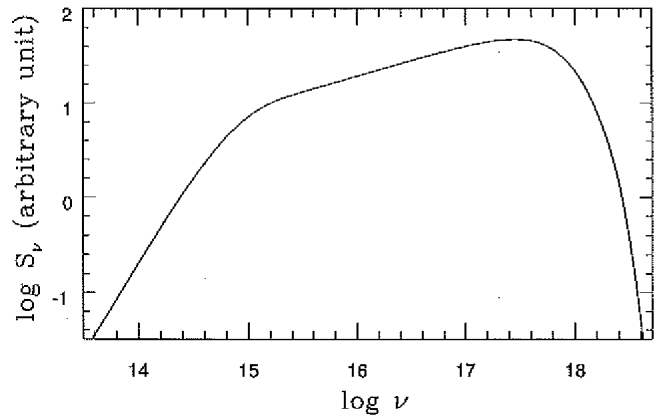}
\caption{Typical spectrum of a thin accretion disk ($p=3/4$) in a soft X-ray transient. Figure after \citem{KatoBook}.}
\label{f.spectrum.kato}
\end{figure}

Useful quantitative results may be obtained if we neglect the variations of $T_{\rm eff}$ near to the inner edge which have little impact on the shape of the total spectrum. Let us then approximate the radial profile of the effective temperature with
\be
T_{\rm eff}=T_{\rm eff,in}\left(\frac{r}{r_{\rm in}}\right)^{-p},\ee
where $T_{\rm eff,in}$ is the effective temperature at the inner edge ($r_{\rm in}=r_{\rm ISCO}$ for thin disks). According to Eq.~\ref{e.spectra.F} the exponent $p$ for radiatively efficient disks is $p=3/4$, while for advection-dominated flows (e.g., slim disks in the limit of high accretion rates) it is equal $p=1/2$. Under this approximation, the spectral flux is given by,
\bea\nonumber
S_\nu&=&\frac{\cos i}{D^2}2\pi\left(-\frac{r_{\rm in}^2}{p}\right)T_{\rm eff,in}^{2/p}\int_{T_{\rm eff,in}}^{T_{\rm eff,out}}B_\nu T_{\rm eff}^{-(2/p)-1}dT_{\rm eff}\\
&=&\frac{\cos i}{D^2}\frac{4\pi h}{c^2}\frac{r_{\rm in}^2}{p}\left(\frac{k_B T_{\rm eff,in}}{h\nu}\right)^{2/p}\nu^3\int_{x_{\rm in}}^{x_{\rm out}}\frac{x^{(2/p)-1}}{e^x-1}dx,
\eea
with
\be x_{\rm in}=\frac{h\nu}{k_B T_{\rm eff,in}}\ee
\be x_{\rm out}=\frac{h\nu}{k_B T_{\rm eff,in}}\left(\frac{r_{\rm out}}{r_{\rm in}}\right)^p.\ee
The observed spectrum consists therefore of a power-law component
\be S_\nu\propto \nu^{3-(2/p)}\ee
limited by the low- and high-energy cutoffs:
\be \frac{k_BT_{\rm eff,in}}{h}\left(\frac{r_{\rm in}}{r_{\rm out}}\right)^p \ll \nu \ll \frac {k_B T_{\rm eff,in}}{h}.\ee
For a standard thin disk with $p=3/4$, the spectrum is given by $S_\nu\propto \nu^{1/3}$, while for advection dominated accretion ($p=1/2$) it is $S_\nu\propto \nu^{-1}$. In Figure~\ref{f.spectrum.kato} we show an exemplary spectrum for $p=3/4$ case. Its middle part ($\nu=10^{15}\div10^{18}\,\rm Hz$) follows the $S_\nu\propto \nu^{1/3}$ profile while the low and high energy tails follow the corresponding limits of the Planck function (Eq.~\ref{e.Bnu}) for the effective temperatures at $r_{\rm out}$ and $r_{\rm in}$, respectively.

\subsection{Spectral hardening}
\label{s.hardening}
Real accretion disks do not radiate as a perfect blackbody.
The emission is affected by such factors as the absorption opacity, density structure 
of the plasma, optical depth, rate of electron scattering, etc. Out of these effects most 
important in case of disks in X-ray binaries is the Compton scattering on hot electrons in 
the disk atmosphere. It distorts
 the spectrum towards temperatures 
that are higher than the effective temperature at given radius. 
However, when integrated over the disk, the total 
emergent disk spectrum can still be approximated to a good accuracy by a multicolor disc blackbody, but with different characteristic
temperatures \citepm{shimuratakahara-95, merlonietal-00}
\be T=f_{\rm c}T_{\rm eff},\ee
where $f_{\rm c}$ is called the color temperature correction, or spectral hardening factor, since it is always greater then unity. In general, $f_{\rm c}$ is a function of disk luminosity and its particular value depends on the disk structure close to its photosphere.

To calculate the proper, hardened spectrum of an accretion disk one has to perform full radiative transfer calculations. \citem{davisomer05} calculated disk atmosphere models (BHSPEC) spanned on a grid of effective temperatures, effective gravity and surface densities relevant to disks in X-ray binaries. In the spectral model described in the following section, we either assume some particular value of the hardening factor $f_{\rm c}$ or take the local spectra directly from the BHSPEC model.

\begin{figure}[h]
\centering
\includegraphics[width=.8\textwidth, angle=0]{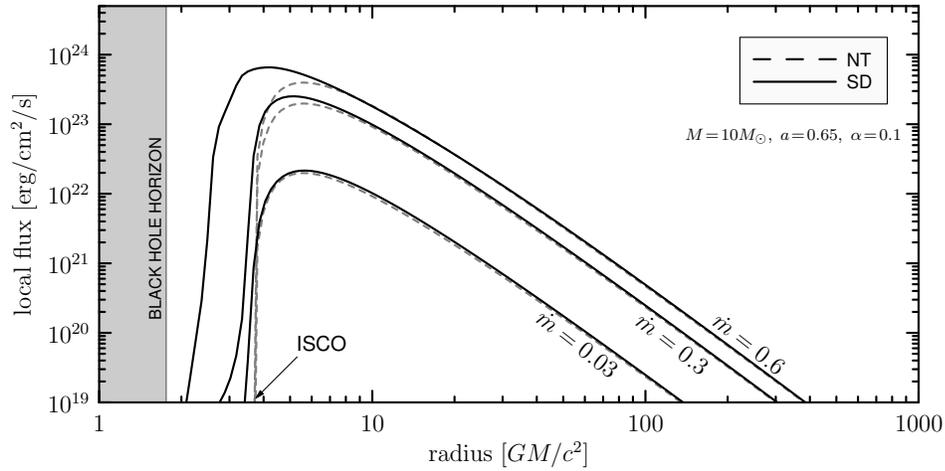}
\caption{Examples of local flux profiles emitted from the surface of slim disk (solid) and \citem{nt} model (dashed lines) for BH spin $a_*=0.65$. Fluxes are plotted for three different mass accretion rates $\dot m=0.03$, $0.3$ and $1.0$.}
\label{f.plotflux}
\end{figure}

In Fig.~\ref{f.plotflux} we present emission profiles of slim disk solutions for $a_*=0.65$ and three accretion rates: $\dot m=0.03,\,0.3$ and $0.6$. For the highest accretion rate (corresponding to luminosity $\sim 1.0L_{\rm Edd}$) the inward shift of emission caused by advection is visible. The dashed lines present profiles of the relativistic standard thin disk solutions \citepm{nt}. The following figures (Figs.~\ref{f.spectra.hf17} and \ref{f.spectra.bhspec}) present spectra calculated basing on these emission profiles. The spectra were obtained using ray-tracing routines developed by \citem{bursa-05} assuming either constant value of the hardening factor ($f_{\rm c}=1.7$, Fig.~\ref{f.spectra.hf17}) or using BHSPEC atmospheric models (Fig.~\ref{f.spectra.bhspec}). The slim disk solutions give in general slightly harder spectra, as they reach higher maximal effective temperatures in comparison to \citem{nt} model (Fig.~\ref{f.plotflux}). Detailed discussion of the slim disk spectra will be given in \citem{bursa-slimbb}.

\begin{figure}[h]
\centering
\includegraphics[width=.8\textwidth, angle=0]{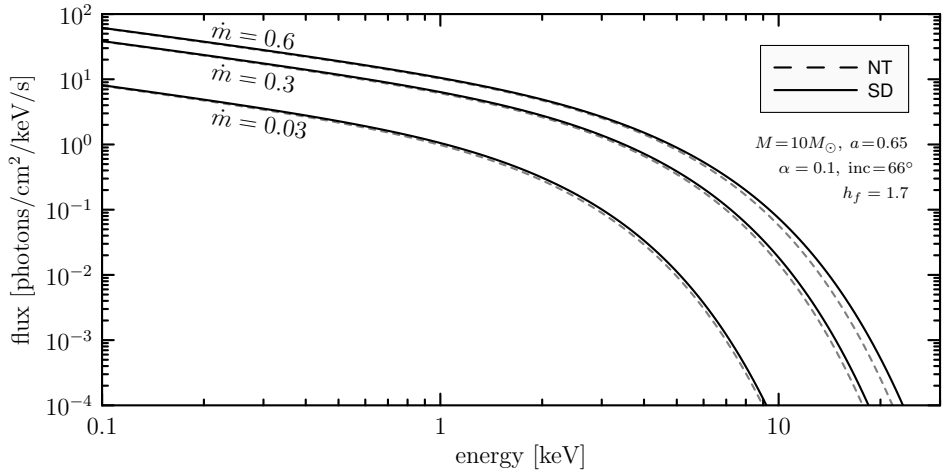}
\caption{
Calculated theoretical continuum spectra corresponding to the emission profiles presented in Fig.~\ref{f.plotflux}. The plot shows spectra of slim disk (solid) and \citem{nt} (dashed lines) models calculated assuming constant value of the hardening factor ($f=1.7$) for an observer at inclination $66^o$, and distance $10 {\rm kpc}$.}
\label{f.spectra.hf17}
\end{figure}

\begin{figure}[h]
\centering
\includegraphics[width=.8\textwidth, angle=0]{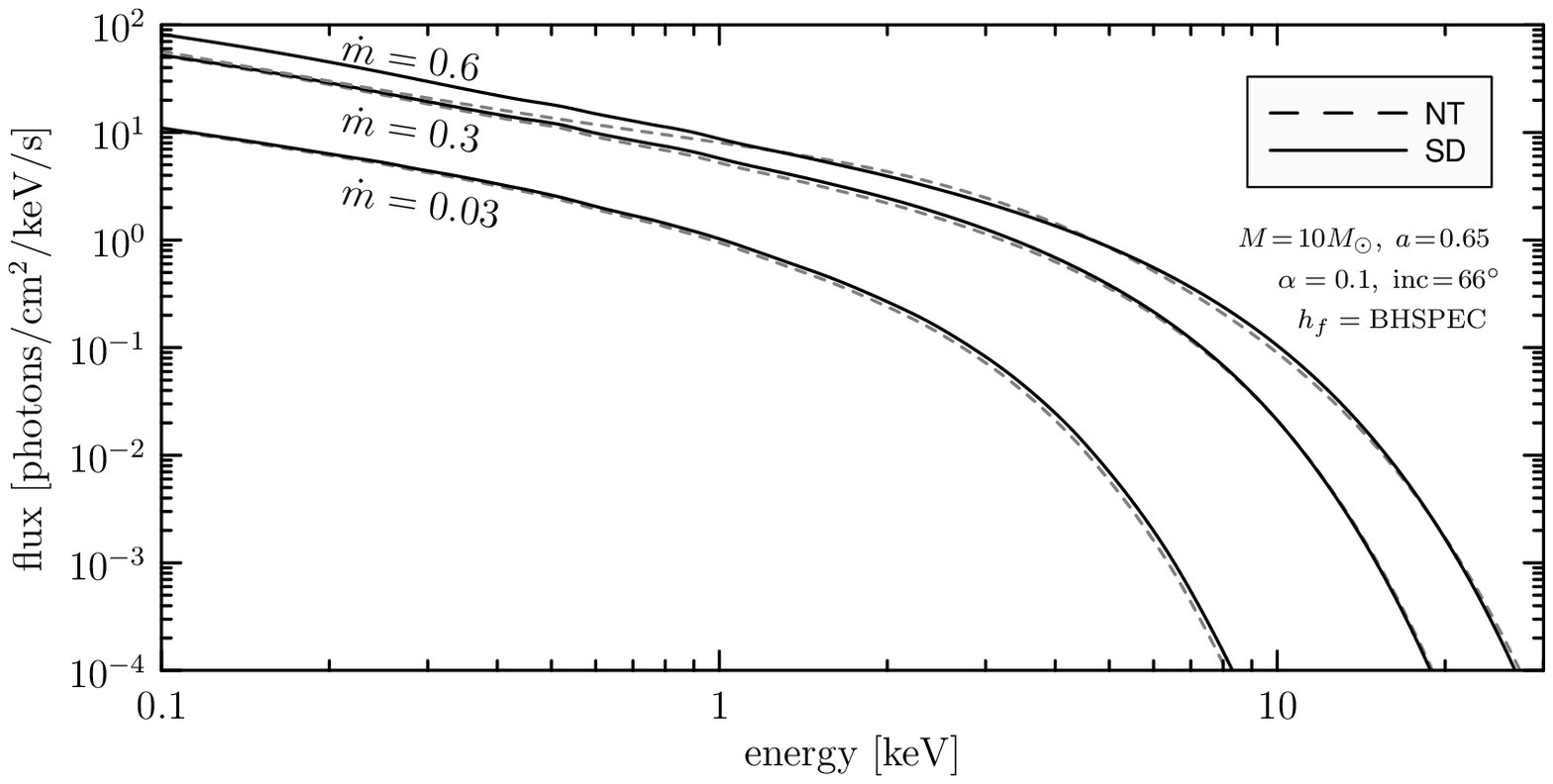}
\caption{
Same as Fig.~\ref{f.spectra.hf17} but calculated using the BHSPEC atmospheric model instead of assuming constant value of the hardening factor.}
\label{f.spectra.bhspec}
\end{figure}

\subsection{\texttt{slimbb} XSPEC fitting routine}
\label{s.slimbb}

Using the numerical solutions of the slim accretion disk equations and an advanced ray-tracing technique \citepm{bursa-05}, 
we have calculated tables of spectra and developed a new model called \texttt{slimbb} for
use with the X-ray spectral 
fitting package XSPEC \citepm{arnaud-96}. This model provides spectra of slim disks on a three-dimensional 
grid of the black-hole spin parameter, disk luminosity and inclination angle. They were calculated assuming
either constant hardening factor $f_{\rm c}$, or using the BHSPEC disk atmosphere model \citepm{davisomer05,davishubeny-06}. In the following subsections we describe most important properties of the model.

\subsubsection{Disk model}

Slim disk solutions (Section~\ref{s.stationarysolutions}) provide radial profiles of the following quantities: total amount of emitted flux $F$, radial velocity $V$,  specific angular
momentum $\ell$, surface density $\Sigma$ and disk height $\cos\Theta_H$.
 They are used to construct and 
tabularize spectra of slim accretion disks at a grid spanned on accretion rate, BH spin and inclination angle.

\subsubsection{disk spectrum as observed by a distant observer}

The specific flux density of the disk radiation as observed by a distant
observer at energy $E_{\rm obs}\!=\!\planckh\nu$ is given by
\be
    F_{\rm obs}(\nu) = \int I_{\rm em}(\nu/g, r, \mue) \, g^3 \, d\Omega \;,
\ee
where $g$ is the photon redshift (Eq.~\ref{e.gfactor}) and $d\Omega$ is the
element of the solid angle subtended by the image of the disk on the observer's
detector plane. $I_{\rm em}$ is the specific intensity of the radiation coming
out from a point on the disk in the right direction to reach the observer as
measured in the fluid comoving frame. Generally, it is a function of energy,
place of emission and emission angle. \texttt{slimbb} model assumes for $I_{\rm em}$
either a diluted blackbody or takes the emerging radiation spectrum from BHSPEC.

In the case of a diluted blackbody radiation, the specific intensity is given by Planck's
law with color-modified temperature
\be
    I_{\rm em}(\nu) = 
        \frac{2 \planckh \nu^3 \speedoflight^{-2} \fcolor^{-4}}
        {\exp(\planckh\nu/\boltzk \fcolor T_{\rm eff})-1} \; 
        \flimbdk(\mue) \;.
\ee
The effective temperature $T_{\rm eff}$ can be obtained from the
total energy flux $F$:
\be
    T_{\rm eff} = \left( \frac{F_{\rm out}}{\sigma} \right)^{1/4}.
\ee

The color hardening factor $\fcolor$ accounts for the effect of free electrons
in the disk atmosphere on photons and the fact that the locally observed color
temperature of the emitted radiation is generally higher than the effective
temperature of the disk related to the locally dissipated energy (see Section~\ref{s.hardening}).
$\flimbdk(\mue)$ describes the dependence on the photon emission angle (the
angle between the emitted photon wavevector and the disk surface normal), which
is $\flimbdk\!\equiv\!1$ for isotropic emission or 
\be
    \flimbdk = \frac{1}{4}(2 + 3\,\mue)
\ee
if the emission is assumed to be limb-darkened. $\mue$ is the cosine of the emission angle.

\subsubsection{g-factor and disk emision angle} 
\label{s.gfactor}

Most of the flux emitted by disk comes from the region of 
strong gravity and high velocities, and is subject to intense Doppler boosting and gravitational 
redshift.

The combined effect of gravitational and Doppler energy shift is described by $g$-factor, 
which is the ratio between energy at which the photon is observed and  energy at which it 
has been emitted. It is calculated using the general formula,
\be
\label{e.gfactor}
    g = \frac{E_{\rm obs}}{E_{\rm em}} = \frac{p_t}{(p_\mu\,u^\mu)_{\rm em}} \,,
\ee
where $p^\mu$ is photon \mbox{four-momentum} and we obtain $u^\mu$ by a
transformation from the co-rotating frame, $u^\mu = e^\mu_{(a)}\,u^{(a)}$. The fluid velocity
measured by a co-rotating observer has only the radial component $V$ and we have
\be
    u^{(a)} = (\sqrt{1+V^2},-V,0,0) \,.
\ee
Here, $V$, although negative, is taken with the minus sign as the radial vector
$e^\mu_{(r)}$ of the orthonormal frame points in the inward direction.

The emission angle is the angle between the direction of the normal to the disk surface
and the direction of escaping photon momentum. The cosine of the emission angle is given 
by the general formula
\be
    \mue = -\frac{p_\mu \, n^\mu}{p_\nu \, u^\nu} \,,
\ee
where $n^\mu$ is the normal vector to the disk surface. The disk normal can be
easily expressed in the co-rotating frame, where it is
\be
    n^{(a)} = e_\mu^{(a)} n^\mu = (0,\sin{\gamma},\cos{\gamma},0) \,.
\ee
The angle $\gamma$ between the disk surface normal and the direction of the poloidal
basis vector $e^{(\theta)}$ is
\be
    \gamma = \frac{\pi}{2} - (\vartheta + \alpha) \,,
\ee
where $\vartheta$ is the poloidal coordinate angle of the place of emission and $\alpha$ is the disk height
derivative with respect to the cylindrical radius $r$,
\be
    \alpha = \tan^{-1} \left( \frac{dh}{dr} \right) \,.
\ee
In Boyer-Lindquist frame, the disk normal vector, normalized to unity, is then
\be
    n^\mu = \left( 0,\frac{1}{\sqrt{\grr}}\cos{(\vartheta+\alpha)},
\frac{1}{\sqrt{\ghh}}\sin{(\vartheta+\alpha)}, 0 \right) \,.
\ee

\begin{sidewaysfigure}
\centering
\includegraphics[width=.8\textwidth]{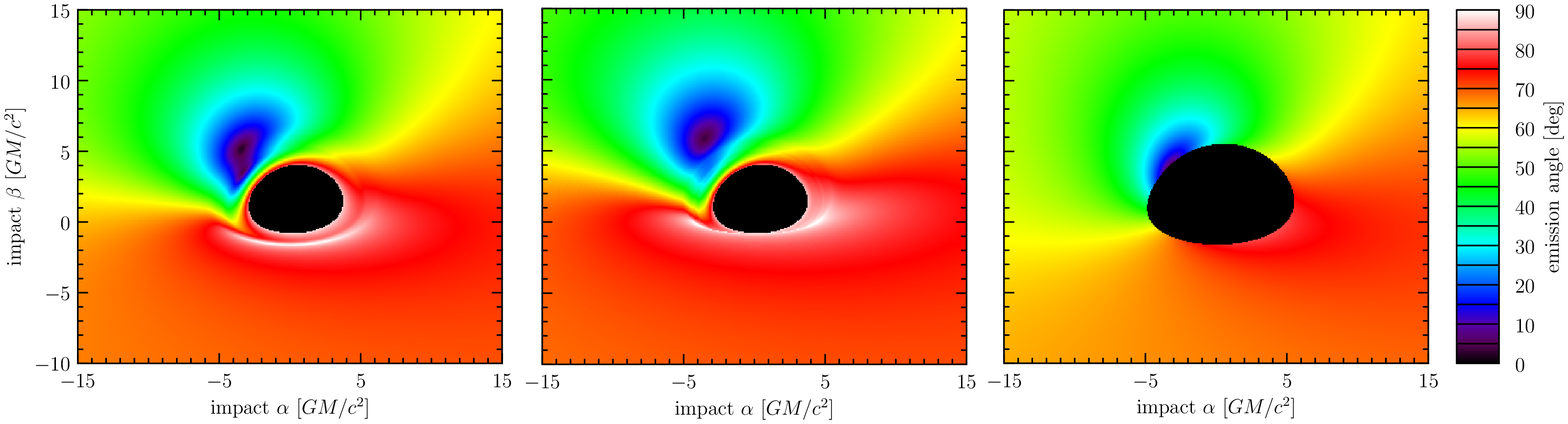}\\
\includegraphics[width=.8\textwidth]{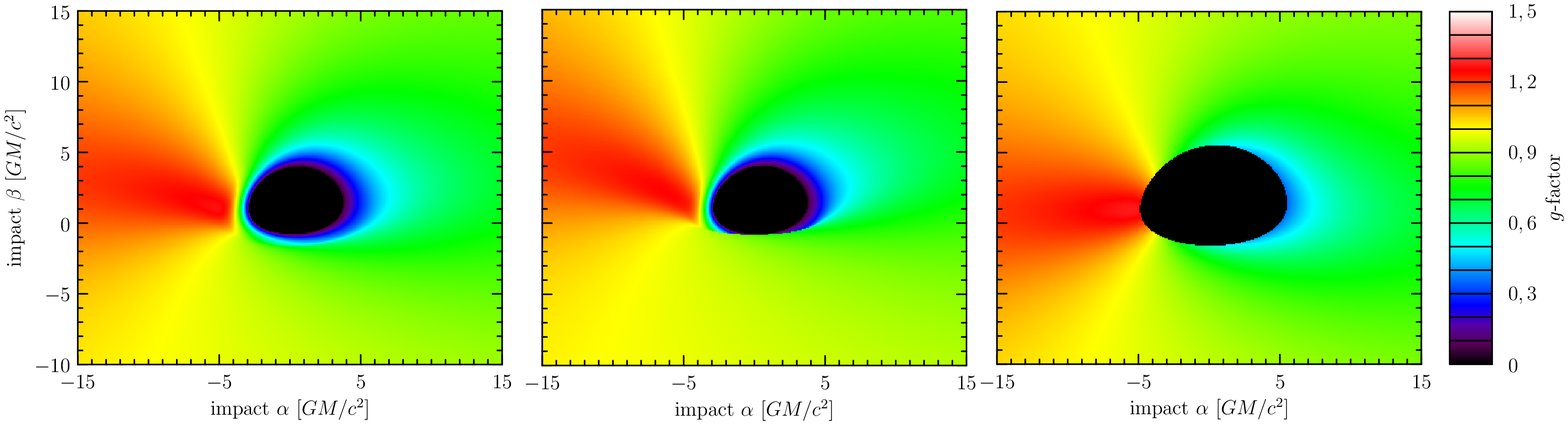}
\caption{Profiles of emission angle ({\sl top row}) and $g$-factor ({\sl bottom row}) for two mass 
accretion rates $\dot m\!=\!0.3$ and $0.6$. Other parameters like spin $a_*\!=\!0.6$, alpha viscosity 
$\alpha\!=\!0.1$ and inclination $i\!=\!66^o$ are fixed. The two leftmost columns represent slim disk, while the 
right column show for comparison corresponding thin disk picture (NT solution), which looks the same 
for any accretion rate. For slim disk the mass accretion rates correspond to luminosities $0.5\,\Ledd$ 
and $1.0\,\Ledd$, respectively. Plots show the image plane view of the disk.}
\label{fig:mue+gfactor}
\end{sidewaysfigure}

An example of how slim disk changes the profiles of $g$-factor and emission angle is given 
in Fig.~\ref{fig:mue+gfactor}. For a chosen observer's viewing angle of $66^o$
and BH spin $a_* = 0.6$ we compare these two quantities for two different mass accretion rates.

For luminosities less than $0.3\,L_{\rm Edd}$ (mass accretion rates $\dot M\!<\!0.2$ for $a_* \= 0.6$) the 
two models differ only slightly. For higher luminosities, presented in Fig.~\ref{fig:mue+gfactor}, the increasing difference between these
models comes from the fact that the radiation inner edge of slim disks shifts to smaller 
radii, that the disk thickness increases, and due to more and more significant departure of angular momentum from the Keplerian profile (see Section~\ref{s.stationarysolutions}).

\subsubsection{Self-obscuration}

We may notice an interesting consequence, unknown to thin disks, namely that at this inclination the inner regions of 
slim disks may be obscured by its outer parts. This fact is noticeable in the plot of the emission angle, 
where we see regions from which the radiation comes almost parallel to the disk surface. Full obscuration of the central 
BH takes place only for super-Eddington luminosities. This effect has also obvious and important implications 
for the temperature--luminosity relations.

\begin{figure*}[t!]
\centering
\includegraphics[width=0.7\textwidth]{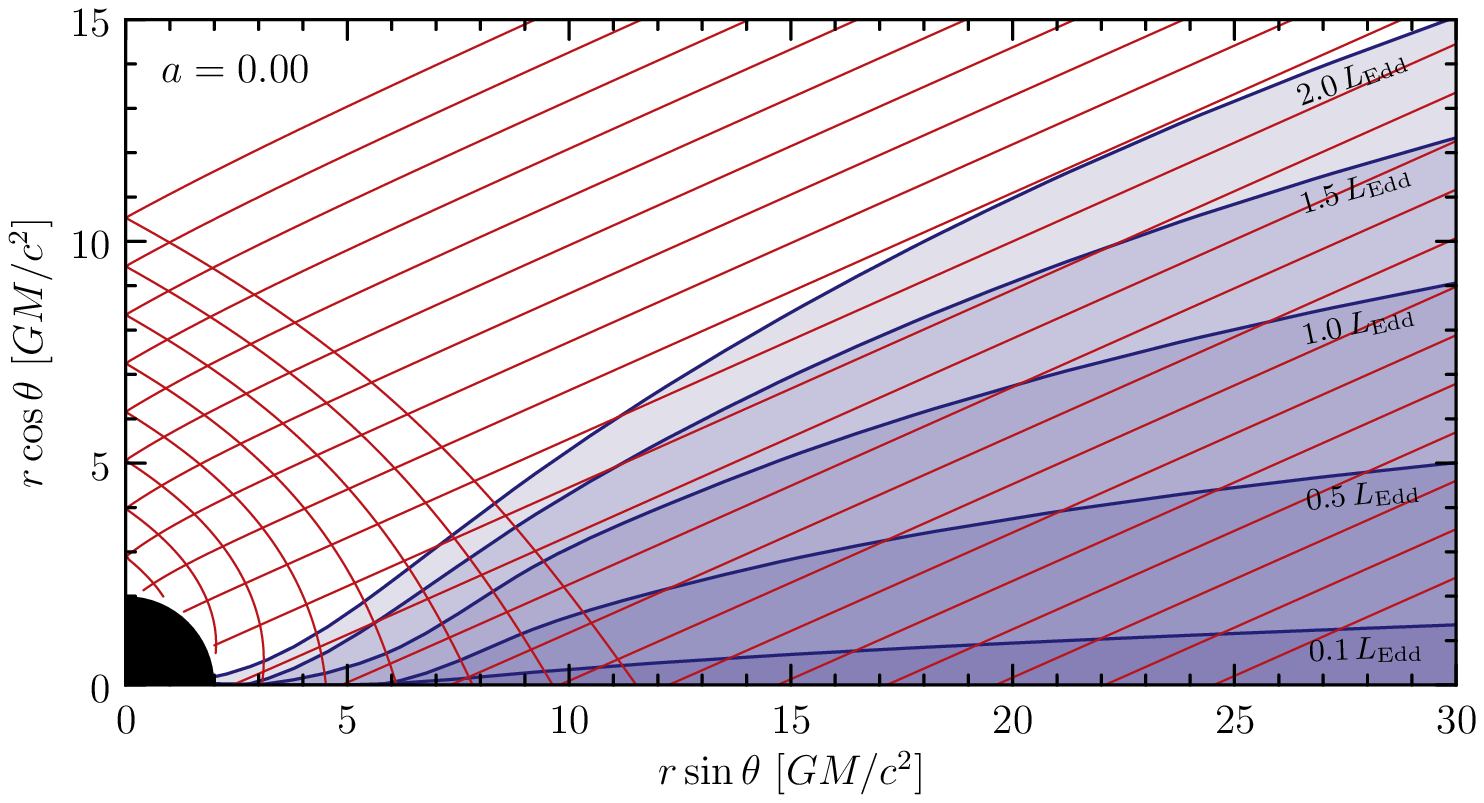}
\includegraphics[width=0.7\textwidth]{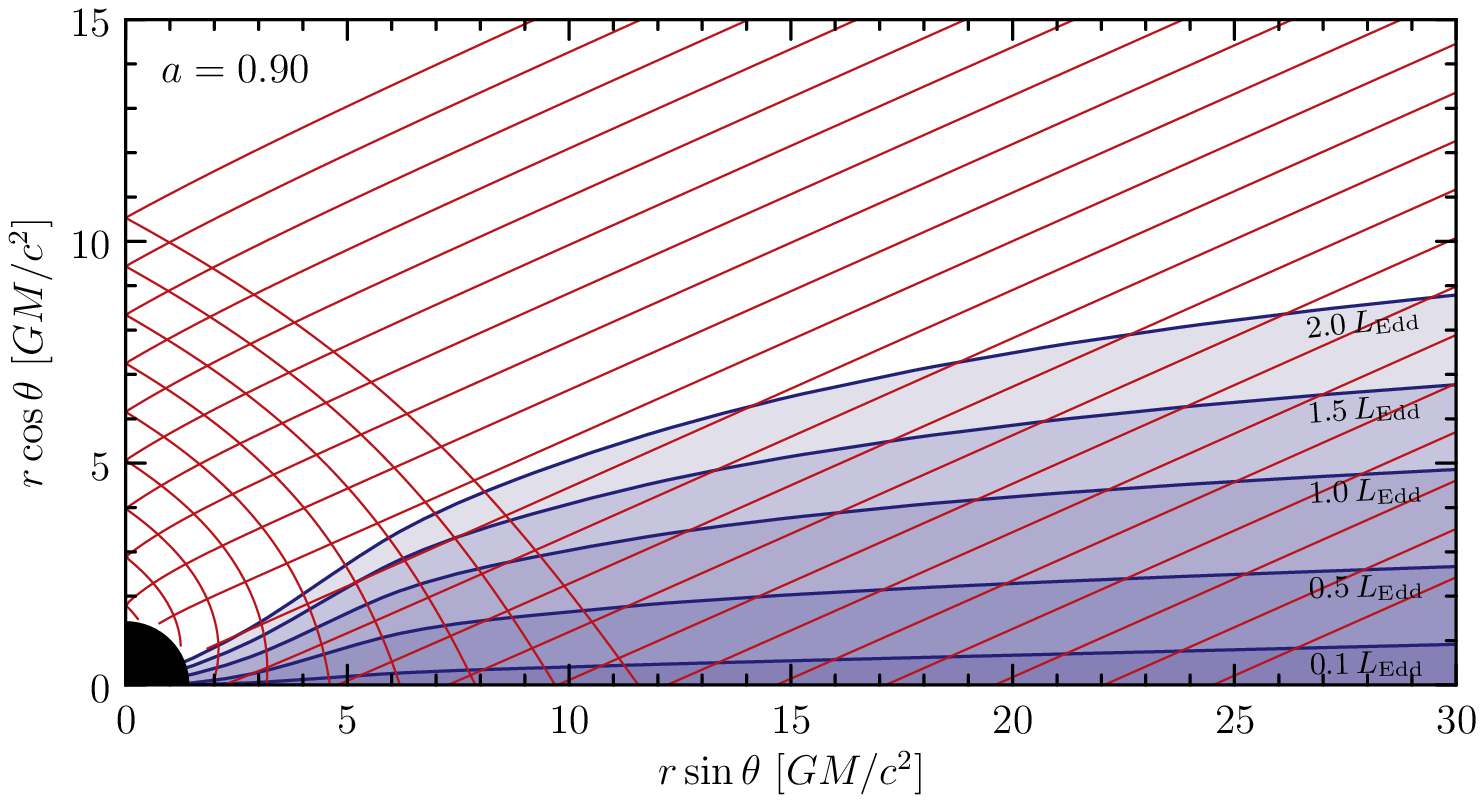}
\caption{Illustration of disk self-obscuration in the $z$-$x$ plane. The figure shows vertical cuts through the accretion disk, 
location of disk photosphere for different luminosities and trajectories of photons 
with zero azimuthal angular momentum at infinity reaching an observer at inclination $66\deg$. 
Top panel corresponds to spin $a_*\!=\!0.0$, bottom panel to spin $a_*\!=\!0.9$. }
\label{fig:self-obscuration}
\end{figure*}

The condition for self-obscuration is roughly that $h/r$ somewhere in the disk is 
larger than the cosine of inclination. If for higher luminosities the radiation 
pressure becomes important, the $h/r$ ratio typically has a maximum near radius 
where radiation pressure dominates over gas pressure. The most luminous part of 
the disk, however, lies still inside this radius and thus if the condition 
$H/R\!\gtrsim\!\cos i$ is met, the strongly radiating part of the disk disk becomes 
partially or even fully obscured by an outer part. This, of course, has important 
observational consequences.


Let us consider our model example of a source with disk inclination $66^o$. The critical 
$h/r$ ratio is $0.4$ and the situation is illustrated in Figure~\ref{fig:self-obscuration}. 
The plot shows the profiles of the disk photosphere for several luminosities and compares 
them with trajectories of photons sent from infinity with zero azimuthal angular momentum.
For zero spin case (top panel), we see that already at luminosity $0.5\,\Ledd$ photons from the 
part of the disk near the BH need to escape almost parallel to disk surface to reach the observer. 
For even larger luminosities this part of the disk starts to be completely obscured and at 
$2\,\Ledd$ the large part of the inner disk is eclipsed including the BH itself.
For high spin case (bottom panel), the situation is quite similar, only the luminosity has to be higher. 

\subsubsection{Model parameters}

\begin{table}[tb]
\begin{center}
\caption{\texttt{slimbb} model parameters.\label{tab:model-parameters}}
\begin{tabular}{lcccc}
\hline\hline
Parameter & Units & Default value & Min value & Max value \\
\hline
M                   &   $\Msun$  & 10.    &    0.      &  100.     \\
a                   &   $GM/c$   &  0.    &    0.      &    0.999  \\
L                   &   $\Ledd$  &   0.1  &    0.0003  &    1.0    \\
inc                 &   deg      &  0.0   &    0.0     &   85.0    \\
D                   &   kpc      & 10.    &    0.      &    1.0e4  \\
f\_hard             &   $-$      &  1.    &  -10.      &   +10.    \\
alpha               &   $-$      &  0.1   &    0.005   &    0.1    \\
lflag               &   $-$      &  0.    &   -1.      &   +1.     \\
vflag               &   $-$      &  0.    &   -1.      &   +1.     \\
\hline
\end{tabular}
\end{center}
\end{table}

The SLIMBB model comes in a form of a FITS table with precalculated spectra and a small routine 
which reads the data and produces the final spectrum in terms of linear interpolation in the 
parameter space. For five different values of $\alpha$ viscosity the table contains extensive three 
dimensional grids of spectra spanned on the spin, luminosity and inclination. These parameters 
vary within the limits given in Table~\ref{tab:model-parameters} with steps chosen to ensure that 
the spectra differ roughly equally between any two adjacent steps.

The model has all together nine parameters with the following meaning (see Table~\ref{tab:model-parameters} 
for their limit values):\\
\indent M -- mass of the central BH in units of solar masses,\\
\indent a -- specific angular momentum of the central BH in units of $GM/c$,\\
\indent L -- disk luminosity in units of Eddington luminosity $\Ledd$,\\
\indent inc -- inclination of the observer with respect to the symmetry axis in degrees,\\
\indent D -- distance of the source in kiloparsecs,\\
\indent f\_hard -- hardening factor (if positive) or a switch for BHSPEC (if negative),\\
\indent alpha -- the value of $\alpha$-viscosity parameter,\\
\indent lflag -- switch for isotropic (0) or limb-darkened (1) emission,\\
\indent vflag -- switch for equatorial plane (0) or photospheric (1) ray-tracing.\\
Most up-to-date version of SLIMBB and its detailed description may be found at \\\texttt{http://astro.cas.cz/bursa/projects/slimbb}.

\chapter{Relativistic, non-stationary slim accretion disks}
\label{chapter-nonstationary}
X-ray binaries and AGN exhibit different types of variability taking place at different timescales. Some of them may result from intrinsic instabilities of accretion disks. Other may result e.g., from the orbital motion. Instabilities caused by the viscous and non-adiabatic processes in disks may be classified into three types \citepm{KatoBook}: thermal, secular instabilities and overstabilities of waves and oscillations. The first one takes place on the thermal timescale what makes it most profound. Its origins will be discussed in the following paragraph, while in the next sections we will present numerical model of non-stationary slim accretion disks which is able to reproduce the limit cycles resulting from the thermal instability.

\section{Thermal instability}
\label{s.thermalinstability}

Thermal instability takes place when a temperature perturbation over the equilibrium is amplified by a thermal process. Therefore, the characteristic timescale is the thermal timescale \citepm{KatoBook},
\be
\tau_{\rm th}=\frac{c_s^2}{V_\phi^2}\frac {r^2}\nu,
\ee
over which the disk temperature changes. The following relations are satisfied,
\be
\tau_{\rm hyd}=\frac{h}{c_s}\ll\tau_{\rm th}\ll\tau_{\rm vis}=\frac{r^2}{\nu},
\ee
where $\tau_{\rm hyd}$ is the hydrodynamical timescale, over which the vertical disk structure varies, while $\tau_{\rm vis}$ is the viscous timescale determining the radial disk structure changes. 

When studying the secular and thermal instabilities one has to filter out the acoustic oscillations occurring on timescales shorter than both thermal and viscous ones. It can be done by neglecting the time derivatives in the momentum equations. 
On the contrary, as the instability involves temperature perturbations, the energy equation should be taken into account in a time-dependent form. The dynamical equilibrium may be adopted as it is achieved over the dynamical timescale, much shorter than the thermal one. Finally, since we consider the force balance between the gravitational and centrifugal forces, we may assume the gas motion resulting from a temperature change occurs only in the vertical direction ($\Sigma={\rm const}$). 

Under these assumptions we may perform a simple thermal stability study. Full mathematical derivation of the stability criterion for both thermal and secular instabilities has been given e.g., in \citem{shakura-76} and \citem{piran-78}. Recently, \citem{ciesielski-11} have generalized the perturbative analysis of the stability criterion to account for modified (delayed, advanced or with different stress-to-pressure response factors) viscosity prescriptions.

Let us write the vertical equilibrium equation in the following form (compare Eq.~\ref{sh_vert2}),
\be
p=\frac12\Omega^2\Sigma h.
\label{e.tstab1}
\ee
In the innermost region of an accretion disk we have $p\approx p_{\rm rad}$ and $\kappa\approx\kappa_{\rm es}={\rm const}$. The viscous heating rate is given by,
\be
Q^{\rm vis}=-\frac32\Omega t_{r\phi}h.
\ee
Under the $\alpha p$ assumption (Section~\ref{standard-alpha-prescription}) and taking Eq.~\ref{e.tstab1} into account, we have,
\be
Q^{\rm vis}\propto ph\propto h^2.
\ee
The radiative cooling rate per unit surface is,
\be
Q^{\rm rad}=16\frac{c p_{\rm rad}}{\kappa\Sigma}\propto h,
\ee
where we used again Eq.~\ref{e.tstab1}.
If the temperature exceeds slightly the equilibrium ($Q^{\rm vis}=Q^{\rm rad}$) value, the disk expands in the vertical direction as $h\propto p\propto T^4$ (Eq.~\ref{e.tstab1}). As a result, both $Q^{\rm vis}$ and $Q^{\rm rad}$ increase, but the rate is larger for the heating component. This means that the specific entropy of the gas (and temperature) increases, which leads to a further increase of $H$. In this way, the positive feedback occurs and the disk becomes unstable.

In the gas dominated case, the heating rate is still given by,
\be
Q^{\rm vis}\propto h^2.
\ee
However, the radiative cooling is now expressed as,
\be
Q^{\rm rad}\propto h^8,
\ee
since $p_{\rm rad}\propto T^4$, $p=p_{\rm gas}\propto T\rho$ and hence $p_{\rm rad}\propto (p/\rho)^4\propto (ph)^4\propto h^8$. For such dependence of the heating and cooling on $H$, the reasoning given above will not be valid. The disk in the gas pressure-dominated case is thermally stable.

The general criterion for the thermal stability may be put in the form \citepm{pringle-76},
\be
\left.\pder{}h(Q^{\rm vis}-Q^{\rm rad})\right|_\Sigma > 0.
\label{e.tstab2}
\ee
Using Eq.~\ref{e.tstab1} we can relate the small-amplitude perturbation of pressure $p_1$ superposed on the equilibrium value $p_0$ as,
\be
\frac{p_1}{p_0}=\frac{h_1}{h_0},
\ee
where similar perturbations of the disk thickness have been introduced. Using the general equation of state (Eq.~\ref{e.shak.pressure}), we may also write,
\be
\frac{T_1}{T_0}=\frac{1+\beta}{4-3\beta}\frac{h_1}{h_0},
\ee
where $\beta=p_{\rm gas}/p$ is the gas pressure to total pressure ratio. The heating rate, $Q^{\rm vis}$, follows $Q^{\rm vis}\propto ph$ and therefore,
\be
\frac{Q^+_{1}}{Q^+_{0}}=2\frac{h_1}{h_0}
\ee
Similarly, the perturbation of $Q^{\rm rad}$ follows $Q^{\rm rad}\propto T^4$ and thus,
\be
\frac{Q^{\rm rad}_{1}}{Q^{\rm rad}_{0}}=\frac{4(1+\beta)}{4-3\beta}\frac{h_1}{h_0}.
\ee
Putting the equations given above into Eq.~\ref{e.tstab2} we obtain the stability criterion,
\be
2-5\beta<0.
\label{e.tstab3}
\ee
The regions of accretion disks with $\beta<0.4$ are thermally unstable. This is the case if only the accretion rate exceeds some critical value, e.g. $\dot m\approx 0.06$ for a non-rotating BH.

\begin{figure}
  \centering\resizebox{.85	\textwidth}{!}{\includegraphics[angle=0]{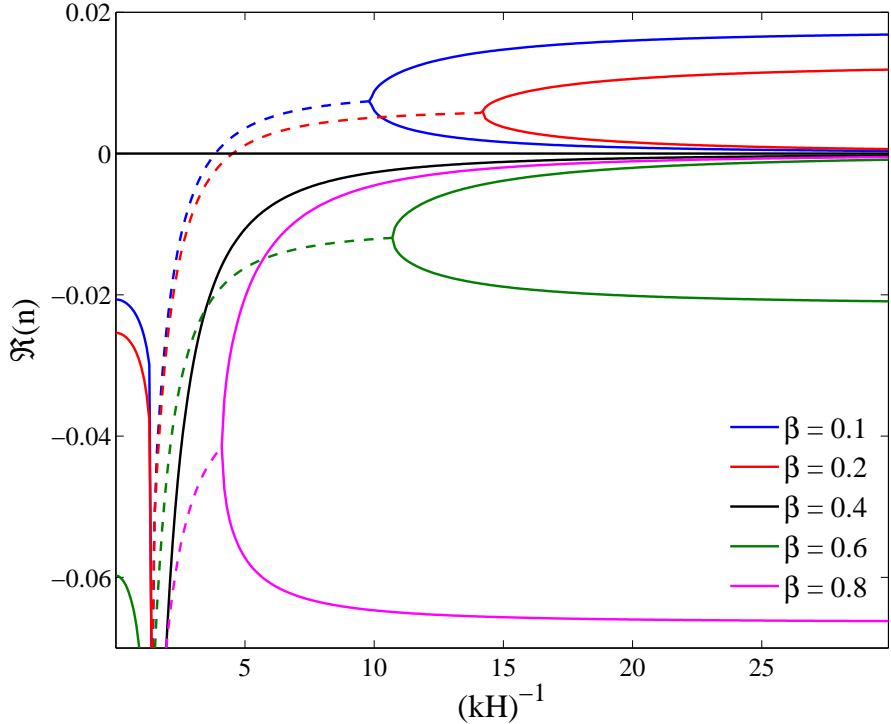}}
\caption{ Growth rate as a function of the radial wave-length of perturbations for $\alpha=0.1$, $\gamma=5/3$ and different values of $\beta$. The branch closer to the horizontal axis denotes the secular mode, while the other represents the thermal one. Dotted curves denote the complex conjugate solutions of the dispersion relation. }
  \label{f.khrstability}
\end{figure}

Fig.~\ref{f.khrstability} summarizes results of a more detailed, based on the perturbative analysis, study of thermal and secular types of instabilities in accretion disks. The growth rates of perturbations (solutions of the dispersion relation) for some arbitrary values of $\beta$ are plotted against the perturbation wave number $k=2\pi/\lambda$. Positive values correspond to the unstable case, while negative prove stability. It is clear that only for $\beta<0.4$ the solutions lie above the x-axis. These curves have two branches: the lower (approaching zero for $\lambda\rightarrow \infty$) corresponds to the secular instability, while the upper to the thermal one. For $\beta=0.4$ the disk is marginally stable in the long wavelength limit and stable for finite wavelengths. Higher values of $\beta$ give solutions with the negative sign reflecting their stability.

The stability criterion (Eq.~\ref{e.tstab3}) is very strong. It predicts that only disks with very low accretion rates are stable. One could therefore expect strong variability of this type in most accretion disks. However, it is not the case since most of observed microquasars exhibit in their high/soft states stable emission for periods of time much longer than the corresponding thermal timescales. Moreover, most recent MHD simulations \citepm{hirose-09-a} show that thermal instability may not occur in real accretion disks. This inconsistency has not yet been resolved. Recently, \citem{ciesielski-11} have addressed this issue, showing that time delays in heating or reduced stress-to-pressure response may quench the thermal instability.

Considering the unique limit-cycle behavior of GRS 1915+105, which contains near-extreme Kerr BH and exhibits limit-cycle-like luminosity variations, we suggest that the classical theory may be indeed correct in predicting the radiation pressure instability, but under some, so far unknown, conditions this behavior is suppressed, as observed in most X-ray binaries. Since the proper description of viscosity is still developing, the limited classical theory is the only reasonable choice for our study before a widely accepted new theory is presented. In the next sections we investigate the time-dependent behavior of relativistic slim accretion disks that exhibit limit cycles resulting from the classical thermal instability. 

\section{Equations}

In this Chapter, we consider axisymmetrical relativistic accretion flows onto Kerr BHs. We use the Boyer-Lindquist spherical coordinates ($t$, $r$, $\theta$, $\phi$) to describe the spacetime around a BH. Putting all of the complicated derivations into Appendix \ref{ap.nonstat}, we give here the basic equations governing the dynamical behavior of the flow. Assuming $\partial_t=0$, they all reduce, upto the coefficients related to the vertical integration, to the equations given in Section~\ref{s.stateq}.

\begin{eqnarray}
&&\frac{\partial\Sigma}{\partial t} = -\frac{r\Delta^{1/2}}{\gamma A^{1/2}}
\left[\Sigma\frac{\partial u^t}{\partial t}+\frac{1}{r}\frac{\partial}{\partial r}\left(r\Sigma\frac{V}{\sqrt{1-V^2}}\frac{\Delta^{1/2}}{r}\right)\right],\label{e.nonstat.continuity}  \\
&&\frac{\partial V}{\partial t} = \frac{\sqrt{1-V^2}\Delta}{\gamma A^{1/2}}\left[-\frac{V}{1-V^2}\frac{\partial V}{\partial r}+\frac{{\cal A}}{r}-\frac{1-V^2}{\rho}\frac{\partial p}{\partial r}\right], \label{e.nonstat.radial} \\
&&\frac{\partial\cal L}{\partial t} = -\frac{V\Delta}{\gamma\sqrt{1-V^2}A^{1/2}}\frac{\partial\cal L}{\partial r}+
\frac{\Delta^{1/2}}{\gamma\Sigma A^{1/2}}\frac{\partial}{\partial r}\left(\frac{\nu\Sigma A^{3/2}\Delta^{1/2}\gamma^3}{r^4}
\frac{\partial\Omega}{\partial r}\right),\label{e.nonstat.angular}\\
&&\frac{\partial h}{\partial t} = -U\cos\Theta_H -\frac{1}{\gamma}\frac{V}{\sqrt{1-V^2}}\frac{\partial h}{\partial r},\label{e.nonstat.thickness}\\
&&\frac{\partial U}{\partial t} = \frac{\Delta^{1/2}}{\gamma^2 A^{1/2}\cos\Theta_H}{\cal R}-
\frac{U}{\gamma^2}\left(\frac{V}{(1-V^2)^2}\frac{\partial V}{\partial t}+
\frac{{\cal L}r^2}{A}\frac{\partial\cal L}{\partial t}\right)-
\frac{U}{h}\frac{\partial h}{\partial t}, \label{e.nonstat.vertical}\\
\nonumber &&\frac{\partial T_C}{\partial t} = \frac{1}{\Sigma}\frac{r\Delta^{1/2}}{\gamma A^{1/2}}\left[
\frac{Q^{\rm vis}-Q^{\rm rad}}{c_V}+(\Gamma_3-1)T_C\Sigma\left(-\frac{\partial u^t}{\partial t}-\frac{1}{r^2}\frac\partial{\partial r}(r^2u^r)\right)\right]\\
&&~~~~~-\frac{V\Delta}{\gamma\sqrt{1-V^2}A^{1/2}}\frac{\partial T_C}{\partial r}, \label{e.nonstat.energy}
\end{eqnarray}
where $\Delta\equiv r^2-2Mr+a^2$, $A\equiv r^4+r^2a^2+2Mra^2$ and $\gamma$ is the Lorentz factor. Equations \ref{e.nonstat.continuity}, \ref{e.nonstat.radial}, \ref{e.nonstat.angular}, and \ref{e.nonstat.energy}, defining the time derivatives of the surface density, radial velocity, angular momentum and central temperature, respectively, represent corresponding conservation laws. Equations \ref{e.nonstat.thickness} and \ref{e.nonstat.vertical} determine the evolution of the disk thickness and the vertical acceleration. We adopt the diffusive form of viscosity (Eq.~\ref{visc.stress2}) in equations \ref{e.nonstat.angular} and \ref{e.nonstat.energy}. In these equations, surface density $\Sigma$, radial velocity $V$, specific angular momentum $\cal L$, half thickness $h$, vertical velocity $U$, and temperature $T$, are six essential quantities  describing the structure of disk. 

The other symbols denote the following auxiliary quantities,
\begin{eqnarray}
&&\nu=\frac{2}{3}\alpha h \sqrt{\frac{p}{\rho}}, \label{nu}\\
&&\rho=\frac{\Sigma}{2h},\\
&&p=\frac{k_B}{\mu m_p}\rho T+p_{\rm rad},\\
&&\cos\Theta_H=\frac{h}{r},\\
&&Q^{\rm vis}=\nu\gamma^4\Sigma\frac{A^2}{r^6}\left(\frac{\partial\Omega}{\partial r}\right)^2, \\
&&Q^{\rm rad}=\frac{8\sigma T^4}{3\tau_R/2+\sqrt{3}+1/\tau_P}, \label{F^-}\\
&&p_{\rm rad}=\frac{Q^{\rm rad}}{2}\left(\tau_R+\frac{2}{\sqrt{3}}\right),
\end{eqnarray}
where equation (\ref{F^-}) gives a bridging formula that is valid for both optically thick and thin regimes \citepm{hubeny-90}, $\alpha$ is the viscosity parameter, $p_{\rm rad}$ is the radiation pressure, and $\tau_R$ and $\tau_P$ are the Rosseland and Planck mean optical depths \citepm{szuszkiewiczmiller-98}.

The properties of an accretion flow depend on three crucial parameters:
\begin{itemize}
\item Dimensionless BH spin $a_*$:
\begin{equation}
a_*\equiv\frac{a}{M},~~~0\leq a_* < 1;
\end{equation}
\item Diffusive viscosity parameter $\alpha$ (Eq.~\ref{nu}):\\
 \be
0<\alpha<1;
\ee
\item Dimensionless mass supply rate $\dot m$ (compare Eq.~\ref{e.mdotcritical}):\\
\begin{equation}
\dot m\equiv\frac{\dot M(r_{\rm out})}{\dot M_{\rm Edd}}.
\end{equation}
\end{itemize}
Where $M$ is the BH mass and $\dot M(r_{\rm out})$ is the accretion rate at the outer boundary ($r=r_{\rm out}$).

\section{Numerical methods}

The given set of equations may be put in the following, general form,
\be
\label{e.nonstat.1}
\pder{u_i(r,t)}{t}=L_i(u_k(r,t))
\ee
where $u_i(r,t)$ stands for all six functions involved ($\Sigma$, $V$, $\cal L$, $H$, $U$ and $T$) and $L$ is a partial differential operator. Such a set of equations may be solved numerically by discretizing in time and radius. For the purpose of solving the non-stationary equations of slim disks we applied the standard Chebyshev pseudospectral method for the spatial, and combined Runge-Kutta and third-order backward differentiation explicit schemes for the time discretization, respectively.

\subsection{Spatial discretization}

A physical quantity $u(r)$ is approximated by the following series with a finite number of terms:
\be
u(r_k)=u(g(\bar r_k))=\Sigma_{n=0}^N\hat u_nT_n(\bar r_k)=\Sigma_{n=0}^N\hat u_n\cos\left(\frac{nk\pi}N\right),
\ee
where $T_n(\bar r_k)$ is the $\rm n^{\rm th}$-order Chebyshev polynomial; $\bar r_k=\cos(k\pi/N)$ ($k = 0,1,...,N$)
represents the Chebyshev-Gauss-Lobatto collocation points, with $N$ being the number of collocation points;
$r_k = g(\bar r_k)$ is the mapping from the Chebyshev-Gauss-Lobatto collocation points $-1\le\bar r_k\le1$
to the physical collocation points $r_{\rm min}\le r_k\le r_{\rm max}$ and is a strictly
increasing function that satisfies both $g(-1) = r_{\rm min}$ and $g(1) = r_{\rm max}$; and the $\bar u_n$ are the spectral coefficients and can be calculated
from the physical values $u(r_k )$ with a fast discrete cosine transform (FDCT; \citem{numericalrecipes}).

In a similar way one may approximate radial derivatives,

\be
\pder{u(r_k)}{r}=\frac 1{{\rm d}g/{\rm d}\bar r}\pder{u(g(\bar r_k))}{\bar r}=\frac 1{{\rm d}g/{\rm d}\bar r}\Sigma_{n=0}^N\hat u_n'T_n(\bar r_k),
\ee
where the spectral coefficients $\hat u'_n$ are evaluated basing on $\hat u_n$ using the following recursive relations,
\bea \nonumber
\hat u'_N&=&0,\\ 
\hat u'_{N-1}&=&2 N \hat u_N,\\ \nonumber
c_n\hat u'_m &=&\hat u'_{n+2} + 2(n+1)\hat u_{n+1},
\eea
with $c_0=2$ and $c_n=2$ for $n=1,2,...,N$. Calculation of these coefficients is followed by an inverse FDCT of $\hat u'_n$ providing $\partial u(g(\bar r_k))/\partial \bar r$ and, finally, discrete spatial derivatives $\partial u(r_k)/\partial r$.

\subsection{Time discretization}

The integration in time is performed using two algorithms. First two time-steps are integrated using the third order total variation diminishing (TVD) Runge-Kutta scheme \citepm{shu1988}, while the further time-evolution using the third
order backward-differentiation explicit scheme \citepm{peyret2002} which provides good enough accuracy and is less CPU-consuming.

The third order TVD Runge-Kutta scheme may be expressed as,
\begin{eqnarray}\label{RK3}
  \nonumber u^{(1)}&=& u^n+\Delta t \tilde{L}(u^n),\\
  u^{(2)}&=& \frac{3}{4}u^n+\frac{1}{4}u^{(1)}+\frac{1}{4}\Delta t\tilde{L}(u^{(1)}),\\\nonumber 
  u^{n+1}&=& \frac{1}{3}u^n+\frac{2}{3}u^{(2)}+\frac{2}{3}\Delta
  t\tilde{L}(u^{(2)}),
\end{eqnarray}
where $\Delta t$ is the time-step; $u^n$ and $u^{n+1}$ are the
values of the physical quantity $u$ at the $n$-th and $(n + 1)$-th
time-levels, respectively; and $u^{(1)}$ and $u^{(2)}$ are two
temporary variables.

The third order backward-differentiation explicit scheme is defined as follows,
\begin{equation}\label{BDE3-1}
    \frac{1}{\Delta t}\sum\limits_{j=0}^{3} a_j
    u^{n+1-j}=\sum\limits_{j=0}^{2} b_j \tilde{L}(u^{n-j}),
\end{equation}
where
\begin{eqnarray}\label{BDE3-2}
  \nonumber a_0&\equiv& 1+\frac{1}{1+k_n}+\frac{1}{1+k_n+k_{n-1}},\\
  \nonumber a_1&\equiv& -\frac{(1+k_n)(1+k_n+k_{n-1})}{k_n(k_n+k_{n-1})},\\
  \nonumber a_2&\equiv& \frac{1+k_n+k_{n-1}}{k_nk_{n-1}(1+k_n)},\\
 a_3&\equiv&
 -\frac{1+k_n}{k_{n-1}(k_n+k_{n-1})(1+k_n+k_{n-1})};\\
 \nonumber\\
  \nonumber b_0&\equiv& \frac{(1+k_n)(1+k_n+k_{n-1})}{k_n(k_n+k_{n-1})},\\
  \nonumber b_1&\equiv& -\frac{1+k_n+k_{n-1}}{k_nk_{n-1}},\\\nonumber 
  b_2&\equiv& \frac{1+k_n}{k_{n-1}(k_n+k_{n-1})};
\end{eqnarray}
and
\begin{eqnarray}\label{BDE3-3}
  k_n&\equiv& \frac{t^n-t^{n-1}}{\Delta t},\\\nonumber 
  k_{n-1}&\equiv& \frac{t^{n-1}-t^{n-2}}{\Delta t};
\end{eqnarray}
with $t^n$, $t^{n-1}$, and $t^{n-2}$ being the values of time of the
$n$-th, $(n-1)$-th, and $(n-2)$-th time-steps, respectively.
This scheme consumes about three times less computational time than the Runge-Kutta scheme but is not able to start the time-integration by itself. Therefore, we combine these approaches and start integration with the latter, followed by the faster third order backward-differentiation explicit scheme. In such a way we are able to obtain a sufficient high order accuracy with minimal CPU-time
consumption.

\subsection{Specific techniques and assumptions}

Besides discretizing the equations describing non-stationary evolution of an accretion disk (Eqs.~\ref{e.nonstat.continuity} - \ref{e.nonstat.energy}) one has also to treat properly several numerical issues which may affect the integration. In order
to obtain a physical, transonic and numerically stable solution in a
finite domain, it is  necessary to impose appropriate
boundary conditions and to apply some filtering techniques to
overcome the inevitable spurious nonlinear numerical instabilities
in the code. For the purpose of this project we applied the following techniques:

(i) domain decomposition into 7 subdomains so that each sub-domain contains at most one single region of rapid variation,

(ii) spectral filtering by using an exponential filer in space to filter out the high-frequency
modes in each time step,

(iii) small numerical viscosity which makes the lowest density regions stable.
Details of these algorithms are given in \citem{lietal-07}.

\section{Limit cycles}

In this section we briefly discuss the limit-cycle evolution of an accretion disk for a non-rotating BH and mass accretion rate high enough to trigger the instability (we choose $\dot m=0.06$) assuming $\alpha=0.1$. Such a disk is radiation pressure dominated in its inner parts and is expected to exhibit the thermal instability.

\begin{figure}
  \centering\resizebox{.9	\textwidth}{!}{\includegraphics[angle=0]{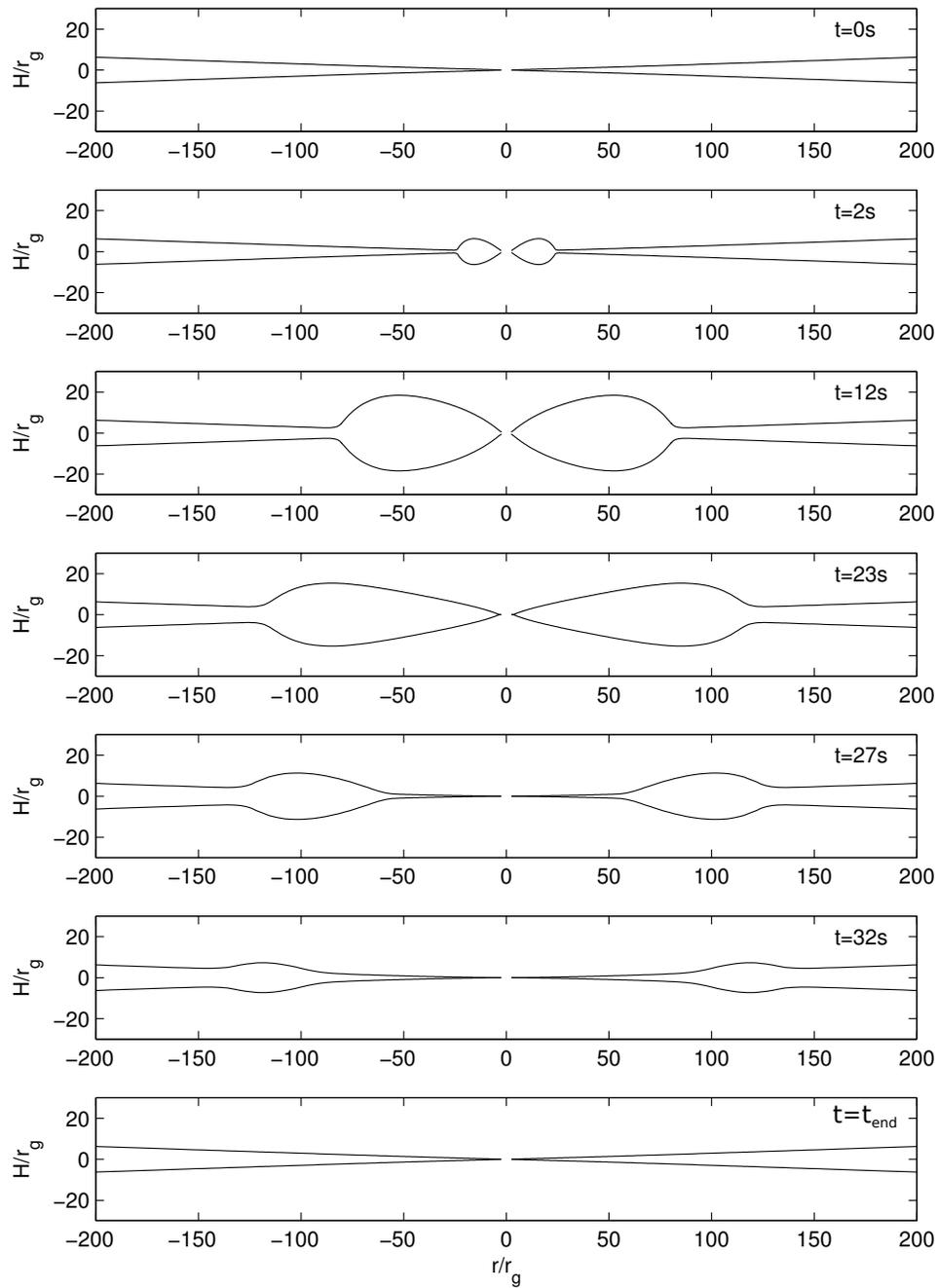}}
\caption{Evolution of the disk thickness during one full cycle for $\dot m=0.06$, $\alpha=0.1$ and $a_*=0$; $r_g=2M$. Figure after \citem{lietal-07}. }
  \label{f.xue_f1}
\end{figure}

\begin{figure}
\centering
 \subfigure
{
\includegraphics[height=.5\textwidth]{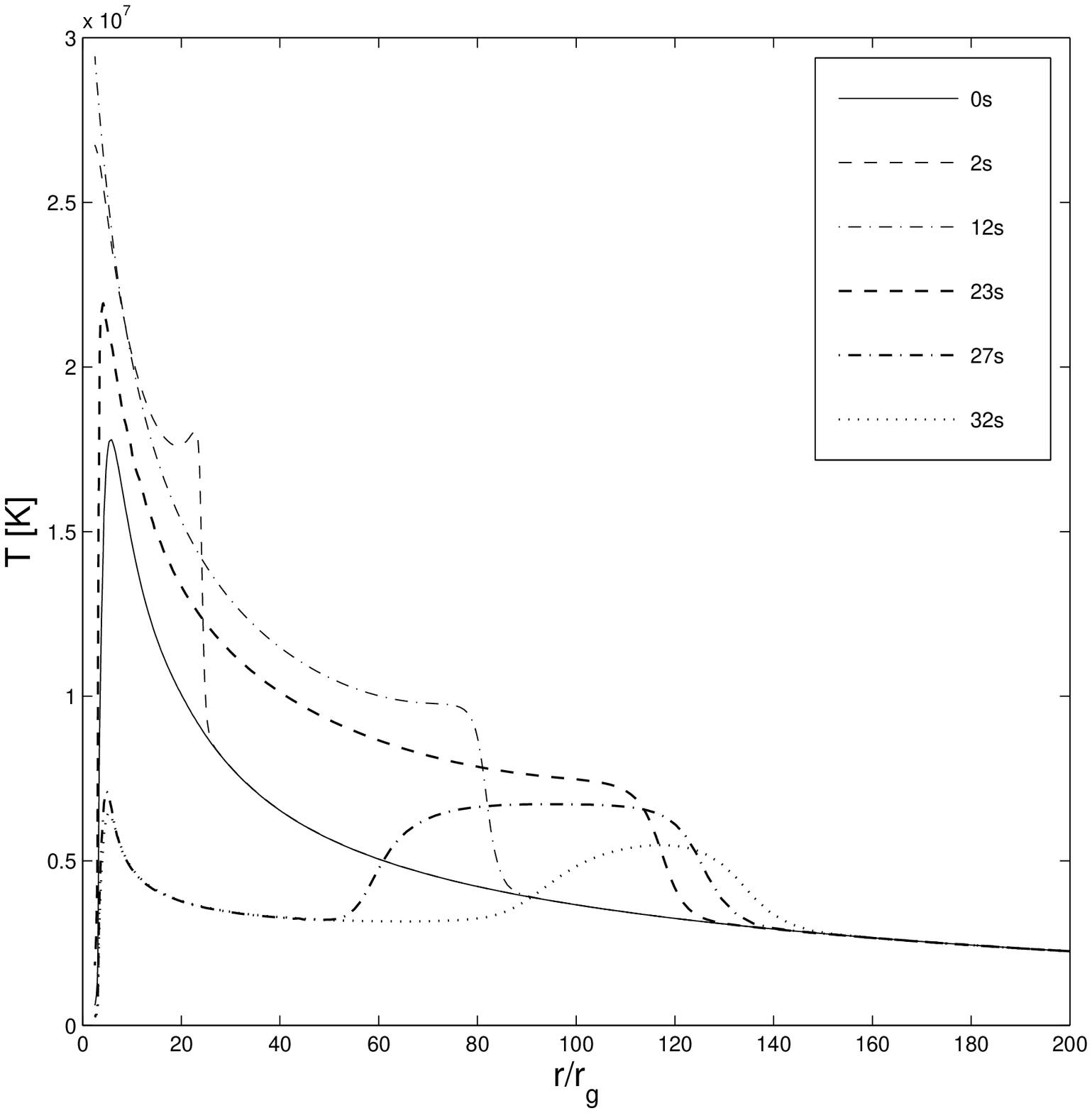}
}
\hspace{-1.2cm}
 \subfigure
{
\includegraphics[height=.5\textwidth]{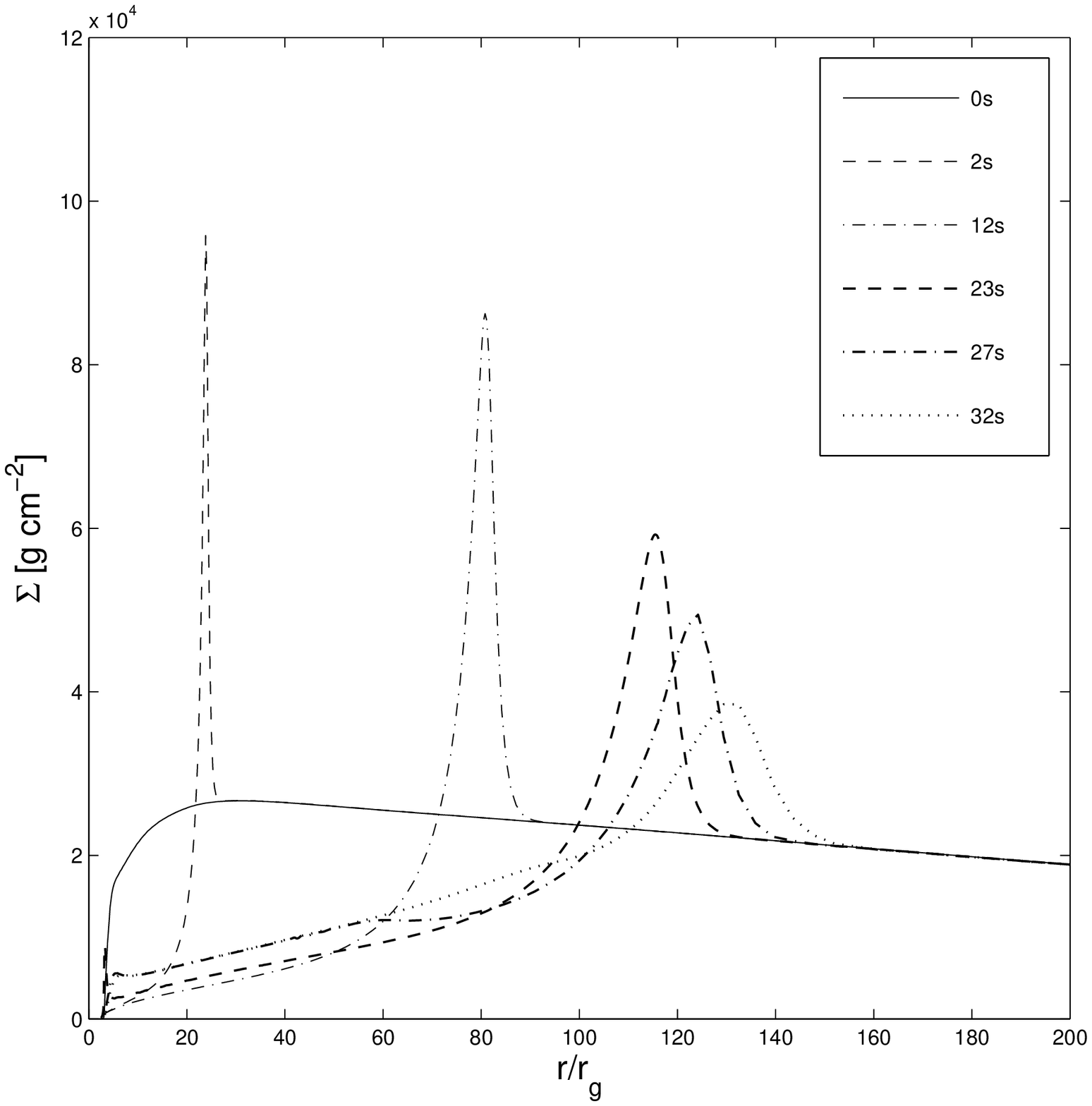}
}
\\
\vspace{-.031\textwidth}
 \subfigure
{
\includegraphics[height=.5\textwidth]{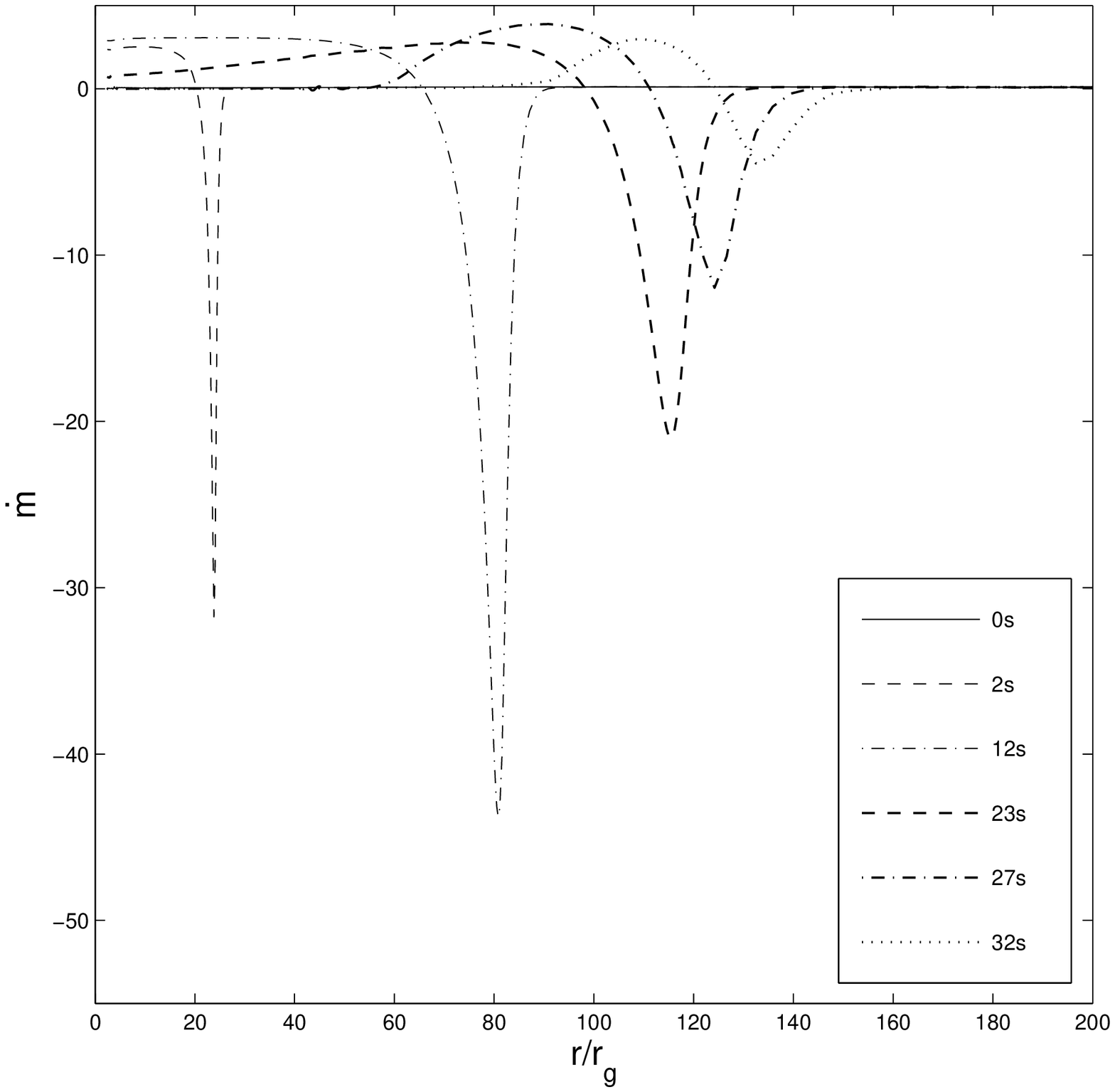}
}
\hspace{-1.2cm}
 \subfigure
{
\includegraphics[height=.5\textwidth]{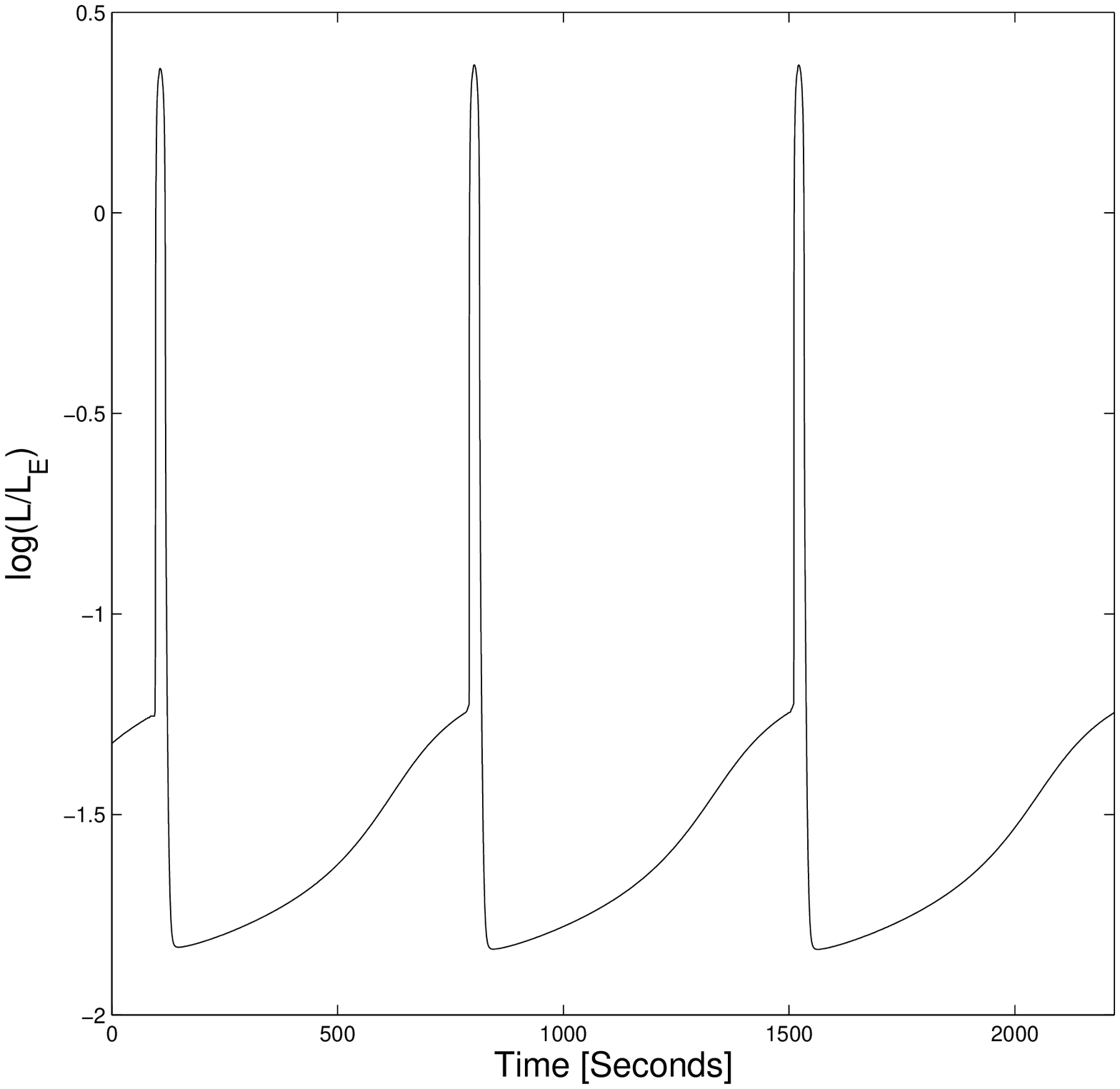}
}
\caption{Evolution of the temperature (top-left), surface density (top-right) and radial velocity (bottom-left panel) during one cycle. The bottom-right panel presents the outcoming light curve covering 3 cycle durations. A non-rotating BH, $\dot m=0.06$ and $\alpha=0.1$ were assumed. Figure after \citem{lietal-07}. $r_g=2M$.}
\label{f.xue_fN}
\end{figure}

The top-most panel of Figure \ref{f.xue_f1} presents the disk thickness just before the
start of the cycle ($t=0\rm s$). As the accretion rate is relatively low, the disk is geometrically thin ($h/r\ll1$). The disk structure does not correspond precisely to the stationary solution of a slim disk with the same model parameters, as the fluid configuration is not stationary and the disk is in the process of reformation which is terminated by the onset of the instability.
As it sets in ($t=2\rm s$, the second panel of Figure \ref{f.xue_f1} and corresponding lines in Figure ~\ref{f.xue_fN}) the temperature
rises rapidly, the disk expands in the vertical direction and a shock appears in the surface density and accretion rate profiles. These shocks move outward with time, forming an
expansion wave, heating the inner material, pushing it into the
BH, and perturbing the outer gas. It is worth mentioning that the accretion rate at the shocked region (the border between the expanding medium and the quasi-stationary disk) is negative which means radial outflows, causing the outer, incoming gas to further pile up.

At $t=12\rm s$ (e.g., the third panel of Fig. \ref{f.xue_f1}) the disk thickness and the negative spike of the mass accretion rate in the 
expanding region reach their maximum values. The local $\dot{m}$ inside the wavefront, however, exceeds $3$ which is far above the initial value
$\dot{m}=0.06$ and reflects the rapid inflow of gas.

Once the wavefront has moved beyond the unstable region
($r\lesssim120r_g=240\rm M$ for this case) the disk deflates as the instability is no longer supported by radiation pressure dominated parts of the disk. However, the expanded region still moves outward. The temperature and surface density in the inner parts rapidly drop down. The inner region becomes geometrically thin, with the temperature and surface density 
lower than the corresponding values at $t=0\rm s$

The outburst is essentially over after the
thermal timescale (e.g., $t=32\rm s$). It is followed
by a much slower process (taking place on the viscous timescale) of
refilling and reheating of the inner part of the disk.
 Finally, after $\sim 976\rm s$, the disk returns to
the same state as at $t=0\rm s$ and the thermal instability triggers again leading to a new cycle.

The bottom-right panel of Fig.~\ref{f.xue_fN} presents the bolometric luminosity of the disk, obtained by integrating the
radiated flux per unit area over the disk at successive
times for three complete cycles.
The light curve exhibits a burst with a duration of about $20$
seconds followed by a quiescent phase corresponding to the disk reformation process.
During the outburst the disk may reach super-Eddington luminosities ($1.5L_{\rm Edd}$ for this model).

\section{Results}
\label{Sec_NumExplo}
\subsection{Parameter space and characteristic quantities}
\label{Sec_NumExplo2}
In order to understand the effects of BH spin on the limit-cycle behavior, it is necessary to investigate the parameter space ($a_*,\alpha,\dot m$) with a series of models. In this section, we describe and discuss results of twelve runs assuming different sets of the input parameters span on the grid: $a_*=(0, 0.5, 0.95)$, $\alpha=(0.07, 0.1)$, and $\dot{m}=(0.06, 0.1)$ (cf. Table \ref{tab1}). All of the cases are thermally unstable and undergo recursive limit-cycle evolution.

\begin{sidewaystable*}
\begin{center}
\begin{tabular}{ccccccccc}
\hline \hline
$a_*$ & $\alpha$ & $\dot{m}$ & OD [s] &CD [s] &OD/CD & ML [$L_{\rm{Edd}}$]\\
\hline
0 & 0.07 & 0.06 &31.21$\pm$0.12 (0.38$\%$)& 1282$\pm$3 (0.23$\%$) &(2.43$\pm$0.02)$\times10^{-2}$ (0.61$\%$)& 1.281$\pm$0.001 (0.07$\%$)\\
0.5 & 0.07 & 0.06 &44.76$\pm$0.32 (0.71$\%$)& 1630$\pm$4 (0.25$\%$) &(2.75$\pm$0.03)$\times10^{-2}$ (0.96$\%$) & 1.568$\pm$0.003 (0.17$\%$)\\
0.95 & 0.07 & 0.06 &35.72$\pm$0.14 (0.38$\%$)& 1276$\pm$10 (0.79$\%$) &(2.80$\pm$0.04)$\times10^{-2}$ (1.17$\%$) & 2.219$\pm$0.018 (0.79$\%$)\\
0 & 0.1 & 0.06 &33.78$\pm$0.12 (0.35$\%$)& 976$\pm$3 (0.31$\%$) &(3.46$\pm$0.03)$\times10^{-2}$ (0.66$\%$) & 1.384$\pm$0.003 (0.18$\%$) \\
0.5 & 0.1 & 0.06 &40.67$\pm$0.21 (0.50$\%$)& 1136$\pm$5 (0.44$\%$) &(3.58$\pm$0.04)$\times10^{-2}$ (0.94$\%$) & 1.609$\pm$0.003 (0.18$\%$) \\
0.95 & 0.1 & 0.06 &54.68$\pm$0.45 (0.81$\%$)& 1517$\pm$10 (0.66$\%$) &(3.60$\pm$0.06)$\times10^{-2}$ (1.47$\%$) & 2.415$\pm$0.008 (0.32$\%$)\\
0 & 0.07 & 0.1 &44.80$\pm$0.15 (0.33$\%$)& 1246$\pm$4 (0.33$\%$) &(3.40$\pm$0.03)$\times10^{-2}$ (0.66$\%$) & 1.385$\pm$0.002 (0.14$\%$)\\
0.5 & 0.07 & 0.1 &46.86$\pm$0.12 (0.24$\%$)& 1233$\pm$3 (0.25$\%$) &(3.80$\pm$0.02)$\times10^{-2}$ (0.49$\%$) & 1.596$\pm$0.003 (0.17$\%$)\\
0.95 & 0.07 & 0.1 &50.87$\pm$0.27 (0.52$\%$) & 1322$\pm$3 (0.23$\%$) &(3.85$\pm$0.03)$\times10^{-2}$ (0.75$\%$) & 2.336$\pm$0.006 (0.23$\%$)\\
0 & 0.1 & 0.1 &53.73$\pm$0.53 (0.98$\%$)& 1070$\pm$4 (0.38$\%$) &(5.02$\pm$0.07)$\times10^{-2}$ (1.36$\%$) & 1.456$\pm$0.002 (0.08$\%$)\\
0.5 & 0.1 & 0.1 &55.47$\pm$0.70 (1.25$\%$)& 1086$\pm$9 (0.83$\%$) &(5.11$\pm$0.11)$\times10^{-2}$ (2.08$\%$) &  1.692$\pm$0.003 (0.17$\%$)\\
0.95 & 0.1 & 0.1 &53.94$\pm$0.47 (0.87$\%$)& 1084$\pm$5 (0.47$\%$) &(4.98$\pm$0.07)$\times10^{-2}$ (1.34$\%$) & 2.482$\pm$0.004 (0.16$\%$)\\

\hline
\end{tabular}
\caption{Limit-cycle characteristic quantities. See Section~\ref{Sec_NumExplo2} for the definitions of OD, CD, and ML. The values showed in the columns, OD, CD, OD/CD and ML are collected as "mean value"$\pm$"standard deviation" (relative deviation). \label{tab1}
}
\end{center}
\end{sidewaystable*}

\begin{figure}
\centering\includegraphics[width=.7\textwidth]{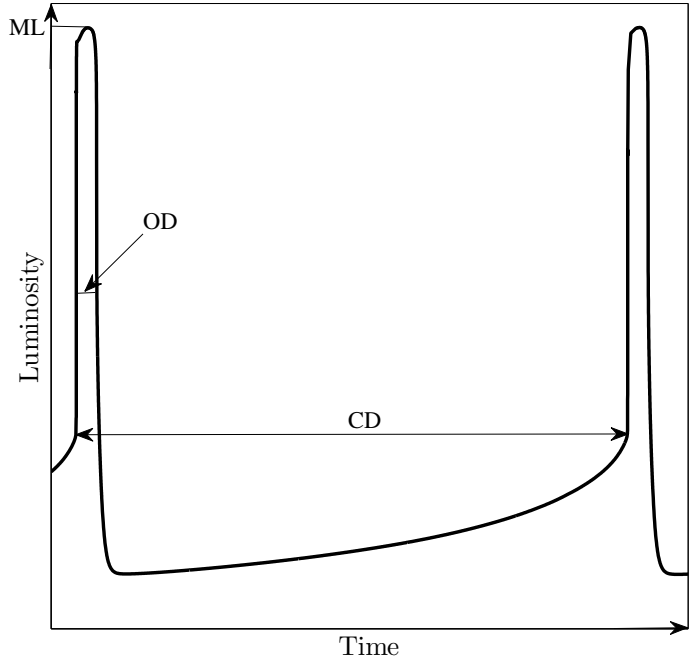}
\caption{Sketch for explaining the definitions of characteristic quantities. \label{fig:sketch}}
\end{figure}

To describe the limit-cycle behavior, we define four characteristic quantities, which are the Outburst Duration (the full width at half maximum of light-curve, hereafter OD), Cycle Duration (time interval between two outbursts, hereafter CD), ratio of OD to CD, and Maximal Luminosity (the peak value of the light curve during outburst, hereafter ML; calculated intrinsically, with no ray-tracing), respectively. In Figure \ref{fig:sketch}, we visualize these definitions on a sketched light-curve.

The computations for each one of those twelve cases were continued until the CD converged to a constant value. The constancy of CDs implies that the computations have reached the state uniquely determined by the crucial parameters ($a_*,\alpha,\dot m$), without any impact of the initial conditions. In practice, the required level of convergence is achieved after three or five cycles. We performed computations for each case for more than seven cycles (some cases reached ten cycles) and we based our subsequent analysis on the last four cycles only.

In Table \ref{tab1} we list the mean values, standard and relative deviations of each characteristic quantity for our twelve models. These values were calculated basing on the retained cycles only. Since the relative deviations (the percentages in the brackets) are all very low, we do not show the error bars for the data points in the subsequent figures.

\subsection{Impact of BH spin}
\begin{figure}
\centering\includegraphics[width=.95\textwidth]{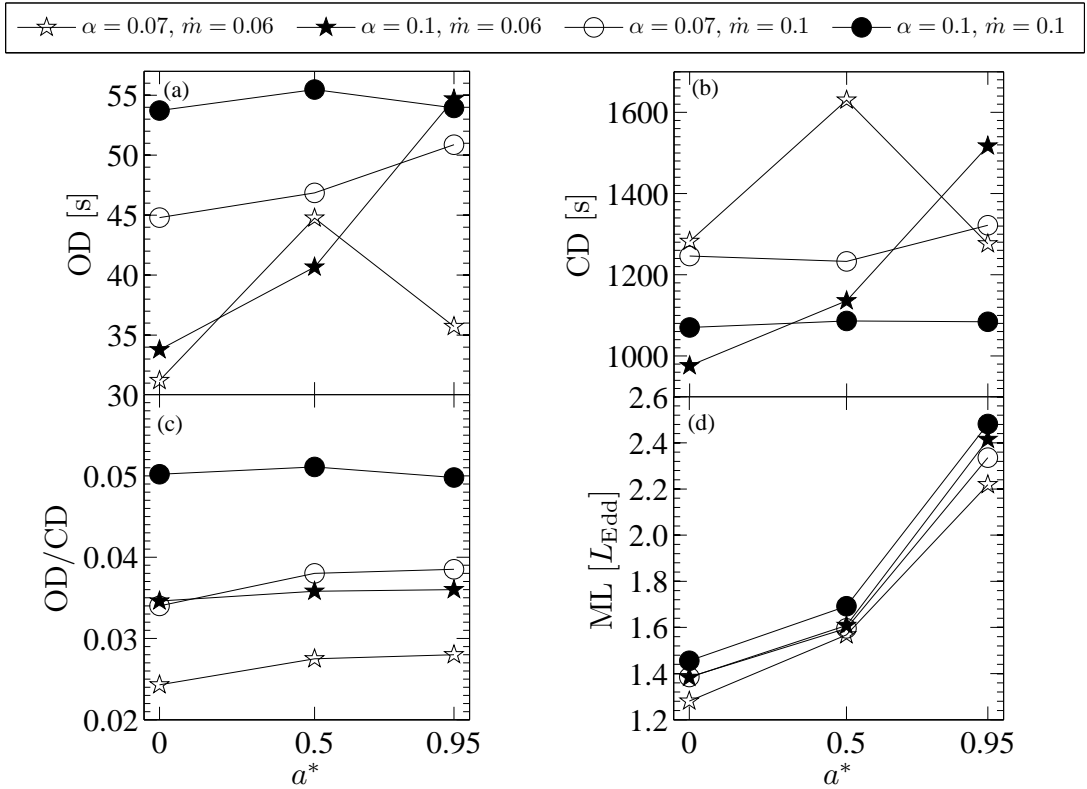}
\vspace{-1cm}
\caption{Correlation between the characteristic quantities and $a_*$. \label{fig:4plots}}
\end{figure}

\begin{figure}
\centering\includegraphics[width=.8\textwidth]{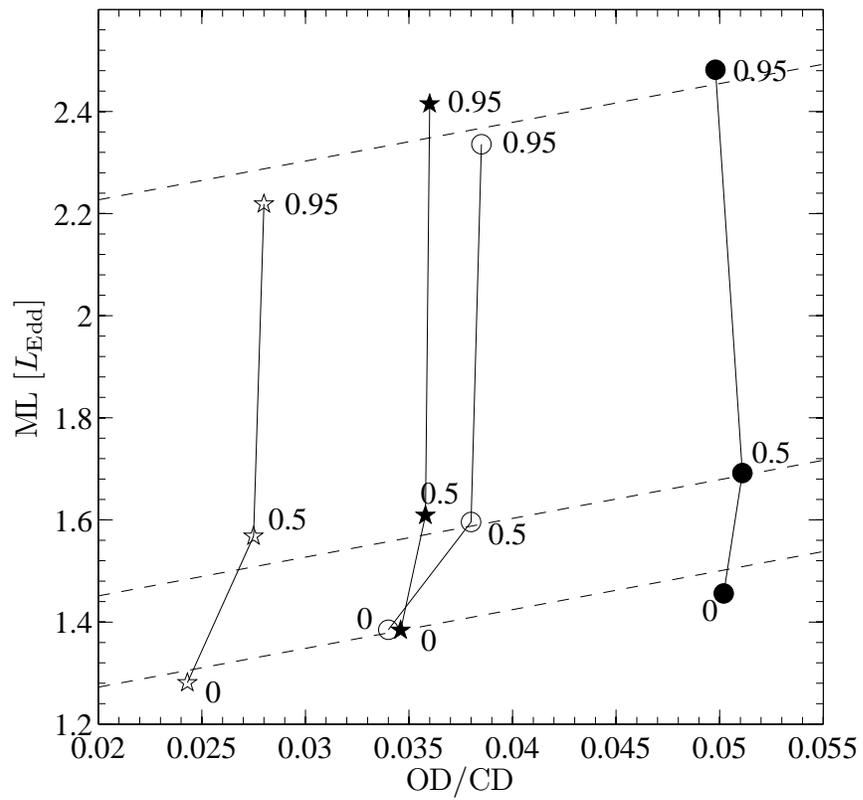}
\caption{Plot of maximal luminosity versus ratio of OD to CD. The numbers near each data point are the values of $a_*$. The meanings of markers are the same as those in Figure \ref{fig:4plots}. The dashed lines are the OD/CD-ML fitted lines for different $a_*$ cases. \label{fig:ML_ODC}}
\end{figure}

In order to reveal the effects of BH spin on limit-cycle behavior, we collect our twelve models into four groups to reflect the different effects of viscosity and mass supply (i.e., mass accretion rate at the outer boundary). The models with ($\alpha=0.07, \dot m=0.06$) and different spins are represented by empty stars, ($\alpha=0.1, \dot m=0.06$) by filled stars, ($\alpha=0.07, \dot m=0.1$) by empty circles, and ($\alpha=0.1, \dot m=0.1$) by filled circles in the Figures \ref{fig:4plots} and \ref{fig:ML_ODC} (both are based on the data from Table \ref{tab1}).

In the four panels of Figure \ref{fig:4plots} we plot OD, CD, OD/CD and ML versus $a_*$.
It is clear that there is no distinct and consistent dependence on $a_*$ neither for OD nor CD (panels (a) and (b)). The ratio of OD to CD (panel (c)) is almost independent of $a_*$ for all sets of input parameters. On the contrary, ML exhibits a perfect correlation with $a_*$ (panel (d)), though this relation is still slightly distorted by the impact of $\alpha$ and $\dot m$.

These facts may be easily understood in the framework of the general relativity. The effects of BH spin are restricted to the innermost parts of accretion disks. Therefore, the quantity ML, which corresponds mostly to the radiation emerging from the inner part of an accretion disk, must display a strong dependence on $a_*$. The other quantities (OD, CD and OD/CD) cannot have such strong dependence on $a_*$ as they depend on the disk structure and its evolution at wide range of radii (from the inner part up to more than $200\rm M$), where the effects of viscosity and mass-supply dominates over the effect of BH spin.

\subsection{Possibility of estimating the black hole spin with limit-cycle}
As has been discussed in the previous section, the effects of BH spin on the quantities OD, CD and OD/CD are easily disturbed and even obscured by the complex effects of viscosity and mass-supply. Thus, they cannot serve as proper probers of BH spin. Even ML, which has monotonic and positive correlation with $a_*$, cannot be used directly to estimate the BH spin, because the ML dependence on BH spin is significantly affected by other factors (see panel (d) of Figure \ref{fig:4plots}).

However, it is reasonable to base both on ML and the OD to CD ratio. In Figure \ref{fig:ML_ODC}, we show the OD/CD-ML diagram for all the models. The OD/CD ratio raises with increasing values $\alpha$ and $\dot m$ and is hardly dependent on $a_*$ (see panel (c) of Figure \ref{fig:4plots}). It is remarkable, that the combined effects of viscosity and mass-supply make ML scale almost linearly with OD/CD ratio. For a given BH spin there is a single straight line reflecting the ML dependence on OD/CD ratio. These lines do not intercept and have similar slopes. The OD/CD-ML diagram may be, therefore, used for rough BH spin estimation if only the limit cycle parameters are read from observational data.

We find that the dependence of ML on OD/CD may be approximated by the following formula
\begin{equation}
\label{mlvsodcd_fit}
 {\rm ML}\,/\,L_{\rm{Edd}}=7.59\,\, {\rm OD/CD} + 0.71 + 7.18\, \tilde\eta,
\end{equation}
where $\tilde\eta$ is the efficiency of accretion for a thin disk and is given by (Eq.~\ref{e.etaorg2}),
\be
\tilde\eta=1-\sqrt{1-\frac{2 M}{3 r_{\rm ISCO}}},
\ee
with $r_{\rm ISCO}$ being the radius of the marginally stable orbit. In Figure \ref{fig:ML_ODC} we plot with dashed lines the fits obtained with these formulae for $a_*=0,\,0.5$ and $0.95$. Knowing the values of ML and OD/CD, one can easily obtain from Eq.~\ref{mlvsodcd_fit} the radius of the marginally stable orbit and, subsequently, the rough estimate of BH spin.

There are, however, at least few limitations for the application of this method. The first one is related to the fact that the OD/CD-ML relation is still subject to significant dispersion caused by the combined effects of $\alpha$ and $\dot m$ (see Fig. \ref{fig:ML_ODC}) as well as to ray-tracing effects we neglect, related to e.g., the inclination angle and limb darkening. Another one is the model-dependence of our method. We assume the limited classical theory to describe the accretion flow. We neglect mass outflows, magnetic fields and apply the $\alpha$ prescription of viscosity with $\alpha$ independent of radius. These factors, especially the viscosity treatment, may have significant impact on the ML vs OD/CD relation that we have obtained. \citem{Nayakshin00} showed that reproducing the light curves of GRS 1915+105 is possible only if much more complicated models are considered. Nevertheless, if the limit-cycle behavior is the actual reason for the observed variability of some microquasars, and if the extraction of ML and OD/CD parameters from the light curves is possible, then, under the limitations of the model presented in this work, the BH spin can be roughly estimated using the method presented here, before other accurate methods, e.g., fitting spectral energy distribution, are applied.

\subsection{Evolution of the temperature-luminosity relation for the limit-cycle in classical theory}
\begin{figure}
\centering\includegraphics[width=.9\textwidth]{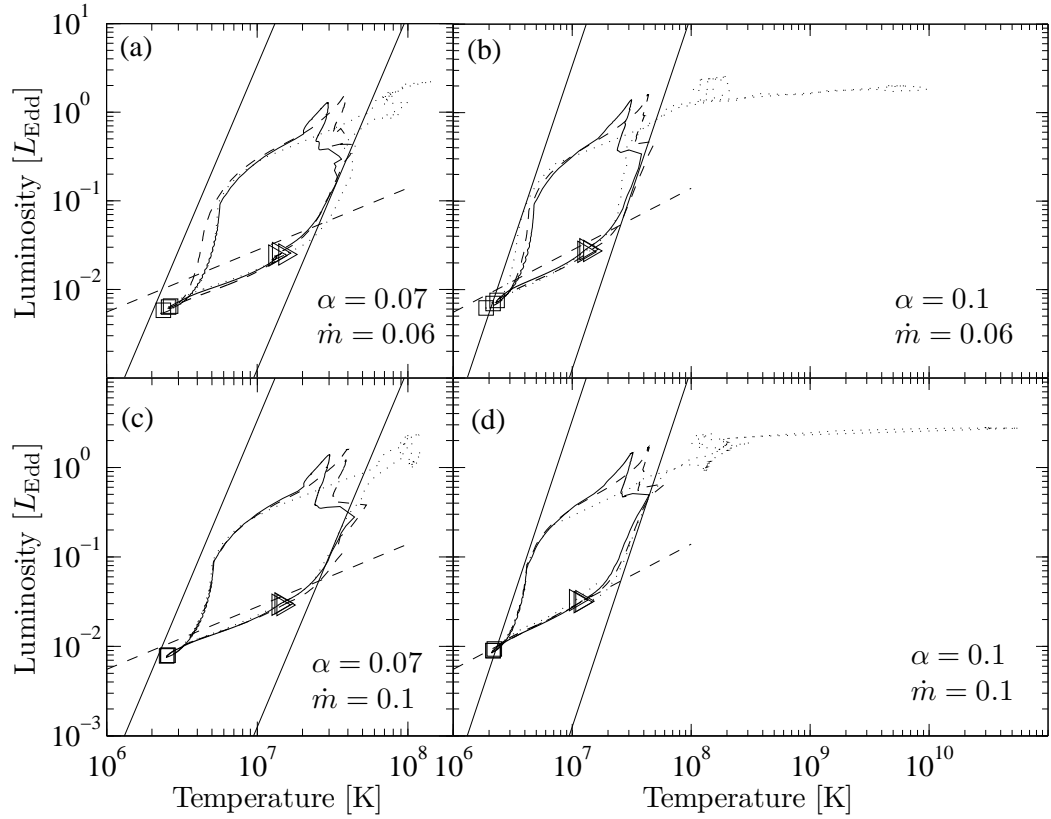}
\caption{Evolution of the last cycle for the groups with different $\alpha$ and $\dot m$ in the maximal disk temperature - luminosity ($T$-$L$) plane. All cycles begin at the triangle points, and then evolve along the curve in the anti-clockwise direction. The square points are used to denote the quarter of cycle durations. The solid straight lines in each panel represent $L_{\rm disk}\propto T^4_{\rm max}$ for a non-spinning BH (see Eq. (3) of \citem{Gierlinski04}). They are calculated with the color temperature correction $f_{\rm col}=1$ for the left lines and $f_{\rm col}=7.2$ for the right ones. The dashed straight lines denote the $L_{\rm disk}\propto T^{0.7}_{\rm max}$ relation. In each panel, the solid, dashed, and dotted curves are for the different BH spins with $a_*=0, 0.5, 0.95$, respectively. \label{fig:L_MT}}
\end{figure}

The maximal disk temperature-luminosity ($T$-$L$) diagrams have been used to compare the observations with the predictions of the classical  theory of limit-cycles (e.g., \citem{Gierlinski04,Kubota04,Kubota01}). In Figure \ref{fig:L_MT}, we show the $T$-$L$ evolution for the last calculated cycle of each model. The panels present the evolution for a given set of $\alpha$ and $\dot m$ parameters and different BH spins ($a_* = 0,\,0.5$ and $0.95$ marked by solid, dashed and dotted lines, respectively).
These curves reveal several common features: (1) All of the cycles begin at the closely-located triangle points and spend a quarter of CD evolving along the curve in the anti-clockwise direction before they reach the square points, and finally spend the other $3/4$ of CD to return back to the starting location (triangles); (2) The evolution of all the cycles fits (with exception to the outbursts of the $a_*=0.95$ case) between two straight lines corresponding to $L_{\rm disk}\propto T^4_{\rm max}$ relations with different color correction factors ($f_{\rm col}=1$ and $7.2$, see \citem{Gierlinski04}); (3) The $T$-$L$
 curves of all the cycles are similar for most of CD except for the outburst state --- the BH spin impact is revealed at the outburst of limit-cycle
 only; (4) During the mass restoring process (prior to the next outburst) the disk obeys $L_{\rm disk}\propto T^{0.7}_{\rm max}$  (see the dashed straight
 lines in all panels).

\section{Discussion}
The theory
based on the standard viscosity prescription
\begin{equation}
{\cal W} = -\alpha P,
\label{standard-alpha-prescription-eq}
\end{equation}
where $\cal W$ and $P$ are vertically integrated stress and total pressure, respectively, predicts that
radiation pressure-supported regions of radiatively-efficient accretion disks
are thermally and viscously unstable.
The unstable luminosity range is smaller for an (ad hoc)
alternative viscosity prescription
\begin{equation}
{\cal W} = -\alpha \sqrt {P^{2 -\mu} P^{\mu}_{\rm gas}} ,
\label{alternative-alpha-prescription}
\end{equation}
where $0 \le \mu \le 1$ is a constant.  The value $\mu =
0$ corresponds to the standard prescription, and $\mu = 1$ to the
often considered ``geometrical mean'' prescription. The
instability disappears for $\mu > 8/7$. From the recent important
paper of \citem{hirose-09-b} it follows that the standard
\citem{shakura-73} viscosity prescription
($\mu=0$) fits much better results of
their MHD simulations of accretion disks than the alternative
prescriptions (\ref{alternative-alpha-prescription}) with $\mu >
0$. However, \citem{hirose-09-a} also found that the radiation
pressure supported MHD disks were thermally {\it stable} in their
simulations (it is not known whether they are viscously stable).
This result contradicts the hydrodynamical stability analysis, and
therefore needs an explanation --- what are the MHD effects that
stabilize the disk? \citem{hirose-09-a} argue that although the
viscosity prescription has the form
(\ref{standard-alpha-prescription-eq}), it should nevertheless be
modified in a subtle way, because the ``viscous heating''
${Q}^{\rm vis}$ that occurs in the disk is {\it delayed} with
respect to the MHD stress ${\cal W} = {\cal W}_{\rm MHD}$,
\begin{equation}
{ Q}^{\rm vis}(t) = {\cal W}(t - \Delta t)
\frac{d\Omega}{dr}.
\label{delayed-heating}
\end{equation}
Here $\Delta t$ is the delay, found to be about 10 dynamical
times, and $\Omega$ is the angular rotation of matter. It should
be obvious that this modification is irrelevant in the stationary
case.

In an exhaustive {\it Newtonian} analytic stability analysis
\citepm{ciesielski-11} our group confirmed (in general) the suggestion made
by \citem{hirose-09-a} that the delayed heating
(\ref{delayed-heating}) stabilizes, in some parameter range, the
radiation pressure thermal instability. However, we also found a
variety of different parameter ranges with interesting, and
complex, oscillatory behaviors that need to be further
investigated. 

On the observational front, spins of BHs in several galactic BH
binaries have been recently measured by fitting their spectral
energy distribution to the relativistic geometrically thin
accretion disk model (e.g., \citem{shafee-06,McClintock06}). From among
these objects, the GRS 1915+105 is the most special one. It is
believed to have a near-extreme Kerr BH with the spin very close
the theoretical maximal value $a_*=1$
(see, \citem{McClintock06}). It is also the only object that
exhibits the quasi-regular luminosity variations \citepm{Belloni97,
Nayakshin00, Janiuk02, Watarai03, Ohsuga06, Kawata06, janiukczerny11}, similar to
the limit-cycle predicted by the standard, Newtonian,
hydrodynamical models. There are obvious questions to investigate
here. May one explain the luminosity variations observed in GRS
1915+105 as the classic ``slim disk'' limit-cycle behavior? What
is the role of the nearly extreme spin of GRS 1915+105 in the
context of the observed variability? Could the case of GRS
1915+105 help to find the answer about the proper viscosity
prescription?

Our research shows that the shape of light curves is weakly dependent on BH spin. The only parameter that feels the impact of the BH angular momentum is the maximal luminosity. Therefore, under the assumptions adopted in this paper (e.g., constant $\alpha$ in radius), limit cycles produce light curves only qualitatively similar to those observed in GRS1915+105, independently of BH spin. To obtain better agreement one has to construct more complicated models (e.g., \citem{Nayakshin00}), but the mechanism of the instability may remain intact. We prove that the cycle duration weakly depends on the rotation of
BH. Thus, the observed value for GRS1915+105 most probably results
from combined effects of viscosity and accretion rate and is not
affected by the spin of the central BH. The very presence of limit cycle variability only in this object challenges our understanding of the mechanisms of viscosity.

\chapter{Vertical structure of slim accretion disks}
\label{chapter-vertical}
\label{s.vertical}
Slim disks, just like \citem{shakura-73} model,
are based on the assumption that the dissipation mechanisms
operating in accretion flows may be described by a viscous stress
tensor whose leading component is proportional to the pressure.
 
Slim disk solutions, again just like the \citem{shakura-73} and \citem{nt}
solutions, are obtained by solving vertically averaged (or
height-integrated) radial equations of motion. Thus, steady slim
disks neglect the vertical structure of
flow, and describe essentially flat fluid configurations. Although an
expression for the radial dependence of disk thickness is obtained,
the slope of the disk surface is usually neglected in the discussion
of emergent spectra, where a plane-parallel atmosphere is typically
considered for the purposes of computing radiative transfer in the
vertical direction.

In the case of slim disks, the radial equations of structure
are a set of ordinary differential equations (Section~\ref{s.stateq}).
The set of equations is closed by including relations describing
vertical hydrostatic equilibrium and vertical transport of energy. By
solving these equations with appropriate boundary (or regularity)
conditions one obtains the radial profiles of the central temperature,
$T_C(r)$, surface density, $\Sigma(r)$, height-integrated
pressure, $P(r)$, half-thickness of the disk, $h(r)$, a radial
velocity $V(r)$, emerging flux of radiation, $F(r)$, and certain
other physical quantities. 

In this Chapter we are taking a first step towards constructing truly
three-dimensional slim disk models, by solving a set of differential
equations describing the vertical structure of the disk. The resulting
$z$-dependence of physical quantities is used to compute certain
coefficients that enter the radial equations, and that up till now
have been estimated with algebraic expressions (Eqs.~\ref{def.eta1} - \ref{def.eta4}).
We will refer to the resulting models as
``two-dimensional`` (2-D),
although it has to be understood that in contrast to the thin-disk solutions of 
\citepm{urpin,kita}, the actual meridional flow has not
been computed in this paper---the average radial velocity
alone has been considered in the structure equations.
However, the models are now
self-consistent in the sense that the vertical averages of physical
quantities that form the coefficients of the radial ODEs do correspond
to the vertical structure considered in the radiative transfer
calculation --- in Chapter~\ref{chapter-stationary} the vertical structure of the radiative
atmosphere was considered {\it a posteriori}, and it had no influence
on the radial structure of the disk.

\section{Radiative transfer}
\label{s.radiative}
Accretion disks are not simple rotating fluid configurations which may be described by the hydrostatic balance only. The turbulent motions, magnetic fields, mean radial velocity, shear and radiation make the description of the disk interior very complicated. The simplest approach, which has been successfully used for a long time, is to consider disk annuli separately, neglecting the radial gradients, radial net flux of radiation and accounting only for the mean flow. Under these conditions, disk layers may be described as plane-parallel and are subject to the vertical hydrostatic equilibrium and radiative transfer along that direction only. Thus, the problem is reduced to one-dimensional and may be relatively easy solved.

The equation of hydrostatic equilibrium reflects the balance between the gradients of pressure and gravitational potential. For accretion disks annuli and cylindrical coordinates $(r,\phi,z)$ it may be put in the form (Eq.~\ref{sh_vert1}),

\be
\frac 1\rho\der pz=-\frac {GM}{r^3}{z},
\ee
where $p=p_{\rm gas}+p_{\rm rad}$.

The equations describing the radiative part are more complicated. In the following sections we briefly summarize the theory of radiative transfer and introduce the Rosseland approximation which will be used in the model described in this Chapter.

\subsection{Transfer equation}

The general radiative transfer equation \citepm{rybickilightman},

\be\der {I_\nu}s=-\alpha_\nu I_\nu + j_\nu,
\label{e.radtr.1}\ee
describes the variation of specific intensity $I_\nu$ along the distance $s$. $\alpha_\nu$ and $j_\nu$ stand for the absorption and emission coefficients, respectively. Eq.~\ref{e.radtr.1} takes a particularly simple form if another variable, $\tau_\nu$ - the optical depth, is introduced. It is defined by

\be d\tau_\nu=\alpha_\nu ds\ee
and
\be\tau_\nu(s)=\int_{s_0}^s\alpha_\nu(s){\rm d}s'.\ee

If $\tau_\nu$ integrated throughout the medium satisfies the condition $\tau_\nu>1$, we say that medium is optically thick. On the contrary, when $\tau_\nu<1$, the medium is optically thin --- the typical photon of frequency $\nu$ can traverse an optically thin medium without being absorbed, while it is not possible in an optically thick medium.

Introducing the optical depth, the radiative transfer equation simplifies to
\be\der{I_\nu}{\tau_\nu}=-I_\nu+S_\nu,\ee
where the source function $S_\nu$, defined as the ratio of the emission and absorption coefficients,
\be S_\nu\equiv \frac{j_\nu}{\alpha_\nu}\ee,
has been introduced.

In general, the intrinsic emission comes from two processes: the reemission after absorption (thermal emission) and scatterings. If the former dominates, according to the Kirchoff's Law for thermal emission \citepm{rybickilightman}, we have,
\be j_\nu=\alpha_\nu B_\nu(T)\ee
and
\be S_\nu=B_\nu(T),\ee
where $B_\nu(T)$ is the Planck function describing the black body radiation (Eq.~\ref{e.Bnu}).
The transfer equation is, then,
\be\der{I_\nu}{s}=-\alpha_\nu (I_\nu-B_\nu(T))\ee
or
\be\der{I_\nu}{\tau_\nu}=-I_\nu+B_\nu(T).\ee

If the medium is dominated by scatterings, we have
\be j_\nu=\sigma_\nu J_\nu,\ee
where $\sigma_\nu$ is the absorption coefficient of the scattering process (the scattering coefficient) and $J_\nu$ is the mean intensity of radiation:
\be J_\nu=\frac1{4\pi}\int I_\nu{\rm d}\Omega.\ee
It is then straightforward to get the following equality,
\be S_\nu=J_\nu\ee
and the following form of the radiative transfer equation,
\be \der {I_\nu}{s}=-\sigma_\nu(I_\nu - J_\nu).\ee

If we take into account combined scattering and absorption we have the following general form of the transfer equation,
\bea\nonumber
\der{I_\nu}{s}&=&-\alpha_\nu (I_\nu-B_\nu(T))-\sigma_\nu(I_\nu-J_\nu) =\\
&=&-(\alpha_\nu+\sigma_\nu)(I_\nu-S_\nu),
\label{e.transferequation}
\eea
with the source function given by,
\be S_\nu=\frac{\alpha_\nu B_\nu + \sigma_\nu J_\nu}{\alpha_\nu+\sigma_\nu}.\ee

\subsection{Optical depths}

The net absorption coefficient is now equal to $\alpha_\nu+\sigma_\nu$ and is called the total extinction coefficient. Thus, we should define the optical depth as,
\be d\tau_\nu=(\alpha_\nu+\sigma_\nu)ds.\label{e.radtran.tautota}\ee
The mean free path of a photon before scattering or absorption is
\be l_\nu=\frac1{\alpha_\nu+\sigma_\nu}.\label{e.radtran.lnu}\ee
Following the random walk arguments \citepm{rybickilightman} we also may obtain an expression for the net displacement between the points of creation and destruction of a typical photon,
\be l_*=\frac{1}{\sqrt{\alpha_\nu(\alpha_\nu+\sigma_\nu)}},\label{e.radtran.lnustar}\ee
which is also called the \textit{effective mean path} or the \textit{thermalization length}.

From Eqs.~\ref{e.radtran.lnu} and \ref{e.radtran.lnustar} it follows that the mean path of a photon before scattering or absorption may be significantly shorter than the effective mean path. This is the case for scattering-dominated media where absorptions are less likely than scatterings. The regular total optical depth defined in Eq.~\ref{e.radtran.tautota} reflects only the more important process. It is therefore convenient to introduce, basing on Eq.~\ref{e.radtran.lnustar}, the \textit{effective optical depth},
\be\tau_*\approx\sqrt{\tau_a(\tau_a+\tau_s)},\ee
where $\tau_a$ and $\tau_s$ stand for the optical depths for absorption and scattering, respectively.

The medium is \textit{effectively optically thin}, $\tau_*\ll1$, when the effective mean path is large compared with the size of the medium --- most photons escape out by random walking before being absorbed. On the other hand, in \textit{effectively thick} medium most photons thermally emitted at large depths will be absorbed before they get out. The medium at large effective optical depths is close to the thermal equilibrium and we have both $S_\nu\rightarrow B_\nu$ and $I_\nu\rightarrow B_\nu$ (e.g., \citem{MihalasBook}).

\subsection{Rosseland approximation}

From the previous section we know that deep inside an effectively thick medium the radiation is in thermal equilibrium. Let us assume that the properties of the medium depend only on the geometrical depth. Symmetry requires that the intensity of radiation may depend only on the angle $\theta$ between the given direction and the distinguished coordinate. Introducing $\mu=\cos\theta$, so that
\be ds=\frac{dz}{\cos\theta}=\frac{dz}{\mu},\ee
we may put the transfer equation (Eq.~\ref{e.transferequation}) in the following form,
\be I_\nu(z,\mu)=S_\nu-\frac{\mu}{\alpha_\nu+\sigma_\nu}\pder{I_\nu}z.\ee
Since at large effective optical depths we have $I_\nu=S_\nu=B_\nu$, we may approximate this formula with
\be I_\nu(z,\mu)\approx B_\nu-\frac{\mu}{\alpha_\nu+\sigma_\nu}\pder{B_\nu}z.\ee
The vertical flux of radiation is given by the following integral,
\be F_\nu(z)=\int I_\nu(z,\mu)\mu {\rm d}\Omega=2\pi\int_{-1}^1I_\nu(z,\mu)\mu {\rm d}\mu,\ee
which may be easily calculated. Since only the angle-dependent part contributes, we have
\be F_\nu(z)=-\frac{4\pi}{3(\alpha_\nu+\sigma_\nu)}\pder{B_\nu(T)}{T}\pder Tz.\ee
The total flux may be obtained by integration over all frequencies:
\be F(z)=\int_0^\infty F_\nu(z){\rm d}\nu=-\frac{4\pi}{3}\pder Tz\int_0^\infty\frac1{\alpha_\nu+\sigma_\nu}\pder{B_\nu(T)}{T}{\rm d}\nu.\ee
Taking into account the following equality,
\be \int_0^\infty\pder{B_\nu(T)}{T}{\rm d}\nu=\frac{4\sigma T^3}{\pi}\ee
and introducing the \textit{Rosseland mean absorption coefficient} $\alpha_R$:
\be \frac1{\alpha_R}\equiv \frac{\int_0^\infty\frac1{\alpha_\nu+\sigma_\nu}\pder{B_\nu(T)}{T}{\rm d}\nu}{\int_0^\infty\pder{B_\nu(T)}{T}{\rm d}\nu},\ee
we finally have:
\be\label{e.radtran.ross}
F(z)=-\frac{16\sigma T^3}{3\alpha_R}\pder Tz.
\ee
This is the \textit{Rosseland} or \textit{diffusive} approximation for the energy flux. It shows that the radiative energy transport deep in effectively thick media is of the same nature as heat diffusion. It is valid not only in plane-parallel media but everywhere where the medium properties change slowly on the scale of the radiation mean free paths. The Rosseland approximation says that the vector flux of energy is opposite to the temperature gradient and gives its magnitude through Eq.~\ref{e.radtran.ross}.

Heat in gases may be transported not only through radiative processes, but also through convection. The latter occurs due to temperature differences which affect the density, and thus relative buoyancy, of the fluid.  Heavier (more dense) components will fall while lighter (less dense) components rise, leading to bulk fluid movement and, as a result, transport of heat. The onset of convection may be determined by comparing the local values of the thermodynamical radiative and convective gradients. In the model described in this Chapter we account for convection by applying the mixing length approximation \citepm{paczynski69}.

\subsection{Dissipation}

Accretion disk annuli are not conservative in the sense that the flux of energy changes with the vertical coordinate due to viscous dissipative processes. Basing on Eq.~\ref{sh_dotQ} we expect the whole annuli to generate the following flux of energy at its surface,
\be F = {\cal F}(h)=\int_{0}^{+h}t^{r\phi}\,\sigma_{r\phi}\,{\rm dz},\ee
where $t^{r\phi}$ and $\sigma_{r\phi}$ are components of the stress and shear tensors, respectively. Using the $\alpha$ approach and assuming the heat is not advected, we may write in the non-relativistic limit 
\be F= -\frac 12 \alpha P r \der\Omega r\ee
with $P$ being the vertically integrated pressure. The models based on vertically averaged quantities (like \citem{shakura-73} or slim disks) are able to provide the total emission but say nothing about the internal distribution of dissipation. So far, no analytical model has been established and one has to make \textit{ad hoc} assumptions or base on the results of numerical simulations (e.g., \citem{hirose-09-a}). However, it has been shown \citepm{sadowski.photosphere} that the profile of the internal dissipation hardly affects the disk observables. 

The simplest and physically justified choice for the dissipation profile is to assume that the $\alpha$ prescription is valid throughout the disk interior. Thus, we may take
\be\der {\cal F}z = \frac 32 \alpha p \Omega_K,\ee
where $p$ is now the local pressure. In the following section we will use the relativistic version of this formula.

\begin{figure}
 \centering \includegraphics[angle=0,width=.7\textwidth]{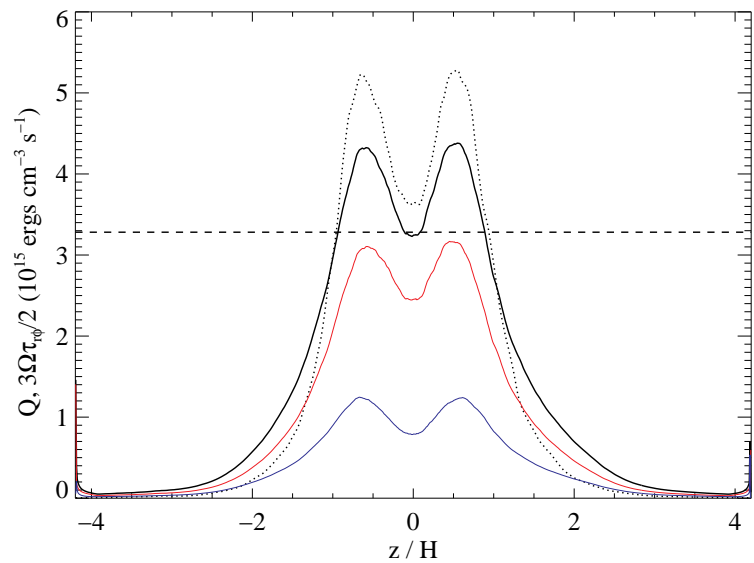}
\caption{ Horizontally and time-averaged total dissipation rate (black) and
stress times shear (dotted) in simulation 1112a of \citem{hirose-09-a}. The total dissipation rate
is the sum of magnetic energy dissipation (red) and kinetic energy dissipation
(blue). The difference between the stress times shear product and the total dissipation rate reflects the 
fact that some part of the generated heat goes into the mechanical energy of gas motion instead of increasing the radiative flux. Figure after \citem{hirose-09-a}.}
  \label{f.hirose.f11}
\end{figure}

However, most recent MHD shearing-box simulations show that the dissipation in accretion disks may not follow the local $\alpha p$ dissipation. Surprisingly, the resulting maximum of the dissipation rate is not even at the equatorial plane (where both the density and the total pressure have maxima). Fig.~\ref{f.hirose.f11} shows the dissipation rate in one of the simulations performed by \citem{hirose-09-a}. The solid black line presents the dissipation rate giving rise to the radiative flux. Its integral is equal to the total emitted radiative flux. It is clear that the dissipation profile has two maxima at $z=\pm 0.5h,$ where $h$ is the disk scale height. As the total (gas, radiation and magnetic) pressure peaks at the equatorial plane, the regular $\alpha p$ prescription cannot describe such dissipation profile. Nevertheless, as has already been discussed in Section~\ref{s.mri}, the $\alpha$ formalism works for the total, vertically integrated, quantities.

\section{Model}

\subsection{Basic assumptions, parameters, and coefficients}
\label{s.assump}

We make the same assumptions as for the one-dimensional model (Section~\ref{s.stateq}), namely:
(i) axisymmetry, (ii) stationarity, (iii) Kerr metric, with fixed values of the BH mass, $M$, 
and spin, $a$, parameters, (iv) symmetry under reflection in the equatorial plane, (v) steady mass supply rate, $\dot M$, through a
boundary ``at infinity'', (vi) zero torque at the BH horizon, and (vii) we neglect the loss of angular momentum to both wind and radiation, self-irradiation and magnetic pressure.

A fraction, $1/(1+f^{\rm adv}(r))$, of the entropy
generated locally by dissipative processes is released into the
radiation field, while the remainder is advected by the gas.

A unique solution to the slim disk model can only be found if certain
additional assumptions are made. We make the following arbitrary
choice.  We neglect the vertical variation ($z$ dependence) of the
velocity field, considering only its height-averaged value; thus, the
velocity is always directed radially inwards and is a function of the
radial coordinate alone. Similarly, we assume there is no $z$
variation of the advection factor $f^{\rm adv}(r)$.  Dissipation and
angular momentum transport are given by the $\alpha$ prescription
\citepm{shakura-73}, with a constant value of $\alpha$. We assume that
the local dissipation rate is proportional to the total pressure,
$p$. For a more detailed statement, see Eq.~\ref{vs.dFdz}
and the comment following it.
Calculations are carried out for the value $\alpha=0.01$.

We are looking for 2-D
slim disk solutions at a definite value of mass
accretion rate for a given Kerr metric. Thus, for a fixed value of
$\alpha$, there are three fundamental parameters describing a given
slim disk solution: $M$, $a_*$, and $\dot M$.

\subsection{Vertical structure equations}
\label{s.verticaleq}
We describe the vertical structure of an accretion disk in the
optically thick regime by the following equations:

(i) Hydrostatic equilibrium \citepm{KatoBook},
\begin{equation}
 \frac 1{\rho}\der pz=-\Omega_\perp^2z,
\label{vs.dpdz}
\end{equation}
where $\Omega_\perp$ is defined in Eq.~\ref{eq.omegatilde}.
Although other expressions for the right-hand side of
Eq.~\ref{vs.dpdz} can be found in the literature (e.g., \citem{vertical}
includes $v^z\neq 0$), the form above is appropriate for our scheme, in which the vertical structure is precalculated before any information about the radial variables becomes available. 

(ii) The energy generation equation. 
We assume that the vertical flux of energy inside the disk
$\cal F$ is generated
according to
\begin{equation}
\label{vs.dFdz}
 \der {\cal F}z=\frac{3\cal D}{2\cal C}\cdot
\left(\frac{\alpha p}{1+f^{\rm adv}}\right)\left(\frac{M}{r^3}\right)^{1/2}.
\end{equation}
Strictly speaking, this does not correspond to a constant $\alpha$ prescription,
as the term $(3/2)(M/r^3)^{1/2}$, which is derived from Keplerian strain,
departs somewhat from the value that would follow from the actually computed
distribution of angular momentum (Fig.~\ref{f.angmom.a0}). However,
as the departure for sub-Eddington accretion rates is small,
we expect Eq.~(\ref{vs.dFdz}) to afford a good approximation.
 Note that
$f^{\rm adv}=0$ corresponds to the NT disk
(going over into the Shakura-Sunyaev disk
in the nonrelativistic limit of thin disks),
$f^{\rm adv}>1$
characterizes advection-dominated disks, while $f^{\rm adv}<0$ describes
those regions where the advected heat is being released. The amount
of heat advected $Q^{\rm adv}=f^{\rm adv}{\cal F}(h)$.

(iii) Energy transport. The structure of the disk
has to be such that the actual value of the divergence
of the flux corresponds to Eq.~(\ref{vs.dFdz}).
Radiative transport is computed in the diffusive approximation (Eq.~\ref{e.radtran.ross}),
\begin{equation}
 {\cal F}(z)=-\frac{16\sigma T^3}{3\kappa\rho}\der Tz,
\end{equation}
while convective transport is computed in the mixing-length
approximation.
Energy is transported in the vertical direction
through diffusion of radiation or convection according
to the value of the thermodynamical gradient,
which can be either radiative or convective.
 Accordingly, we take
\begin{equation}
\derln Tp=\left\{\begin{array}{lll}
\nabla_{\rm rad}, & {\rm for} & \nabla_{\rm rad}\le\nabla_{\rm ad} \\
\nabla _{\rm conv}, & {\rm for} & \nabla_{\rm rad}>\nabla_{\rm ad} \\
\end{array}\right.
\label{vs.gradient}
\end{equation}
with the adiabatic gradient given by a derivative at constant entropy:
$\nabla_{\rm ad} \equiv (\partial\, {\rm ln}\,T/\partial\, {\rm ln}\,p)_S$.

The radiative gradient $\nabla_{\rm rad}$ is calculated in the
diffusive approximation,
\begin{equation}
\nabla_{\rm rad}=\frac{3p\kappa_R {\cal F}}{16\sigma T^4\Omega_\perp^2z},
\label{vs.gradrad}
\end{equation}
where $\kappa_R$ is the Rosseland mean opacity, and
$\sigma$ is the Stefan-Boltzmann constant. 

When the temperature gradient exceeds the
value of the adiabatic gradient, we have to consider
the convective energy flux. The convective gradient
$\nabla _{\rm conv}$ is calculated using the mixing length theory
introduced by \citem{paczynski69}. We take the following
mixing length,
\begin{equation}
H_{\rm ml}=1.0 H_p,
\label{vs.hml}
\end{equation}
with pressure scale height $H_{\rm p}$ defined as \citepm{hameury98}
\begin{equation}
H_{\rm p}=\frac p{\rho\Omega_\perp^2 z+\sqrt{p\rho}\Omega_\perp}
\label{vs.hP}.
\end{equation}
The convective gradient is defined by the formula
\begin{equation}
\nabla _{\rm conv}=\nabla_{\rm ad}+(\nabla_{\rm rad}-\nabla_{\rm ad})y(y+w)
\label{vs.gradconv}
\end{equation}
where $y$ is the solution of the equation
\begin{equation}
\frac94\frac{\tau_{\rm ml}^2}{3+\tau_{\rm ml}^2}y^3+wy^2+w^2y-w=0,
\label{vs.eqy1}
\end{equation}
with the typical optical depth for convection
$\tau_{\rm ml}=\rho\kappa_R H_{\rm ml}$, and $w$ given by

\begin{equation}
\frac{1}{w^2}=\left(\frac{3+\tau_{\rm ml}^2}{3\tau_{\rm ml}}\right)^2
\frac{\Omega_\perp^2zH_{\rm ml}^2\rho^2C_p^2}{512\sigma^2T^6H_P}
\left(\pderln\rho T\right)_p
\left(\nabla_{\rm rad}-\nabla_{\rm ad}\right)
\label{vs.eqy2}.
\end{equation}
The thermodynamical quantities $C_p$, $\nabla_{\rm ad}$ and $(\partial\,{\rm ln}\,\rho/\partial\,{\rm ln}\,T)_p$ are calculated using
standard formulae (e.g., \citem{stellarstructure}) assuming solar abundances
($X=0.70$, $Y=0.28$) and, when necessary, taking the
effect of partial ionization of gas on the gas mean molecular
weight into account.

We use Rosseland mean opacities $\kappa_R$ (including 
the processes of absorption and scattering) taken from
\citem{alexander} and \citem{seaton}. Following other authors
(e.g., \citem{idanlasota08}), we neglect expansion opacities,
in agreement with our neglect of vertical velocity gradients.

(iv) We set the following boundary conditions. At the equatorial plane
($z=0$) we set ${\cal F}(0)=0$, in accordance with the assumption of the
reflection symmetry, while at the disk surface, $\tau(h)=0$, we follow
the Eddington approximation \citepm{MihalasBook} and require
${\cal F}(h)\equiv\sigma T_{\rm eff}^4=2\sigma T^4(h)$.

In practice, for a fixed $r$, and prescribed values of $T_C$ and
$f^{\rm adv}$, a trial value of the central density, $\rho_C$, is assumed
and the equations are integrated in $z$ until
$\rho(z_*)=10^{-16}\,{\rm g\,cm^{-3}}$ (as a stand-in for the disk
surface, $z_*=h$). If ${\cal F}(h)$ and $T(h)$ fail to satisfy the surface
boundary condition, the assumed value of $\rho_C$ is adjusted, and the integration
is repeated, until the condition ${\cal F}(h)=2\sigma T^4(h)$ is
met. Convergence is usually attained in a few iterations.
The emergent flux of radiation at any given $r$ is then
$F={\cal F}(h)$.

\subsection{Radial structure equations}
\label{s.radialeq}

The radial sector of the model is described by four laws of
conservation (of mass, angular and radial momenta, and energy), and a regularity condition at the sonic point. These equations have exactly the same form as in Section~\ref{s.stateq} with exception to the amount of heat advected $Q^{\rm adv}$ which is now calculated precisely basing on the solutions of the vertical structure (see the previous section).

\section{Numerical methods}

\label{s.numerical}
\subsection{Vertical structure}
\label{s.numerical.vertical}

The set of ordinary differential equations describing the vertical
structure, i.e., Eqs.~\ref{vs.dpdz}, \ref{vs.dFdz} and
\ref{vs.gradient} together with appropriate boundary conditions are
solved for a given BH spin on a three-dimensional grid spanned  by the radius $r$, the
central temperature $T_C$, and the advection  factor $f^{\rm adv}$. For
a given set of these parameters, we start the integration from the
equatorial plane ($z=0$), and the solution satisfying the outer
boundary condition is found as described at the end of Section~\ref{s.verticaleq}.  The
resulting quantities describing the vertical structure 
($T_C$, $\Sigma$, $Q^{\rm adv}$, $P$, $\eta_1$, $\eta_2$, $\eta_3$,
and $\eta_4$),
together with $r$, are printed out to tables for subsequent
use in interpolation routines. As it turns out, the first two of these
parameters, $T_c$ and $\Sigma$, can be used to uniquely
determine all the other quantities characterizing the vertical structure,
including $f^{\rm adv}$ (see Fig.~\ref{f.scurve}).

Calculating the
full grid of vertical solutions for a single value of BH spin takes
about 5 hours on a 4-CPU workstation.

\subsection{Radial structure}
\label{s.numerical.radial}

The radial sector is described by exactly the same differential equations as given in Section\,\ref{sect.numerical}. However, the coefficients $\eta_1$ to $\eta_4$ and the amount of advected heat $Q^{\rm adv}$ are now taken from the precise solutions of the vertical structure.

Similarly like for the one-dimensional case, to obtain a solution one has to solve this system of two ordinary
differential equations, together with the following regularity
conditions at the sonic radius $r_{\rm son}$, defined by the same conditions as before:
\be \left.{\cal N}\right|_{r_{\rm son}}=\left.{\cal D}\right|_{r_{\rm son}}=0,
\label{num_regcond}
\ee and with outer boundary conditions given at some large radius
$r_{\rm out}$. The solution between the outer boundary and the
sonic point is found using the relaxation technique
\citepm{numericalrecipes}, with ${\cal L}_{\rm in}$ treated as the
eigenvalue of the problem. 
The sonic point for $a_*=0$ is located at $5.9M$ for accretion rate $0.01\dot M_{\rm Edd}$ and at $5.0M$
   for $2.0\dot M_{\rm Edd}$.
We use 25 mesh
points spaced logarithmically in the radius on the section between the sonic point 
and $r_{\rm out}=1000M$. This particular number of grid
points is enough to resolve all disk features. We have verified that the results are accurately reproduced with a denser grid.

The parameters linked to the vertical structure ($Q^{\rm adv}$,
$\eta_1$, $\eta_2$, $\eta_3$, and $\eta_4$) for given $\Sigma$ and
$T_C$ are linearly interpolated from precalculated tables of the
vertical structure solutions
(for given value of $V$, $\Sigma$ is determined directly from the mass conservation). 
The radial derivatives ${\rm d}\ln
\eta_1/{\rm d}\ln r$, ${\rm d}\ln \eta_2/{\rm d}\ln r$, ${\rm d}\ln
\eta_3/{\rm d}\ln r$, and ${\rm d}\ln \eta_4/{\rm d}\ln r$ are
evaluated numerically from the $\eta_1$, $\eta_2$, $\eta_3$, and
$\eta_4$ profiles in the previous iteration step. A relaxed solution
is obtained in a few iteration steps.  Once a solution outside the
sonic point is found we numerically estimate the radial derivatives of
$V$ and $T_C$ at the sonic point using values given at $r>r_{\rm son}$ and use
these derivatives to start direct integration inside the sonic point.

The solution thus obtained may then be used as a trial solution when
looking for the relaxation solution of another slim disk, i.e., when
one of the three fundamental parameters ($\dot M$, $a_*$, $M$) has a
slightly different value.
Each relaxation step takes approximately 5
seconds on a single-CPU workstation.

\section{Disk structure}

In the following two subsections we present and discuss both the
radial and vertical structure of the two-dimensional model of slim accretion disks.  All the
solutions, if not stated otherwise, were computed assuming
$\alpha=0.01$ and $M=10~{\rm M_\odot}$.

\subsection{Disk radial structure}
\label{ss.radialstructure}

The radial profiles are very similar to those described in Section~\ref{s.stationarysolutions} (detailed comparison is given in Section~\ref{s.comparison}). However, to make this section complete, we shortly discuss them again. In addition to figures presented before, we show disk thickness and surface density dependence on BH spin (Fig.~\ref{f.surfdens.a0}) as well as profiles of the total and effective optical depths (Figs.~\ref{f.optdepth.a0} and \ref{f.1alpha}).

\begin{figure}
 \centering \includegraphics[angle=0,width=.7\textwidth]{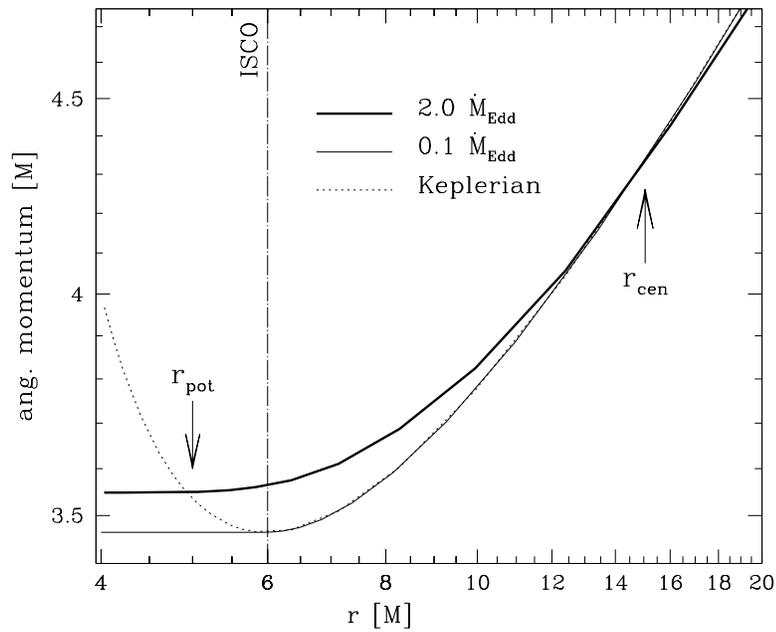}
\caption{ Angular momentum ($u_{\phi}$) in a Schwarzschild slim disk for 
two accretion rates. For very low accretion rates the
angular momentum follows the Keplerian profile (dotted line) down to the ISCO. 
For high accretion rates the flow is
super-Keplerian between the ``center'' of the disk at $r_{\rm cen}$ 
and the ``potential spout''  at $r_{\rm pot}$. The vertical dot-dashed
line on this and subsequent figures denotes the location of the ISCO.}
  \label{f.angmom.a0}
\end{figure}

\subsubsection*{Angular momentum}

The angular momentum profiles for our accretion disk solutions near a
non-rotating BH are presented in Fig.~\ref{f.angmom.a0}. 
Results for two values of mass accretion rate are shown, $\dot M=0.1M_{\rm Edd}$
and $2.0M_{\rm Edd}$.
 For the
lowest accretion rates the profiles follow the Keplerian profile and
reach its minimal value (${\cal L}_{\rm in}$ in Eq.~\ref{eq_ang6}) at
the marginally stable orbit (ISCO). The higher the accretion rate, the
stronger the departure from the Keplerian profile. For low values of $\alpha$ (e.g., $0.01$) the disk is
sub-Keplerian at large distances and super-Keplerian at moderate
radii.  The Keplerian profile is crossed again at a point located
inside the marginally stable orbit, and corresponding to what is
usually called ``the cusp'' or ``the potential spout''.  For a
detailed study of the physics of differently defined inner edges of accretion disks see
Section~\ref{s.inneredges}.

\subsubsection*{S-curves}

\begin{figure}

  \centering\resizebox{.75\hsize}{!}{\includegraphics[angle=270]{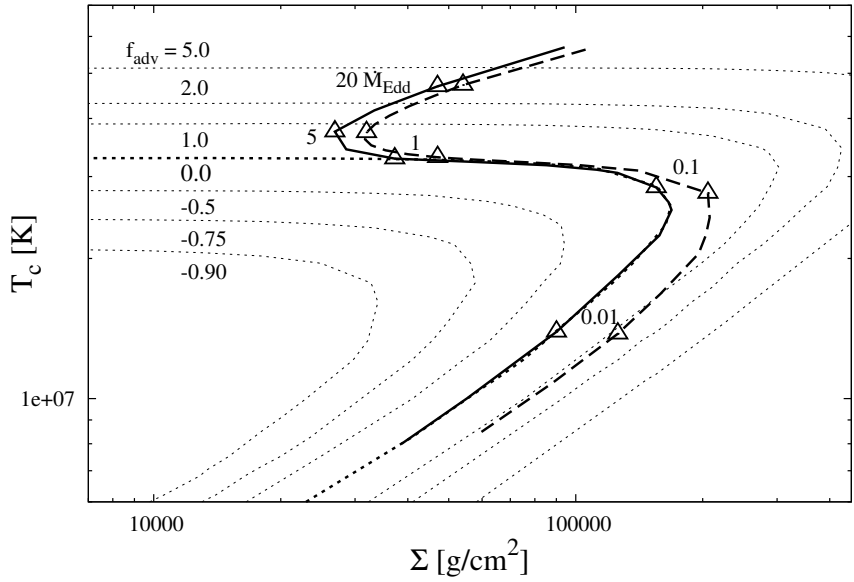}}
\caption{The $T_C$-$\Sigma$ plane at $r=20M$ for a non-rotating BH ($a_*=0$).
 The dotted lines connect
solutions for the vertical structure of slim disks that have the same value
of the advection parameter $f^{\rm adv}$. The locus of standard
(radiatively efficient, $f^{\rm adv}=0$)
 disk solutions is shown with the thick dotted line.
The solid thick line represents the vertical slim disk solutions for
different accretion rates
(indicated by triangles), and the dashed line presents corresponding solutions
of the conventional polytropic slim disk model
(see Section~\ref{s.comparison}). The difference between the two lines in the low
$\dot M$ limit corresponds to the difference in $\Sigma$ between the two models
(see Fig.~\ref{f.3comp}).
}
  \label{f.scurve}
\end{figure}

Figure~\ref{f.scurve} presents slim disk
solutions at $r=20 M$ on the $T_C$-$\Sigma$ plane, for a non-rotating BH. Solutions of the 
polytropic, height-averaged models are presented for comparison 
(for detailed discussion see Section~\ref{s.comparison}). The locus of solutions
for various values of the mass accretion rate has the shape of the
so-called ``S-curve'' \citepm{slim}.  The lower, gas-pressure dominated
branch accurately follows the track of radiatively efficient solutions
($f^{\rm adv}=0$).  The middle, radiation-pressure dominated branch is
reached at $\mdot\approx0.1\mdot_{\rm Edd}$. As advection becomes
significant, the slim disk solution leaves the $f^{\rm adv}=0$ track
and moves to higher advection rates. Around
$\mdot=5\mdot_{\rm Edd}$, the solutions enter the upper advection-dominated branch corresponding to $f^{\rm adv}>1.0$  (more than 50\%
of heat stored in the accreted gas).  At $\mdot=20\mdot_{\rm Edd}$
this rate increases almost up to 80\% ($f^{\rm adv}\approx4.0$).

\subsubsection*{Surface density}

\begin{figure}
\centering
 \subfigure
{
\includegraphics[height=.6\textwidth]{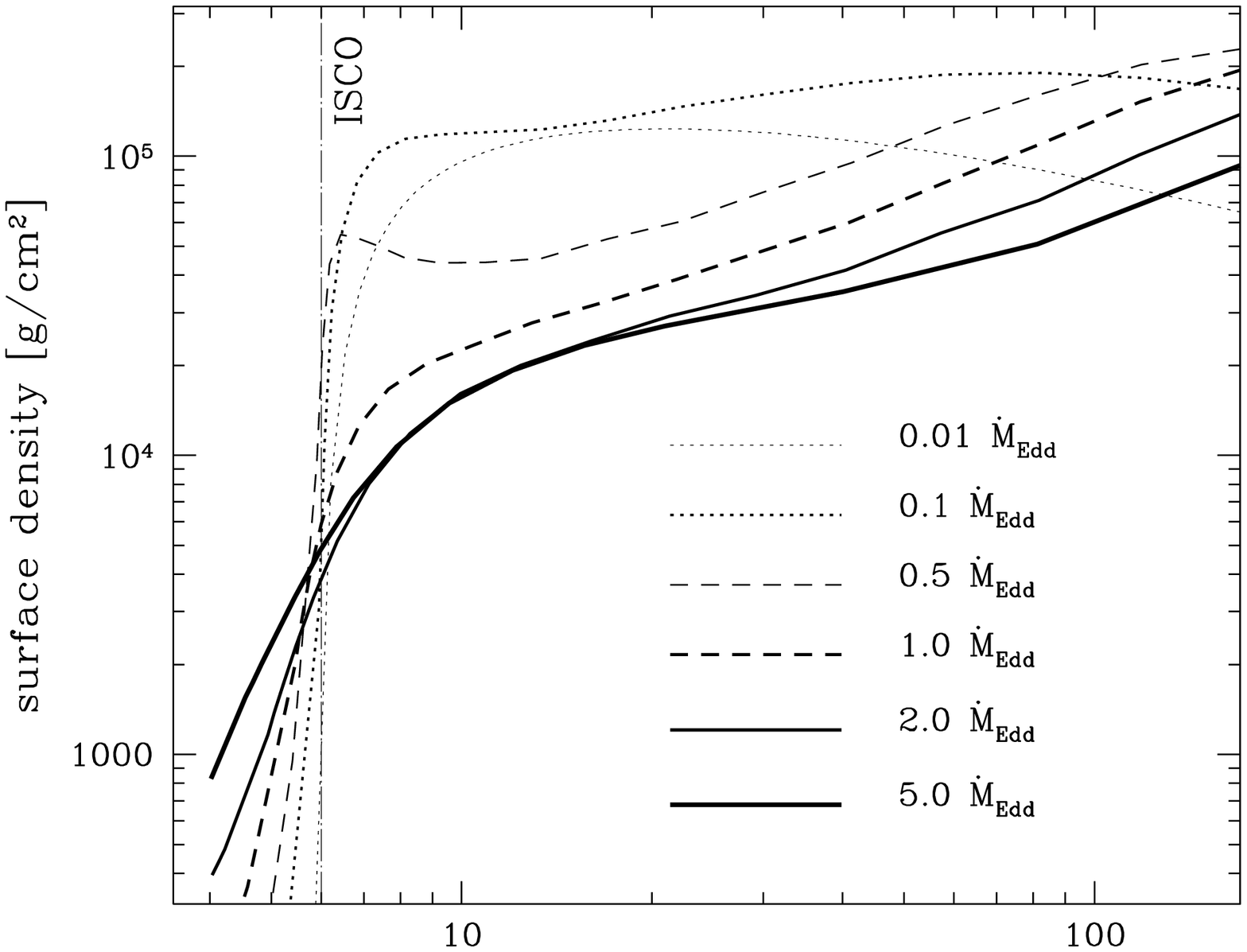}
}
\\
\vspace{-.07\textwidth}
 \subfigure
{
\includegraphics[height=.6\textwidth]{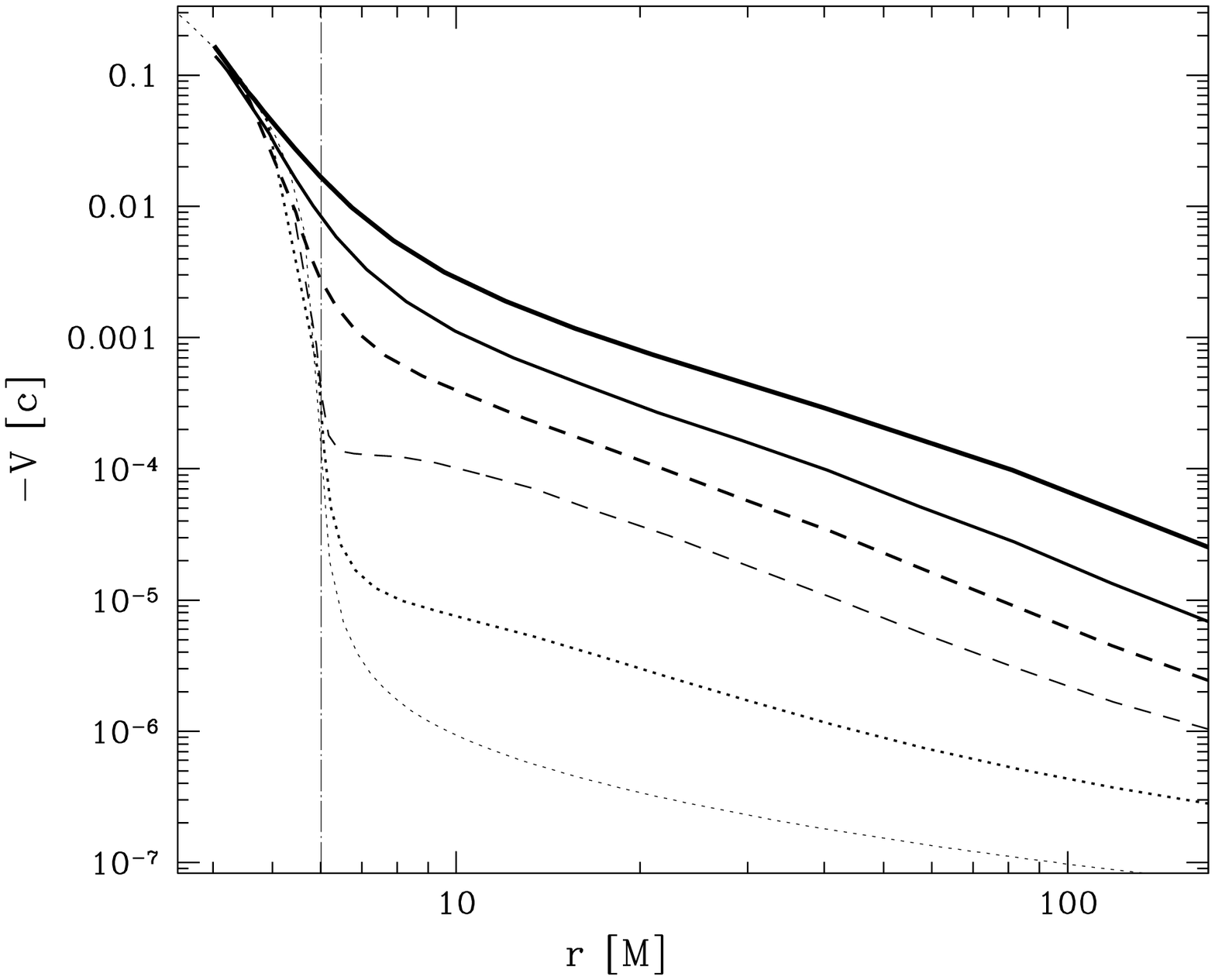}
}
\caption{ Profiles of the surface density (upper panel)
and corresponding values of radial velocity $V$ (bottom
panel) of a slim disk for
a non-rotating BH. Solutions for different accretion rates
are presented.}
\label{f.surfdens.a0}
\end{figure}

Profiles of the surface density for 2-D slim disk solutions for
a non-rotating BH are presented in the upper panel of Fig.~\ref{f.surfdens.a0}. 
Different 
regimes, corresponding to different branches of the ``S-curve''
on the ($\Sigma$, $T_C$) plane are visible.
For large radii the surface density increases with increasing 
accretion rate (the lower gas-pressure dominated branch), while this relation is opposite for
moderate radii (the middle radiation-pressure
dominated branch). For accretion rates $\mdot>5.0\mdot_{\rm Edd}$ the 
upper advection-dominated branch would be reached. The local maxima
in the surface density profiles (discussed in detail
 in, e.g., \citem{sadowski.slim})
 are visible for moderate accretion rates ($\sim 0.5\mdot_{\rm Edd}$).
The bottom panel of Fig.~\ref{f.surfdens.a0} presents corresponding 
profiles of the radial velocity $V$ as measured by an observer co-rotating
with the fluid. 

\begin{figure}
  \centering\resizebox{.7\hsize}{!}{\includegraphics[angle=0]{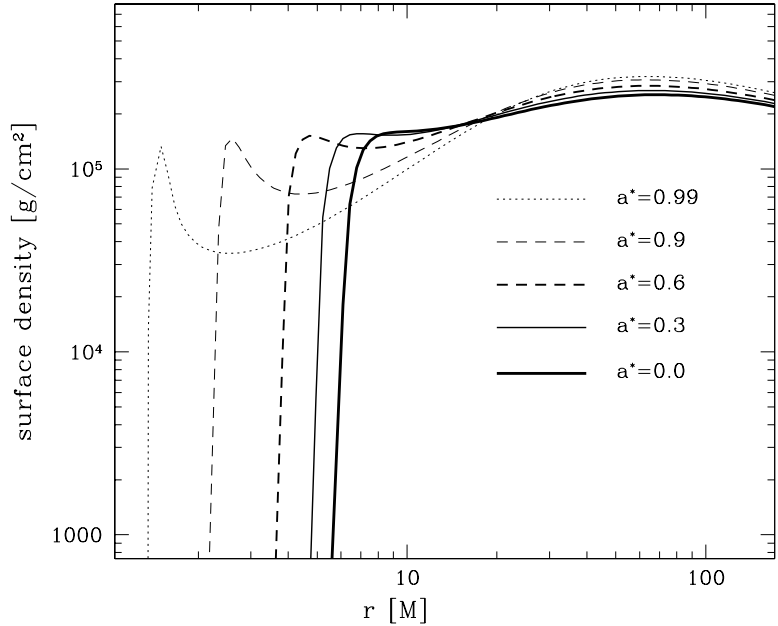}}
\caption{ Profiles of the surface density for slim disks 
at a constant accretion rate ($\dot M = 0.1\dot M_{\rm Edd}$) 
and various BH spins. }
  \label{f.surfdens.aN}
\end{figure}

The surface density dependence on BH rotation is presented in 
Fig.~\ref{f.surfdens.aN}. The profiles are shifted to lower radii as the inner 
edge of the disk moves inward for higher BH spins. The outer
parts of the accretion disk are insensitive to the metric.



\subsubsection*{Optical depth}

\begin{figure}
\centering
 \subfigure
{
\includegraphics[height=.6\textwidth]{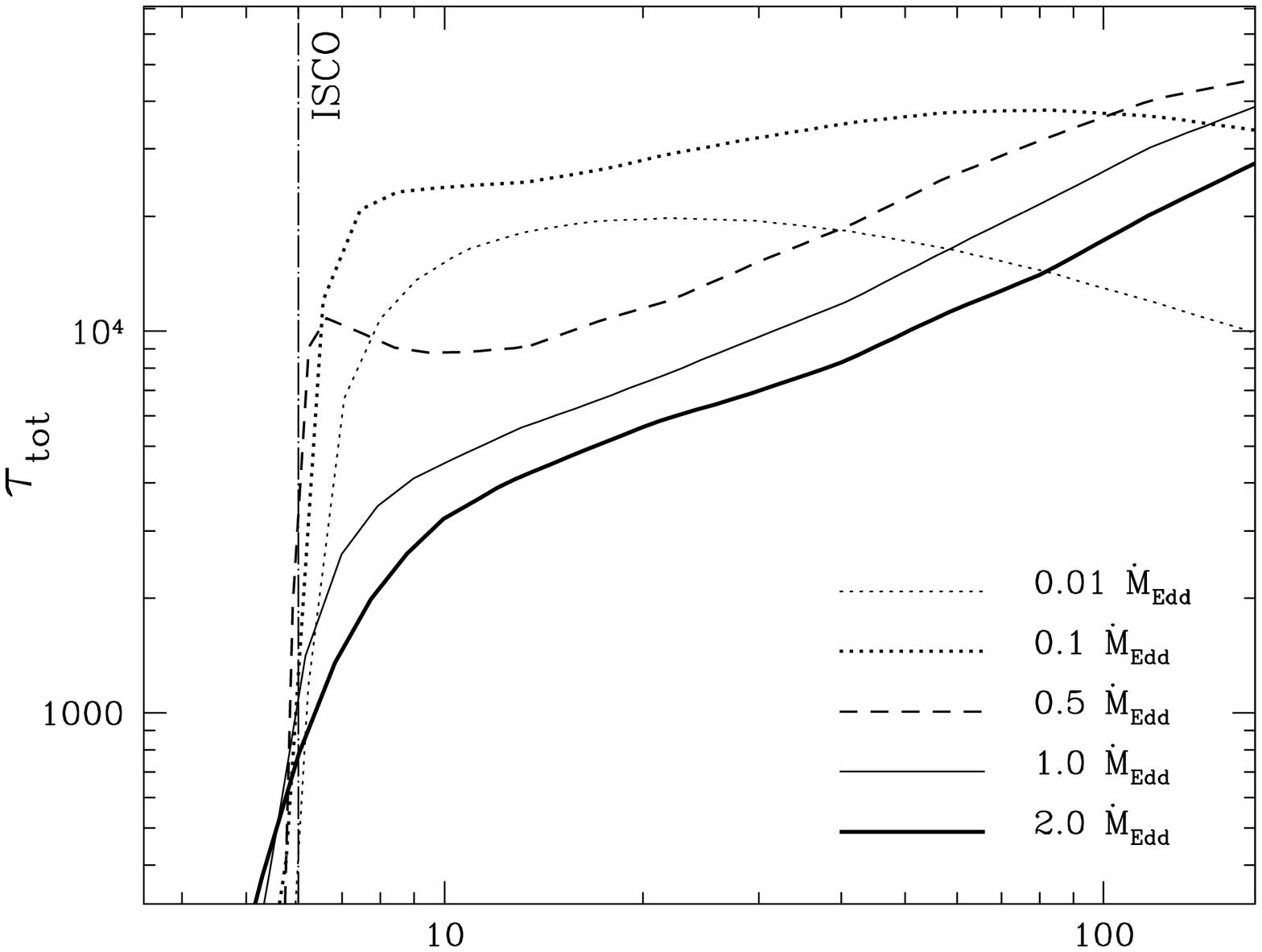}
}
\\
\vspace{-.07\textwidth}
 \subfigure
{
\includegraphics[height=.6\textwidth]{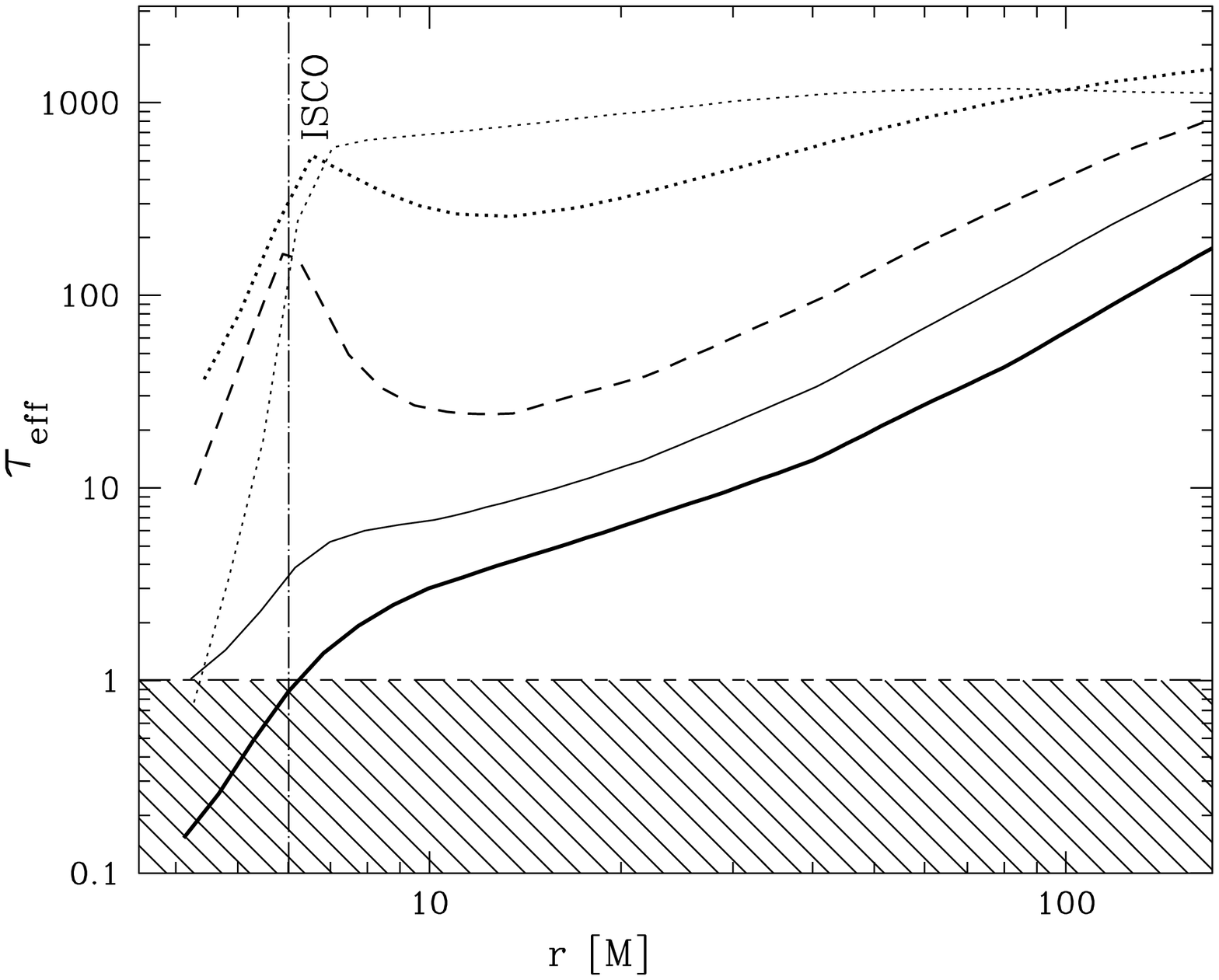}
}
\caption {Optical depth for $\alpha=0.01$ slim disks
around a  Schwarzschild black hole at different accretion rates. 
{\sl Top panel:} The total optical depth as a function of the radius.
{\sl Bottom panel:} The effective optical depth as a function of the radius. 
The ISCO is shown at $r=6M$. The shaded region in the bottom plot
indicates the region where the diffusive approximation is invalid.
}
 \label{f.optdepth.a0}
\end{figure}

In the top panel of Fig.~\ref{f.optdepth.a0} we plot the total
optical depth of the vertical slim disk solutions for different accretion 
rates and $a_*=0$. The total optical depth,
\be \tau_{\rm tot}=\int_0^h\kappa_R\rho~{\rm dz},\ee
where the total opacity coefficient $\kappa_R$,
which includes the processes of absorption and scattering,
is closely related to the surface density.
Indeed, the radial profiles of the optical depth 
shown in Fig.~\ref{f.optdepth.a0} follow the corresponding profiles of 
surface density. Any differences in the profiles
come from the dependence of the opacity coefficient 
on local density and temperature. Outside the ISCO the total
optical depth is always large ($\tau_{\rm tot}>10^3$).

%
%

\begin{figure}
 \centering \includegraphics[angle=0,width=.7\textwidth]{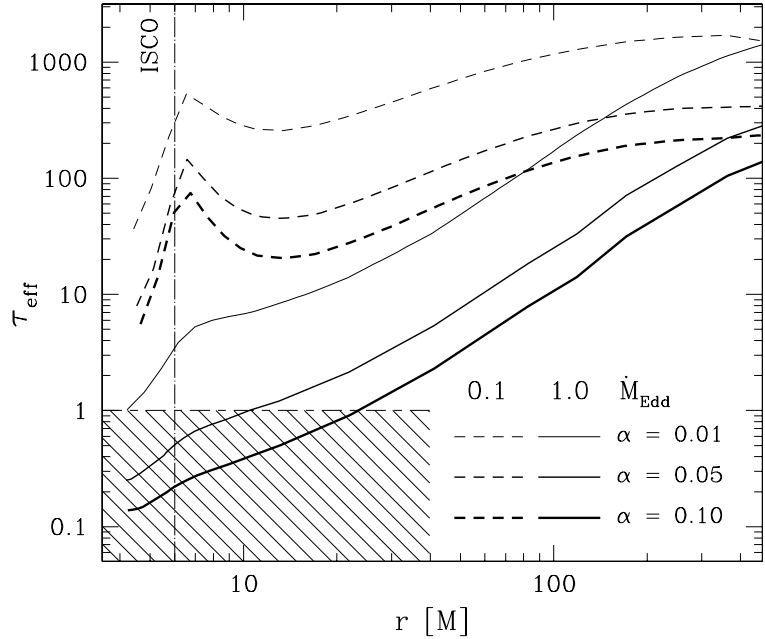}
\caption{ Profiles of the effective optical depth of a Schwarzschild
slim disk for three values of viscosity ($\alpha=0.01$, $0.05$, and $0.1$),
calculated for two accretion rates, $0.1\mdot_{\rm Edd}$ (dashed lines)
and $1.0 \mdot_{\rm Edd}$ (solid lines). }
  \label{f.1alpha}
\end{figure}

The diffusive approximation for radiative transport may only be used if photons are absorbed, otherwise LTE cannot be established. In a scattering-dominated atmosphere, the effective optical depth is then the relevant quantity to be used in checking for the self-consistency of the diffusive approximation. 
The bottom panel of Fig.~\ref{f.optdepth.a0} presents corresponding profiles
of the effective optical depth, which is
 estimated in the following way:
\be \tau_{\rm eff}=\int_0^h\sqrt{(\kappa_R-\kappa_{\rm es})\kappa_R}~\rho\,{\rm dz}\ee
where $\kappa_{\rm es}=0.34~{\rm cm^2 g^{-1}}$ 
is the mean opacity for Thomson electron scattering.
For $\dot M>0.1\dot M_{\rm Edd}$ the inner region of the disk 
becomes radiation-pressure dominated 
(it enters the middle branch on the corresponding S-curve),
 the surface density decreases with increasing
accretion rate, and electron scattering begins to dominate
absorption. Therefore, the effective optical depth decreases with
increasing
accretion rate and reaches values  $\tau_{\rm eff}<1$ 
in the inner parts of the disk for accretion rates above $1.0\dot M_{\rm Edd}$. 
As a result, for $\dot M>1.0\dot M_{\rm Edd}$, 
the diffusive approximation can no longer be applied, 
and our present approach to solving for the disk structure breaks down.
 In Fig.~\ref{f.1alpha}
we exhibit the dependence of the effective optical depth on the value
of $\alpha$.
In general, $\tau_{\rm eff}$ is inversely proportional to  $\alpha$
(as is the surface density). At lower accretion rates ($\dot M<0.1\mdot_{\rm Edd}$) 
the effective optical depth remains large even for high values of $\alpha$.

\subsubsection*{Flux profiles}
\begin{figure}
 \centering \resizebox{.7\hsize}{!}{\includegraphics[angle=0]{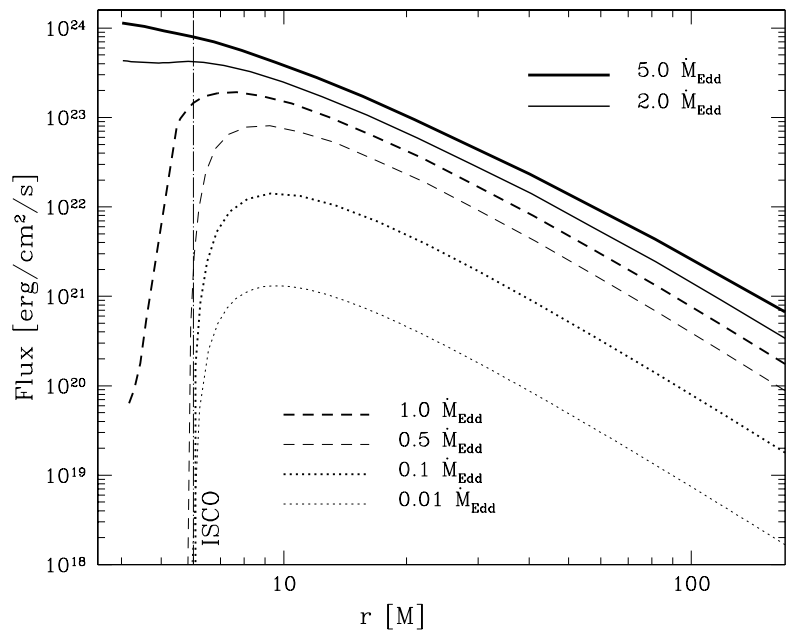}}
\caption{Flux emitted from the surface of a slim disk at five
accretion rates onto a Schwarzschild black hole.
At high accretion rates significant emission from within the
ISCO is clearly visible in the figure.}
  \label{f.flux.a0}
\end{figure}

Profiles of the flux emitted from the disk surface ($F=\sigma
T_{\rm eff}^4$)  in the case of a non-rotating BH are presented in
Fig.~\ref{f.flux.a0}. Results corresponding to accretion rates
from $0.01$ up to $5.0\mdot_{\rm Edd}$ are shown. For the lowest
rates the emission from inside the marginally stable orbit is
negligible as expected in the standard accretion disk
models. This is no longer true for higher accretion rates,
and the advection of energy causes  significant emission from smaller
radii. For super-Eddington accretion rates the
emitted flux continues to grow with a decreasing radius even inside the
marginally stable orbit. Radiation coming from the direct vicinity of the BH is
suppressed by the gravitational redshift (the $g$-factor, see Section~\ref{s.gfactor}).
Therefore, an observer at infinity will observe a
maximum in the profile of the effective temperature even for
the highest accretion rates.

\begin{figure}
  \centering\resizebox{.7\hsize}{!}{\includegraphics[angle=0]{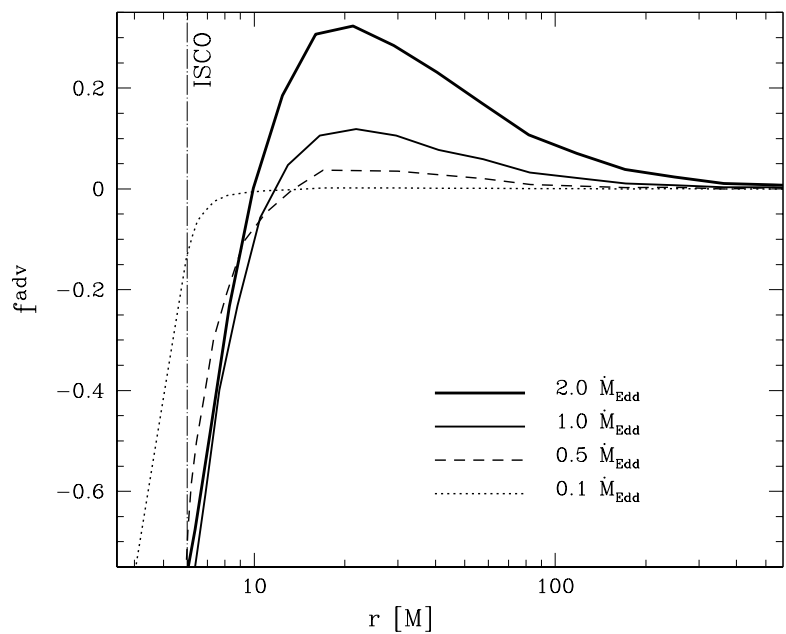}}
\caption{ Profiles of the advection coefficient $f^{\rm adv}$
for different accretion rates (Schwarzschild black hole).}
  \label{f.fadv.a0}
\end{figure}

The increase of the advective flux with increasing accretion rate
is clearly visible in Fig.~\ref{f.fadv.a0}. The ratio of the
heat advected to the amount of energy emitted 
is presented for different accretion rates. For very low accretion rates 
these profiles approach the limit of a radiatively efficient disk
($f^{\rm adv}\equiv 0$), and the advection component becomes significant for higher accretion rates.
Some part (up to $30\%$ at $r=20M$ for $2.0\mdot_{\rm Edd}$)
of the energy generated at moderate radii is advected with gas and radiated 
away at $r<10M$. This
causes the significant change in the emitted flux profile
at the higher accretion rates visible in 
Fig.~\ref{f.flux.a0}.

\begin{figure}
  \centering\resizebox{.7\hsize}{!}{\includegraphics[angle=0]{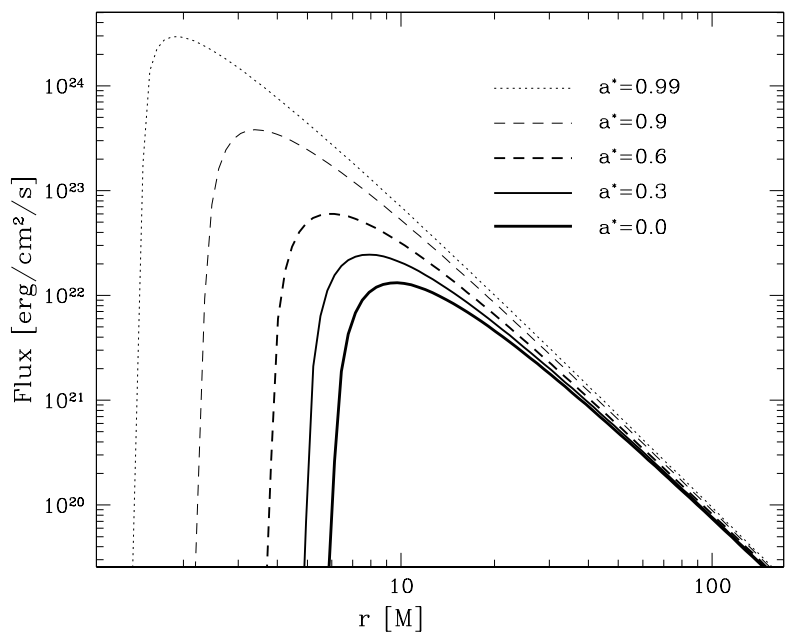}}
\caption{ Flux profiles at fixed accretion rate ($\dot M=0.1\dot M_{\rm Edd}$)
for five values of BH spin.}
  \label{f.flux.aN}
\end{figure}

In Fig.~\ref{f.flux.aN}  we present the emitted flux profiles for
different BH angular momenta at a constant accretion rate
$\mdot=0.1\mdot_{\rm Edd}$. These profiles coincide at large radii
where the influence of the BH  rotation is negligible; however, the
higher the BH spin, the closer to  the horizon the marginally stable
orbit. Therefore, in  the case of rotating BHs, the accreting matter
can move much deeper into the gravity well, compared to non-rotating
BHs. This effect leads to an increase in the disk luminosity and
hardening of its spectrum, which can be inferred from
Fig.~\ref{f.flux.aN} --- the higher the spin, the higher the disk luminosity,
 and the
higher the flux (which corresponds to the effective temperature).

\begin{figure}
  \centering\resizebox{.7\hsize}{!}{\includegraphics[angle=0]{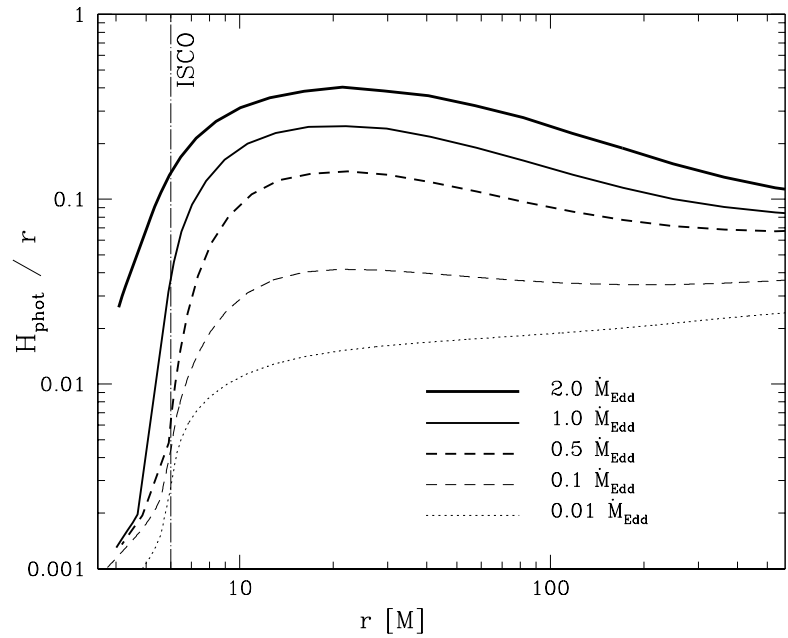}}
\caption{ The height of the photosphere at different accretion rates
onto a Schwarzschild black hole. }
  \label{f.Hphot.a0}
\end{figure}

\begin{figure}
  \centering\resizebox{.7\hsize}{!}{\includegraphics[angle=0]{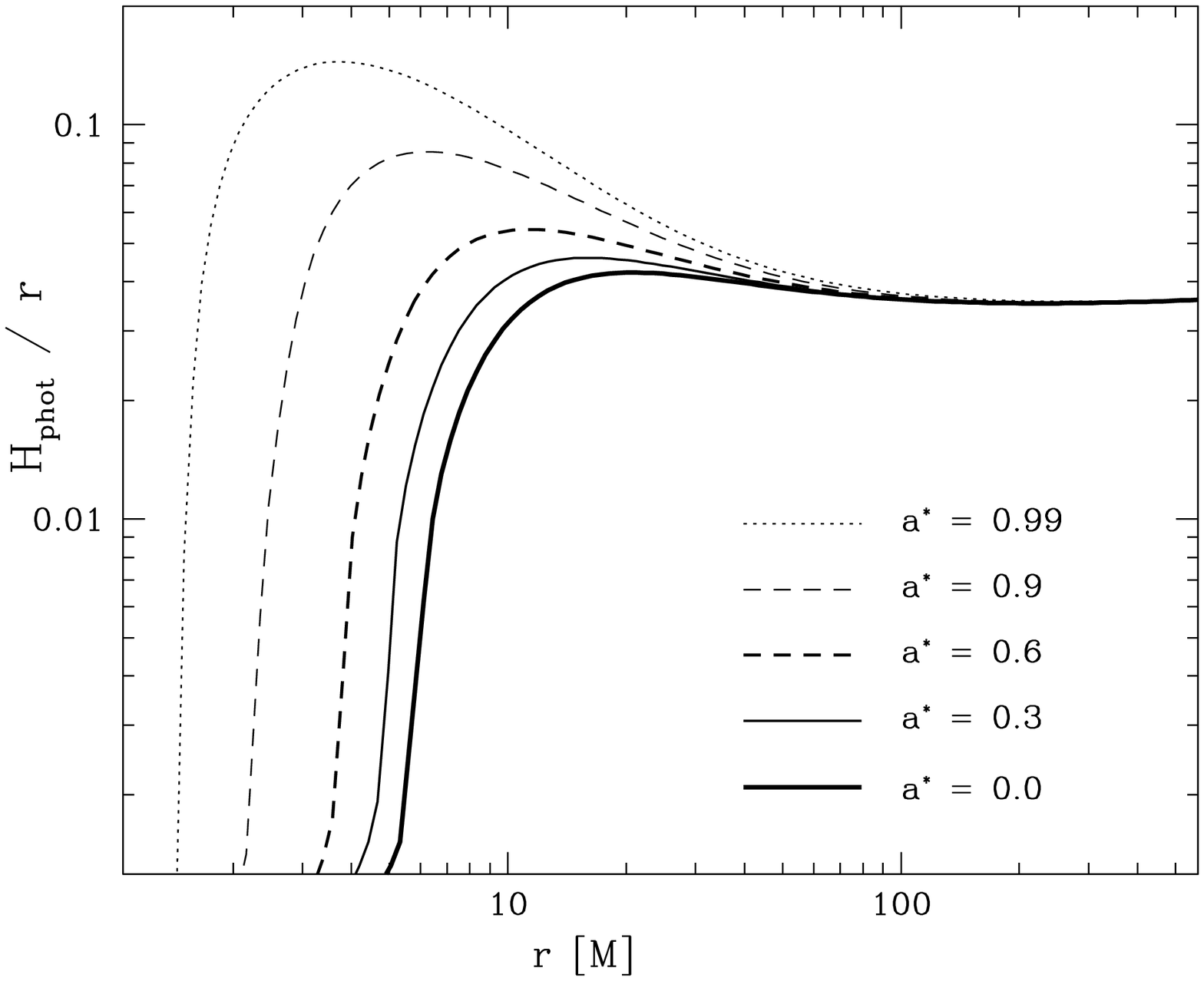}}
\caption{Profiles of photospheric height at a constant accretion rate
($\dot M=0.1\dot M_{\rm Edd}$) and various BH spins. }
  \label{f.Hphot.aN}
\end{figure}

\subsubsection*{Photosphere location}
\label{s.photlocation}
The flux observed at infinity may be obtained by performing ray tracing
of photons emitted from the accretion disk (see Section~\ref{s.comparison}).
Scattering in the layer
above the photosphere must also be taken into account.
An accurate calculation requires detailed knowledge of
atmospheric properties, including the location of the photosphere.
This is particularly
important when accretion rates are high and the disk is no longer geometrically
thin \citepm{sadowski.photosphere}. 

\begin{figure}
\centering
\resizebox{.7\hsize}{!}{\includegraphics[]{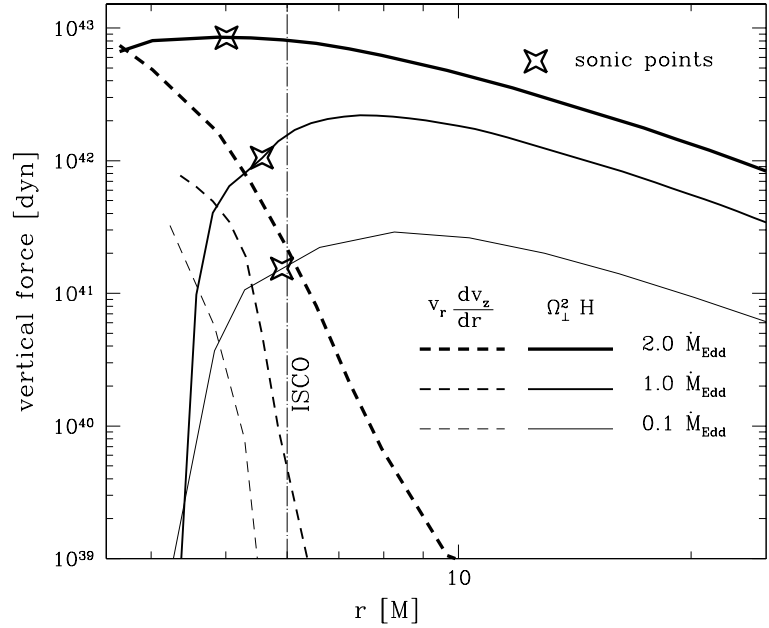}}
\caption{Comparison of the dynamical ($v_r{\rm d}v_z/{\rm d}r\approx V{\rm d}(V{\rm d}H/{\rm d}r)/{\rm d}r$) and gravitational ($\Omega_\perp^2 H$) components
of the vertical equilibrium equation at the photosphere. The stars denote locations of the sonic
radii.}
\label{f.vertbal}
\end{figure}

In Fig.~\ref{f.Hphot.a0} we plot the profiles for the
$z$-location of the photosphere, $h_{\rm phot}$, obtained in our
model at different accretion rates for a non-rotating BH.
Clearly, for high accretion rates, $\mdot>0.1\mdot_{\rm Edd}$,
 the inner regions become thicker (effects of
radiation pressure). For $\mdot = 1.0\mdot_{\rm Edd}$,
 the ratio $h_{\rm phot}/r$ reaches values as high as $0.25$. 
Close to the ISCO the height of the photosphere rapidly
  decreases because of vigorous cooling.
Although the rapid change in disk thickness violates the assumption of the hydrostatic equilibrium that we make when solving for the disk vertical structure, the accelerations connected with the vertical motions
involved are much lower than the vertical component of gravity.
Indeed, the vertical accelerations are close to
$v_r dv_z/dr\sim v_rd(v_rdh/dr)/dr\sim v_s^2 d(h/r)/dr\sim r\Omega_\perp^2 (h/r)^2\sim\Omega_\perp^2 h (h/r) $.
In deriving this estimate we liberally assumed that  $dh/dr\sim 1$ and
used the fact that the rapid decrease in disk height occurs near the sonic point, 
while the speed of sound $v_s$ is approximately $r\Omega_\perp (h/r)$.
Thus the acceleration terms modifying Eq.~\ref{vs.dpdz} would be
smaller than
the gravitational acceleration by a factor of a few percent: $(h/r)\sim10^{-1}$.
In Fig.~\ref{f.vertbal} we plot the dynamical and gravitational components 
of the vertical equilibrium equation at the photosphere. It is clear that the former
is at least $10$ times smaller than the latter at the sonic radii for $\dot M\le\dot M_{\rm Edd}$.
Therefore, in all likelihood, our results correctly describe the disk structure that would be obtained without assuming strict hydrostatic equilibrium. 

In Fig.~\ref{f.Hphot.aN} we present radial photosphere profiles,
at a fixed accretion rate and different
values of the BH spin. The photospheric heights
coincide for large radii.  In the inner regions of the disk, the height of the
photosphere increases with BH spin, reflecting the increased
luminosity and radiation pressure.

\subsection{Vertical structure}
\label{ss.vertstructure}
\begin{figure}
\centering
\resizebox{.8\hsize}{!}{\includegraphics[]{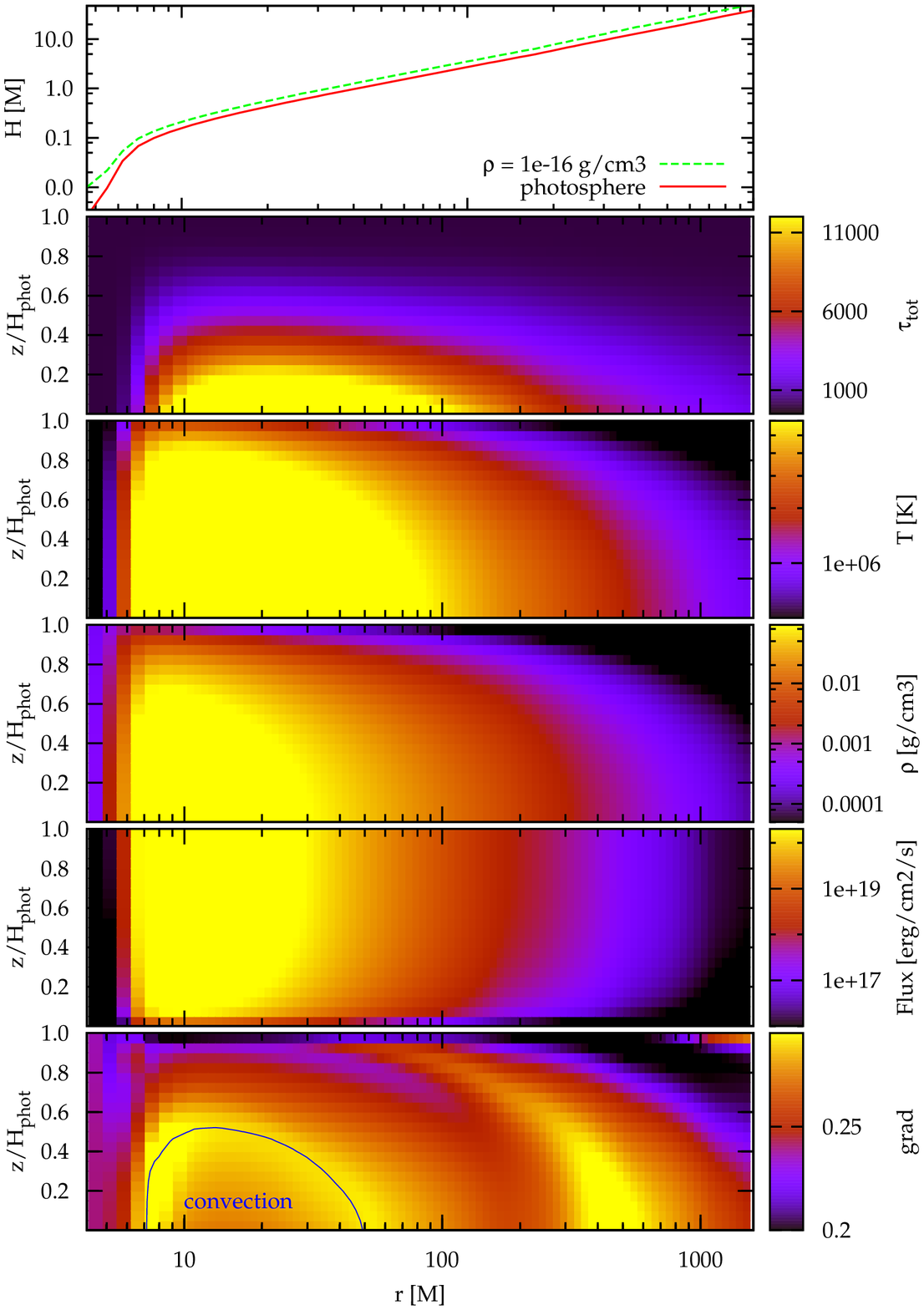}}
\caption{Vertical structure of a slim disk
for $\mdot=0.01\dot M_{\rm Edd}$ and $a_*=0$. 
The top panel presents the surface of the disk (green dashed line),
and the photospheric surface (red solid line). 
The other panels present the structure of the disk below the 
photosphere. {\sl Top to bottom:} total optical thickness, temperature, density,
vertical flux of energy, and the thermodynamical gradient.
 The blue solid line in the bottom panel delimits
the convective region.}
\label{f.rv.md0.01.a0}
\end{figure}

In Fig.~\ref{f.rv.md0.01.a0} we present a
vertical cross-section of a Schwarzschild slim disk
for $\mdot=0.01\mdot_{\rm Edd}$.
At this accretion rate the disk is radiatively efficient and no advection
of entropy is expected. The top panel presents the radial profiles
of the photosphere and the disk surface (defined as a layer
with $\rho=10^{-16}\,$g/cm$^3$).

\begin{figure}
\centering
\resizebox{.8\hsize}{!}{\includegraphics[]{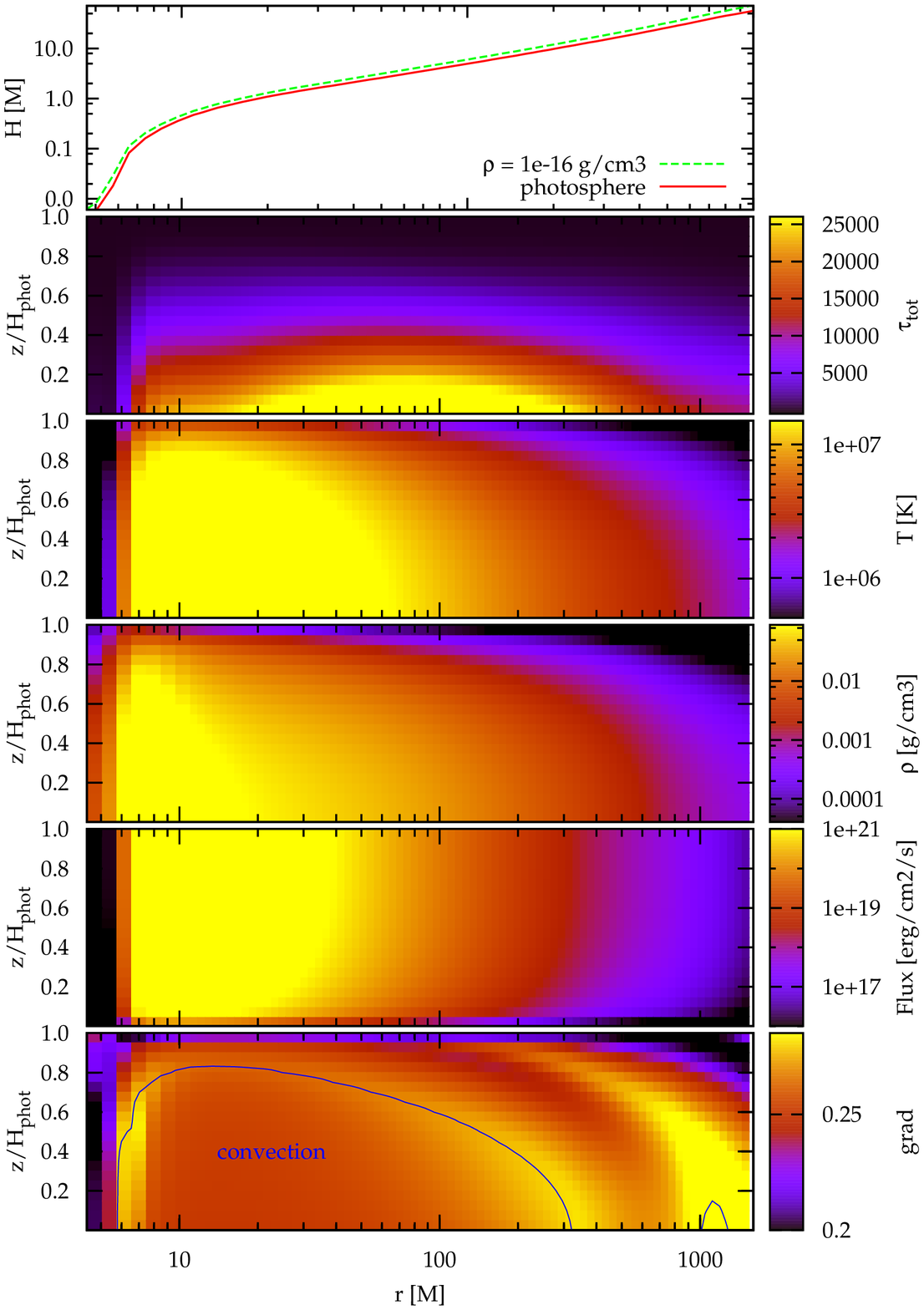}}
\caption{Same as Fig.~\ref{f.rv.md0.01.a0} but for 
$\mdot=0.1\dot M_{\rm Edd}$.}
\label{f.rv.md0.1.a0}
\end{figure}
\begin{figure}
\centering
\resizebox{.8\hsize}{!}{\includegraphics[]{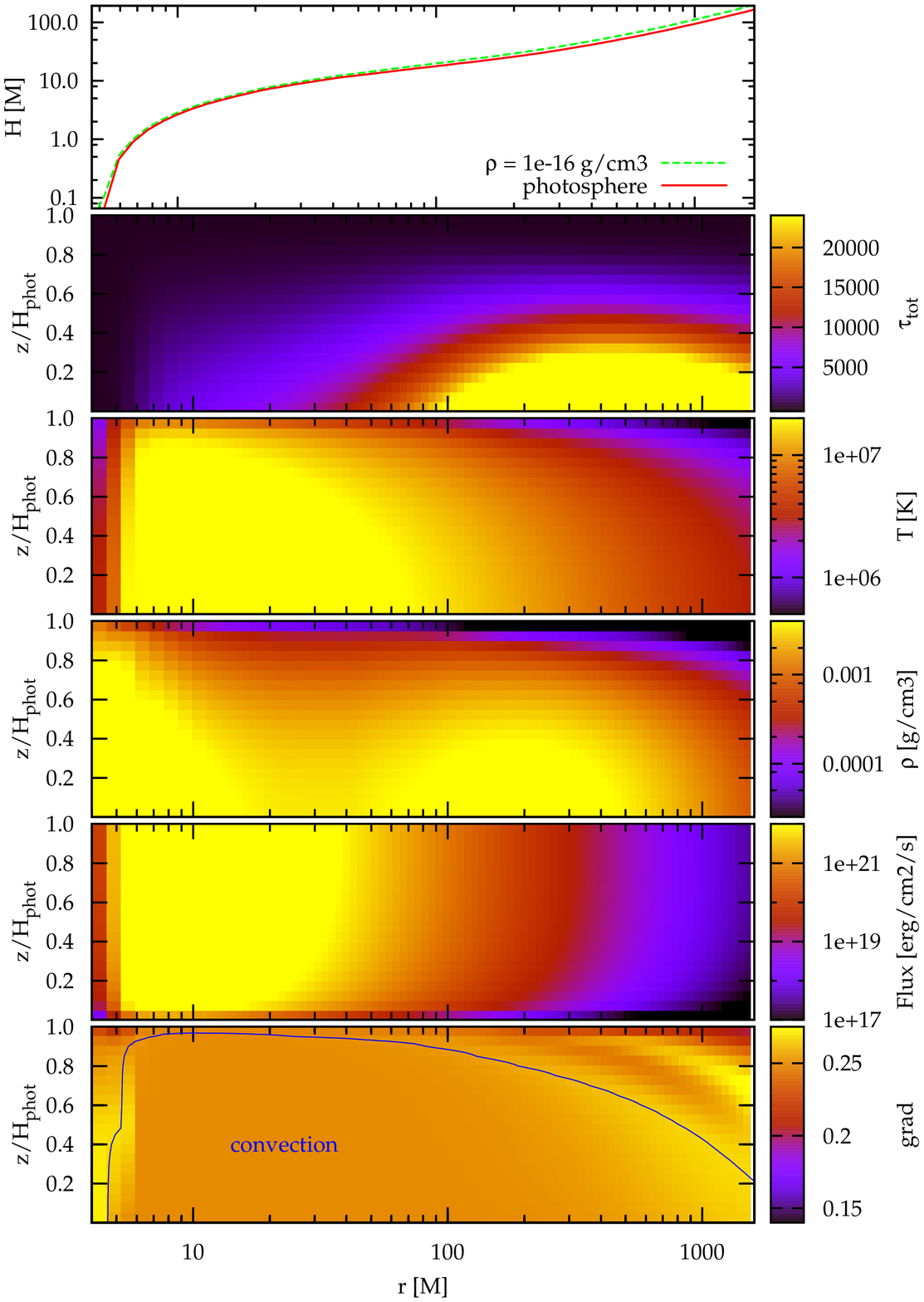}}
\caption{Same as Fig.~\ref{f.rv.md0.01.a0} but for 
$\mdot=1.0\dot M_{\rm Edd}$.}
\label{f.rv.md1.0.a0}
\end{figure}

 The total optical depth reaches
values as high as $\sim20000$ on the equatorial plane and decreases
monotonically towards the photosphere at $\tau=2/3$
(Fig.~\ref{f.rv.md0.01.a0}, second panel from the top).
Within the marginally stable orbit, the total optical depth
significantly decreases, as shown in Fig.~\ref{f.optdepth.a0}.

The third panel of Fig.~\ref{f.rv.md0.01.a0} presents the
temperature, which decreases with height from $T_C$ at the equatorial plane
to $T_{\rm eff}$ at the photosphere. The maximum value of the temperature is 
attained on the equatorial plane at $r\sim10M$.

Density is presented in the fourth panel,
and the next panel presents the vertical flux that is generated
inside the disk according to Eq.~\ref{vs.dFdz}. It is set to zero on
the equatorial plane by reflection symmetry, and then rapidly increases,
because in an alpha disk the dissipation is proportional to the pressure,
which reaches its maximum on the $z=0$ plane. Close to the disk
surface, where  pressure is almost negligible, the flux slowly settles
down to the emitted value. At the accretion rate chosen for the figure the flux
rapidly decreases inside the marginally stable orbit.

The bottom panel of Fig.~\ref{f.rv.md0.01.a0} presents the non-monotonic 
distribution of the thermodynamical gradient
(Eq.~\ref{vs.gradient}), which ranges between
$0.2$ and $0.4$. Therefore, the disk's vertical structure cannot be
described by a simple polytropic relation. Moreover, in the region of the
highest temperature
(close to the equatorial plane at moderate radii, $r\approx 20M$),
the heat is transported upward through convection.

The vertical structure of an accretion disk with ten times higher accretion
rate, $\mdot=0.1\mdot_{\rm Edd}$, 
is presented in Fig.~\ref{f.rv.md0.1.a0}. The
general picture remains the same, since the advection of heat is still 
insignificant. However, as the temperature increases,
 the inner disk regions become dominated by radiation pressure. 
 For $0.1\mdot_{\rm Edd}$ the 
convective region extends from the marginally stable orbit up to $300M$ and
covers more than half of the disk thickness.


The disk structure is significantly different in the case of
a high accretion rate (e.g., $1.0\mdot_{\rm Edd}$), with a 
significant amount of advection. The 
vertical cross-sections of the slim disk  are presented in 
Fig.~\ref{f.rv.md1.0.a0}. The inner regions are dominated by radiation 
pressure, so the disk geometrically
 thickens and the photosphere is now higher.
The total optical depth mostly follows the surface density
 (compare Fig.~\ref{f.optdepth.a0}), and therefore decreases 
considerably  towards the BH in the inner parts of the disk.

The temperature maximum is again
 located at the equatorial plane close to $r\approx 10M$. 
The maximum of the effective temperature 
(corresponding to the vertical flux shown
in the fifth panel (compare also Fig.~\ref{f.flux.a0}) 
is shifted inwards, down to $r\approx 8M$.

The fourth panel of Fig.~\ref{f.rv.md1.0.a0} presents the density distribution.
Despite the fact that the surface density monotonically increases
outward (see Fig.~\ref{f.surfdens.a0}), $\rho$ has
 two maxima in the equatorial plane: at $r\approx200M$
 and $r\approx 6M$. Finally, the bottom panel presents the thermodynamical gradient distribution.
 For such a high accretion rate, the convective zone extends
nearly to the photosphere for $r<200M$ and is present up to $r=2000M$.

\begin{figure}
\centering
\resizebox{.8\hsize}{!}{\includegraphics[angle=270]{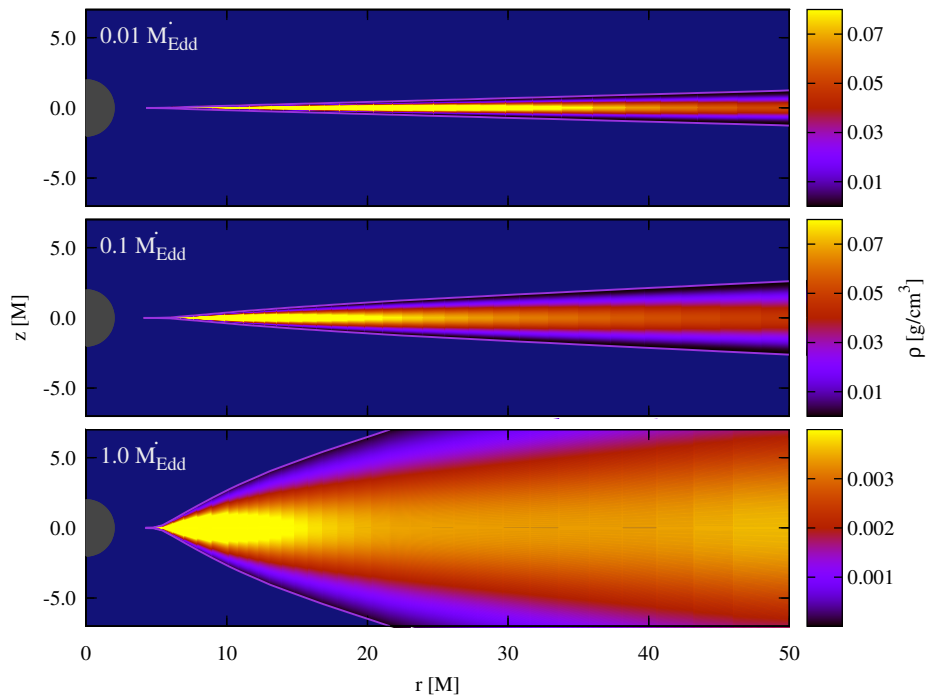}}
\caption{Meridional profiles of density for three accretion rates: $
\mdot=0.01$ (top), $0.1$ (middle) and $1.0 \mdot_{\rm Edd}$ (bottom
panel) in (r,z) coordinates. The violet boundaries show the location of
the photosphere. The black hole is described by
$M_{\rm BH}=10\msun$, and $a_*=0$.}
\label{f.slice}
\end{figure}

In Fig.~\ref{f.slice} we present the density in the meridional plane
for three different accretion rates. The
equatorial plane lies in the middle of each plot. The violet boundaries
denote the photosphere. The two maxima of density at the equatorial plane
for $\mdot=1.0\mdot_{\rm Edd}$ are clearly visible also in this representation,
as are some other features discussed above.

\section{Comparison with height-averaged slim disk solutions}
\label{s.comparison}

In this section we compare the two-dimensional (2-D) slim disk solutions,
 where the radial structure
equations are coupled to those for the vertical structure, with the
standard polytropic slim disk model, introduced in Chapter~\ref{chapter-stationary}, in which the slim disk equations and properties are
averaged over the thickness of the disk. The latter may be extended to mimic the more detailed 2-D solutions by introducing two additional parameters: $f_F$ and $f_H$.

%

Under the one-zone approximation and assuming that radiation is transported in the vertical direction
only through diffusion, one obtains the following formula
for the total flux emitted from disk surface (compare Eq.~\ref{e.radial.Qadv}),
\be
Q^{\rm rad}=f_F\frac{64\sigma T_C^4}{3\Sigma\kappa},
\ee
where factor $f_F$ has been introduced to account for inaccuracies 
arising from this approximation, as well as from the dominance
of the disk convection in certain regions (as discussed in \S\ref{ss.vertstructure}). 
Now, advective cooling takes the form
\be
Q^{\rm adv}=-\alpha P\frac{A\gamma^2}{r^3}\der\Omega r - f_F\frac{64\sigma T_C^4}{3\Sigma\kappa}.
\ee

One has to remember that disk thickness, $h$, introduced in Eq.~\ref{eq_vertbalance}, 
is not the exact location of the photosphere but rather disk pressure scale. Therefore, we introduce a
factor $f_h$ relating these quantities,
\be
h_{\rm phot}=f_h h.
\ee

Usually (like in Chapter~\ref{chapter-stationary}), the following values of $N$ (the polytropic index), $f_H$ and $f_F$ are assumed:
\begin{eqnarray}\nonumber\label{poli.param}
N&=&3.0,\\
f_H&=&1.0,\\\nonumber
f_F&=&1.0.
\end{eqnarray}

\begin{figure}
 \centering \includegraphics[angle=0,width=.7\textwidth]{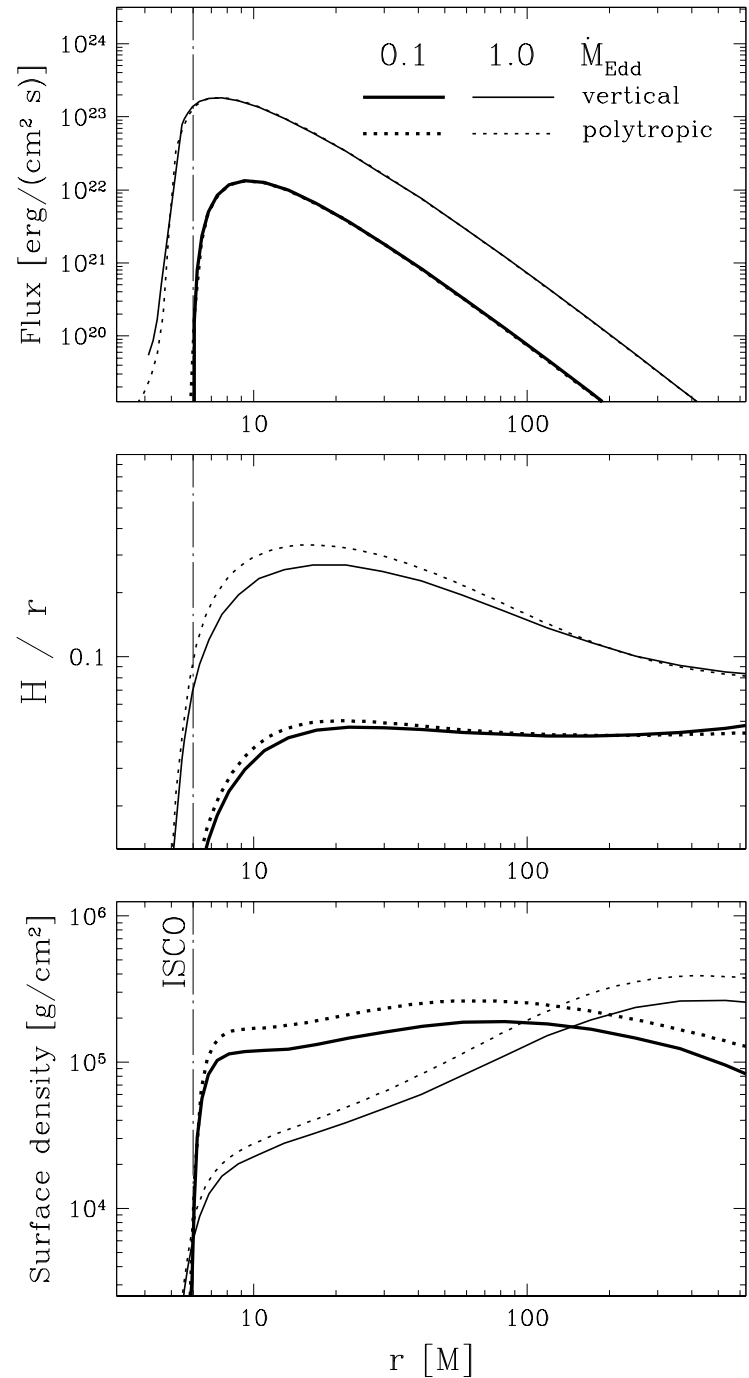}
\caption{ Comparison of the flux, disk thickness 
 and surface
density profiles calculated using the 2-D (this Chapter, solid lines)
 and the usual polytropic (Chapter~\ref{chapter-stationary}, dotted lines)
slim disk models for $\alpha=0.01$.
The solutions for two accretion rates ($0.1$ and $1.0 \mdot_{\rm Edd}$)
are presented with thick and thin lines, respectively.}
  \label{f.3comp}
\end{figure}

In Fig.~\ref{f.3comp} we compare radial profiles of the flux, photospheric
height and surface density of the 2-D
slim disk model described in this Chapter (consistently
taking its vertical structure into account)
with profiles obtained for the polytropic, height-averaged, slim disk, 
as presented in Chapter~\ref{chapter-stationary}.
The comparison was carried out for two accretion rates
($0.1$ and $1.0\mdot_{\rm Edd}$) and $\alpha=0.01$. 
 For the lower accretion rate ($0.1\mdot_{\rm Edd}$), the disk is 
radiatively efficient (advective cooling is negligible), and therefore 
both flux profiles almost coincide and correspond to the
\citem{nt} solutions. 
However, the 
disk photosphere location in the polytropic model turns out
to be overestimated by more than 20\% in the region corresponding
to the maximal emission.
 The profiles of the surface density do not coincide either,
the 2-D solutions giving values
 $\sim25\%$ lower than the corresponding polytropic solutions
(compare Fig.~\ref{f.scurve}).

For the higher accretion rate ($1.0 \mdot_{\rm Edd}$),
the flux profiles remain similar (up to 1\%).
However,  as advection becomes important, the emission is shifted
inwards with respect to the Novikov \& Thorne profile.
The photosphere location in the polytropic model is overestimated
by $\sim30\%$ 
and the surface density by $\sim20\%$.

\begin{figure}
 \centering \includegraphics[angle=0,width=.7\textwidth]{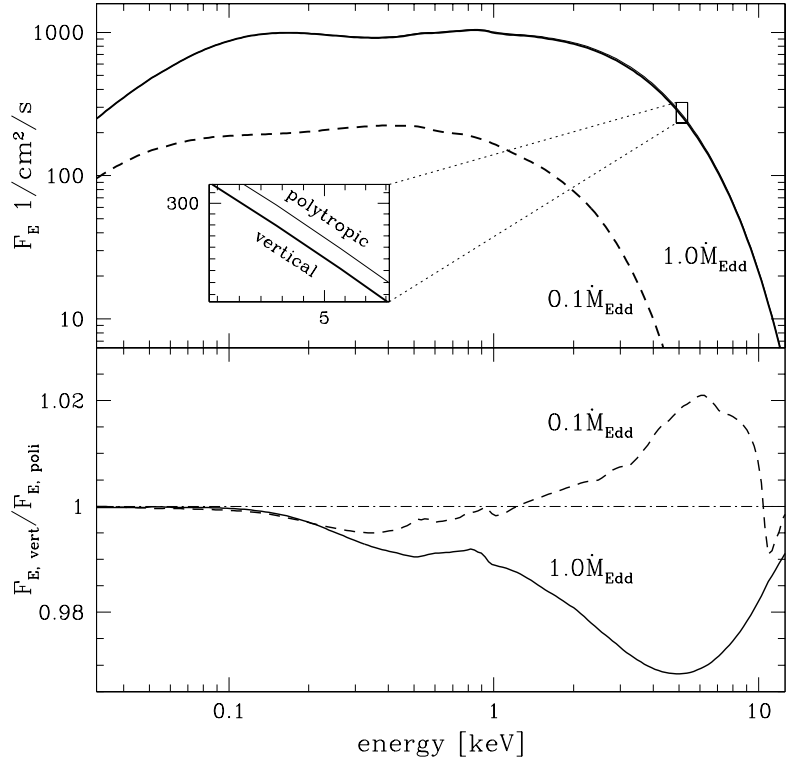}
\caption{ 
The upper panel presents spectral profiles of the 2-D and polytropic
solutions for two accretion rates 
($0.1$ and $1.0 \mdot_{\rm Edd}$), at inclination angle 
$i=70^o$ and distance $d=10{\rm kpc}$.
The bottom panel presents ratios of the corresponding spectra
(of 2-D to polytropic solutions)
for both accretion rates.}
  \label{f.spectra}
\end{figure}

A question arises as to whether such differences in the flux, photosphere, and surface
density profiles affect the resulting disk spectrum. In Fig.~\ref{f.spectra}
we present spectral profiles and their ratios 
(2-D to polytropic) for two accretion rates.
The spectra were calculated with ray-tracing routines \citepm{bursa.raytracing} 
 using the BHSPEC package \citepm{davisomer05}, assuming the inclination angle $i=70^o$ 
and distance to the observer $d=10\,{\rm kpc}$.
As BHSPEC gives tabulated solutions of the full,
frequency-dependent, radiative transfer
equations for the disk vertical structure taking the Compton scatterings into account
 in the disk atmosphere, the spectra
presented in  Fig.~\ref{f.spectra}
are not those of a simple multicolor blackbody. However, one should
be aware that using BHSPEC for calculating spectral color 
correction is not consistent with the assumed vertical structure
(calculated or height-averaged) because it is based on a stand-alone
disk model.

The general shape of the spectra is similar for both types of slim
disk models, because the emission profiles nearly coincide. However, the
spectra are not identical. 
The bottom panel of Fig.~\ref{f.spectra} presents ratios
of the spectral profiles of the corresponding solutions for each accretion rate.
They all coincide att low energies ($<0.1{\rm keV}$), while for 
higher energies the discrepancies
are as large as $3\%$ for  $1.0 \mdot_{\rm Edd}$ at $5{\rm keV}$. These
differences are attributed to (slightly) different profiles of the flux, 
the photosphere, and the surface density. We conclude that the proper
treatment of the vertical structure hardly affects the spectra of 
slim disks for this range of accretion rates ($\dot M<\dot M_{\rm Edd}$) and 
the viscosity parameter ($\alpha\leq0.01$).


We end this section by giving fitting formulae for $N$, $f_H$, and $f_F$,
which approximate the full numerical 2-D solutions described in this paper 
with a polytropic slim disk model. An advantage of using these formulae
lies in avoiding the need to perform time-consuming calculations
of the vertical structure and avoiding numerical problems connected to
interpolation in the vertical solutions grid. The formulae
for the polytropic model parameters for $\alpha=0.01$ are
\begin{eqnarray}\nonumber
\label{poli.fit}
 N&=&3.25\times{\cal S}_N,\\ 
 f_H&=&0.63\times{\cal S}_H,\\\nonumber
 f_F&=&0.94\times{\cal S}_F,
\end{eqnarray}
where the spin correction coefficients ${\cal S}_N$, ${\cal S}_H$, and ${\cal S}_F$ are given by
\begin{eqnarray}\nonumber\label{poli.fit2}
 {\cal S}_N&=&1+0.002~(6-r_{\rm ISCO}/M),\\\nonumber
 {\cal S}_H&=&1+0.003~(6-r_{\rm ISCO}/M),\\\nonumber
 {\cal S}_F&=&1+0.064~(6-r_{\rm ISCO}/M).
\end{eqnarray}
Here, $r_{\rm ISCO}$ is the radius of the marginally stable orbit.
For a non-rotating BH, at the radius $r=7M$
(corresponding to the highest disk effective 
temperature at $\dot M=\Medd$), the fitting formulae are accurate to 1\% for
the emitted flux, the photospheric height and the surface density. 

\section{Discussion}

Motivated by a desire to explain and fit
the observed spectra of accreting BHs in binary systems to theoretical models,
we have developed a 2-D model of optically thick
slim disks. These should be particularly relevant to transient
binaries. 
In quiescence, their inner disk regions are described well by optically thin
advection-dominated accretion flows 
(ADAFs; see e.g., \citem{lasotaetal-96,dubusetal-01}).
However, a few BH systems (e.g., GRS 1915+105 and LMC X-3)
have been observed in 
thermal states corresponding to disk luminosities higher than 
$0.3 L_{\rm Edd}$ \citepm{mcclintockremillard03, steineretal-10},
and modeling these require optically thick models going beyond the standard
thin disks.

In this Chapter we presented
a 2-D slim disk model, in which the radial and vertical structures
are coupled. 
Such an approach eliminates arbitrary factors that influence
solutions of the usual polytropic slim 
disk model. 
The results were obtained under two key assumptions:
an alpha disk was assumed (dissipation proportional to pressure),
with a uniform value of $\alpha$,
and the fraction of the generated entropy that is advected
was computed at every radius under the assumption that
this fraction does not vary with the height above the disk
plane (Eq.~\ref{vs.dFdz}).
Both of these assumptions seem arbitrary, and we can offer no physical
motivation for the (conventional) choice we made.

Under these assumptions and for the value $\alpha=0.01$ of the viscosity
parameter, we computed and presented the detailed
structure of 2-D slim disks, parametrized by the
mass accretion rate, and the two Kerr metric parameters, $M$ and $a$.
Somewhat surprisingly, the spectra observed at infinity from such disks
differ by only a few percent (and only at high energies) from those obtained from previously considered
slim disk models (in which the equations and structure
correspond to a height average over a polytropic atmosphere).
Such differences are unlikely to introduce any large corrections
to spin measurements based on X-ray continuum fits 
made with corresponding height-averaged
polytropic models of slim disks. Therefore, applying the spectral routines based on regular, polytropic slim disk solutions (e.g.,  \texttt{slimbb}, described in Section~\ref{s.slimbb}) is justified.

One has to be aware that the model of vertical
structure presented here is only an approximation
of the real physical processes taking place in disk interiors.
The diffusion approximation and the convection
treatment in the mixing length approach are known to successfully
describe media with large effective optical depths but break down when
the disk becomes optically thin. We have shown that the effective optical depth of slim accretion
disks may drop below unity for super-Eddington luminosities 
and sufficiently high values of $\alpha$. For such conditions,
a more sophisticated model of radiation transfer should be 
implemented. However, for $\alpha\leq0.01$ and $L\leq L_{\rm Edd}$
the assumptions of this work are self-consistent.
 For higher values of $\alpha$ their range of
applicability is limited to lower luminosities
(e.g., to $0.5 L_{\rm Edd}$ for $\alpha=0.1$).
Kerr-metric slim disks with low effective optical depth
were discussed by \citem{beloborodov-98}, who finds them to be significantly
hotter than the optically thick ones.

Another remark is connected to the fact that one can expect winds to be blown out
of the disk surface at super-Eddington luminosities. Such a phenomenon
may significantly change the disk structure, e.g., its thickness. This feature of slim disks, not described in
our calculations, has recently been recently cleverly modeled by  \citem{nir}.

\chapter{Self-irradiated slim accretion disks}
\label{chapter.selfirradiated}
The slim disk models presented and discussed in Chapters~\ref{chapter-stationary}, \ref{chapter-nonstationary} and \ref{chapter-vertical} make a few fundamental assumptions which make the problem of solving the hydrodynamical equations simpler, e.g., the negligence of the angular momentum taken away by photons in the angular momentum equation (Eq.~\ref{e.ang1}) makes its direct integration possible and puts the hydrodynamics into the form of ordinary differential equations instead of an integral-differential problem. The same physical process may be accounted for consistently in the standard model of a relativistic thin and radiatively efficient disk \citepm{nt}.

The returning flux of radiation and angular momentum due to the self-irradiation in a curved spacetime is another factor which has been so far neglected. Again, only in the case of thin disks one can account for this effect in a (relatively) simple way. It has been shown (\citem{lietal-05}, \citem{KatoBook}) that for thin disks with no torque at their inner edge the returning radiation has little influence on the inner disk regions and therefore on the emerging flux and high-energy disk spectrum. However, as was shown in the previous Chapters, accretion disks with high accretion rates are geometrically thick and the effect of self-irradiation may be much more significant. Unfortunately, including the appropriate terms into the disk equations once again leads to a complicated set of integral-differential equations. Therefore, a different approach for solving them is necessary. 

In this Chapter we present a state-of-art numerical method of solving slim disk equations including all of the above-mentioned factors. We limit our discussion to the case of a non-rotating BH. More detailed study will be included in a separate paper.

\section{Equations}

We start with presenting modified forms of the vertically integrated equations describing an advective accretion disk which include the flux of angular momentum taken away by photons and the fluxes of radiation and angular momentum coming back with the returning radiation. Only two equations need to be modified: the angular momentum conservation and the energy balance. Deriving them we assume that the disk albedo is zero, i.e., all the returning radiation is absorbed and thermalized by the disk. We note that this assumption may not be satisfied in the disk regions which are effectively optically thin (e.g., deep inside marginally stable orbit). Our approach gives the upper limit on the impact of the returning radiation on the disk structure and neglects the reflected radiation.

\subsection{Angular momentum conservation}

Let us take the angular momentum conservation law in the general form (Eq. \ref{e.ang1}) assuming stationarity ($\partial_t=0$) and leaving the general form of the $(r,\phi)$ component of the stress-energy tensor,

\be \frac{\Mdot}{2\pi}\der{\cal L}{r}=-\der{}{r}(r^2T_{r\phi}).\ee
This expression neglects the angular momentum fluxes due to the emerging and returning radiation which are equal,
\be F_{{\cal L},\rm out}={\cal L} F,\ee
\be F_{{\cal L},\rm in}=\int_{{\rm disk\,surface}} \int_{{\rm hemisphere}}{\cal L}({\bf r'})F({\bf r'}){\cal R}({\bf r},{\bf r'},{\bf p}) {\rm d}\Omega\, {\rm d}{\bf r'},\ee
respectively. $F=F({\bf r})$ denotes the emerging flux at given point on the disk surface, ${\cal L}={\cal L}({\bf r})$ is the corresponding angular momentum, ${\cal R}({\bf r},{\bf r'},{\bf p})$ is a ray-tracing function determining if a photon emitted from $\bf r'$ along the direction $\bf p$ returns to the disk at radius $\bf r$, and which accounts for the change of the photon energy (expressed by the g-factor, see Section~\ref{s.slimbb} or \citem{lietal-05}).

Including these terms leads to,
\be\frac{\Mdot}{2\pi}\der{\cal L}{r}=-\der{}{r}(r^2T_{r\phi})+rF_{{\cal L},\rm out}-rF_{{\cal L},\rm in}\ee
and further, after radial integration, to,
\be
\label{e.sirr.angmom}
\frac{\Mdot}{2\pi}({\cal L}-{\cal L}_{\rm in})=r^2(-T_{r\phi}+T_{\rm out}-T_{\rm in}),
\ee
where
\be\label{e.sir.tout}
T_{\rm out}=\frac1{r^2}\int_{r_{\rm in}}^r r'F_{{\cal L},\rm out}\,{\rm d}r'
\ee
and
\be\label{e.sir.tin}
T_{\rm in}=\frac1{r^2}\int_{r_{\rm in}}^r r'F_{{\cal L},\rm in}\,{\rm d}r'.
\ee
Calculating these integrals requires {\it a priori} knowledge about the emission and angular momentum profiles between the horizon and given radius $r$. We will solve this problem using the relaxation method where the integrals are calculated basing on previous relaxation steps. Details are given in Section~\ref{s.sirr.numerical}.

\subsection{Energy balance}

The energy balance equation (Eq.~\ref {e.radial.Qadv}) relates the amount of the advected heat (left hand side) to the difference between the heating and cooling (right hand side). Originally, the latter involved the viscous heating and radiative cooling only. When accounting for the returning radiation one has to include an additional heating term as the incoming flux of radiation brings its energy and stores it in the disk heating it up. The expression for the advected heat takes the following form,

\be
\label{e.sirr.Qadv}
Q^{\rm adv}=Q^{\rm vis}-Q^{\rm rad}+Q^{\rm ret}
\ee
where 
\be
Q^{\rm vis}=-\alpha P\frac{A\gamma^2}{r^3}\der\Omega r 
\ee
is the viscous heating, 
\be
Q^{\rm rad}= \frac{64\sigma T_C^4}{3\Sigma\kappa}
\ee
is the radiative cooling, and the returning flux $Q^{\rm ret}$ is given by,
\be\label{e.sir.qret}
Q^{\rm ret}=\int_{{\rm disk\,surface}} \int_{{\rm hemisphere}}{F}({\bf r'}){\cal R}({\bf r},{\bf r'},{\bf p}) {\rm d}\Omega\, {\rm d}{\bf r'}.\ee
The integral will be calculated using the emission profile from the previous iteration step as described in detail in the next section.

\section{Numerical methods}
\label{s.sirr.numerical}

Including self-irradiation into the slim disk model introduces integrals which cannot be directly evaluated because they depend, e.g., on the emission and angular momentum profiles which are not known at the time of solving the equations. However, the exact solution may be found using an iteration scheme where these integrals are calculated basing on the profiles from previous iteration steps. The solution is found when and if such a scheme converges. Fortunately, this is the case for self-irradiated slim disks. In this paragraph we give details of the numerical methods used to solve the problem.

We base on the equations for the radial structure of the polytropic slim accretion disk described in Chapter~\ref{chapter-stationary}. We take without any modifications the equation of continuity (Eq.~\ref{eq_poly_cont2}), the radial momentum balance (Eq.~\ref{eq_poly_rad3}), the vertical equilibrium formula (Eq.~\ref{eq_vertbalance}) and the expression for the advected heat (Eq.~\ref{eq.qadv}). We modify the formula for the angular momentum conservation and the heating-cooling balance and take the forms derived in the previous two paragraphs,

\be
\frac{\Mdot}{2\pi}({\cal L}-{\cal L}_{\rm in})=r^2\left(-T_{r\phi}+\xi (T_{\rm out}- T_{\rm in})\right),
\label{e.angsir}
\ee
\be
Q^{\rm adv}=Q^{\rm vis} - Q^{\rm rad}+\xi Q^{\rm ret}
\ee
introducing a dimensionless parameter $\xi$ satisfying $0\le \xi \le1$.

In the beginning we put $T_{\rm out}=T_{\rm in}=Q^{\rm ret}=0$ and solve the set of regular slim disk equations. Basing on this solution, we calculate the integrals in Eqs.~\ref{e.sir.tout} and \ref{e.sir.tin} as well as the returning radiation profile $Q^{\rm ret}$ (Eq.~\ref{e.sir.qret}). We set $\xi$ to some small number (typically $0.001$) and relax the equations (in the same way as described in Section~\ref{sect.numerical}) to obtain a transonic solution. Once the transonic solution is found we slightly (not to lose convergence) increase value of $\xi$ making the impact of the self-irradiation effects stronger. During the relaxation in $\xi$ we keep the calculated values of $T_{\rm out}$, $T_{\rm in}$ and $Q^{\rm ret}$ fixed. This procedure is repeated until the solution with $\xi=1$ is found. Then, new profiles of the integrals and returning radiation are calculated. Basing on them, we relax the solution from the one obtained in the previous set of iterations to a new one, following the most recent estimate of $T_{\rm out}$, $T_{\rm in}$ and $Q^{\rm ret}$. This scheme is repeated until a converged solution is found.

Usually, around $50$ iterations in $\xi$ are required and the whole scheme has to be repeated few times. Due to the numerical evaluation of the self-irradiation, the required computational time increased significantly (up to 10 minutes on a single-CPU workstation) in comparison with the regular, non-irradiated disk model.

\section{Results}

In this section we discuss the most important features of self-irradiated solutions of slim disks for a non-rotating BH. More detailed study, including discussion of the impact of self-irradiation on accretion disks around highly-spinning BHs, will be put in a forthcoming paper. 

\subsection{Emission profiles}

\begin{figure}
\centering

\includegraphics[width=1.\textwidth]{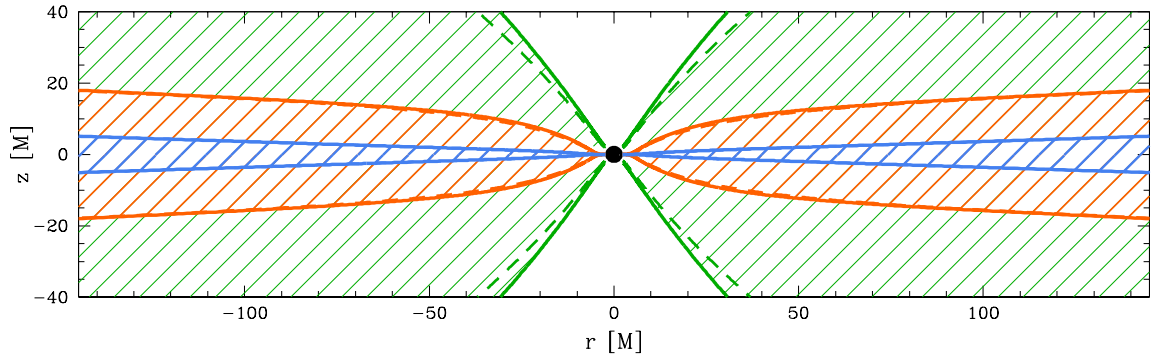}

\caption{
Disk thickness profiles of the self-irradiated solutions (solid lines) for three accretion rates: $0.1$ (blue), $1.0$ (orange) and $10.0\Medd$ (green), $\alpha=0.1$ and a non-rotating BH. The dashed lines (which significantly differ from the solid lines only for the highest accretion rate) present corresponding profiles for the standard slim disk model (Fig.~\ref{f.sir.height.stand}).
}
\label{f.sir.height}
\end{figure}

The profile of emission, corresponding to the rate of radiative cooling, is the most important factor determining the accretion disk spectrum. Other factors which are important for the spectrum are: location of the disk photosphere (the location where the ray-tracing starts) and disk surface density (related to the intrinsic hardening of the spectrum due to photon up-scattering by hot electrons, see Section~\ref{s.hardening}). These additional factors, as our study shows (e.g., Fig~\ref{f.sir.height}), are slightly affected by self-irradiation, and the changes in the emission profile dominate. Thus, we put most attention to discussing the impact of returning radiation on the emergent flux profiles.


\begin{figure}
\centering
\includegraphics[angle=0,width=.8\textwidth]{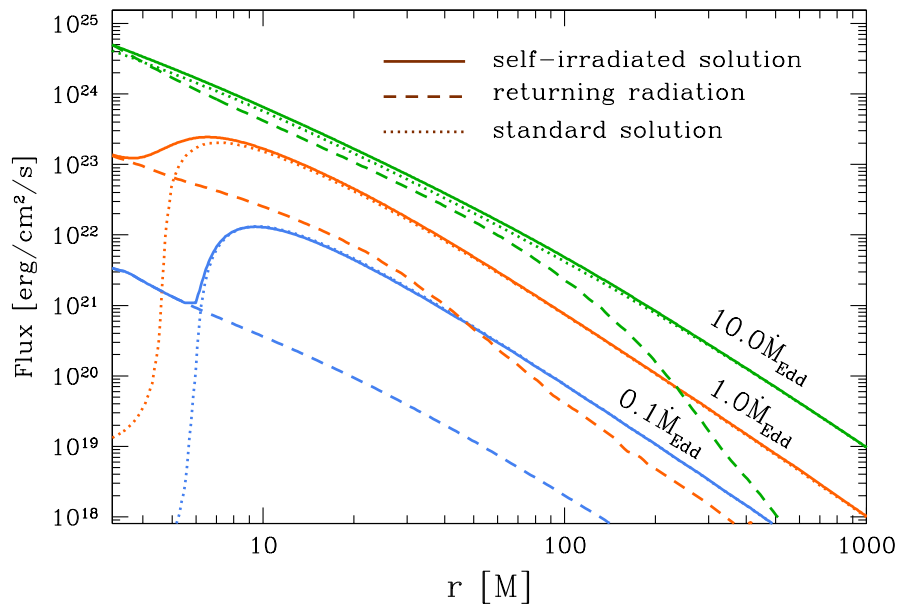}
\caption{Profiles of the emitted and returning fluxes of radiation (solid and dashed, respectively) predicted by the self-irradiated slim disk model for a non-rotating BH and $\alpha=0.1$. Profiles of the emitted flux for regular (non-irradiated) disks are also presented for comparison (dotted line). Colors correspond to different accretion rates: $0.1$ (blue), $1.0$ (orange) and $10.0\Medd$ (green).}
\label{f.sir.fluxes}
\end{figure}

In Fig.~\ref{f.sir.fluxes} we plot profiles of the emitted and returning fluxes of radiation for slim disks with three accretion rates: $0.1$, $1.0$ and $10.0\Medd$ (marked with different colors). The dotted lines denote solutions of the regular, non-self-irradiated model presented in Chapter~\ref{chapter-stationary}. The dashed lines give the profiles of the returning radiation, while the solid lines show the emission from the new, self-irradiated solutions.

The disk with the lowest accretion rate presented ($0.1\Medd$) is very thin ($h/r<0.1$, see Fig.~\ref{f.sir.height}) and therefore, the returning radiation is a result of the light bending only --- there is hardly any direct irradiation by the disk itself and curved geodesics are required to direct photons back to the disk. Thus, the returning radiation for this accretion rate (blue, dashed line) with good accuracy corresponds to the profile of returning radiation calculated for the standard thin disk \citepm{lietal-05}. For the case of a non-rotating BH, the profile of returning radiation is roughly a power law with the same exponent as the emission at $r\gg r_{\rm ISCO}$, i.e., $F_{\rm ret}\propto r^{-3}$. At large radii the returning radiation is equal to roughly $3\%$ of the radiation emitted. The relation between the emitted and returning fluxes changes rapidly inside the marginally stable orbit. Inside this particular radius, there is little viscous heating, and therefore the standard solutions emit hardly any radiation. On the other hand, the radiation emitted from outside the marginally stable orbit is efficiently focused by the BH on the regions close to the horizon. Therefore, the returning flux increases with decreasing radius, and dominates over the viscosity-generated emission.

The solid blue line presents the ultimate solution for $0.1\Medd$, which takes into account the impact of self-irradiation and the angular momentum taken away by photons, and which has relaxed following the numerical procedure described in Section~\ref{s.sirr.numerical}. Outside the marginally stable orbit the disk is heated up by viscosity, while the self-irradiation dominates inside. One can expect that an accretion disk at such low accretion rate is radiatively efficient. This is in fact the case, not only outside the marginally stable orbit, but also inside, where heating by the returning radiation dominates. As a result, all the heat generated is released at the same radius, and the outcoming flux is the sum of the profiles of viscous and returning radiation heating rates (the blue solid line follows the standard solution outside the marginally stable orbit, and the returning radiation profile inside this radius). The maximal disk temperature, which determines the spectral hardness and is crucial for estimating the BH spin using X-ray continuum fitting (see Section~\ref{s.spindetermination}), does not change --- the self-irradiation affects the emission profile only inside the marginally stable orbit and the modified emission does not exceed the maximal value reached at $r\approx 10\rm M$.

The behavior discussed above changes when the accretion rate becomes high. For $1.0\Medd$ the disk is no longer thin ($h/r\approx 0.3$), and the profile of emission of the standard solution slightly shifts inward due to advection. Furthermore, the disk surface is no longer straight ($\theta=\rm const$), but the innermost region is puffed up by the pressure of radiation. As a result, the inner region of this radiation pressure-related hump captures back radiation emitted from disk inner regions on the other side of the central BH. What follows, the radiation from the innermost disk does not reach the most distant ($r>100\rm M$) regions, being eclipsed by the inner hump. The increase of the amount of returning radiation due to the radiation pressure-induced disk thickening around $r=30\rm M$ is visible in Fig.~\ref{f.sir.fluxes}. The orange dashed line presents its profile for $1.0\Medd$. At large distances it contributes to $\sim 3\%$ of the viscous heating, exactly as in the case of a thin disk. In the region inside $r=30\rm M$, however, its amount, compared to the standard solution, increases to $\sim 20\%$, due to the photon capture by the hump. The profile of the returning flux further increases towards the BH horizon due to efficient light bending close to the BH. All the captured radiation, as will be discussed in one of the following paragraphs is reemitted, and the profile of emission of the self-irradiated solution is, similarly as for the $0.1\Medd$ case, the sum of the viscosity and returning fluxes. This time, however, the maximal effective temperature of the emitted flux is affected and increases by $\sim 5\%$ when compared to the standard solution.

The impact of the radiation pressure-induced inner hump, and, in general, increased disk thickness is even more profound for accretion disks with super-critical accretion rates. The profile of the estimated location of the disk photosphere for $10.0\Medd$ is presented in Fig.~\ref{f.sir.height}. It is clear that disk inner region irradiates itself, and only small fraction of radiation is able to escape to the observer at infinity. The photons emitted from the most energetic, inner regions are not able to reach the most distant regions, obscured by the thickening. The green curves in Fig.~\ref{f.sir.fluxes} show this behavior in terms of the returning and emitted radiation. The flux re-captured by the disk in the inner region ($r<50\rm M$) is equal to $\sim 80\%$ of the emerging one --- only fifth part of the emitted radiation escapes to infinity. If the disk was radiatively efficient, this flux of returning radiation would increase the emitted one almost by a factor of two. However, at such high accretion rate the disk is advection dominated, and most of the heat returning with radiation to the disk is advected towards BH instead of being instantaneously reemitted. As a result, the flux emitted by the self-irradiated solution is higher than the flux of the standard one only by $\sim 25\%$, what corresponds to the same increase of the effective temperature as in the case of $1.0\Medd$.

\subsection{Returning radiation}

\begin{figure}
\centering

\includegraphics[height=.8\textwidth,angle=270]{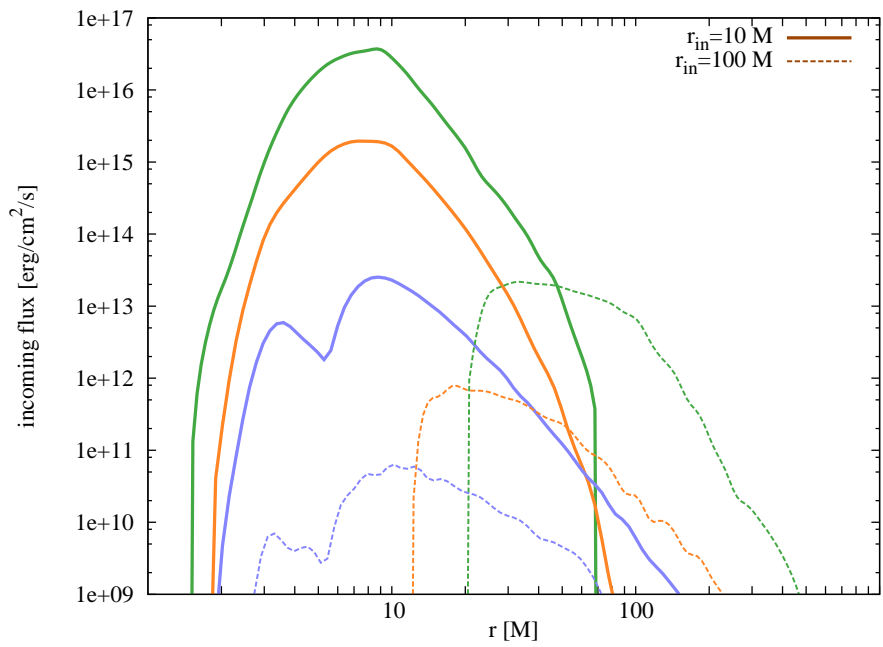}

\caption{
Radial profiles of radiation coming back to $r=10\rm M$ (solid) and $r=100\rm M$ (dashed lines). The curves show where the flux returning to a given radius comes from. Their integrals are equal to the returning radiation flux at these radii plotted  in Fig.~\ref{f.sir.fluxes}. Profiles for three accretion rates: $0.1$ (blue), $1.0$ (orange) and $10.0\Medd$ (green lines) are presented.
}
\label{f.sir.srad}
\end{figure}

In Fig.~\ref{f.sir.srad} we plot the radial distribution of the radiation returning back to the disk at two radii: $r=10\rm M$ (solid) and $r=100\rm M$ (dashed lines) for the same three accretion rates. The blue curves correspond to the lowest, $0.1\Medd$. The profiles of the returning flux for both radii reflect the profile of the emitted radiation with the peak around $10\rm M$ and a minimum slightly inside the marginally stable orbit. The returning flux, however, does not increase, like the emitted flux does, towards the horizon, as the photons coming back from the vicinity of BH are subject to very strong redshift making their contribution to the returning flux at larger radii negligible. Even for large distances from the BH (e.g., $r=100\rm M$, blue dashed line), the photons from the most energetic, inner regions around $r=10\rm M$ contribute to the majority of the returning flux. 

Once the disk thickens and the inner hump appears, the radial distribution of the returning radiation changes. Most of the flux returning to $r=10\rm M$ still comes from the hottest region inside this radius, the radiation from the vicinity of BH is also strongly redshifted and does not contribute significantly to the returning flux. However, the distribution does no longer extend uniformly towards large radii. The region outside $r\approx80\rm M$ is obscured by the inner hump and no single photon from this region is able to return to the disk at $r=10\rm M$. This fact is clearly visible also on the profiles of the radiation returning to $r=100\rm M$ (dashed lines). For $1.0\Medd$ the distribution terminates at $r=10\rm M$ meaning that the region inside this radius is hidden for an observer located at the disk photosphere at $r=100\rm M$. For $10.0\Medd$ this obscuration takes place starting from $r=20\rm M$.


\subsection{Energy balance}

At each radius the balance between heating and cooling must be satisfied. In the standard approach, only three mechanisms were involved: viscous heating, radiative cooling and advective heating (decreasing radial advective flux heats the disk at given radius up) or cooling (some fraction of heat generated by viscosity increases the advective flux). When the self-irradiation is taken into account, an additional heating mechanism resulting from the returning radiation must be considered. 

\begin{figure}
\centering
 \subfigure
{
\includegraphics[height=.45\textwidth]{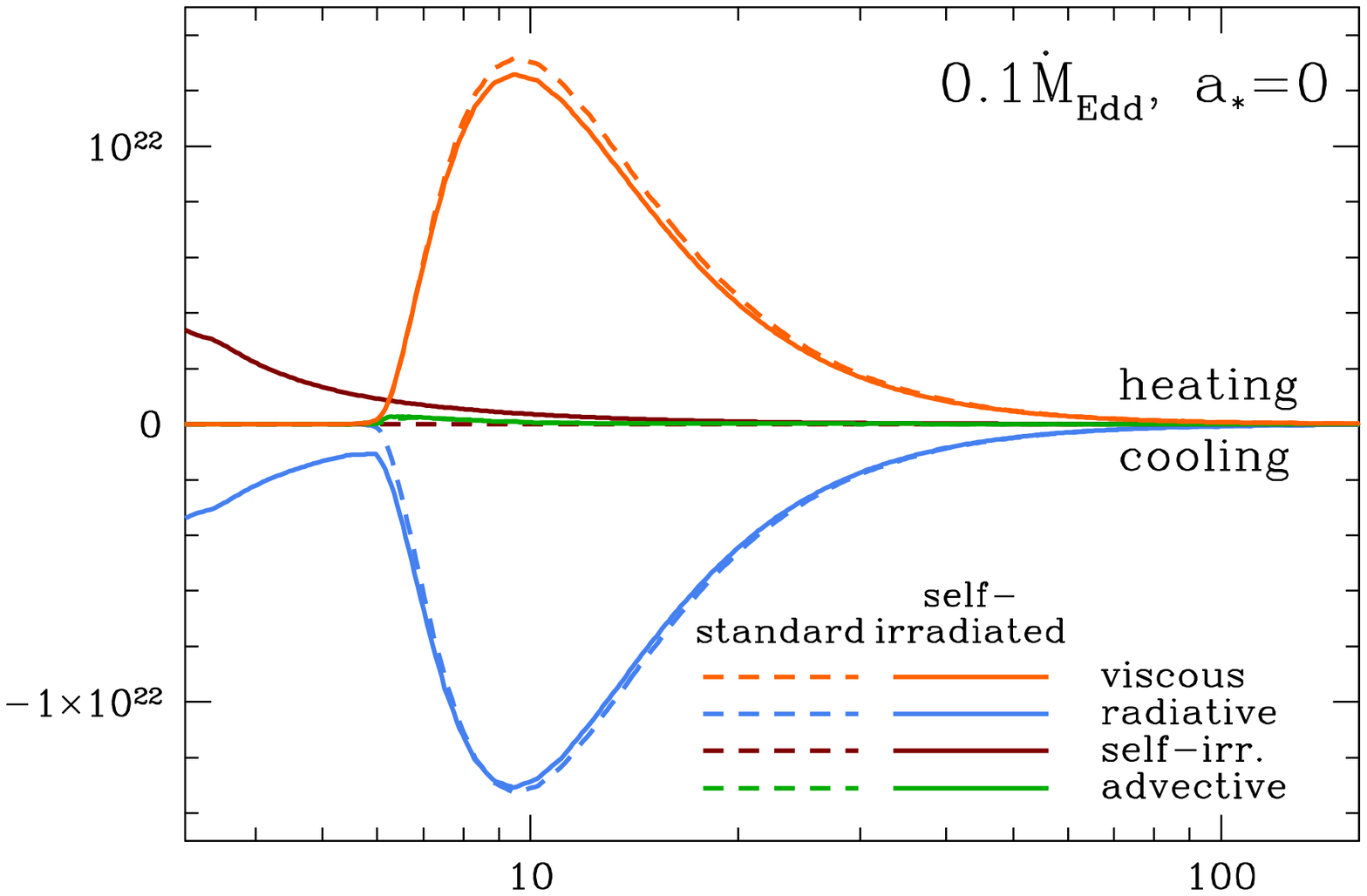}
}
\\
\vspace{-.07\textwidth}
 \subfigure
{
\includegraphics[height=.45\textwidth]{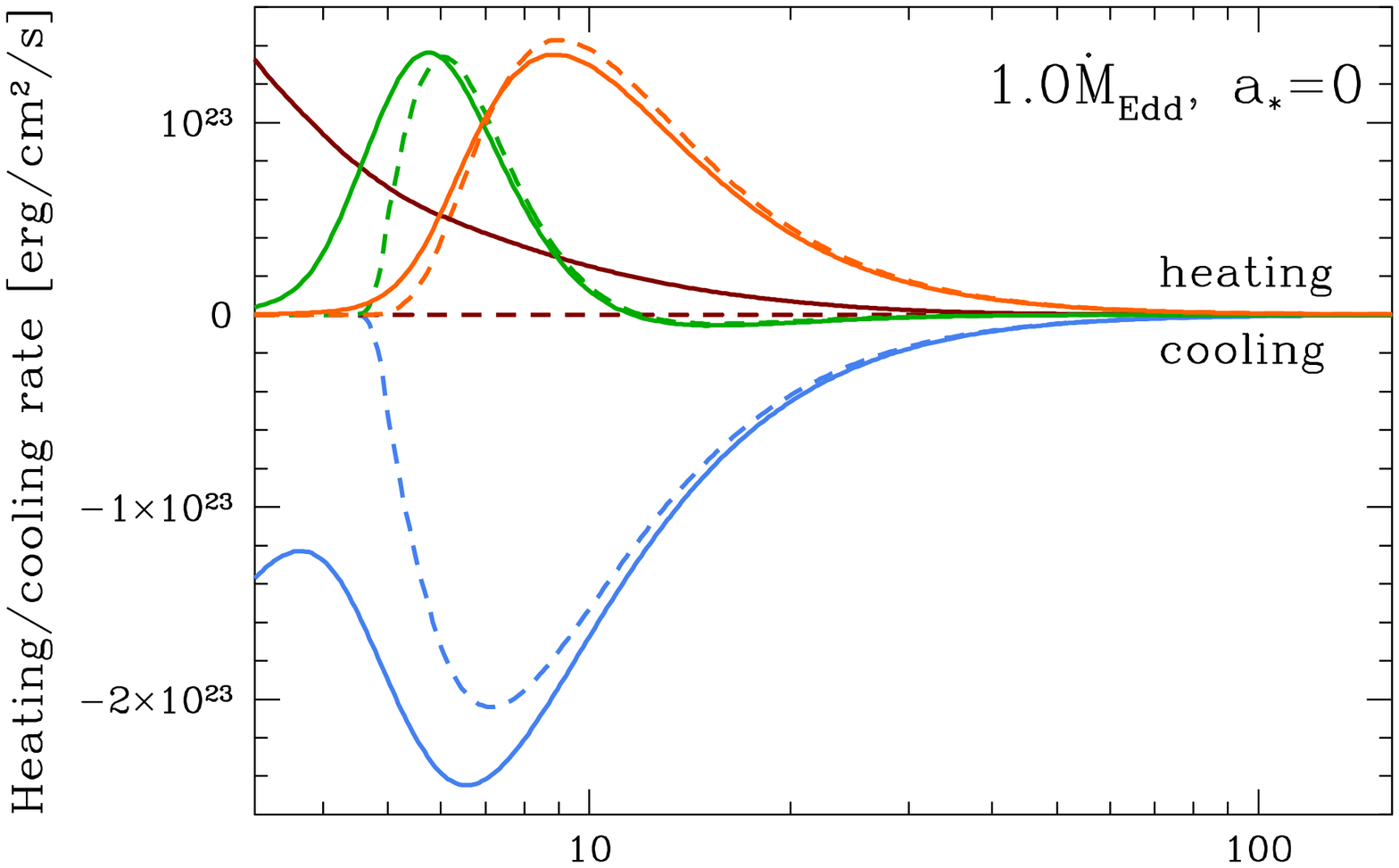}
}
\\
\vspace{-.07\textwidth}
 \subfigure
{
\includegraphics[height=.45\textwidth]{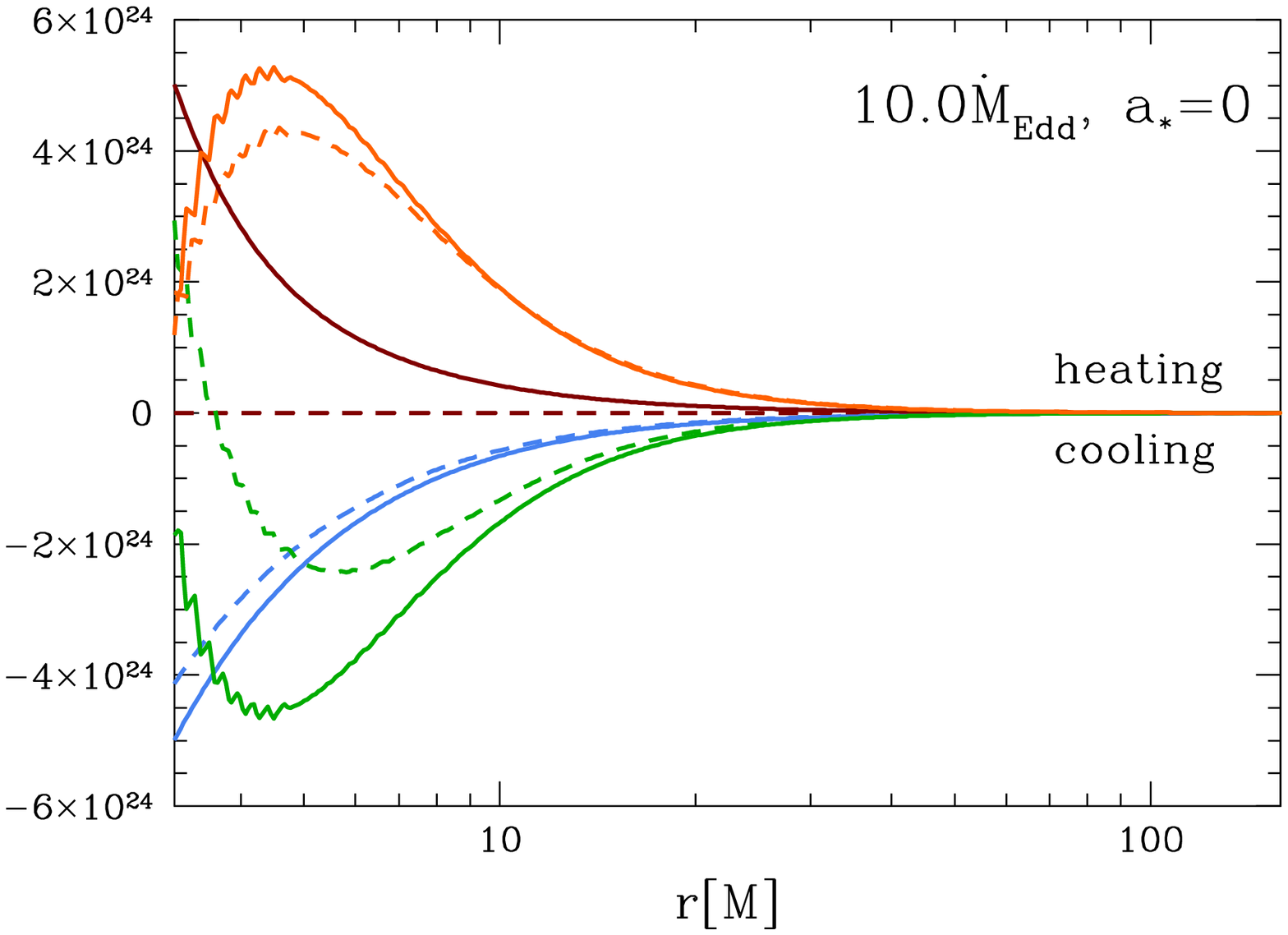}
}
\caption{Energy balance as a function of radius for three accretion rates: $0.1$ (top), $1.0$ (middle) and $10.0\Medd$ (bottom panel). Positive values denote heating, while negative stand for cooling. Profiles for the self-irradiated solutions are marked with solid lines, while dashed lines correspond to the regular disk model. Viscous heating, radiative cooling and advective transport are marked with orange, blue and green lines, respectively. The brown lines denote heating by the returning flux of radiation. }
\label{f.sir.balans.a0}
\end{figure}

In Fig.~\ref{f.sir.balans.a0} we compare the heating/cooling profiles of the standard (dashed) and self-irradiated (solid lines) slim disk solutions for three accretion rates: $0.1$ (top), $1.0$ (middle) and $10.0\Medd$ (bottom panel). Advection for the lowest accretion rates (top panel), corresponding to radiatively efficient disks, is negligible. Therefore, in the standard model the heat generated by viscosity is completely balanced by the radiative cooling. When the returning radiation is considered, another heating mechanism occurs. The excess of heating at $r\lesssim 6\rm M$ must be balanced by the radiative emission. This is in fact the case --- the rate of heating by the returning radiation (brown solid line) is equal to the absolute value of the radiative cooling rate in this region (blue solid line).


Advection becomes an important factor in the energy balance when the accretion rate is large. The middle panel of Fig.~\ref{f.sir.balans.a0} presents the heating and cooling rates for both models for $1.0\Medd$. In the standard approach, viscosity is the only heating mechanism at $r>10\rm M$. Small fraction (compare Fig.~\ref{f.fadv}) of the heat generated in this region is stored in the advective flux and is released, contributing to the total heating rate, inside this radius. At $5{\rm M} < r < 7\rm M$ the advective heating dominates over the viscosity-induced mechanism, leading to slight inward shift of the emission. 

This characteristic changes, mostly in the innermost region, when the returning radiation is considered (solid lines). The heating caused by this process is most efficient near the BH, where it dominates over the other mechanisms. For the whole region inside $r=10\rm M$, the advection contributes to heating only, and all the heat coming with the returning radiation must be balanced by the increase of radiative cooling. As a result, the amount of the emitted flux increases by the rate of the captured radiation, leading to the modified emission profile presented in Fig.~\ref{f.sir.fluxes}.

The solutions for the highest accretion rates (e.g., $10.0\Medd$, bottom panel) are significantly different, as the advective cooling is prominent at most radii. For the standard model, the advection contributes to cooling at $r>4\rm M$ and may be twice as efficient as the radiative mechanism at moderate ($r\approx 10\rm M$) radii. This leads to an inefficient emission (blue line) equal to only a small fraction of the heating rate (orange). Disks at such high accretion rates are very thick and capture significant part of the emitted radiation. However, most of this radiation is not radiated away at the same radius due to the efficient advective cooling mechanism. Instead, the captured heat is advected towards the BH. As a result, the emission profile does not change significantly. Only the rate of advective cooling significantly increases.

\section{Discussion}

We have introduced a model of an advective, optically thick accretion disk which properly accounts for the heating impact of the returning radiation, as well as the angular momentum taken away and brought in by photons. In the previous section we described in detail solutions for disks around non-rotating BHs and compared them with the solutions of the standard model presented in Chapter~\ref{chapter-stationary}. This analysis leads to the following conclusions:

(i) The impact of self-irradiation onto disks with sub-Eddington accretion rates is negligible, as the returning radiation is comparable with the viscous heating only inside the marginally stable orbit. Despite the fact that the emission profile is significantly modified in this region, the resulting spectrum will be hardly affected, as the radiation coming from the vicinity of BH is strongly redshifted and  hardly contributes to the total emission.

(ii) For disks with accretion rates close to the Eddington value, the relative amount of the captured radiation in the innermost region increases due to radiation pressure-induced disk thickening. For the Eddington accretion rate, the returning radiation around $r=10\rm M$ may be equal up to $20\%$ of the flux emitted there. At these accretion rates, all the returning radiation contributes to the emission, increasing the amount of the emerging flux. As a result, the maximal effective temperature increases by a few percent, leading to a spectrum slightly harder than the predicted by the standard model.

(iii) The relative amount of the captured radiation is much larger for super-critical accretion rates. For $10\Medd$, the returning flux may reach as much as $80\%$ of the emitted one. However, at this regime of luminosities accretion disks are efficiently cooled by advection, and most of this returning heat is advected onto the BH. Only small fraction ($20\%$ for $10\Medd$) contributes to the radiative emission. The emerging spectrum is slightly hardened in comparison with the standard model.

The study presented in this Chapter was limited to the case of a non-rotating BH. In a forthcoming paper we will extend this approach to study the impact of returning radiation on accretion disks around rotating BHs. It has been shown by \citem{lietal-05} that its impact on disk spectra may be more significant for rapidly rotating BHs\footnote{They prove, that for the thin disk case, the impact of self-irradiation on the spectrum may be simulated to quite high accuracy by a slight increase of the accretion rate.}. Our results show, that for slowly rotating central objects, only solutions with low accretion rates are unaffected by the returning radiation, and the standard models may be applied to model spectra. For higher accretion rates, however, they have to be used with care, as they underestimate the amount of emitted radiation and the spectrum hardness. This discrepancy may be more profound for rotating BHs. Nevertheless, the uncertainty related to the model inadequacy is most likely negligible when compared to other unknown factors, e.g., the value of the viscosity parameter $\alpha$ at different luminosities (see Section~\ref{s.spindetermination}).

\chapter{Applications}
\label{chapter-applications}

\section{Spinning-up black holes with super-critical accretion flows}
\label{s.spinningup}

Astrophysical BHs are very simple objects -- they can be described by just two parameters: mass and angular momentum. In isolation, BHs conserve the birth values of these parameters but often, e.g., in close binaries or in active nuclei, they are surrounded by accretion disks and their mass and angular momentum change. Accretion of matter always increases BH's irreducible mass and may change its angular momentum, usually described by the dimensionless spin parameter $a_*=a/M$. The sign of this change and its value depend on the (relative) sign of accreted angular momentum and the balance between accretion of matter and various processes extracting BH's rotational energy and angular momentum.

The question about the maximal possible spin of an object represented by the Kerr solution of the Einstein equation is of fundamental and practical (observational) interest. First, a spin $a_* > 1$ corresponds to a {\it naked} singularity and not to a BH. According to the Penrose cosmic censorship conjecture, naked singularities cannot form through actual physical processes, i.e. singularities in the Universe (except for the initial one in the Big Bang) are always surrounded by event horizons \citepm{wald-84}. This hypothesis has yet to be proven.

In any case, the {}``third law'' of BH thermodynamics \citepm{bardeenetal-73} asserts that a BH cannot be spun-up in a finite time to the extreme spin value $a_* = 1$.  
Determining the maximum value of BH spin is also of practical interest because the efficiency of accretion through a disk depends on the BH's spin value. For example, for the {}``canonical'' value $a_*=0.998$ (see below) it is about $\tilde\eta\approx 32\%$, while for $a_*\rightarrow 1$ one has $\tilde\eta\rightarrow 42\%$ (Section~\ref{s.efficiency}). Recently, \citem{banadosetal-09} showed that the energy $E_{\rm COM}$ of the center-of-mass collision of two particles colliding arbitrary close to the BH horizon, grows to infinity ($E_{\rm COM} \rightarrow \infty$) when $a_* \rightarrow 1$\footnote{Of more fundamental interest is the fact that the proper geodesic distances $D$ between the ISCO and several other special Keplerian orbits relevant to accretion disk structure tend to infinity $D \rightarrow \infty$ when $a_* \rightarrow 1$ \citepm{bardeen-72}.}.

A definitive study of the BH spin evolution will only be possible when reliable, non-stationary models of accretion disks and jet emission mechanisms are established. For now, one has to use simplified analytical or numerical models.

\citem{thorne-74} used the model of a radiatively efficient, geometrically thin accretion disk \citepm{nt} to evaluate BH spin evolution taking into account the decelerating impact of disk-emitted photons. The maximum value so obtained, $a_*=0.9978$, has been regarded as the canonical value for the maximal BH spin. In this Chapter we generalize Thorne's approach, using models of advective, optically thick accretion disks (slim disks, Chapter~\ref{chapter-stationary}) to calculate maximum BH spin values for a large range of accretion rates. We show that for sufficiently large accretion rates they differ from the canonical value.

We start with a short discussion of previous works devoted to the BH spin evolution. In Section~\ref{s.tetrad} we give formulae for a general tetrad of an observer comoving with the accreting gas along the arbitrary photosphere surface. In Section~\ref{s.spinevolution} we give basic equations describing the BH spin evolution. In Section~\ref{s.results} we present and discuss the terminal spin values for all the models we consider. Finally, in Section~\ref{s.summary} we summarize our results.

\subsection{Previous studies}

A number of authors have studied the evolution of the BH spin
resulting from disk accretion. \citem{bardeen-70} initiated this field of research
by stating the problem and solving equations describing
the BH spin evolution for
accretion from the marginally stable orbit. Neglecting the effects of radiation he proved that such a process
may spin-up the BH up to $a_*=1$. Once the classical models of accretion disks were formulated
\citepm{shakura-73,nt}, it was possible to account properly
for the decelerating impact of radiation (frame dragging makes counter-rotating photons more likely to be captured by the BH). As mentioned above, \citem{thorne-74} performed this study
and obtained the terminal spin value for an isotropically emitting thin disk of
$a_*=0.9978$, independently of the accretion rate.
The original study by Thorne was followed by many papers, some of which are briefly mentioned below.

The first to challenge the universality of Thorne's limit were \citem{abramowiczlasota-80} who showed that geometrically thick accretion disks may spin up BHs to terminal spin values much closer to unity than the presumably canonical $a_*= 0.9978$. Their simple argument was based on models by \citem{koz-1978} who showed that for high accretion rates the inner edge of a disk may be located inside the marginally stable orbit; with increasing accretion rate, arbitrarily close to the marginally bound orbit. However, this conclusion assumed implicitly a low viscosity parameter $\alpha$, whereas for high viscosities the situation is more complicated (see \citem{leavingtheisco} and references therein).

\citem{moderski-98} assessed the impact
of possible interaction between the disk magnetic field and the BH through
the Blandford-Znajek process. They showed that the terminal spin value may be decreased
to any, arbitrarily small value, if only the disk magnetic field is strong enough.
Given the current lack of knowledge about the strength of magnetic fields (or the magnetic transport of angular momentum in the disk, see \citem{ghoshabramowicz-97})
and processes leading to jet emission, a more detailed study cannot be performed. The situation may further be complicated by energy extraction from the inner parts of accretion disks \citepm{livioetal-99}.

\citem{pophamgammie-98} studied the spinning-up
of BHs by optically thin advection dominated accretion flows (ADAFs). They
neglected the contribution of radiation to BH spin as such disks are radiatively
inefficient. They found that the terminal value of BH spin is very sensitive to
the assumed value of the viscosity parameter $\alpha$ and may vary between
$0.8$ and $1.0$. \citem{gammie-04}, besides making a comprehensive summary of different
ways of spinning up supermassive BHs, presented results based on a single run of a general relativistic magnetohydrodynamical (GRMHD)
simulation (with no radiation included) obtaining terminal spin $a_*=0.93$.

Cosmological evolution of spins of supermassive BHs due to hierarchical mergers and thin-disk accretion episodes has been recently intensively studied. Although \citem{volonterietal-05} arrived to the conclusion that accretion tend  to spin-up BHs close to unity, as opposed to mergers which, on the average, do not influence the spin evolution, the following studies by e.g.,
\citem{volonterietal-07,kingetal-08,bertivolonteri-08} showed that the situation is more complex, the final spin values depending on the details of the history of the accretion events (see also \citem{fanidakisetal-11}).

\citem{lietal-05} included the returning radiation into the thin-disk model of \citem{nt} and calculated the spin-up limit for the BH assuming the radiation crossing the equatorial plane inside the marginally stable orbit to be advected onto the BH. Their result ($a_*=0.9983$) slightly differs from Thorne's result, thus showing that returning radiation has only a slight impact on the process of spinning-up BHs. In our study we use advective, optically thick solutions of accretion disks and account for photons captured by the BH in detail. However, we neglect the impact of the radiation returning to the disk on its structure.

\subsection{Tetrad}
\label{s.tetrad}
We base this work on slim accretion disks, which are not razor-thin and have angular momentum profile that is not Keplerian (for details on the assumptions made and the disk appearance see Chapter~\ref{chapter-stationary}). Therefore, photons are not emitted from matter in Keplerian orbits in the equatorial plane and the classical expressions for photon momenta (e.g., \citem{mtw}) cannot be applied. Instead, to properly describe the momentum components of emitted photons, we need a tetrad for the comoving observer instantaneously located at the disk photosphere. Below we give the explicit expression for the components of such a tetrad assuming time and axis symmetries. A detailed derivation is given in Appendix~\ref{ap.tetrad}.

Let us choose the following comoving tetrad,
\be
\label{tetrad}
e^i_{(A)}=[u^i,N_*^i,\kappa_0^i,S^i],
\ee
where
\noindent $ u^i $ is the four-velocity of matter (living in $[ t, \phi, r, \theta]$),

\vskip 0.1truecm

\noindent $ N_*^i$ is a unit vector orthogonal to the photosphere ($[r, \theta]$),

\vskip 0.1truecm

\noindent $ \kappa_0^i$ is a unit vector orthogonal to $u^i$ ($[t, \phi]$),

\vskip 0.1truecm

\noindent $ S^i$ is a unit vector orthogonal to $u^i$, $N^i$ and
$\kappa_0^i$ ($[t, \phi, r, \theta]$).

The tetrad components are given by,
\be N_*^r=\der{\theta_*}r(g_{\theta\theta})^{-1/2}\left[1+\frac{g_{rr}}{g_{\theta\theta}}\left(\der{\theta_*}r\right)^2\right]^{-1/2},\ee
\be N_*^\theta=(g_{\theta\theta})^{-1/2}\left[1+\frac{g_{rr}}{g_{\theta\theta}}\left(\der{\theta_*}r\right)^2\right]^{-1/2},\ee

\vspace{.5cm}

\be u^i=\frac{\eta^i+\Omega\xi^i+vS^i_*}{\sqrt{-g_{tt}-\Omega g_{\phi\phi}(\Omega-2\omega)-v^2}},\ee

\vspace{.5cm}
\be \kappa_0^i=\frac{(l\eta^i+\xi^i)}{\left[g_{\phi\phi}(1-\Omega l)(1-\omega l)\right]^{1/2}},\ee

\vspace{.5cm}
\be S^i=(1+\tilde A^2v^2)^{-1/2}(\tilde A vu^i+S^i_*),\ee
where $\theta=\theta_*(r)$ defines the location of the photosphere, $\eta_i$ and $\xi_i$ are the Killing vectors, $l=u_\phi/u_t$, $\Omega=u^\phi/u^t$, $\omega$ is the frequency of frame-dragging and the expressions for $S^i_*$ and $v$ are given in Eqs.~\ref{e.Sstar} and \ref{e.V}, respectively.

\subsection{Spin evolution}
\label{s.spinevolution}
\subsubsection{Basic equations}
The equations describing the evolution of BH dimensionless spin
parameter $a_*$ with respect to the BH energy $M$ and the accreted
rest-mass $M_0$ are \citepm{thorne-74},
\be
\label{eq.spinevolution1}
\der{a_*}{{\,\rm ln} M}=\der{J/M^2}{\,{\rm ln} M}=\frac 1M\frac{\dot M_0
  u_{\phi}+\left(\der Jt\right)_{\rm rad}}{\dot M_0 u_{t}+\left(\der
    Mt\right)_{\rm rad}}-2a_*,
\ee
\be
\label{eq.spinevolution2}
\der M{M_0}=u_{t}+\left(\der Mt\right)_{\rm rad}/\dot M_0.
\ee
The energy and angular
momentum of BH increases due to the capture of photons according to
the following formulae,
\be
\label{e.dM}
(dM)_{\rm rad}=\int^{}_{\rm disk} T^{ik}\eta_k N_i{\rm d}S,
\ee
\be
\label{e.dJ}
(dJ)_{\rm rad}=\int^{}_{\rm disk} T^{ik}\xi_k N_i{\rm d}S,
\ee
where $\eta_k$ and $\xi_k$ are the Killing vectors connected with time
and axial symmetries, respectively, $T^{ik}$ is the stress-energy
tensor of photons, which is non-zero only for photons crossing the BH
horizon and ${\rm d}S$, the ``volume element'' in the hypersurface
orthogonal to $N^i$, is given by Eq.~\ref{ap.dS}.

From Eqs.~\ref{e.dM} and \ref{e.dJ} it follows that
\be
\left(\der  Mt\right)_{\rm rad}=\int^{2\pi}_0\int_{r_{\rm in}}^{r_{\rm out}}T^{ik}\eta_k
N_i {\rm d}\tilde S,
\label{e.dMdt1}
\ee
\be
\left(\der
  Jt\right)_{\rm rad}=\int^{2\pi}_0\int_{r_{\rm in}}^{r_{\rm out}}T^{ik}\xi_k N_i {\rm d}\tilde S,
\label{e.dJdt1}
\ee
where
\be
 {\rm d\tilde S} = {\rm d}\phi\,{\rm d}r\,\left( g_{t\phi}^2 -
g_{tt}\,g_{\phi\phi}\right )^{1/2}\,\sqrt{ g_{rr} + g_{\theta
\theta}\left(\frac{d\theta_*}{dr}\right)^2}.
\ee
\subsubsection{Stress energy tensor in the comoving frame}

Let us choose the tetrad given in Eq.~\ref{tetrad}:
\begin{eqnarray}
\label{e.comtetrad}
e^i_{(0)}= u^i&&e^i_{(1)}=N^i\\\nonumber
e^i_{(2)}=\kappa^i&&e^i_{(3)}= S^i
\end{eqnarray}
The disk properties,
e.g., the emitted flux, are usually given in the comoving frame defined
by Eq.~\ref{e.comtetrad}. The stress tensor components in the two
frames (Boyer-Lindquist and comoving) are related
in the following way,

\be
T^{ik}=T^{(\alpha)(\beta)}e^{i}_{(\alpha)}e^{k}_{(\beta)}.
\ee
The stress tensor in the comoving frame is
\be
T^{(\alpha)(\beta)}=2\int_0^{\pi/2}\int_0^{2\pi}I_0SC\pi^{(\alpha)}\pi^{(\beta)}\sin\tilde
a\,d\tilde a\,d\tilde b,
\ee
where $I_0S=I_0(r)S(\tilde a,\tilde b)$ is the intensity of the emitted
radiation, $\tilde a$ and $\tilde b$ are the angles between the emission vector and the
$N^i$ and $S^i$ vectors, respectively, $C$ is the capture function
defined in Section~\ref{s.capture}, the factor $2$ comes from the fact that the disk emission comes from both sides of the disk and $\pi^{(\alpha)}=p^{(\alpha)}/p^{(0)}$ are the
  normalized components of the photon four-momentum in the comoving
  frame. The latter are given by the following simple relations \citepm{thorne-74},
\begin{eqnarray}\nonumber
\pi^{(0)}&=&1,\\
\pi^{(1)}&=&\cos \tilde a,\\\nonumber
\pi^{(2)}&=&\sin{\tilde a}\cos \tilde b,\\\nonumber
\pi^{(3)}&=&\sin{\tilde a}\sin \tilde b.
\end{eqnarray}
Eqs.~\ref{e.dMdt1} and \ref{e.dJdt1} take the form,
\be
\left(\der Mt\right)_{\rm rad}=\int^{}_{\rm disk}
T^{(\alpha)(\beta)}e^{i}_{(\alpha)}e^{k}_{(\beta)}\eta_k N_i{\rm
  d}\tilde S,
\label{e.dMdt2}
\ee
\be
\left(\der Jt\right)_{\rm rad}=\int^{}_{\rm disk}
T^{(\alpha)(\beta)}e^{i}_{(\alpha)}e^{k}_{(\beta)}\xi_k N_i{\rm
  d}\tilde S,
\label{e.dJdt2}
\ee
where $e^i_{(a)}$ is our local frame tetrad given by
Eq.~(\ref{e.comtetrad}). Taking the following relations into account,
\bea
\pi^{(\alpha)}&=&\pi^je^{(\alpha)}_j,\\\nonumber
e^{(\alpha)}_je_{(\alpha)}^i&=&\delta_j^i,
\eea
we have,
\vspace{.3cm}
\be \pi^{(\alpha)}e^{i}_{(\alpha)}N_i=\pi^{(\alpha)}\delta^{(1)}_{(\alpha)}=\pi^{(1)}=\cos \tilde a,\ee
\be \pi^{(\beta)}e^{k}_{(\beta)}\eta_k=\pi^je^{(\beta)}_ju^k_{(\beta)}\eta_k=\pi^j\delta_j^k\eta_k=\pi^k\eta_k=\pi_t,\ee
\be \pi^{(\beta)}e^{k}_{(\beta)}\xi_k=\pi^je^{(\beta)}_ju^k_{(\beta)}\xi_k=\pi^j\delta_j^k\xi_k=\pi^k\xi_k=\pi_\phi,\ee
where,
\bea
\pi_t&=&\pi^{(i)}e^t_{(i)} g_{tt}+\pi^{(i)}e^\phi_{(i)}
g_{t\phi},\\
\pi_\phi&=&\pi^{(i)}e^t_{(i)} g_{t\phi}+\pi^{(i)}e^\phi_{(i)} g_{\phi\phi}.
\eea
Therefore, Eqs.~\ref{e.dMdt2} and \ref{e.dJdt2} may be finally expressed as,
\begin{eqnarray}\nonumber
\left(\der
  Mt\right)_{\rm rad}&=&4\pi\int_{r_{\rm in}}^{r_{\rm out}}\int_0^{2\pi}\int_0^{\pi/2}I_0SC\pi_t\times
\\
&\times&\cos\tilde
a\sin\tilde a\,d\tilde a\,d\tilde b \sqrt{\tilde g}\,{\rm d}r,
\end{eqnarray}
\begin{eqnarray}\nonumber
\left(\der
Jt\right)_{\rm rad}&=&4\pi\int_{r_{\rm in}}^{r_{\rm out}}\int_0^{2\pi}\int_0^{\pi/2}I_0SC\pi_\phi\times
\\
&\times&\cos\tilde
a\sin\tilde a\,d\tilde a\,d\tilde b \sqrt{\tilde g}\,{\rm d}r,
\end{eqnarray}
with
\be \sqrt{\tilde g}\equiv\left( g_{t\phi}^2 -
g_{tt}\,g_{\phi\phi}\right )^{1/2}\,\sqrt{ g_{rr} + g_{\theta
\theta}\left(\frac{d\theta_*}{dr}\right)^2}.\ee

\subsubsection{Emission}
The intensity of local radiation may be identified with the flux
emerging from the disk surface $F(r)$ (presented and discussed for the case of slim disks in Section~\ref{s.diskappearance})
\be I_0=F(r).\ee
The angular emission factor $S$ is given by \citepm{thorne-74},
\begin{equation}
S(\tilde a,\tilde b)=\left\{\begin{array}{ll}
1/\pi & {\rm isotropic}\\
(3/7\pi)(1+2\cos\tilde a)& {\rm limb\,darkening}\\
\end{array}\right.
\end{equation}
for isotropic and limb-darkened cases, respectively. In this work we assume that the
radiation is emitted isotropically.
\subsubsection{Capture function}
\label{s.capture}
The BH energy and angular momentum are affected only by photons
crossing the BH horizon. Following \citem{thorne-74}, we define
the capture function $C$,
\begin{equation}
C=\left\{\begin{array}{ll}
1& {\rm if\ the\ photon\ hits\ the\ BH,}\\
0& {\rm in\ the\ opposite\ case.}\\
\end{array}\right.
\end{equation}
Herein, we calculate $C$ in two ways. First, we use the original
\citem{thorne-74} algorithm modified to account for emission out
of the equatorial plane. For this purpose we calculate
the constants of motion, $j$ and $k$, for a geodesic orbit of a photon in
the following way,
\be j=a_*^2+a_*(\pi_\phi/M\pi_t),\ee
\be k=\frac1{(M\pi_t)^2}\left[\pi_\theta^2-(\pi_\phi+a_*M\pi_t \sin\theta_*)^2/\sin^2\theta_*\right]\ee
which replaces Thorne's Eqs.~A10. This approach does not take into
account possible returning radiation, i.e., a photon hitting the disk
surface is assumed to continue its motion. Such a treatment is not appropriate
for optically thick disks - returning photons are most likely
absorbed or advected towards the BH.

To assess the importance of this inconsistency we adopt two additional
algorithms for calculating $C$. Using photon equations of motion we determine if the photon hits the disk surface \citepm{bursa.raytracing}. Then we make two assumptions, either the angular momentum and energy of all returning photons are advected onto the BH ($C_1$), or all are reemitted carrying away their original angular momentum and energy, and never hit the BH ($C_2$). In this way we establish two limiting cases allowing us to assess the impact of the returning radiation.

We note here that for fully consistent treatment of the returning
radiation (as in \citem{lietal-05} for geometrically thin disks) it is not enough to modify the capture function ---
solving for the whole structure of a self-irradiated accretion disk (like in Chapter~\ref{chapter.selfirradiated})
is necessary. The impact of self-irradiation onto the BH spin evolution will be studied in detail in the future.

\subsection{Results for BH spin evolution}
\label{s.results}
Using the slim disk solutions described in Chapter~\ref{chapter-stationary} we solve Eqs.~\ref{eq.spinevolution1} and \ref{eq.spinevolution2} using regular Runge-Kutta method of the 4th order. To calculate the integrals (Eqs.~\ref{e.dMdt1} and \ref{e.dJdt1}) we use the alternative extended Simpson's rule \citepm{numericalrecipes} basing on 100 grid points in $\tilde a$, $\tilde b$ and radius $r$. We have made convergence tests proving this number is sufficient.

\begin{figure}
\centering
  \includegraphics[width=.75\textwidth,angle=0]{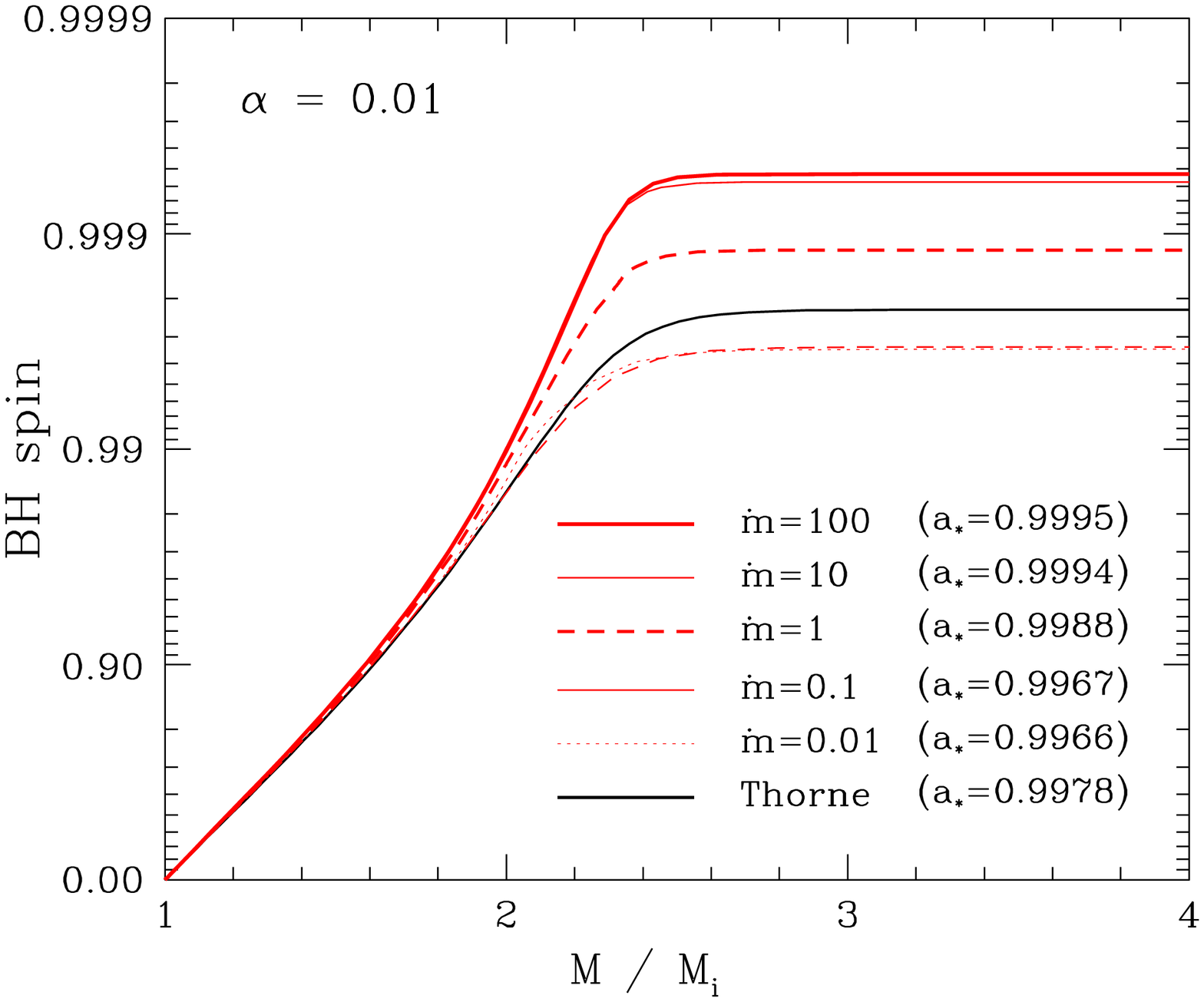}

  \caption{Spin evolution for $\alpha=0.01$.}
  \label{f.sev0.01}
\end{figure}

\begin{figure}
\centering
  \includegraphics[width=.75\textwidth,angle=0]{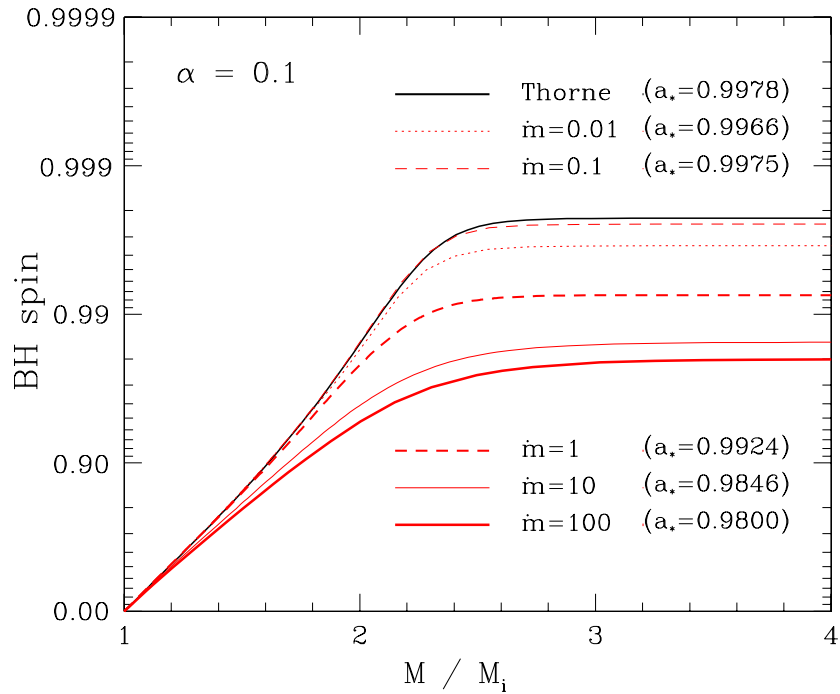}
  \caption{Spin evolution for $\alpha=0.1$.}
  \label{f.sev0.1}
\end{figure}

In Figs.~\ref{f.sev0.01} and \ref{f.sev0.1} we present the BH spin evolution for $\alpha=0.01$ and $0.1$, respectively. The red lines show the results for different accretion rates while the black line shows the classical \citem{thorne-74} solution based on the \citem{nt} model of thin accretion disk. The accretion rates are given in the units of the critical accretion rate defined in Eq.~\ref{e.mdotcritical}. Our low accretion rate limit does not perfectly agree with the black line as the slim disk model does not account for the angular momentum carried away by radiation. As a result, the low-luminosity slim disk solutions slightly overestimate (by no more than few percent) the emitted flux leading to stronger deceleration of the BH by radiation --- the \citem{thorne-74} result is the proper limit for the lowest accretion rates. When the accretion rate is high enough (e.g., $\dot m>0.1$), the impact of the omitted angular momentum flux is overwhelmed by the modification of the disk structure introduced by advection.

It is clear the spin evolution is quantitatively different for different values of the viscosity parameter. For the lower value ($\alpha = 0.01$) the BH spin can reach values significantly higher than $0.998$ for the highest accretion rates ($a_*=0.9995$ for $\dotm=100$), while for the higher viscosity ($\alpha = 0.1$) the terminal spin value decreases with increasing accretion rate down to $a_*=0.9800$ for $\dotm=100$. This behavior is connected with the impact of viscosity on the angular momentum profile. Generally speaking, the higher the value of $\alpha$, the lower the angular momentum of the flow at BH horizon for a given accretion rate (see Fig.~\ref{fig:horizon-angular-momentum} in Section~\ref{s.diskappearance}), leading to a slower acceleration of the BH rotation.

\begin{figure}
\centering
  \includegraphics[width=.75\textwidth,angle=0]{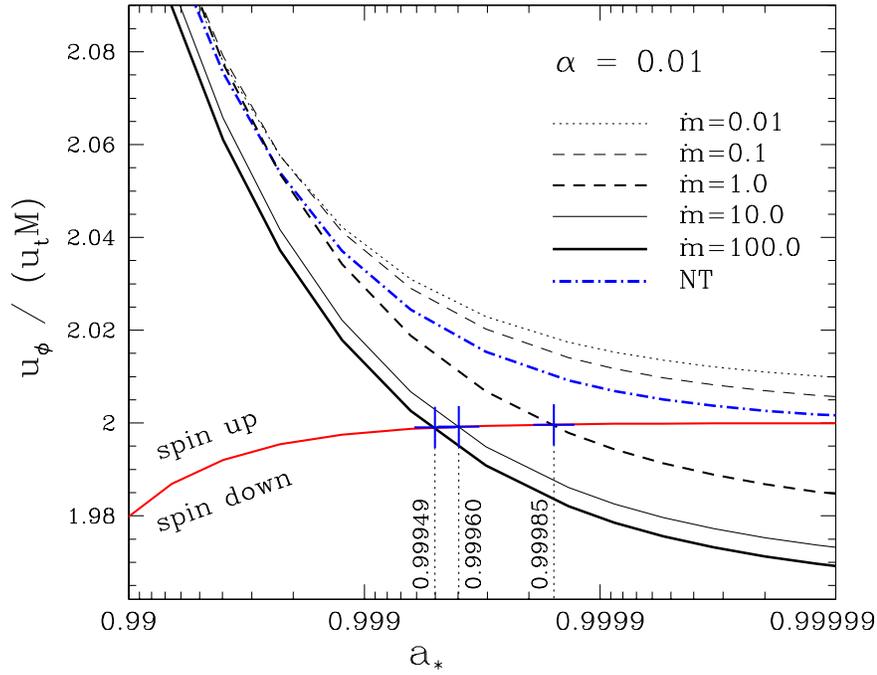}
  \caption{The rate of spin-up or spin-down by ''pure'' accretion
    (radiation neglected) for $\alpha=0.01$. Profiles for five
    accretion rates are presented. Their intersections
 with the red line (marked with blue
    crosses) correspond to equilibrium states. For the two lowest
    accretion rates the equilibrium state is never reached
    ($a_*\rightarrow 1$).}
  \label{f.eqalp0.01}
\end{figure}

\begin{figure}
\centering
  \includegraphics[width=.75\textwidth,angle=0]{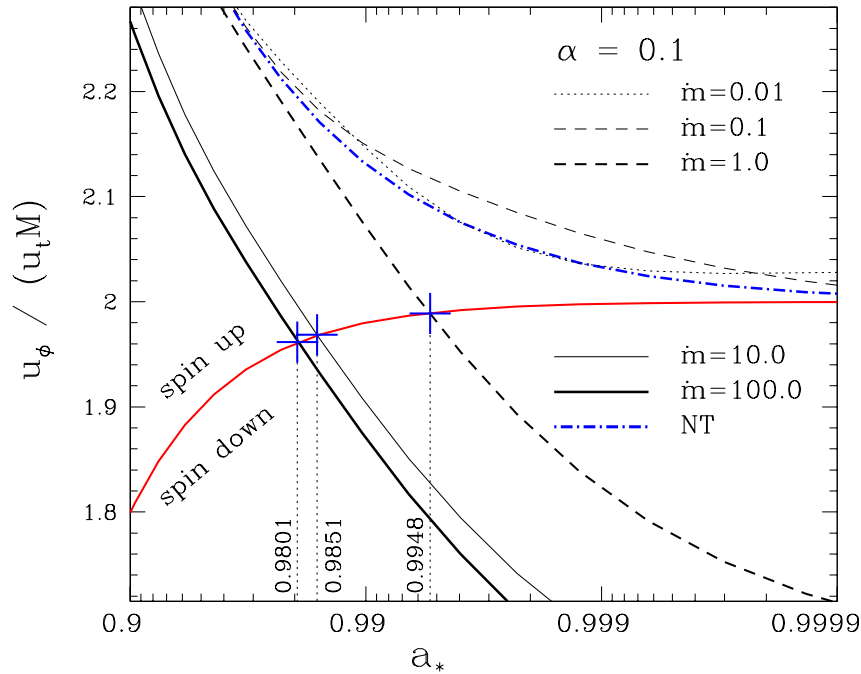}
  \caption{Same as Fig.~\ref{f.eqalp0.01} but for $\alpha=0.1$.}
  \label{f.eqalp0.1}
\end{figure}

\begin{figure}
\centering
  \includegraphics[width=.75\textwidth,angle=0]{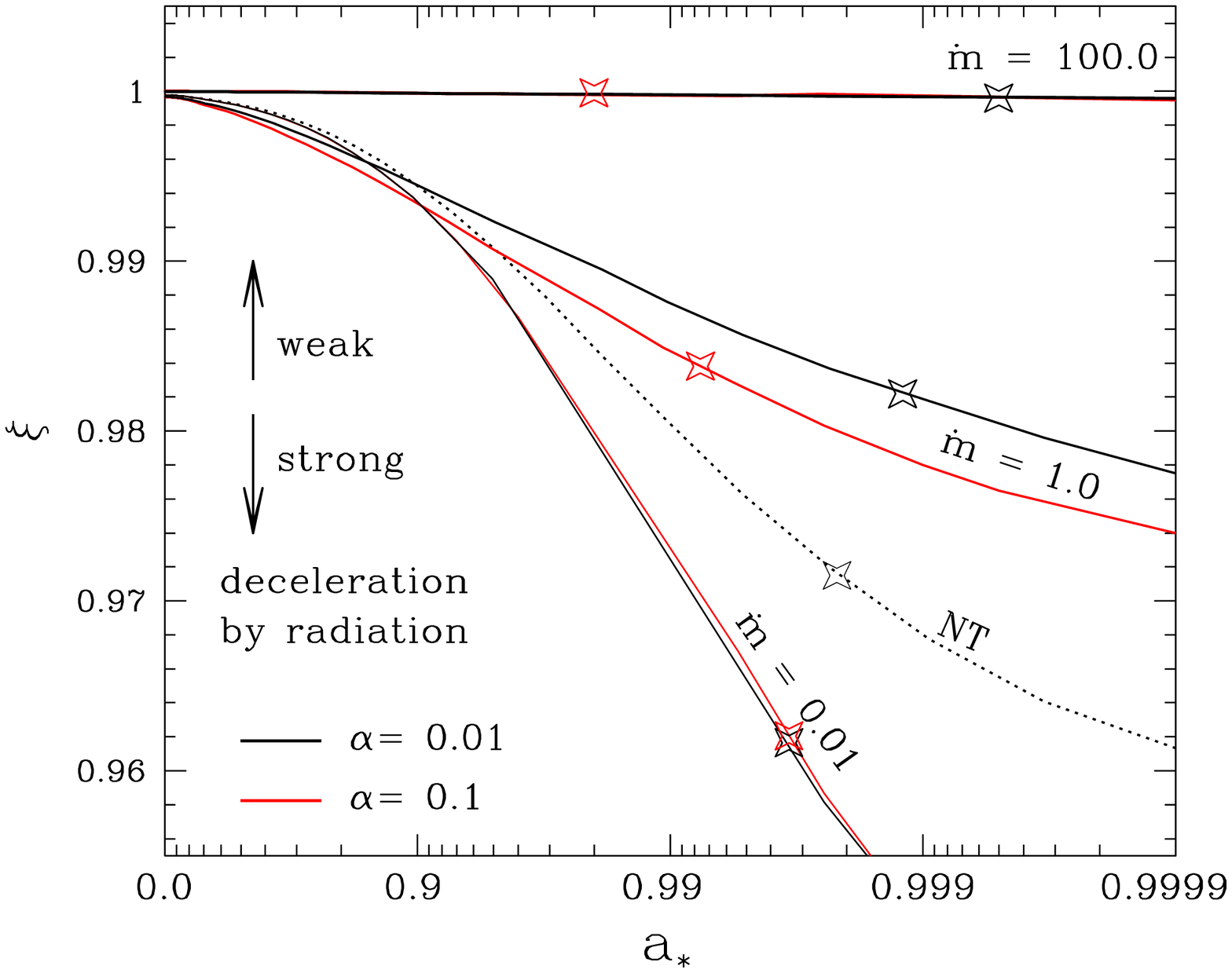}
  \caption{Radiation impact factor $\xi$ (Eq.~\ref{e.xi}) for different accretion rates and values of $\alpha$. The dotted line corresponds to the thin-disk induced spin evolution. For $\xi\approx1$ spin evolution is not affected by radiation. Stars denote the equilibrium states (compare Table~\ref{t.astars}).}
  \label{f.radimp}
\end{figure}

To study this fact in detail we calculated the rate of BH spin-up for ''pure'' accretion of matter (without accounting for the impact of radiation). For that case the BH spin evolution is given by (compare Eq.~\ref{eq.spinevolution1}):
\begin{equation}
\label{eq.spinevpure}
\der{a_*}{{\rm ln} M}=\frac 1M\frac{
  u_{\phi}}{ u_{t}}-2a_*.
\end{equation}
In Figs.~\ref{f.eqalp0.01} and \ref{f.eqalp0.1} we plot with black lines the first term on the right hand side of the above equation for different accretion rates and values of $\alpha$. The red lines on these plots show the absolute value of the second term. The intersections of the black and red lines denote the equilibrium states i.e., the limiting values of BH spin for pure accretion. These values differ significantly from the previously discussed results only for low accretion rates. In contrast, for high accretion rates radiation has little impact on the spin evolution and the value of terminal spin is mostly determined by the properties of the flow. In Fig.~\ref{f.radimp} we plot the radiation impact parameter $\xi$, defined as the ratio of the disk-driven terms on the right hand sides of Eqs.~\ref{eq.spinevolution1} and \ref{eq.spinevpure},
\be
\label{e.xi}
  \xi = \left.\frac{\dot M_0
  u_{\phi}+\left(\der Jt\right)_{\rm rad}}{\dot M_0 u_{t}+\left(\der
    Mt\right)_{\rm rad}}\right/ \frac{
  u_{\phi}}{ u_{t}}
\ee
 If the captured radiation significantly decelerates BH spin-up, this ratio drops below unity. On the other hand, it is close to unity for BH spin evolution which is not affected by the radiation. According to Fig.~\ref{f.radimp} the latter is in fact the case for the highest accretion rates independently of $\alpha$.

In Table~\ref{t.astars} we list the resulting values of the terminal BH spin for all the models considered. The first column gives results for our fiducial model ($A$) including Thorne's capture function, emission from the photosphere at the appropriate radial velocity.

The second column presents results obtained assuming the same (Thorne's) capture function and profiles of emission, angular momentum and radial velocity as in the model A, but assuming the emission takes place from the equatorial plane instead of the photosphere. The resulting terminal spin values are equal, up to 4 decimal digits, to the values obtained with the fiducial model. This result is expected for the lowest accretion rates, where the photosphere is located very close to the equatorial plane. For the highest accretion rates the location of the emission has no impact on the BH spin-up, as the spin evolution is driven by the flow itself and the effects of radiation are negligible. However, for moderate accretion rates one could have expected significant change of the terminal spin. We find that the location of the photosphere has little impact on the resulting BH spin regardless of the accretion rate.

Our third model ($V$) neglects the flow radial velocity when the radiative terms are evaluated. Similar arguments to those given in the previous paragraph apply.  For the lowest accretion rates, the radial velocity is negligible and therefore it should have no impact on the resulting spin. For the highest accretion rates, the spin-up process depends only on the properties of the flow. Once again, however, the impact of this assumption on moderate accretion rates is not obvious. The radial velocity turns out to be of little importance for the calculation of the terminal spin (only for $\alpha=0.1$ and moderate accretion rates the difference between models $A$ and $V$ is higher than $0.01\%$).

In the fourth and fifth columns of Table~\ref{t.astars}, results for models with the same assumptions as the fiducial model, but with different capture functions are presented. The first alternative capture function ($C_1$) assumes that the angular momentum and energy of all photons returning to the disk are added to that of the BH. This assumption has a strong impact on the spin evolution --- the terminal spin values are higher, sometimes approaching $a_*=1$. This may seem surprising because in the classical approach the captured photons are responsible for decelerating the spin-up. This effect comes from the fact that the cross-section (with respect to the BH) of photons going ``against`` the frame dragging is larger than of photons following the BH sense of rotation. As frame dragging is involved, this effect is significant only in the vicinity of the BH horizon. For our model $C_1$, however, the probability of photons returning to the disk does not appreciably differ for co- and counter-rotating photons, as they both hit the disk surface mostly at large radii.


The other capture function ($C_2$) assumes, on the contrary, that all returning photons are re-emitted from the disk with their original angular momentum and energy (and never fall onto the BH). This assumption cuts off the photons which would hit the BH in the fiducial model after crossing the disk surface. Thus, we may expect decreased radiative deceleration and increased values of the terminal spin parameter. However, these changes are not significant, reflecting the fact that most of the photons hit the BH directly, along slightly curved trajectories. Only for $\alpha=0.1$ and moderate accretion rates do the terminal spin values differ in the $\rm 4^{\rm th}$ decimal digit.

Neither of the models with a modified capture functions is self-consistent. To account properly for the returning radiation one has to modify the disk equations by introducing appropriate terms for the outgoing and incoming fluxes of angular momentum and additional radiative heating. No such model for advective, optically thick accretion disk has been constructed. The emission profile should be significantly affected (especially inside the marginally stable orbit) by the returning radiation leading to different rates of deceleration by photons. In view of our results for models $C_1$ and $C_2$, as well as the results of \citem{lietal-05}, one may expect the final spin values for super-critical accretion flows to be slightly higher than the ones obtained in this work.

The last column of Table~\ref{t.astars} gives terminal spin values for ''pure'' accretion (radiation neglected). Under such assumptions the BH spin could reach $a_*=1$ for sub-Eddington accretion rates as there are no photons which could decelerate and stop the spin-up process. As discussed above, for the highest accretion rates the resulting BH spin values agree with the values obtained for the fiducial model as radiation has little impact on spin evolution in this regime.

\begin{table}
\centering
\begin{tabular}{lccccccc}
\hline \hline
\multicolumn{2}{l}{capture function:} & $C$ & $C$ & $C$ & $C_1$ & $C_2$ & - \\
\hline
\multicolumn{2}{l}{model:} & $\bf{A}$ & $T$ & $V$ & $A$ & $A$ & \textit{NR} \\
\hline \hline
\multicolumn{2}{l}{thin disk}   & \textbf{0.9978} & 0.9978 & 0.9978 & 0.9981 & 0.9978 & $\rightarrow 1$\\
\hline
\multirow{5}{*}{$\alpha=0.01$} &
  $\dot m=0.01$  & \textbf{0.9966} & 0.9966 & 0.9966 & 0.9975 & 0.9966 & $\rightarrow 1$\\
& $\dot m=0.1$   & \textbf{0.9967} & 0.9967 & 0.9967 & $\rightarrow 1$ & 0.9967 & $\rightarrow 1$\\
& $\dot m=1$     & \textbf{0.9988} & 0.9988 & 0.9988 & $\rightarrow 1$ & 0.9988 & 0.9998 \\
& $\dot m=10$    & \textbf{0.9994} & 0.9994 & 0.9994 & $\rightarrow 1$ & 0.9994 & 0.9996 \\
& $\dot m=100$   & \textbf{0.9995} & 0.9995 & 0.9995 & $\rightarrow 1$ & 0.9995 & 0.9995 \\
\hline
\multirow{5}{*}{$\alpha=0.1$} &
  $\dot m=0.01$  & \textbf{0.9966} & 0.9966 & 0.9966 & 0.9975 & 0.9966 & $\rightarrow 1$\\
& $\dot m=0.1$   & \textbf{0.9975} & 0.9975 & 0.9975 & $\rightarrow 1$ & 0.9975 & $\rightarrow 1$\\
& $\dot m=1$     & \textbf{0.9924} & 0.9924 & 0.9923 & $\rightarrow 1$ & 0.9927 & 0.9948 \\
& $\dot m=10$    & \textbf{0.9846} & 0.9846 & 0.9845 & 0.9901 & 0.9847 & 0.9951 \\
& $\dot m=100$   & \textbf{0.9800} & 0.9800 & 0.9800 & 0.9803 & 0.9800 & 0.9801 \\
\hline \hline
\end{tabular}
\caption{BH spin terminal values. $\, C$ - Thorne's capture function, $C_1$ - all
returning photons advected onto the BH, $C_2$ - all returning photons neglected;
$A$ - our fiducial model, $T$ - emission from the equatorial plane, $V$ - zero radial
velocity, \textit{NR} - pure accretion, radiation neglected.}
\label{t.astars}
\end{table}

\subsection{Summary}
\label{s.summary}
We have studied BH spin evolution due to disk accretion assuming that the angular momentum and energy carried by the flow and the emitted photons is the only process affecting the BH rotation. We generalized the original study of \citem{thorne-74} to high accretion rates by applying a relativistic, advective, optically thick slim accretion disk model. Assuming isotropic emission (no limb darkening) we have shown that

(i) the terminal value of BH spin depends on the accretion rate for $\dot m\gtrsim 1$,

(ii) the terminal spin value is very sensitive to the assumed value of the viscosity parameter $\alpha$ --- for $\alpha\lesssim 0.01$ the BH is spun up to $a_*>0.9978$ for high accretion rates, while for $\alpha\gtrsim 0.1$ to $a_*<0.9978$,

(iii) with a low value of $\alpha$ and high accretion rates, the BH may be spun up to spins significantly higher than the canonical value $a_*=0.9978$ (e.g., to $a_*=0.9994$ for $\alpha=0.01$ and $\dotm =10$) but, under reasonable assumptions, BH cannot be spun up arbitrarily close to $a_*=1$,

(iv) BH spin evolution is hardly affected by the emitted radiation for high ($\dotm\gtrsim 10$) accretion rates (the terminal spin value is determined by the flow properties only),

(v) for all accretion rates, neither the photosphere profile nor the profile of radial velocity significantly affects the spin evolution.

We point out that the inner edge of an accretion disk cannot be uniquely defined for a super-critical accretion (Section~\ref{s.inneredges}), as opposed to geometrically thin disks where the inner edge is uniquely located at the marginally stable orbit ($r_{\rm ISCO}$). In the thin disk case the BH spin evolution is determined by the flow properties at this particular radius (as there is no torque between the marginally stable orbit and BH horizon) and the profile of emission (terminating at $r_{\rm ISCO}$). For super-critical accretion rates, however, one cannot distinguish a particular inner edge which is relevant to studying BH spin evolution. On the one hand, the values of the specific energy ($u_t$) and the angular momentum ($u_\phi$) remain constant within the \textit{stress inner edge}. On the other, the radiation is emitted outside the \textit{radiation inner edge}. These inner edges do not coincide as they are related to different physical processes.

Our study was based on a semi-analytical, hydrodynamical model of an accretion disk which makes a number of simplifying assumptions like stationarity, no returning radiation, $\alpha p$ prescription and no wind outflows. Thus, the results obtained in this work should be considered only qualitative, as the precise values of the terminal spin parameter are very sensitive (e.g., through viscosity) to the flow and emission properties. Moreover, we neglected the impact of magnetic fields and jet ejection mechanisms. Nevertheless, our study shows that Thorne's canonical value for BH spin ($a_*=0.9978$) may be exceeded under certain conditions.

\section{Spin determination via X-ray continuum fitting}
\label{s.spindetermination}

Astrophysical BHs may be uniquely characterized by two parameters: mass and angular momentum. From among those two, the latter is much more difficult to measure, as it has no impact onto the orbital motion of the compact binary. Currently, three methods of estimating BH spin are in use: based on the shape of emission lines (e.g., \citem{rey-2008}), on the continuum fitting (e.g., \citem{steineretal-10}), and on studying the resonances leading to quasi-periodic oscillations (QPOs, \citem{toroketal-07}). First two give consistent results for those few objects that can be studied using both approaches.  

\begin{figure*}[ht]
  \centering
  \begin{minipage}{0.65\textwidth}
  \vspace{0pt}
  \includegraphics[width=\linewidth]{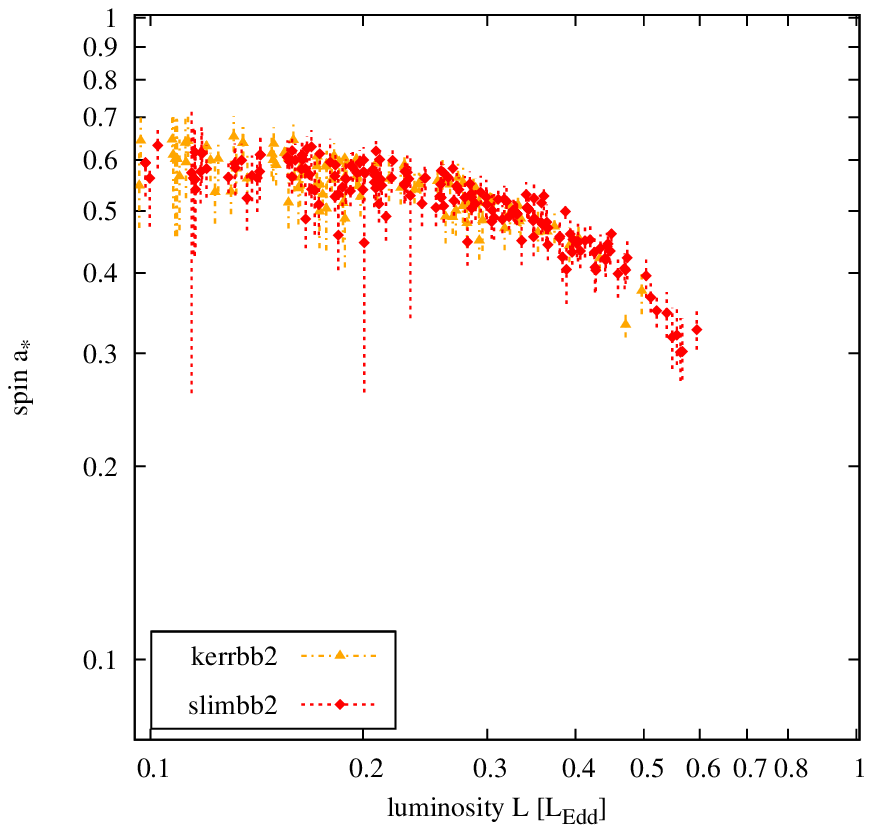}
  \end{minipage}
  \hfill
  \begin{minipage}{0.65\textwidth}
  \vspace{0pt}\raggedright
  \includegraphics[width=\linewidth]{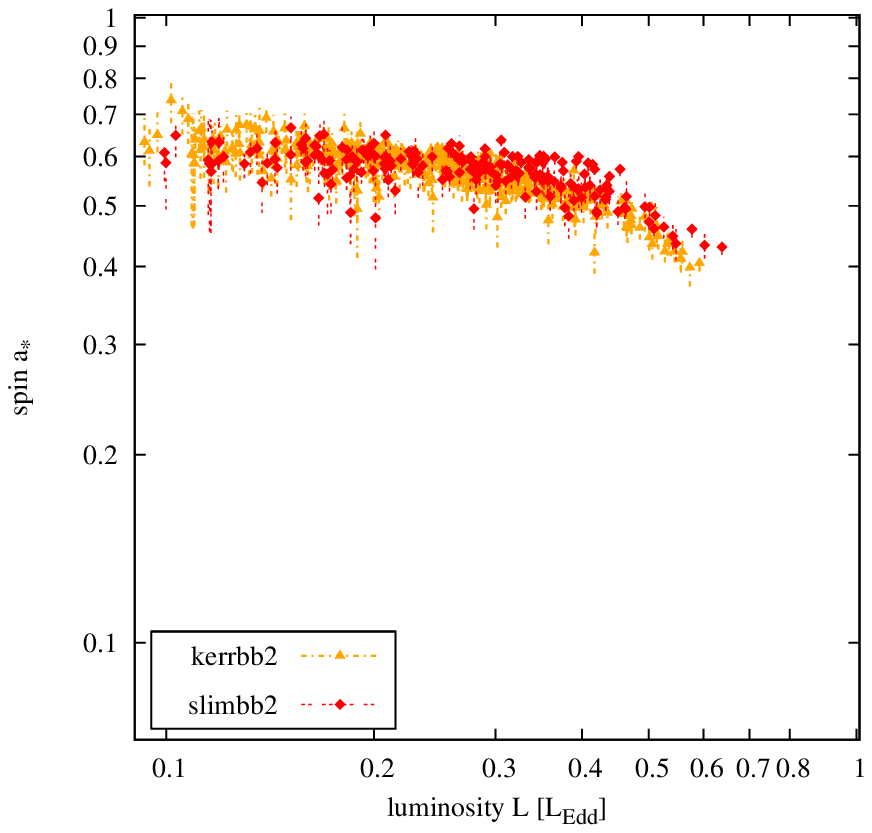}
  \end{minipage}
  \caption{Results of BH spin estimation for LMC X-3 using the continuum fitting method. Disk models {\tt slimbb} (red diamonds) and {\tt kerrbb2} (orange triangles) for viscosity parameters $\alpha = 0.1$ (top) and $\alpha = 0.01$ (bottom) are compared. In all cases we assume LMC X-3 contains a $10\Msun$ black hole, is located at a distance of $48.1$kpc and seen at an inclination of $66^{o}$ (figure after: \citem{straubetal-11}).}
 \label{f.odelespin}
\end{figure*}

The continuum-fitting method is usually based on the \citem{nt} thin disk model and is used for X-ray binaries which exhibit clear high/soft spectra dominated by the disk thermal component (Section~\ref{s.highsoft}). It has been successfully applied for a number of objects. However, its range of applicability seems to be limited to low-luminosity spectra only. As has been shown (e.g., \citem{steineretal-10}), the spin estimates for high luminosity 
spectra are not consistent. The characteristic tail of dropping spin estimates in this regime (Fig.\,\ref{f.odelespin}) is common for all objects. 

This inconsistency emerges for luminosities which are close to or outside the limit of applicability of radiatively efficient disks. We have shown that advective significantly modifies the disk structure for $L\gtrsim 0.6L_{\rm Edd}$ (Section~\ref{s.inneredges}). Therefore, thin disk models should not be applied in this regime, as they cannot produce consistent disk flux profiles which are used for continuum fitting purposes.

In this section we briefly discuss the application of the spectral model \texttt{slimbb} (Section \ref{s.slimbb}) to high/soft state observations of LMC X-3. We report results described in \citem{straubetal-11}.

The data which has been analyzed, consisting of 712 PCU-2 X-ray spectra, was obtained using the large-area Proportional Counter Array (PCA) aboard the {\it Rossi X-ray Timing Explorer} (RXTE). The most important selection criteria for those set of spectra were: (i) disk component exceeding $75\%$ of the total emission, (ii) small variability, (iii) no high-frequency QPOs\footnote{QPOs are observed only in the low/hard state or during state transitions, so the data containing QPOs is most likely not disk-dominated.}. The X-ray spectral fitting software, XSPEC \citepm{arnaud-96} was used
with the following model combination {\tt TBabs * simpl * slimbb}, where  {\tt TBabs\/} stands for the Galactic neutral hydrogen absorption with $n_H=4 \times 10^{20}\,$ cm$^{-2}$ \citepm{pag+03} and {\tt simpl\/} for a power-law component \citepm{ste+09}. The mass $M=10\Msun$, the average distance to the LMC, $D=48.1\,$kpc \citepm{oro+09} and the inclination, $i = 66^\circ$ \citepm{kui+88} were assumed.
The proper shape of the Compton-hardened spectrum was taken  from the BHSPEC atmospheric models \citepm{davisomer05}.

In Fig.\,\ref{f.odelespin} we present a comparison of the fittings obtained with {\tt slimbb} with analogous study conducted with {\texttt kerrbb2} --- the BHSPEC enhanced version of {\tt kerrbb} (with switched off returning radiation, to match the slim disk assumptions). The top panel presents results for $\alpha=0.1$. Both approaches give the same results producing the characteristic, artificial spin-luminosity dependence --- the results are in fact indistinguishable. Only for low luminosities ($L\lesssim0.2L_{\rm Edd}$) one gets roughly constant value of BH spin $a_*\approx 0.6$. For higher luminosities the spin estimates drop down --- as far as to $a_*=0.3$ for $0.5L_{\rm Edd}$. 

The bottom panel presents the corresponding comparison of {\tt slimbb} and {\tt kerrbb2} models for lower value of the viscosity parameter $\alpha=0.01$. The fittings behave qualitatively in the same way. However, the spin-luminosity relation is much flatter --- the spin estimates drop down only to $a_*=0.4$ for $0.5L_{\rm Edd}$. Similarly as in the $\alpha=0.1$ case, the results provided by these models are indistinguishable.

%

In comparison to the standard (based on thin disk models) \texttt{kerrbb2}, the model based on slim disk solutions (\texttt{slimbb}) accounts additionally for the advective cooling. In this way it is able to consistently describe accretion flows with moderate and high luminosities. As the comparison of BH spin estimates for LMC X-3 shows, this single improvement is not able to solve the high-luminosity inconsistency. Despite the fact that advection modifies the emission (however, as discussed in Section \ref{s.stationarysolutions}, its impact is hardly visible for sub-Eddington luminosities), the spectra produced by combined disk and atmospheric models are too hard, when compared to the observed data (the harder the spectrum, the lower the spin estimate). One may conclude that the reason for this discrepancy lies in a different process than advection.

There are at least few shortcomings of the hydrodynamical models in use, which may be responsible for this fact. To start with, these models neglect mass outflows assuming constant (in radius) accretion rate. However, it has been recently shown that radiation driven outflows, while hardly modify the effective temperature, lead to significant decrease of the local surface density which, in turn, would lead to further spectrum hardening \citepm{takeuchietal-09}. Another possible source of this discrepancy may be connected to the treatment of comptonization in the disk atmosphere. Currently, the disk atmosphere model BHSPEC, based on radiative transfer code {\tt TLUSTY}, is the only available. Its approach to comptonization is sophisticated, but still involves some approximations. It would be beneficial, if spectra produced by BHSPEC may be compared to another similar, stand-alone radiative transfer code. It is, most probably, the question of time. 

\begin{figure*}[hb]
  \centering\includegraphics[width=.8\linewidth]{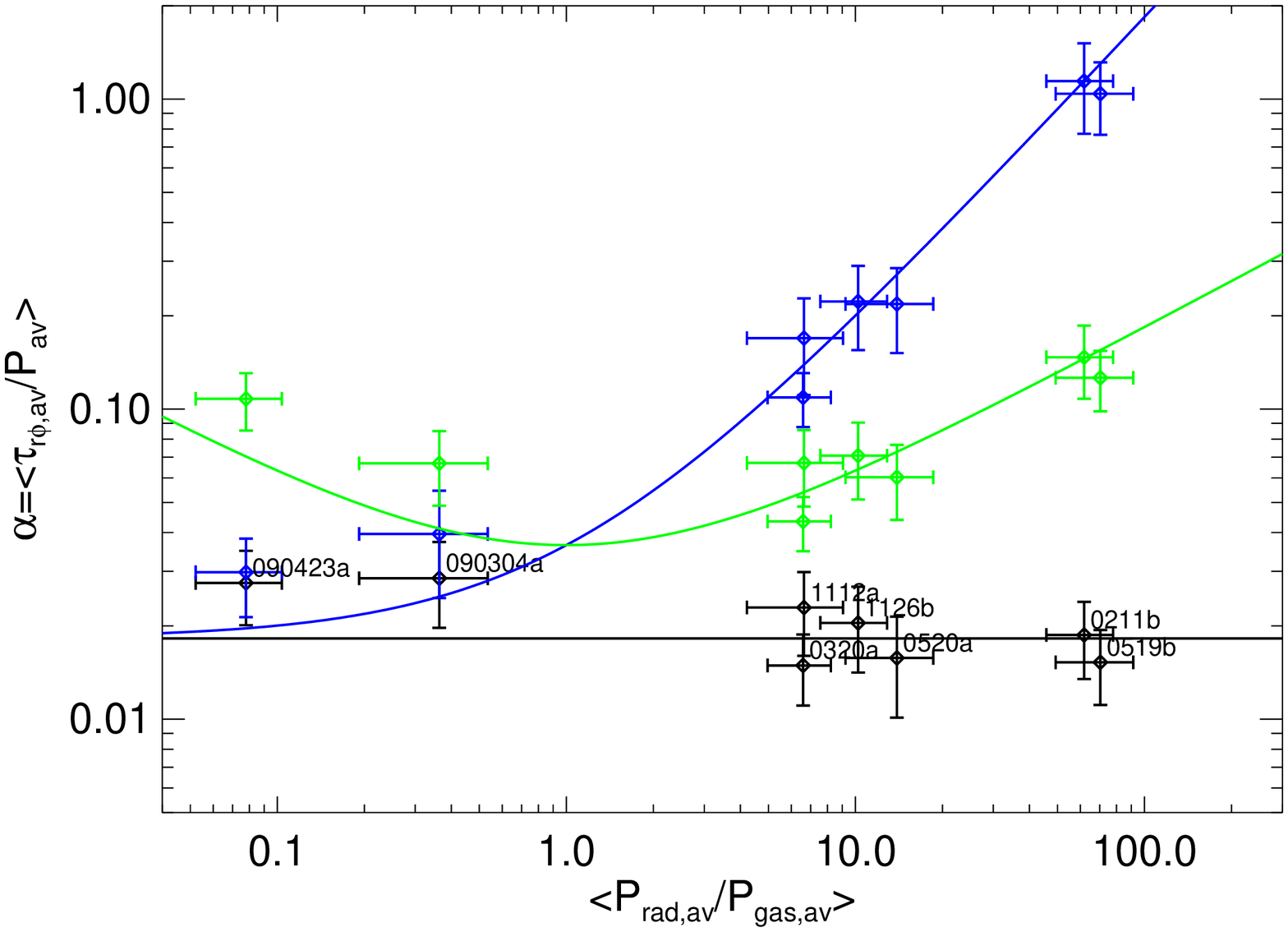}
 \caption{Measured values of the stress parameter $\alpha$ as a function of
the time-averaged ratio of the box-averaged radiation pressure to the
box-averaged gas pressure.  The black points define $\alpha$ as the
time-averaged ratio of the vertically averaged stress to the box-averaged
total thermal pressure.  The blue and green points define $\alpha$
in the same way except with total thermal pressure replaced by gas pressure
and the geometric mean of gas and radiation pressure, respectively.  Both
horizontal and vertical error bars indicate one standard deviation in the
time averages.  The horizontal black line indicates the weighted mean of
$\alpha$ for the total pressure stress prescription.  Assuming that stress
really scales with total thermal pressure, the green and blue curves show
how $\alpha$ would then behave under the other stress prescriptions.
structure parameters. Figure and caption after: \citem{hirose-09-b}}
 \label{f.hirose.alphas}
\end{figure*}

The viscosity in accretion disks most likely results from magnetorotational instability (Section~\ref{s.mri}). Magnetic fields are not explicitly taken into account in hydrodynamical models (e.g., thin or slim disks). One has to include the viscosity in an artificial way. It is usually done by assuming that the stress is proportional to pressure (Section~\ref{standard-alpha-prescription}). Recent MHD shearing-box studies tend to support this assumption for mean accretion flows. These local simulations, however, say little about the dependence of the $\alpha$ parameter on radius or luminosity. Only recently, shearing boxes with different values of radiation to gas pressure ratio have been studied \citepm{hirose-09-b} proving that the value of $\alpha$, defined as the ratio of stress to local pressure, may slightly depend on the luminosity (Fig.~\ref{f.hirose.alphas}).

One may infer about the $\alpha$ dependence on luminosity not only from numerical simulations of MRI, but also from observations of accretion disks at high luminosities. As Fig.~\ref{f.odelespin} shows, the spin estimates are $\alpha$-degenerate at low luminosities ($L\lesssim 0.2L_{\rm Edd}$). At higher accretion rates, however, they strongly depend on $\alpha$ --- the higher the value of $\alpha$, the lower the spin value. This behavior results from the fact that lower $\alpha$ corresponds to higher surface density at given radius, which, in turn, decreases the efficiency of spectrum hardening through comptonization. If one sets the BH spin to a constant value corresponding to the results of the low-luminosity data analysis, then fitting for the value of $\alpha$ which produces a consistent spin estimate, is possible.

\begin{figure*}[hb]
  \includegraphics[width=.9\linewidth]{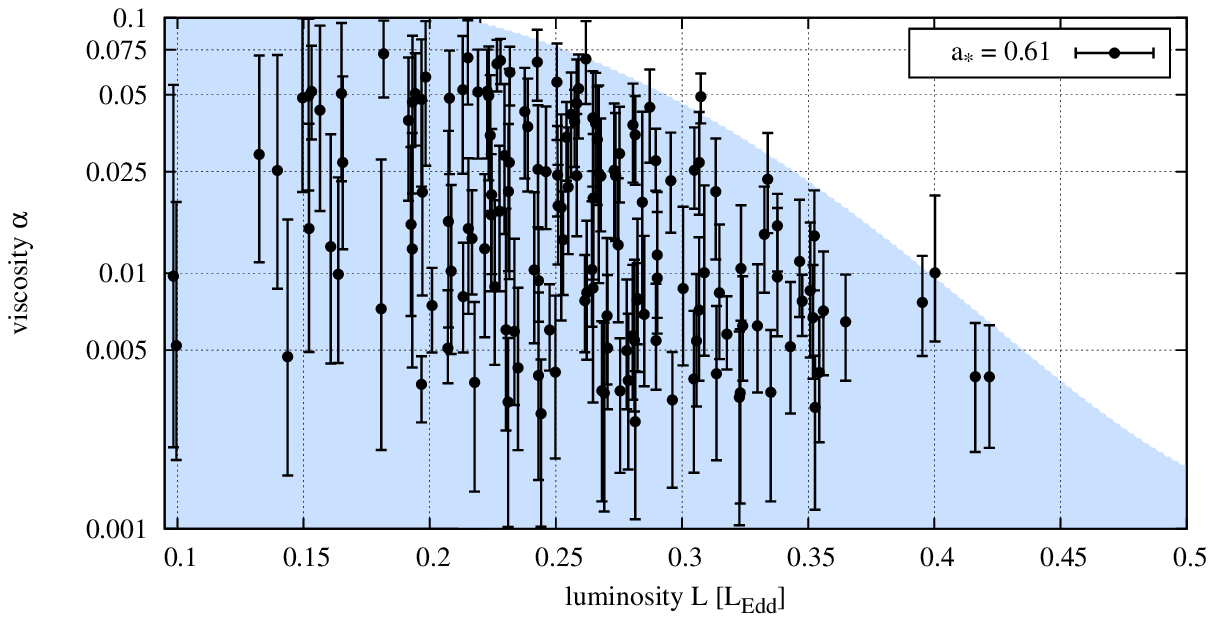}
 \caption{Behavior of the $\alpha$-viscosity parameter with luminosity for LMC X-3. Figure after: \citem{straubetal-11}.}
 \label{f.odele.alpha}
\end{figure*}

In Fig.~\ref{f.odele.alpha} we present results of such a study for LMC X-3 assuming BH spin $a_*=0.61$ (the mean value for fittings based on the $L\lesssim 0.2L_{\rm Edd}$ data). The large error bars for low-luminosity data reflect the fact that spin estimates are $\alpha$-degenerate in this regime. For $L\gtrsim 0.3L_{\rm Edd}$, however, the magnitude of viscosity may be determined. This study shows that high ($\alpha\gtrsim 0.05$) values of $\alpha$ are forbidden when the disk luminosity is not small. Therefore, either the effective $\alpha$ in LMC X-3 is independent of luminosity and small ($\alpha \lesssim 0.01$), or its value is larger for low-luminosity states and decreases with increasing luminosity down to the allowed values in the high luminosity regime. It is worth noting, that MHD shearing-box simulations of radiation dominated disks (Fig.~\ref{f.hirose.alphas}) suggest weak decrease of $\alpha$ with luminosity (from $0.03$ down to $0.01$) what is both in qualitative and quantitative agreement with the results previously discussed.

\section{Normalization of global MHD disk simulations}
\label{s.normalizing}
The general relativistic MHD models that are currently in use 
(e.g., \citem{pennaetal-10})
often make use of
dimensionless quantities and do not include detailed radiation
transfer. Hence,	 there is no direct way of estimating the physical mass accretion rate (${\rm g/s}$) or the true radiative luminosity
(${\rm erg/s}$) of these models. To estimate these quantities one may use an
indirect method, in which the vertical thicknesses
of the simulated disks is compared against physical disk models that include radiation transfer and radiation pressure and solve for the
vertical disk structure, e.g., two-dimensional slim disks introduced in Chapter~\ref{chapter-vertical}.

Our solutions of the two-dimensional slim disk model (Chapter~\ref{chapter-vertical}) 
were recently used, together with a non-LTE, radiatively efficient model of \citem{davisomer05},
to estimate the luminosities that the set of global MHD simulations presented in \citem{kulkarni-10} corresponds to. In this section we briefly compare their profiles of the disk thickness with the predictions of our two-dimensional slim disk model.

\begin{figure}
\centering
\includegraphics[angle=270,width=.9\textwidth]{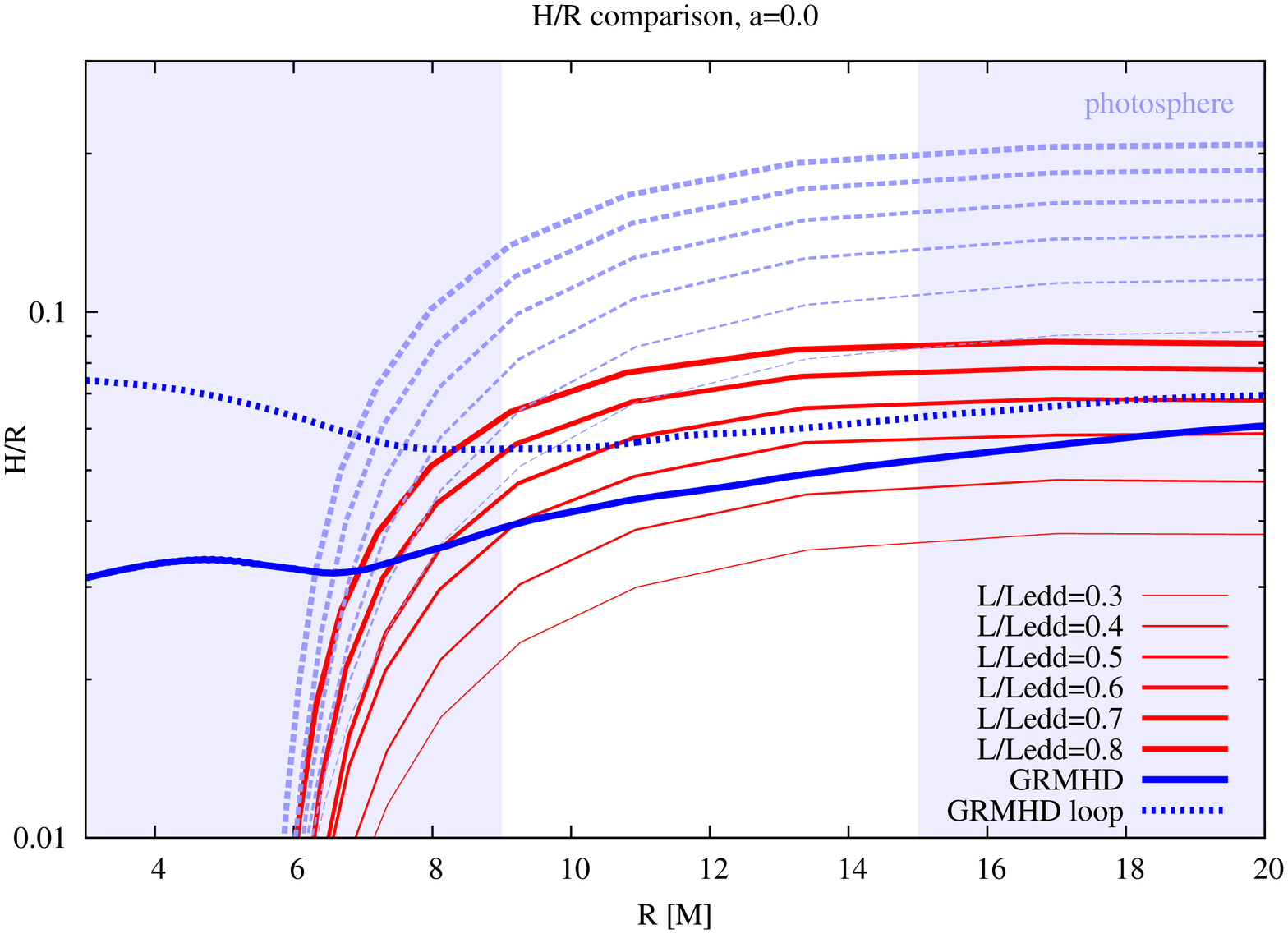}
\caption{Comparison of the density-weighted disk thickness predicted by the two-dimensional slim disk model (red lines) with the results of two GRMHD simulations for a non-rotating BH. Photosphere profiles of the slim disk solutions are presented with violet dotted lines.}
\label{f.grmhdcomp1}
\end{figure}

In Fig.~\ref{f.grmhdcomp1} we plot the density-weighted disk thickness, defined as,
\be|h|=\frac{\int_0^\infty \rho z\,{\rm dz}}{\int_0^\infty \rho\,{\rm dz}}\ee
as predicted by the slim disk model (red solid lines, corresponding to $L = 0.3\div0.8 L_{\rm Edd}$, bottom to top) for a non-rotating BH. These curves should
be compared with the disk thickness in the GRMHD simulations (deep blue lines). Results for two different configurations of the initial magnetic filed, purely-toroidal and loop, are shown with solid and dotted lines, respectively. The slim disk photosphere profiles are also presented (light blue, broken lines). 

At radii close to the ISCO, the slim disk and BHSPEC models
indicate that the disk thickness plunges to small values whereas
the GRMHD simulation shows a much smaller decrease. We believe there are at least three reasons for this discrepancy: (i) the
GRMHD simulations cool the gas by forcing it towards a constant entropy, which may not be justified in the plunging region; (ii) the simulated GRMHD disk includes magnetic fields
whose pressure provides additional support in the vertical direction whereas the other two models do not; and (iii) the simulated disc begins to deviate from hydrostatic equilibrium as the
radial velocity becomes large near the ISCO and the gas has less
time to reach equilibrium, whereas the other models hardwire
the condition of hydrostatic equilibrium at all radii. We estimate
that the last two are only important well inside the ISCO (compare Fig.~\ref{f.vertbal}). These
are interesting issues which we hope to explore in the future.

For the purpose of comparing the disk thickness profiles, the region of the simulation near
the ISCO is simply ignored. The comparison is based on radii 
$r = 9 \div 15 {\rm M}$, which are well outside the ISCO and determines the luminosities at which the slim disk model
gives the same disk thickness as obtained in the simulated GRMHD disk.
We see from Fig.~\ref{f.grmhdcomp1} that the thickness measure $|h|/r \sim 0.05$ in the
simulated GRMHD disk corresponds to $L/L_{\rm Edd}\approx 0.45$ according to the slim disk model. Similar analysis for $a_∗ =
0.7, 0.9\, {\rm and}\, 0.98$ shows that $L/L_{\rm Edd}\approx0.35, 0.5\, {\rm and}\, 0.5$ respectively.

Performing such comparison \citem{kulkarni-10} were able to draw the following conclusion: the observational errors
in current measurements of BH spin by the continuum-fitting method dominate over the errors incurred by using the
idealized NT model for $L/L_{\rm Edd}<0.3$. Comparison of the results of GRMHD simulations of accretion disks with the slim disk model was a crucial point in justifying this result.

\chapter{Conclusions}

Accretion of matter onto compact objects is one of the most efficient ways of extracting gravitational energy. Most luminous objects in the Universe: Active Galactic Nuclei, quasars, X-ray binaries are fed by accretion. Physics of accretion has, therefore, been one of the primary interests of astrophysicists for many years.

Currently, two approaches to modeling accretion disks are in use: (semi-)analytical and numerical. The former was pioneered by the seminal work of \citem{shakura-73}. In this approach, the most general equations of magnetohydrodynamics are reduced, thanks to a number of simplifying assumptions, to a set of ordinary differential or algebraic equations. The most obvious advantage of such treatment is the simplicity of the final equations and, what follows, small computational complexity. The slim disk models described in this thesis are based on such an approach.

The other way of modeling accretion disks is based on solving the equations describing them in the full, unreduced form, using sophisticated numerical methods. In this way one is able to follow the magnetorotational instability in detail, as well as to discard all of the simplifying, but only approximate, assumptions. Unfortunately, such an approach is extremely computational time-consuming and therefore, its applicability is limited. A global (3D), relativistic MHD simulation of a thin disk with simplified treatment of cooling, similar to the one described in Section~\ref{s.normalizing}, takes currently months to reach the quasi-equilibrium state. As a result, it is not possible to obtain a full grid (spanned on BH mass, BH spin, and accretion rate) of disk solutions which might be applied for fitting to particular X-ray sources. Nevertheless, numerical modeling, either global or shearing-box, gives the insight into the real turbulent behavior of disk interior.

The analytical, standard model of relativistic thin accretion disks \citepm{nt} has been in use for almost forty years. It successfully describes the disk component of the high/soft state spectra of AGN and microquasars. BH spins of a number of objects were estimated applying the \citem{nt} model to their X-ray continuum spectra. Most recent global MHD simulations \citepm{kulkarni-10} prove that, in the limit of thin disks, the emission and angular momentum profiles produced by MRI-generated viscosity are in good agreement with the predictions of the analytical, based on the $\alpha p$ prescription, model of \citem{nt}.

However, as we have proven in the introduction to Chapter~\ref{chapter-stationary}, some of the assumptions behind the standard models of thin disks break down for accretion rates close to the Eddington value. Most importantly, the advective cooling triggers in and starts to affect disk observables. 

The model of slim accretion disks successfully generalizes standard thin disks to the regime of high luminosities. It accounts for the advective transport of heat and radial gradient of pressure which affects the profile of angular momentum.

In this thesis we have presented the slim disk model in detail. We started from introducing the stationary, relativistic, height-averaged  model (Chapter~\ref{chapter-stationary}). We widely discussed the solutions, putting most attention to the locations of various disk inner edges. Basing on these solutions, a spectral model {\tt slimbb} was constructed and described in Section~\ref{s.slimbb}. 

The initial model was generalized to the non-stationary case in Chapter~\ref{chapter-nonstationary}. The equations describing time-evolution of a relativistic slim disk (Appendix~\ref{ap.nonstat}) have not yet been presented in the literature. We solved them using spectral methods and presented the limit cycle behavior of accretion disks around rotating BHs. We put most attention to the impact of BH spin on the disk observables. We showed that the only one significantly affected by BH spin was the maximal luminosity. We suggested a rapid, but very rough, method of estimating BH spin basing on the limit-cycle light curves.

We developed a two-dimensional slim disk model including, in a self-consistent way, detailed treatment of the vertical radiative transfer in the disk interior (Chapter~\ref{chapter-vertical}). We presented and discussed the vertical structure of disks and proved that the regular, one-dimensional models produce spectra which are very similar to the spectra obtained using the more precise, two-dimensional model.

We also constructed a state-of-art model of self-irradiated slim disks (Chapter~\ref{chapter.selfirradiated}). The brief discussion included in the dissertation will be followed by a more detailed study to be published in a separate paper.

In Chapter~\ref{chapter-applications} we presented three possible applications of the slim disk models presented in the thesis. First, we studied the spin evolution of BHs accreting at super-critical accretion rates. We proved that, under reasonable assumptions and for a reasonable set of disk parameters, the canonical spin value $a_*=0.9978$ \citepm{thorne-74}, may be exceeded. We also showed that for very high ($\dot m\gtrsim 10$) accretion rates, the evolution of BH spin is not influenced by the angular momentum and energy of photons captured by the BH. 

We applied the slim disk-based spectral model {\tt slimbb} to high/soft spectra of LMC X-3. We suggested that the inconsistency in BH spin estimates for high luminosity data may be resolved if the viscosity parameter $\alpha$ decreases with luminosity.

Finally, we also presented the way of normalizing global MHD simulations of accretion disks using the disk thickness profiles of slim disks.

Slim disks are a powerful tool for modeling high/soft states of accretion disks at high luminosities. They can be further developed by including additional effects which may affect the real sources, and which are currently neglected or treated in a simplified way, e.g., outflows or modified viscosity prescriptions. As long as global, radiative MHD simulations of disks remain limited by the amount of computational power, slim disks will remain the reference model for luminous disks in AGN and X-ray binaries.

\appendix

\chapter{Kerr metric}
\label{ap.kerr}

The specific angular momentum $a$ of a BH with mass $M$ and angular momentum $J$ is given by ($G=c=1$),
\be
a\equiv \frac JM.
\ee
Often a dimensionless parameter $a_*$ is introduced to express the magnitude of the spin,
\be
a_*\equiv \frac aM=\frac J{M^2}.
\ee
It ranges between $a_*=-1$ (maximally counter-rotating BH), through $a_*=0$ (Schwarzschild BH), to $a_*=1$ (maximally rotating BH).

The space-time metric near rotating BHs was described by \citem{kerr-63}. The general formula for the line element $ds^2$ is given in the Boyer-Lindquist coordinates $[t,\phi,r,\theta]$ by,
\bea
{\rm d}s^2&=&-\left(1-\frac{2Mr}{\rho^2}\right){\rm d}t^2 - 2\frac{2Mr}{\rho^2}a\sin^2\theta {\rm d}\phi {\rm d}t+\\\nonumber
&+&\frac{\rho^2}\Delta {\rm d}r^2 + \rho^2{\rm d}\theta^2+\left(r^2+a^2+\frac{2Mr}{\rho^2}a^2\sin^2\theta\right)\sin^2\theta {\rm d}\phi^2,
\eea
where
\be
\Delta=r^2-2Mr+a^2,
\ee
\be
\rho^2=r^2+a^2\cos^2\theta.
\ee
The metric is stationary and symmetric with respect to the polar axis of $\theta=0$ and the equatorial plane $\theta=\pi/2$. The signature of the metric is $(-+++)$.

For the purpose of studying accretion disks it is convenient to write the Kerr metric in the approximate form valid close to the equatorial plane ($\cos^2\theta \ll 0$) and introduce the vertical cylindrical coordinate $z=r\cos\theta$. The non-zero metric components are:
\begin{center}

$g_{tt}=-\frac{r-2M}{r}$, $\quad g_{t\phi}=-\frac{2Ma}{r}$ ,$\quad g_{\phi\phi}=\frac{A}{r^2}$,$\quad g_{rr}=\frac{r^2}{\Delta}$,$\quad g_{zz}=1$.

$g^{tt}=-\frac{A}{r^2\Delta}$, $\quad g^{t\phi}=-\frac{2Ma}{r\Delta}$, $\quad g^{\phi\phi}=\frac{r-2M}{r\Delta}$, $\quad g^{rr}=\frac{\Delta}{r^2}$, $\quad g^{zz}=1$,
\end{center} where  $A=r^4+r^2a^2+2Mra^2$.  

\section{Killing vectors}
\label{ap.killing}

A vector field $X$ is a Killing field if the Lie derivative with respect to $X$ of the metric $g$ vanishes:
\be\mathcal{L}_{X} g = 0 \,.\ee
If the metric coefficients $g_{\mu \nu} \,$ in some coordinate basis $dx^{a} \,$ are independent of $x^{K} \,$, then $x^{\mu} = \delta^{\mu}_{K} \,$ is automatically a Killing vector, where $\delta^{\mu}_{K} \,$ is the Kronecker delta \citepm{mtw}.

In the Kerr metric there are two obvious Killing vectors:
\be\eta=\pder{}{t},\ee
\be\xi=\pder{}{\phi},\ee
connected with time and axial symmetries of the metric, respectively. Their contravariant components ($[t,\phi,r,\theta]$) are,
\be\eta^i=(1,0,0,0),\ee
\be\xi^i=(0,1,0,0),\ee
while the covariant take the form,
\be\eta_t=\eta^t g_{tt}+\eta^\phi g_{t\phi}=g_{tt},\ee
\be\eta_\phi=\eta^t g_{t\phi}+\eta^\phi g_{\phi\phi}=g_{t\phi},\ee
\be\xi_t=\xi^t g_{tt}+\xi^\phi g_{t\phi}=g_{t\phi},\ee
\be\xi_\phi=\xi^t g_{t\phi}+\xi^\phi g_{\phi\phi}=g_{\phi\phi}.\ee
Products of the Killing vectors yield,
\be\eta^i\eta_i=\eta^t\eta_t=g_{tt},\ee
\be\xi^i\xi_i=\xi^\phi\xi_\phi=g_{\phi\phi},\ee
\be\eta^i\xi_i=\eta^t\xi_t=g_{t\phi},\ee
\be\xi^i\eta_i=\xi^\phi\eta_\phi=g_{t\phi}.\ee

\section{Christoffel symbols}
\label{ap.christoffel}
The Christoffel symbols can be derived from the vanishing of the covariant derivative of the metric tensor $g_{ik}$ :

\be    
0 = \nabla_\ell g_{ik}= \frac{\partial g_{ik}}{\partial x^\ell}- g_{mk}\Gamma^m_{i\ell} - g_{im}\Gamma^m_{k\ell} = \frac{\partial g_{ik}}{\partial x^\ell}- 2g^{~}_{m[k}\Gamma^m_{i]\ell}. 
\ee
By permuting the indices, and resumming, one can solve explicitly for the Christoffel symbols as a function of the metric tensor:
\be
    \Gamma^i_{k\ell}=\frac{1}{2}g^{im} \left(\frac{\partial g_{mk}}{\partial x^\ell} + \frac{\partial g_{m\ell}}{\partial x^k} - \frac{\partial g_{k\ell}}{\partial x^m} \right) = {\frac 12} g^{im} (g_{mk,\ell} + g_{m\ell,k} - g_{k\ell,m}), 
\ee
Below we give the exact formulae for all the non-zero Christoffel symbols calculated for the general Kerr metric close to the equatorial plane (i.e., $\sin\theta=\cos^2\theta=0$).

\bea
\label{e.christoffels}
\Gamma^t_{tr}\ =\ \Gamma^t_{rt}&=&\frac{M(r^2+a^2)}{\Delta r^2}\\\nonumber
\Gamma^\phi_{tr}\ =\ \Gamma^\phi_{rt}&=&\frac{Ma}{\Delta r^2}\\\nonumber
\Gamma^t_{t\theta}\ =\ \Gamma^t_{\theta t}&=&-\frac{2Ma^2\cos\theta}{r^3}\\\nonumber
\Gamma^\phi_{t\theta}\ =\ \Gamma^\phi_{\theta t}&=&-\frac{2Ma\cos\theta}{r^3}\\\nonumber
\Gamma^t_{r\phi}\ =\ \Gamma^t_{\phi r}&=&-\frac{Ma(3r^2+a^2)}{\Delta r^2}\\\nonumber
\Gamma^\phi_{r\phi}\ =\ \Gamma^\phi_{\phi r}&=&\frac{r(r^2-2Mr)-Ma^2}{\Delta r^2}\\\nonumber
\Gamma^t_{\theta \phi}\ =\ \Gamma^t_{\phi\theta}&=&\frac{2Ma^3\cos\theta}{r^3}\\\nonumber
\Gamma^\phi_{\theta\phi}\ =\ \Gamma^\phi_{\phi\theta}&=&\left(1+\frac{2Ma^2}{r^3}\right)\cos\theta\\\nonumber
\Gamma^\phi_{\theta\phi}\ =\ \Gamma^\phi_{\phi\theta}&=&\left(1+\frac{2Ma^2}{r^3}\right)\cos\theta\\\nonumber
\Gamma^r_{tt}&=&\frac{M\Delta}{r^4}\\\nonumber
\Gamma^r_{t\phi}\ =\ \Gamma^r_{\phi t}&=&-\frac{Ma\Delta}{r^4}\\\nonumber
\Gamma^\theta_{tt}&=&-\frac{2Ma^2\cos\theta}{r^5}\\\nonumber
\Gamma^\theta_{t\phi}\ =\ \Gamma^\theta_{\phi t}&=&\frac{2Ma(r^2+a^2)\cos\theta}{r^5}\\\nonumber
\Gamma^r_{rr}&=&\frac1r+\frac{M-r}{\Delta}\\\nonumber
\Gamma^r_{\theta\theta}&=&-\frac{\Delta}{r}\\\nonumber
\Gamma^\theta_{r\theta}\ =\ \Gamma^\theta_{\theta r}&=&\frac1r\\\nonumber
\Gamma^r_{r\theta}\ =\ \Gamma^r_{\theta r}&=&-\frac{a^2\cos\theta}{r^2}\\\nonumber
\Gamma^\theta_{\theta\theta}&=&-\frac{a^2\cos\theta}{r^2}\\\nonumber
\Gamma^\theta_{rr}&=&\frac{a^2\cos\theta}{r^2\Delta}\\\nonumber
\Gamma^r_{\phi\phi}&=&\frac{\Delta(Ma^2-r^3)}{r^4}\\\nonumber
\Gamma^\theta_{\phi\phi}&=&-\frac{\cos\theta\left(\Delta r^4+2Mr(r^2+a^2)^2\right)}{r^6}\\\nonumber
\eea

\section{Characteristic radii}
\label{ap.isco}

The following characteristic radii, important for the accretion disk physics, may be distinguished \citepm{bardeen-72, KatoBook}:

(i) The event horizon:

\noindent The horizon is a boundary in spacetime beyond which events cannot affect an outside observer. Even light emitted from beyond the horizon can never cross it and leave the BH. The radius of the horizon is given by,
\be
r_{\rm h}=M(1+\sqrt{1-a_*^2}).
\ee

(ii) Innermost stable circular orbit (marginally stable orbit):

\noindent Defines the radius of the innermost stable circular geodesic orbit. It is located at,
\be
r_{\rm ISCO}=M(3+Z_2- \sqrt{(3-Z_1)(3+Z_1+2Z_2)},
\ee
where
\be
Z_1=1+(1-a_*^2)^{1/3}\left[(1+a_*)^{1/3}+(1-a_*)^{1'3}\right],
\ee
\be
Z_2=\sqrt{3a_*^2+Z_1^2}.
\ee

(iii) Marginally bound orbit

\noindent Defines the innermost unstable circular orbit. Its radius is given by,
\be
r_{\rm mb}=2M\left(1-\frac{a_*}2+\sqrt{1- a_*}\right).
\ee

(iii) Photon orbit

\noindent At the photon orbit photons follow circular, unstable, trajectories around BH. The radius of such an orbit is,
\be
r_{\rm ph}=2M\left[1+\cos\left(\frac23\arccos(- a_*)\right)\right].
\ee

\begin{figure}
\centering
  \includegraphics[width=.75\textwidth,angle=0]{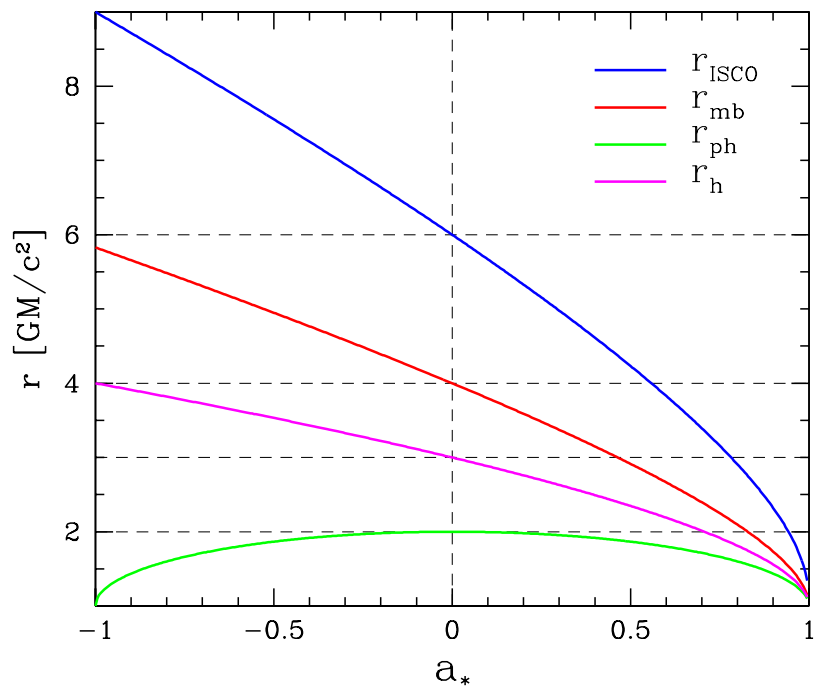}
  \caption{Characteristic orbits of the Kerr metric. The innermost stable circular orbit ($r_{\rm ISCO}$), the marginally bound orbit ($r_{\rm mb}$), the photon orbit ($r_{\rm ph}$) and the BH horizon ($r_{\rm h}$) are presented.}
  \label{f.rrr}
\end{figure}

Fig.~\ref{f.rrr} presents locations of these characteristic radii versus dimensionless BH spin parameter $a_*$.

\chapter{Relativistic, non-stationary equations for slim disks\label{APwork}}
\label{ap.nonstat}

\section{Metric and four-velocity components}
The Kerr metric coefficients close to the equatorial plane are given in Appendix~\ref{ap.kerr}. 

The following form of the stress-energy tensor of the matter in the disk is assumed:
\be T^{ik}=\rho u^iu^k+pg^{ik}-t^{ik}+u^kq^i+u^iq^k,\ee
where $\rho$ is the rest mass density, $t^{ik}$ is the viscous stress tensor and $q^i$ is the radiative energy flux. In the further part of the text we will also use the ''vertically integrated'' pressure $P=2Hp$.

The general form of the four-velocity is:
\begin{equation}\label{eq1}u^i=\gamma\left(e^i_{(t)}+V^{(r)}e^i_{(r)}+V^{(\phi)}e^i_{(\phi)}+V^{(\theta)}e^i_{(\theta)}\right),\end{equation}
where $V^{(j)}$ are velocities as measured in the Local Non-Rotating Frame (LNRF) and $e_{(j)}$ are LNRF basis vectors. Near the equatorial plane they take the following forms \citepm{bardeen-72}:
\begin{eqnarray}
e_{(t)}&=&\frac{A^{1/2}}{r\Delta^{1/2}}\frac{\partial}{\partial t}+\frac{2Ma}{A^{1/2}\Delta^{1/2}}\frac{\partial}{\partial \phi},\\\nonumber
e_{(r)}&=&\frac{\Delta^{1/2}}{r}\frac{\partial}{\partial r},\\\nonumber
e_{(\theta)}&=&\frac{1}{r}\frac{\partial}{\partial \theta},\\\nonumber
e_{(\phi)}&=&\frac{r}{A^{1/2}}\frac{\partial}{\partial \phi}.
\end{eqnarray}
According to Eq. \ref{eq1} the contravariant components of the four-velocity are:
\begin{eqnarray}
u^t&=&\gamma\frac{A^{1/2}}{r\Delta^{1/2}},\\\nonumber
u^r&=&\gamma V^{(r)}\frac{\Delta^{1/2}}{r},\\\nonumber
u^\phi&=&\gamma\frac{2Ma}{A^{1/2}\Delta^{1/2}}+\gamma V^{(\phi)}\frac{r}{A^{1/2}},\\\nonumber
u^\theta&=&\gamma V^{(\theta)}\frac{1}{r}.
\end{eqnarray}

We introduce the following velocity variables describing the flow:
\begin{itemize}
\item the radial velocity of the fluid measured in the co-rotating frame (CRF) $V$ (compare Eq.~\ref{e.vasmeasured}):
\be V^{(r)}=\frac{1}{\gamma}\frac{V}{\sqrt{1-V^2}},\label{eap.vr}\ee
\item the angular momentum per unit mass $\cal{L}$:
\be {\cal L}=u_{\phi},\ee
\item the vertical velocity $U$ in the form:
\be V^{(\theta)}=U\cos{\theta}.\ee
\end{itemize}
By simple calculations we can express $V^{(\phi)}$ in terms of $\cal L$:
\be V^{(\phi)}=\frac{1}{\gamma}\frac{{\cal L}r}{A^{1/2}}.\label{eap.vphi}\ee
The Lorentz factor $\gamma$ at the equatorial plane is defined as following:
\be \gamma^{-2}=1-\left(V^{(r)}\right)^2-\left(V^{(\phi)}\right)^2.\ee
Taking into account Eqs.~\ref{eap.vr} and \ref{eap.vphi} we get:
\be \gamma^{2}=\left(\frac{1}{1-V^2}+\frac{{\cal L}^2r^2}{A}\right)=\frac{1}{1-V^2}+\frac{{\cal L}^2r^2}{A}.\ee
We assume that the Lorentz factor $\gamma$ at the disk surface is given by the same expression, as we expect the vertical velocities to be small compared to the radial and orbital motion.

Now the contravariant components of the four-velocity take following forms:
\begin{eqnarray}u^t&=&\gamma\frac{A^{1/2}}{r\Delta^{1/2}},\\\nonumber
u^r&=&\frac{V}{\sqrt{1-V^2}}\frac{\Delta^{1/2}}{r},\\\nonumber
u^\phi&=&\frac{{\cal L}r^2}{A}+\gamma\omega\frac{A^{1/2}}{r\Delta^{1/2}},\\\nonumber
u^\theta&=&\frac{\gamma}{r}U\cos{\theta}.\end{eqnarray}
The covariant four-velocity components are:
\begin{eqnarray}u_t&=&-\gamma\frac{r\Delta^{1/2}}{A^{1/2}}-\omega{\cal L},\\\nonumber
u_r&=&\frac{r}{\Delta^{1/2}}\frac{V}{\sqrt{1-V^2}},\\\nonumber
u_\phi&=&\cal L,\\\nonumber
u_\theta&=&\gamma rU\cos{\theta}.\end{eqnarray}

\section{Momentum conservation}
Let us consider the r-component of the general equation:
\be \nabla_iT^{ik}=0\Leftrightarrow a^r=-\frac{1}{\rho}g^{rr}\frac{\partial p}{\partial r}-\frac{1}{\Sigma r^2}(r^2T^{ir})_{,i},\ee
\be a^r=u^i\nabla_iu^r=u^t\frac{\partial u^r}{\partial t}+u^r\frac{\partial u^r}{\partial r}+u^i\Gamma^r_{ki}u^k,\ee
\be u^i\Gamma^r_{ki}u^k=\Gamma^r_{tt}u^tu^t+2\Gamma^r_{t\phi}u^tu^\phi+\Gamma^r_{rr}u^ru^r+\Gamma^r_{\theta\theta}u^\theta u^\theta+
2\Gamma^r_{r\theta}u^ru^\theta+\Gamma^r_{\phi\phi}u^\phi u^\phi.\ee
From among all components of the vertically integrated viscous stress tensor $T^{ir}$ we leave only $T^{rr}$.
Neglecting the terms proportional to $\cos^2\theta$ we get:
\be u^t\frac{\partial u^r}{\partial t}+u^r\frac{\partial u^r}{\partial r}+\Gamma^r_{tt}u^tu^t+2\Gamma^r_{t\phi}u^tu^\phi+\Gamma^r_{rr}u^ru^r+\Gamma^r_{\phi\phi}u^\phi u^\phi=-\frac{1}{\rho}g^{rr}\frac{\partial p}{\partial r}+\frac{1}{\Sigma r^2}\frac{\partial}{\partial r}(r^2T^{rr}).\ee
What is equivalent to:
\be (1-V^2)g_{rr}u^t\frac{\partial u^r}{\partial t}+\frac{V}{1-V^2}\frac{\partial V}{\partial r}=\frac{{\cal A}}{r}-\frac{1-V^2}{\Sigma}\frac{\partial P}{\partial r}+\frac{1-V^2}{\Sigma r^2}g_{rr}\frac{\partial}{\partial r}(r^2T^{rr}),\ee
where $\cal A$, $\Omega$, $\Omega_k^+$, $\Omega_k^-$, $\tilde{\Omega}$ and $\tilde{R}$ are defined as in \citem{adafs}:
\begin{eqnarray}
{\cal A}&=&-\frac{MA}{r^3\Delta\Omega_k^+\Omega_k^-}\frac{(\Omega-\Omega_k^+)(\Omega-\Omega_k^-)}{1-\tilde{\Omega}^2\tilde{R}^2},\\
\Omega&=&\frac{u^\phi}{u^t}=\frac{2Mar}{A}+\frac{r^3\Delta^{1/2}{\cal L}}{\gamma A^{3/2}},\\
\Omega_k^\pm&=&\pm\frac{M^{1/2}}{r^{3/2}\pm aM^{1/2}},\\
\tilde{\Omega}&=&\Omega-\omega,\\
\tilde{R}&=&\frac{A}{r^2\Delta^{1/2}}.
\end{eqnarray}
Finally:
\begin{equation}\label{urt}
\frac{\partial u^r}{\partial t}=\frac{\Delta^{3/2}}{\gamma rA^{1/2}(1-V^2)}\left[-\frac{V}{1-V^2}\frac{\partial V}{\partial r}+\frac{{\cal A}}{r}-\frac{1-V^2}{\rho}\frac{\partial p}{\partial r}-\frac{1-V^2}{\Sigma\Delta}\frac{\partial}{\partial r}(r^2T^{rr})\right]
\end{equation}
and
\begin{equation}\label{radial}\frac{\partial V}{\partial t}=\frac{\sqrt{1-V^2}\Delta}{\gamma A^{1/2}}\left[-\frac{V}{1-V^2}\frac{\partial V}{\partial r}+\frac{{\cal A}}{r}-\frac{1-V^2}{\rho}\frac{\partial p}{\partial r}-\frac{1-V^2}{\Sigma\Delta}\frac{\partial}{\partial r}(r^2T^{rr})\right].
\end{equation}
The viscous stress tensor component $T^{rr}$ is given by,
\be
T^{rr}=\nu\Sigma \sigma^{rr},
\ee
where $\nu$ is the kinematic viscosity coefficient and $\Sigma$ is the surface density.
The shear tensor components $\sigma_{\alpha\beta}$ are given in general by,
\be
\sigma_{\alpha\beta}=\frac{1}{2}(u_{\alpha;\mu}h^\mu_\beta+u_{\beta;\mu}h^\mu_\alpha)-\frac{1}{3}\Theta h_{\alpha\beta},\ee
where $\Theta=u^\nu_{;\nu}$ and $h^{\alpha\beta}=g^{\alpha\beta}+u^{\alpha}u^\beta$.

\section{Angular momentum conservation}
Let us write the angular momentum conservation law in the form:
\be \nabla_i(T^i_k\xi^k)=0.\ee
where $\xi^k=\delta^k_{(\phi)}$. After vertical integration we get:
\be \Sigma\left(u^t\frac{\partial u_\phi}{\partial t}+u^r\frac{\partial u_\phi}{\partial r}\right)-\nabla_i(T^i_\phi)=0,\ee
where $T^i_\phi$ is vertically integrated $(i,\phi)$ component of the viscous stress tensor. Let us follow \citem{lasota94} and assume that the only nonvanishing component of $T^i_\phi$ is:
\be T^r_\phi=2\nu\Sigma\frac{A^{1/2}\Delta^{1/2}\gamma}{r^2}\sigma_{\bar{r}\bar{\phi}}=\nu\Sigma\frac{A^{3/2}\Delta^{1/2}\gamma^3}{r^5}\frac{\partial\Omega}{\partial r}.\ee
Finally:
\be \Sigma u^t\frac{\partial u_\phi}{\partial t}=-\Sigma u^r\frac{\partial u_\phi}{\partial r}+\frac{1}{r}\frac{\partial}{\partial r}\left(\frac{\nu\Sigma A^{3/2}\Delta^{1/2}\gamma^3}{r^4}\frac{\partial\Omega}{\partial r}\right),\ee
where:
\be \Omega=\frac{u^\phi}{u^t}=\frac{2Mar}{A}+\frac{r^3\Delta^{1/2}{\cal L}}{\gamma A^{3/2}}\ee 
is the angular velocity with respect to the stationary observer.
Introducing $\cal L$ and $V$ we get:
\begin{equation}\label{angular}
\frac{\partial\cal L}{\partial t}=-\frac{V\Delta}{\gamma\sqrt{1-V^2}A^{1/2}}\frac{\partial\cal L}{\partial r}+
\frac{\Delta^{1/2}}{\gamma\Sigma A^{1/2}}\frac{\partial}{\partial r}\left(\frac{\nu\Sigma A^{3/2}\Delta^{1/2}\gamma^3}{r^4}
\frac{\partial\Omega}{\partial r}\right).
\end{equation}
\section{Disk stability criterion}
Let us consider the vertical acceleration of the disk surface in the Zero Angular Momentum Observer (ZAMO) system of coordinates.
The disk surface is defined in cylindrical coordinates as following:
\be z=H(r,t).\ee 
Therefore, the infinitesimal shift in the vertical direction can be expressed in the following way:
\be {\rm dz}=\frac{\partial H}{\partial r}{\rm dr}+\frac{\partial H}{\partial t}{\rm dt}.\ee 
Dividing both sides by $\rm dt$ we get:
\be \frac{\rm dz}{\rm dt}=-V^{(\theta)}=\frac{\partial H}{\partial r}V^{(r)}+\frac{\partial H}{\partial t}.\ee 
Introducing $U$ and $V$ and substituting $H=r\cos\Theta_H$ we obtain:
\begin{equation}
\frac{\partial\cos\Theta_H}{\partial t}=-\frac{U\cos\Theta_H}{r}-\frac{1}{r\gamma}\frac{V}{\sqrt{1-V^2}}\frac{\partial}{\partial r}(r\cos\Theta_H).
\end{equation}
\section{Vertical equilibrium}
Let us assume the following form of pressure \citepm{vertical}:
\be P(r,\theta,t)=P_0(r,t)\left[1-\frac{\cos^2\theta}{\cos^2\Theta_H}\right].\ee
We will derive the vertical equilibrium equation taking $\theta$ component of the general equation:
\be \nabla_kT^k_i=0.\ee
Simple manipulations give (after vertical integration):
\be 0=\Sigma u^k\nabla_ku_\theta+\frac{\partial P}{\partial\theta},\ee
\be \frac{1}{\Sigma}\frac{\partial P}{\partial\theta}=-u^ku_{\theta,k}+\Gamma^i_{\theta k}u_iu^k.\ee
Since:
\be \Gamma^i_{\theta k}u_iu^k=\frac{\cos\theta}{r^2}\left[u_\phi u_\phi-a^2(u_tu_t-1)\right],\ee
\be u^ku_{\theta,k}=u^tu_{\theta,t}+u^ru_{\theta,r},\ee
\be \left.\frac{\partial P}{\partial \theta}\right|_{\theta=\Theta_H}=\frac{2P_0}{\cos\Theta_H},\ee
we finally obtain (for $\theta=\Theta_H$ and $P_0\equiv P$):
\begin{equation}\label{uthetat}
u^t\frac{\partial u_\theta}{\partial t}=-\frac{2P}{\Sigma\cos\Theta_H}+
\left(u_\phi u_\phi-a^2(u_tu_t-1)\right)\frac{\cos\Theta_H}{r^2}-
u^r\frac{\partial u_\theta}{\partial r}\equiv{\cal R},
\end{equation}
where:
\be u_{\phi}u_\phi={\cal L}^2,\ee
\be u_tu_t=\left[\gamma\frac{r\Delta^{1/2}}{A^{1/2}}+\omega{\cal L}\right]^2,\ee
\be u^r\frac{\partial u_\theta}{\partial r}=\frac{V\Delta^{1/2}}{r\sqrt{1-V^2}}\frac{\partial}{\partial r}
\left(\gamma Ur\cos\Theta_H\right),\ee
and \be u^t\frac{\partial u_\theta}{\partial t}=\gamma\frac{A^{1/2}\cos\Theta_H}{\Delta^{1/2}}\left[
\gamma\frac{\partial U}{\partial t}+
\frac{U}{\gamma}\left(\frac{V}{(1-V^2)^2}\frac{\partial V}{\partial t}+
\frac{{\cal L}r^2}{A}\frac{\partial\cal L}{\partial t}\right)+\frac{\gamma U}{\cos\Theta_H}\frac{\partial\cos\Theta_H}{\partial t}\right].\ee
Solving for $\frac{\partial U}{\partial t}$ we get:
\begin{equation}\label{vertical}
\frac{\partial U}{\partial t}=\frac{\Delta^{1/2}}{\gamma^2 A^{1/2}\cos\Theta_H}{\cal R}-
\frac{U}{\gamma^2}\left(\frac{V}{(1-V^2)^2}\frac{\partial V}{\partial t}+
\frac{{\cal L}r^2}{A}\frac{\partial\cal L}{\partial t}\right)-
\frac{U}{\cos\Theta_H}\frac{\partial\cos\Theta_H}{\partial t},
\end{equation}
where $\cal R$ is the right hand side of the Eq. \ref{uthetat}.\\
\section{Equation of continuity}
The general form of the continuity equation is:
\be \nabla_i(\rho u^i)=0,\ee
what can be written as:
\begin{equation}
\nabla_i(\rho u^i)=\frac{1}{\sqrt{-g}}(\sqrt{-g}\rho u^i)_{,i}=(\rho u^t)_{,t}+\frac{1}{r^2}(r^2\rho u^r)_{,r}.
\end{equation}
After the vertical integration it takes the following form:
\be u^t\frac{\partial\Sigma}{\partial t}=-\frac{1}{r}\frac{\partial}{\partial r}(\Sigma ru^r)-\Sigma\frac{\partial u^t}{\partial t}.\ee
The time derivative $\frac{\partial u^t}{\partial t}$ can be eliminated using the following relation:
\be \frac{\partial u^t}{\partial t}=\frac{A^{1/2}}{r\Delta^{1/2}}\frac{1}{\gamma}\left[\frac{V}{(1-V^2)^2}\frac{\partial V}{\partial t}+
\frac{{\cal L}r^2}{A}\frac{\partial\cal L}{\partial t}\right].\ee
Introducing the physical quantities we get:
\begin{equation}\label{continuity}
\frac{\partial\Sigma}{\partial t}=-\frac{r\Delta^{1/2}}{\gamma A^{1/2}}
\left[\Sigma\frac{\partial u^t}{\partial t}+\frac{1}{r}\frac{\partial}{\partial r}\left(r\Sigma\frac{V}{\sqrt{1-V^2}}\frac{\Delta^{1/2}}{r}\right)\right].
\end{equation}
\section{Energy conservation}
From the general form of the energy conservation equation:
\be \nabla _i(T^{ik}\eta_k)=0,\ee
we obtain, using the most general form of $T^{ik}$ (Eq.~\ref{e.cons2}), in the non-relativistic fluid approximation,
\be 0=\rho\left[u^t\frac{\partial\epsilon}{\partial t}+u^r\frac{\partial\epsilon}{\partial r}\right]+\nabla_i(t^{it})+\nabla_i(u^tq^i+u^iq^t),\ee
where $\epsilon$ is gas internal energy.
Similarly like in \citem{landau} (their Eq. 49.4), after vertical integration, it could be put into the following form:
\be \Sigma T\left[u^t\frac{\partial S}{\partial t}+u^r\frac{\partial S}{\partial r}\right]=Q^{\rm vis}-Q^{\rm rad},\ee
Where $S$ is entropy per unit mass, $\Sigma$ is surface density of the disk, $Q^{\rm vis}$ is the local viscous heat generation rate and $Q^{\rm rad}$ is the radiative cooling rate \citem{adafs}:
\be \label{ap.nonstat.fplus}Q^{\rm vis}=T^{ij}\sigma_{ij}\approx2T^{r\phi}\sigma_{r\phi}\approx\nu\gamma^4\Sigma\frac{A^2}{r^6}\left(\frac{\partial\Omega}{\partial r}\right)^2,\ee
\be Q^{\rm rad}=\frac{32\sigma T^4}{3\Sigma\kappa}.\ee
Thermodynamical relations give (for $p=p_{\rm rad}+p_{\rm gas}$):
\be T\frac{\partial S}{\partial r}=c_VT\left(\frac{1}{T}\frac{\partial T}{\partial r}-(\Gamma_3-1)\frac{1}{\rho}\frac{\partial\rho}{\partial r}\right),\ee
\be T\frac{\partial S}{\partial t}=c_VT\left(\frac{1}{T}\frac{\partial T}{\partial t}-(\Gamma_3-1)\frac{1}{\rho}\frac{\partial\rho}{\partial t}\right).\ee
The appropriate sum of the derivatives of $\rho=\Sigma/2H$ can be substituted using the continuity equation:
\be u^t\frac1\rho\frac{\partial\rho}{\partial t}+u^r\frac1\rho\frac{\partial\rho}{\partial r}\approx-\frac{\partial u^t}{\partial t}-\frac{1}{r^2}\frac\partial{\partial r}(r^2u^r)+\frac{U}{r}.\ee
Taking all together we obtain:
\be c_V\left[\Sigma\left(u^t\frac{\partial T}{\partial t}+u^r\frac{\partial T}{\partial r}\right)-(\Gamma_3-1)T\Sigma\left(-\frac{\partial u^t}{\partial t}-\frac{1}{r^2}\frac\partial{\partial r}(r^2u^r)\right)\right]=Q^{\rm vis}-Q^{\rm rad},\ee
where
\be c_V=\frac{4-3\beta}{\Gamma_3-1}\frac{P}{\Sigma T},\ee
\be \Gamma_3-1=\frac{(4-3\beta)(\gamma_g-1)}{12(1-\beta)(\gamma_g-1)+\beta},\ee
\be \beta=\frac{p_g}{p}.\ee
Ultimately:
\begin{eqnarray}\label{energy}
\frac{\partial T}{\partial t}&=&\frac{1}{\Sigma}\frac{r\Delta^{1/2}}{\gamma A^{1/2}}\left[
\frac{Q^{\rm vis}-Q^{\rm rad}}{c_V}+(\Gamma_3-1)T\Sigma\left(-\frac{\partial u^t}{\partial t}-\frac{1}{r^2}\frac\partial{\partial r}(r^2u^r)\right)\right]\\\nonumber&-&
\frac{V\Delta}{\gamma\sqrt{1-V^2}A^{1/2}}\frac{\partial T}{\partial r}.
\end{eqnarray}

\chapter{Derivation of an arbitrary tetrad in the Kerr spacetime}
\label{ap.tetrad}

Our aim is to derive the tetrad of an observer moving along the photosphere that
would depend only on the quantities which are calculated in accretion disk models, i.e., on
the radial and azimuthal velocities of gas and the location of disk photosphere.

The metric considered here is the Kerr geometry $g_{ik}$
in the Boyer-Lindquist coordinates $[t, \phi, r, \theta]$ (Appendix~\ref{ap.kerr}). Similarly as in Carter's Les Houches
lectures \citep{leshouches}, we will consider two fundamental planes; the symmetry
plane ${\cal S}_0 = [t, \phi]$ and the meridional plane ${\cal
M}_* = [r, \theta]$. (Four)-vectors that belong to the plane ${\cal
S}_0$, will be denoted by the subscript $0$, and vectors that belong
to the plane ${\cal M}_*$, will be denoted by the subscript $*$. For
example, the two Killings vectors are $\eta_0^i$, $\xi_0^i$. Note,
that for any pair $X_0^i, Y_*^i$ one has,
\be X_0^i\,Y_*^k\,g_{ik} \equiv (X_0\,Y_*) = 0.\ee

\section{Stationary and axially symmetric photosphere}

\subsection{Photosphere}

\noindent Numerical solutions of slim accretion disks provide the location of
the photosphere given by $h_{\rm Ph}(r)=r \cos\theta$. This may be put into
$r \cos\theta - h_{\rm Ph}(r) \equiv F(r, \theta) = 0$. The normal vector to the
photosphere surface has the following $[r,\theta]$ components,

\be
N_*^i=\tilde N_* \left[\pder F r,\pder F\theta\right]=\tilde N_*' \left[\der{\theta_*}{r},1\right],\ee
where
\be\der{\theta_*(r)}r=-\pder Fr\left /\pder F\theta\right.\ee
is the derivative of the angle defining the location of the photosphere at a given radial coordinate [$\cos\theta_*(r)=h_{\rm Ph}(r) / r$]. Its non-zero components after normalization [$(N_* N_*)=1$] are

\be N_*^r=\der{\theta_*}r(g_{\theta\theta})^{-1/2}\left[1+\frac{g_{rr}}{g_{\theta\theta}}\left(\der{\theta_*}r\right)^2\right]^{-1/2},\ee
\be\nonumber N_*^\theta=(g_{\theta\theta})^{-1/2}\left[1+\frac{g_{rr}}{g_{\theta\theta}}\left(\der{\theta_*}r\right)^2\right]^{-1/2}.
\ee
There are two unique vectors $S_*$ confined in the $[r,\theta]$ plane which are orthogonal to $N_*$ (and therefore are tangent to the surface). From $(S_*N_*)=0$ and $(S_*S_*)=1$ one obtains the non-zero components of one of them:

\be\label{e.Sstar}
 S_*^r=(g_{rr})^{-1}\left[\frac1{g_{rr}}+\frac1{g_{\theta\theta}}\left(\der{\theta_*}r\right)^2\right]^{-1/2},\ee
\be\nonumber S_*^\theta=-(g_{\theta\theta})^{-1}\left(\der{\theta_*}r\right)\left[\frac1{g_{rr}}+\frac1{g_{\theta\theta}}\left(\der{\theta_*}r\right)^2\right]^{-1/2}.\ee

\subsection{Tetrad}
\label{s.decomposition}
\noindent The four-velocity $u_0$ of an observer with azimuthal motion only is,
\be
u_0^i=\tilde A_0\left(\eta^i+\Omega\xi^i\right).
\ee
The normalization constant $\tilde A_0$ comes from $(u_0u_0)=-1$ and equals
\be\tilde A_0=\left[-g_{tt}-\Omega g_{\phi\phi}(\Omega-2\omega)\right]^{-1/2}.\ee
 It is useful to construct a spacelike vector ($\kappa_0$) confined in
 the $[t,\phi]$ plane, that is perpendicular to $u_0$. From $(\kappa \kappa)=1$ and $(\kappa u_0)=0$ we have,
\be\kappa_0^i=\frac{(l\eta^i+\xi^i)}{\left[g_{\phi\phi}(1-\Omega l)(1-\omega l)\right]^{1/2}},\ee
where $l=-u_\phi/u_t$ is the specific angular momentum and $\omega=-g_{t\phi}/g_{\phi\phi}$ is the frequency of frame dragging. The set of vectors $[u_0^i, ~N_*^i, ~\kappa_0^i, ~S_*^i]$ already forms the desired tetrad valid for the pure rotation ($u^r=0$) case.

Similarly, the four-velocity $u$ of gas moving along the photosphere with non-zero radial velocity may be decomposed into
\be
u^i=\tilde A\left(\eta^i+\Omega\xi^i+vS_*^i\right).
\ee
\noindent The normalization condition $(uu)=-1$ gives,
\be\tilde A=\left[-g_{tt}-\Omega g_{\phi\phi}(\Omega-2\omega)-v^2\right]^{-1/2},\ee
where $v$ is related to the radial component of the gas four-velocity $u^r$ by:
\be v=-\sqrt{\frac{\left(u^r/S^r_*\right)^2(-g_{tt}-\Omega g_{\phi\phi} (\Omega - 2\omega))}{1+\left(u^r/S^r_*\right)^2}}.\label{e.V}
\ee 
Taking into account that $S_*^r$ and $S_*^\theta$ are the only non-zero components of vector $S_*^i$, it is obvious that $u^i$ is orthogonal to $\kappa_0^i$.

The vectors we have just calculated ($u$, $\kappa_0$) are both orthogonal to $N_*$ since $(N_*S_*)=0$. To complete the tetrad we need one more spacelike vector ($S$) that is orthogonal to these three. Let us decompose it into,
\be S^i=\alpha u^i + \beta \kappa_0^i + \gamma N_*^i +\delta S_*^i. \ee
The orthogonality conditions $(\kappa_0 S)=0$ and $(N_* S)=0$ give immediately $\gamma=\beta=0$. The only non-trivial condition is $(u S)=0$. Together with $(SS)=1$ it leads to:

\be S^i=(1+\tilde A^2v^2)^{-1/2}(\tilde A vu^i+S^i_*).\ee

The vectors $u^i,  ~N_*^i, ~\kappa_0^i,~S^i$ form an orthonormal tetrad in the Kerr spacetime:
\begin{equation} \label{velocity-tetrad}
e^i_{~(A)} = [u^i,~N_*^i,  ~\kappa_0^i, ~S^i].
\end{equation}
This tetrad {\it is known}
directly from the slim disk solutions, as it depends on the calculated quantities ($u^r$, $\Omega$, $l$ and $\theta_*(r)$) only.
Any spacetime vector $X^i$, could be uniquely
decomposed into this tetrad with,
$X_{(A)} = X_i\,e^i_{~(A)}$.

\section{General case}

In this section we will assume nothing about the four-velocity of matter
$u^i$ and the location of photosphere. Both may be non-stationary and non-axially symmetric.
Following the same framework as in the previous subsection, we will describe how to obtain the tetrad of an observer
instantaneously located at the photosphere that depends only on the quantities calculated by accretion disk models.

\subsection{Photosphere}

\noindent In the most general case of a
non-stationary and non-axially symmetric photosphere, the location of the photosphere
may be described by the following condition,
\begin{equation} \label{general-photosphere}
F(t, \phi, r, \theta) = 0.
\end{equation}
\noindent The vector $\tilde N$ normal to the photosphere has
the components,
\begin{equation}
\label{normal-photosphere}
{\tilde N}^i = \left[
\frac{\partial F}{\partial t},
\frac{\partial F}{\partial \phi},
\frac{\partial F}{\partial r},
\frac{\partial F}{\partial \theta}
\right]
\end{equation}
\noindent which may be calculated from slim disk solutions.

\noindent Let us project ${\tilde N}$ into the instantaneous
3-space of the comoving observer (\ref{comoving-metric}) and
normalize to a unit vector after the projection,
\begin{equation}
\label{projected-normal}
N^i = \frac{{\hat N}^i}{\vert ({\hat N}{\hat N})\vert^{1/2}},
~~~{\hat N}^i = {\tilde N}^k\,h^i_{~k}.
\end{equation}
In terms of the tetrad~\ref{velocity-tetrad2}, such constructed
vector $N^i$ has the following decomposition,
\begin{equation}
\label{projected-normal-decomposition}
N^i = {\tilde N}\,[\alpha\, (u_0^i) + 1\, (N_*^i) + \gamma\,
(\kappa_0^i) + \delta\, (S_*^i)].
\end{equation}
\noindent The components ${\tilde N}, \alpha, \gamma, \delta$ are
known.

\subsection{Tetrad}

\noindent Similarly as in Section \ref{s.decomposition}, we may always
{\it uniquely} decompose $u^i$, a general timelike unit vector, into
\begin{equation} \label{four-velocity2}
u^i = {\tilde A} (u_0^i + v S_*^i),
\end{equation}
\noindent where $u_0^i$ is a timelike unit vector, and $S_*^i$ is
a spacelike unit vector. Formula~\ref{four-velocity2} uniquely
defines the two vectors $u_0^i, S_*^i$ and the two scalars
${\tilde A}, v$. The vectors and scalars
\begin{equation} \label{known-velocity}
\{ {\tilde A}, v, u_0^i, S_*^i \},
\end{equation}
can be calculated
from {\it known} quantities given by slim disk model solutions.

\noindent The four-velocity (Eq.~\ref{four-velocity2}) defines also the
instantaneous 3-space of the comoving observer with the metric
$\gamma_{ik}$ and the projection tensor $h^{~i}_k$,
\begin{eqnarray}
\gamma_{ik} &=& g_{ik} - u_i\,u_k,\\
\label{comoving-metric}
h^i_{~k} &=& \delta^i_{~k} - u^i\,u_k
\end{eqnarray}

We define two unit vectors $\kappa_0^i$ and $N_*^i$ by
the unique condition,
\begin{equation} \label{kappa-N}
(\kappa_0 u_0) = 0, ~~(S_* N_*) = 0.
\end{equation}

\noindent As before, the four vectors
\begin{equation} \label{velocity-tetrad2}
e^i_{~(A)} = [u_0^i, ~\kappa_0^i, ~N_*^i, ~S_*^i],
\end{equation}
\noindent which form an orthonormal tetrad of an observer with the four-velocity $u_0^i$ can calculated from the solutions of the slim disk equations.

Let us now write decompositions of the four vectors: the
first two we have derived, the next two guessed (but the
guess should be obvious):
\begin{eqnarray}
u^i &=&{\tilde A}\,[1\, (u_0^i) \,+ 0\, (N_*^i) \,\,+ 0\,
(\kappa_0^i) \,+ V\, (S_*^i)], \label{final-tetrad-u}\\
N^i &=&{\tilde N}\,[\alpha\, (u_0^i) + 1\, (N_*^i) \,\,+ \gamma\,
(\kappa_0^i) \,+ \delta\, (S_*^i)], \label{final-tetrad-N}\\
\kappa^i &=&{\tilde \kappa}\,\,\,[0\, (u_0^i) \,+ b\, (N_*^i)
\,\,+
1\,(\kappa_0^i) \,+ 0\, (S_*^i)], \label{final-tetrad-K}\\
S^i &=&{\tilde S}\,\,[A\, (u_0^i) + B\, (N_*^i) + C\,(\kappa_0^i)
+ 1\, (S_*^i)] \label{final-tetrad-S}.
\end{eqnarray}
\noindent The four unknown components, $b, A, B, C$ one calculates
from the four non-trivial orthogonality conditions ($(u\kappa) \equiv 0$ by construction, cf. Eqs. \ref{final-tetrad-u}
and \ref{final-tetrad-K}),
\begin{equation}
\label{orthogonality-conditions}
 (uS) = 0,~~ (NS) = 0,~~
(S\kappa) = 0,~~ (N\kappa) = 0,
\end{equation}
\noindent and the two unknown factors ${\tilde \kappa}, ~~ {\tilde
S}$, from the two normalization conditions
\begin{equation}
\label{normalization-conditions} (\kappa \kappa) = -1, ~~ (SS) =
-1.
\end{equation}
\noindent The conditions \ref{orthogonality-conditions},
\ref{normalization-conditions} are given by linear equations.

Equations
\ref{final-tetrad-u}-\ref{normalization-conditions} define the
tetrad $e_{~(A)}^{~i}$ of an observer comoving with matter,
and instantaneously located at the photosphere:
\begin{equation}
\label{tetrad-final-final}
e_{~(A)}^{~i} = [u^i, ~N^i, ~\kappa^i, ~S^i].
\end{equation}
Both the matter
and the photosphere move in a general manner. The zenithal
direction in the local observer's sky is given by $N^i$.

\chapter{Integration over the world-tube of the
photosphere}
\label{ap.integration}

For stationary and axially symmetric models we have so far (Appendix~\ref{ap.tetrad}):
\vskip 0.2truecm
$ N^i = N^i_* =~$ unit vector orthogonal to the photosphere. It
is in the $[r, \theta]$ plane,

\vskip 0.1truecm

$ S_*^i =~$ unit vector orthogonal to $N^i$ that lives
in the $[r, \theta]$ plane,

\vskip 0.1truecm

$ u^i =~$ four-velocity of matter. It lives in the $[t,
\phi, r, \theta]$ space-time,

\vskip 0.1truecm

$ \kappa^i =\kappa_0^i=~$ unit vector orthogonal to $U^i$ that
lives in the $[t, \phi]$ plane,

\vskip 0.1truecm

$ S^i =~$ unit vector orthogonal to $U^i$, $N^i$ and
$\kappa^i$. It lives in the $[t, \phi, r, \theta]$ space-time,

\vskip 0.1truecm

$ e_{~(A)}^{~i} =~$ $[ u^i, N^i,\kappa^i,  S^i] =~$ the
tetrad comoving with an observer located in the photosphere.

\vskip 0.2truecm

\noindent The integration of a vector $(...)_i$ over the 3-D
hypersurface ${\cal H}$ orthogonal to $N^i$ (i.e. the 3-D
world-tube of the photosphere) may be symbolically written as

\be \int_{\cal H} (...)_i N^i {\rm d}S, \ee  \vskip 0.1truecm

\noindent where ${\rm d}S$ is the ``volume element'' in ${\cal H}$.

\vskip 0.2truecm

\noindent Obviously, the hypersurface ${\cal H}$ is spanned by the
three vectors $[u^i, \kappa^i, S^i]_N$. Each of them is a linear
combination of $[\eta^i, \xi^i, S_*^i]_{N}$, and each of
the three vectors from $[\eta^i, \xi^i, S_*^i]_{N}$ is
orthogonal to $N^i$.

\vskip 0.2truecm

\noindent Therefore, one may say that the hypersurface ${\cal H}$
is spanned by $[\eta^i, \xi^i, S_*^i]_{N}$. It will be
convenient to write

\be {\rm d}S = {\rm d}A {\rm d}R, ~~~{\rm d}R = {\rm d}r\sqrt{ g_{rr} +
g_{\theta \theta}\left(\frac{d\theta_*}{dr}\right)^2}, \ee

\noindent where ${\rm d}R$ is the line element along the vector $S_*^i$,
i.e. along the photosphere in the $[r, \theta]$ plane, with
$\theta = \theta_*(r)$ defining the location of the photosphere,
and ${\rm d}A$ is the surface element on the $[t, \phi]$ plane.

\vskip 0.2truecm

\noindent In order to calculate ${\rm d}A$, imagine an infinitesimal
parallelogram with sides that are located along the $t =$~const
and $\phi =$~const lines. The proper lengths of the sides are  ${\rm d}u
= \vert g_{tt}\vert^{1/2}{\rm d}t$ and ${\rm d}v = \vert g_{\phi
\phi}\vert^{1/2}{\rm d}\phi$ respectively, and therefore ${\rm d}A$, which is
just the area of the parallelogram, is given by

\be {\rm d}A = {\rm d}u\, {\rm d}v \sin \alpha = {\rm d}t\,{\rm d}\phi \,\vert g_{tt}\vert^{1/2}\vert g_{\phi
\phi}\vert^{1/2} \sin \alpha  ,\ee

\noindent where $\alpha$ is the angle between the two sides.
Obviously, the cosine of this angle is given by the scalar product
of the two unit vectors $n_i$ and $x_i$ pointing in the $[t,
\phi]$ plane into $t$ and $\phi$ directions respectively. These
vectors are given by (note that $n_i =~$ZAMO),

\be n_i = \frac{(\nabla_i t)}{\vert g^{jk} (\nabla_j t)( \nabla_k
t)\vert^{1/2}}, ~~~x_i = \frac{(\nabla_i \phi)}{\vert g^{jk}
(\nabla_j \phi)( \nabla_k \phi) \vert^{1/2}}.\ee

\noindent Because $(\nabla_i t) = \delta^t_{~i}$ and $(\nabla_i
\phi) = \delta^{\phi}_{~i}$, one may write,

\be n_i = \frac{\delta^t_{~i}}{\vert g^{tt}\vert^{1/2}}
, ~~~x_i = \frac{\delta^\phi_{~i}}{\vert g^{\phi
\phi}\vert^{1/2}}.\ee
Therefore,
\be  \cos \alpha = n_i x_k g^{ik} = \frac
{g^{t\phi}}{\vert g^{tt}\vert^{1/2}\vert g^{\phi \phi}\vert^{1/2}}
= - \frac {g_{t\phi}}{\vert g_{tt}\vert^{1/2}\vert g_{\phi
\phi}\vert^{1/2}},\ee
and
\be \sin \alpha = \frac{(g_{t\phi}^2 - g_{tt}\,g_{\phi\phi})^{1/2}}{\vert g_{tt}\vert^{1/2}\vert g_{\phi
\phi}\vert^{1/2}}.\ee

\noindent Inserting this into the formula for ${\rm d}A$ we get ${\rm d}A =
{\rm d}t\,{\rm d}\phi\,(g_{t\phi}^2 - g_{tt}\,g_{\phi\phi})^{1/2}$. The final formula for ${\rm d}S$ is,

\be
\label{ap.dS}
{\rm d}S = {\rm d}t\,{\rm d}\phi\,{\rm d}r\,\left( g_{t\phi}^2 -
g_{tt}\,g_{\phi\phi}\right )^{1/2}\,\sqrt{ g_{rr} + g_{\theta
\theta}\left(\frac{d\theta_*}{dr}\right)^2}.\ee


\bibliographystyle{apj}

\end{document}